\RequirePackage{ifpdf}
\documentclass[letterpaper]{article}
\usepackage{jheppub}
\usepackage{xcolor}
\usepackage{amsmath}
\usepackage{epsfig}
\usepackage{caption}
\usepackage{subcaption}
\usepackage{soul}

\usepackage{dsfont}
\usepackage{mathtools}
\usepackage{amssymb}
\usepackage{stmaryrd}
\usepackage{slashed}
\usepackage{bbold}
\usepackage{braket}
\usepackage{bm}
\usepackage{tikz}
\usepackage{color}
\usepackage{multirow}
\usetikzlibrary{decorations.markings}
\usepackage{scalerel}

\pdfoutput=1

\NewDocumentCommand{\dn}{e{_^}}{%
  _{\IfValueT{#1}{#1}\vphantom{\smash[b]{|}}}
  ^{\IfValueT{#2}{#2}\vphantom{\smash[t]{\big|}}}
}

\newcommand{\changelocaltocdepth}[1]{%
  \addtocontents{toc}{\protect\setcounter{tocdepth}{#1}}%
  \setcounter{tocdepth}{#1}%
}

\newcommand{\roughly}[1]{\mathrel{\raise.3ex\hbox{$#1$\kern-0.85em
\lower1ex\hbox{$\sim$}}}}
\newcommand{\lsim}{\roughly<}

\def\nn{\nonumber}

\newcommand{\be}{\begin{equation}}
\newcommand{\bee}{\begin{equation}}
\newcommand{\ee}{\end{equation}}
\newcommand{\beea}{\begin{eqnarray}}
\newcommand{\eea}{\end{eqnarray}}
\newcommand{\bea}{\begin{eqnarray}}

\def\nott#1{\setbox0=\hbox{$#1$}                
   \dimen0=\wd0                                 
   \setbox1=\hbox{/} \dimen1=\wd1               
   \ifdim\dimen0>\dimen1                        
      \rlap{\hbox to \dimen0{\hfil/\hfil}}      
      #1                                        
   \else                                        
      \rlap{\hbox to \dimen1{\hfil$#1$\hfil}}   
      /                                         
   \fi}                                         %
\def\Dsl{\nott{D}}

\def\uxsl{\hbox{/\kern-.4000em$u$}}
\def\uxslsm{\hbox{\smaller/\kern-.5600em$u$}}
\def\pxpsl{\hbox{/\kern-.5000em$p$}}
\def\epssl{\hbox{/\kern-.5600em$\epsilon$}}
\def\delsl{\hbox{/\kern-.7000em$\nabla$}}
\def\lxpsl{\hbox{/\kern-.5600em$l$}}
\def\kxpsl{\hbox{/\kern-.5600em$k$}}
\def\qxpsl{\hbox{/\kern-.3900em$q$}}
\def\psibar{\overline{\psi}}

\def\pref#1{(\ref{#1})}
\def\exd{{\rm d}}
\def\ol#1{{\overline{#1}}}

\def\cA{{\cal A}}
\def\cB{{\cal B}}

\def\cC{{\cal C}}

\def\cF{{\cal F}}

\def\cH{{\cal H}}
\def\cI{{\cal I}}
\def\cJ{{\cal J}}
\def\cK{{\cal K}}
\def\cL{{\cal L}}
\def\cM{{\cal M}}
\def\cN{{\cal N}}
\def\cO{{\cal O}}
\def\cP{{\cal P}}
\def\cQ{{\cal Q}}

\def\cT{{\cal T}}

\def\cV{{\cal V}}

\def\cZ{{\cal Z}}

\def\bfe{{\bf e}}

\def\bfr{{\bf r}}

\def\bfB{{\bf B}}
\def\bfE{{\bf E}}

\def\bfR{{\bf R}}

\def\bmr{{\bm r}}

\def\bmy{{\bm y}}
\def\bmz{{\bm z}}

\def\bmpsi{{\bm \psi}}
\def\bmchi{{\bm \chi}}

\def\mfa{{\mathfrak a}}

\def\mfc{{\mathfrak c}}

\def\mfh{{\mathfrak h}}

\def\mfp{{\mathfrak p}}
\def\mfs{{\mathfrak s}}

\def\mfu{{\mathfrak u}}
\def\mfv{{\mathfrak v}}

\def\mfz{{\mathfrak z}}

\def\mfC{{\mathfrak C}}

\def\mfF{{\mathfrak F}}

\def\ssA{{\scriptscriptstyle A}}
\def\ssB{{\scriptscriptstyle B}}
\def\ssC{{\scriptscriptstyle C}}
\def\ssD{{\scriptscriptstyle D}}
\def\ssE{{\scriptscriptstyle E}}
\def\ssF{{\scriptscriptstyle F}}

\def\ssI{{\scriptscriptstyle I}}

\def\ssL{{\scriptscriptstyle L}}
\def\ssM{{\scriptscriptstyle M}}

\def\ssR{{\scriptscriptstyle R}}

\def\ssU{{\scriptscriptstyle U}}
\def\ssV{{\scriptscriptstyle V}}
\def\ssW{{\scriptscriptstyle W}}

\def\outs{{\rm out}}
\def\ins{{\rm in}}

\def\UV{{\scriptscriptstyle U\hbox{\kern-0.1em}V}}
\def\PPN{{\scriptscriptstyle P\hbox{\kern-0.1em}P\hbox{\kern-0.1em}N}}
\def\MN{{\scriptscriptstyle M\hbox{\kern-0.1em}N}}
\def\MNP{{\scriptscriptstyle M\hbox{\kern-0.1em}N\hbox{\kern-0.1em}P}}
\def\KK{{\scriptscriptstyle K\hbox{\kern-0.1em}K}}
\def\SM{{\scriptscriptstyle S\hbox{\kern-0.1em}M}}
\def\EH{{\scriptscriptstyle E\hbox{\kern-0.1em}H}}

\def\QCD{{\scriptscriptstyle Q\hbox{\kern-0.1em}C\hbox{\kern-0.1em}D}}
\def\IR{{\scriptscriptstyle I\hbox{\kern-0.1em}R}}
\def\TEV{{\scriptscriptstyle T\hbox{\kern-0.1em}E\hbox{\kern-0.1em}V}}
\def\aff{{a\hbox{\kern-0.1em}f\hbox{\kern-0.1em}f}}

\title{On the EFT of Dyon-Monopole Catalysis}

\author[1,2]{S.~Bogojevi\' c}  
\author[1,2,3]{and C.P.~Burgess}

\affiliation[1]{Department of Physics \& Astronomy, McMaster University, 1280 Main Street West, Hamilton ON, Canada.
}
\affiliation[2]{Perimeter Institute for Theoretical Physics, 31 Caroline Street North, Waterloo ON, Canada.
}

\affiliation[3]{School of Theoretical Physics, Dublin Institute for Advanced Studies,
 10 Burlington Road, Dublin, 
Ireland}

\date{\today}

\abstract{Monopole-fermion (and dyon-fermion) interactions provide a famous example where scattering from a compact object gives a cross section much larger than the object's geometrical size. This underlies the phenomenon of monopole catalysis of baryon-number violation because the reaction rate is much larger  in the presence of a monopole than in its absence. It is sometimes claimed to violate the otherwise generic requirement that short distance physics decouples from long-distance observables -- a property that underpins the general utility of effective field theory (EFT) methods. Decoupling in this context is most simply expressed using point-particle effective field theories (PPEFTs) designed to capture systematically how small but massive objects influence their surroundings when probed only on length scales large compared to their size. These have been tested in precision calculations of how nuclear properties affect atomic energy levels for both ordinary and pionic atoms. We adapt the PPEFT formalism to describe low-energy $S$-wave dyon-fermion scattering with a view to understanding whether large catalysis cross sections violate decoupling (and show why they do not). We also explore the related but separate issue of the long-distance complications associated with polarizing the fermion vacuum exterior to a dyon and show in some circumstances how PPEFT methods can simplify calculations of low-energy fermion-dyon scattering in their presence. We propose an effective Hamiltonian governing how dyon excitations respond to fermion scattering in terms of a time-dependent vacuum angle and outline open questions remaining in its microscopic derivation.
}

\begin{document}
\maketitle

\section{Introduction} 

Magnetic monopoles and their electrically charged cousins -- dyons -- are a gift that keeps on giving. First came the observation that they could exist and, if they did, how their presence provides an instance where nature explores nontrivial topology \cite{Dirac}. Next came the realization that they not only could exist but are actually predicted to exist within the semiclassical limit of some classes of nonabelian gauge theories \cite{tHooft:1974kcl, Polyakov:1974ek}. Then dyonic systems provided a first glimpse at the rich structure of weak/strong coupling duality in relativistic dynamics \cite{Montonen:1977sn, Goddard:1976qe, Duality, Schwarz:1993vs}. To this was added the remarkable phenomenon of monopole catalysis wherein monopole-mediated reactions can sometimes proceed with strong-interaction cross sections unsuppressed by the small monopole size \cite{Rubakov:1981rg, Callan:1982ah}. (For a review see \cite{Preskill:1984gd, Rubakov:1988aq, Schnir}.)

Monopole catalysis is the focus of this paper because it is sometimes held up as a rare example where decoupling fails. For the present purposes `decoupling' is the statement that short-distance physics contributes only in a limited number of ways to long-distance physics, and its validity in nature is one of the reasons why science is so successful; nature comes to us with many scales but decoupling allows us to understand each scale on its own terms without needing to understand everything at once. 

Decoupling is also a property of the mathematics -- quantum field theory -- we use to describe nature and the formalism that organizes calculations to exploit this property as efficiently as possible has come to be known as effective field theory (EFT) \cite{Weinberg:1968de, Wilson:1969zs, Weinberg:1978kz} (see \cite{EFTBook} for a review). Our goal here is to understand how monopole catalysis works within an EFT framework and thereby to understand how decoupling can be consistent with the large interaction cross sections predicted by monopole dynamics. Besides being a conceptually interesting question in its own right our hope is to develop tools that can eventually be used to compute low-energy monopole/dyon scattering more efficiently. 

To this end we adapt to dyons the formalism of point-particle effective theories\footnote{PPEFTs are equivalent to EFTs like NRQED \cite{NRQED} or HQET \cite{HQET} but differ from them in that they treat the massive heavy things as first-quantized objects rather than second-quantized fields when describing their interactions with lighter relativistic degrees of freedom.} (PPEFTs) that describe systematically the interactions of individual heavy but small objects with their surroundings.\footnote{For simplicity we do not discuss here multiparticle scattering processes for which the complications of \cite{Csaki:2021ozp, Csaki:2022tvb} might be relevant or those involving {\it e.g.}~the twist operators described in \cite{vanBeest:2023dbu, vanBeest:2023mbs}.} The PPEFT formalism is equivalent to more graphical first-quantized point-particle EFT applications -- such as \cite{Goldberger:2004jt} for instance -- but as used here does not as heavily rely on momentum-space Feynman rules. It instead relates the effective world-line lagrangian for the small source to a set of near-source boundary conditions satisfied by the relativistic fields with which the source interacts. It does so by generalizing to arbitrary interactions the standard Gauss' Law technique that draws a small Gaussian pillbox around the compact source and relates the total charge within the pillbox to the electric flux through the pillbox's surface (which can be regarded as the value of $r^2 \partial A_0/\partial r$ near the source) \cite{PPEFT, PPEFT2, PPEFT3}.\\

Relating the near-source boundary condition to the source action in this way allows a systematic expansion of the source's influence at low energies that includes but extends the usual multipole expansion to include nonlinear interactions. PPEFT methods have been tested extensively in applications for which answers are known by other methods, by using it to describe the influence of finite nuclear size (and various nuclear moments) for atomic energy levels \cite{Burgess:2020ndx, Zalavari:2020tez}, absorption by hot wires in atom traps \cite{Plestid:2018qbf} and to describe the gravitational back-reaction of codimension-two branes in various dimensions \cite{Burgess:2008yx, Bayntun:2009im}. Applying these same techniques to dyons allows a comparison between how $S$-wave scattering with no angular momentum barrier differs for a spinless particle (a pion, say) interacting with a nucleus and a spin-half particle interacting with a dyon. This comparison shows why scattering cross sections in one case are set by the size of the nucleus but in the other can be much larger. 

The key difference is kinematic. For spinless particles interacting with a nucleus there are two types of mode functions for each angular-momentum quantum number $l$. One typically varies as $r^l$ for small $r$ and the other varies as $r^{-l-1}$. Because one solution grows and the other shrinks as $r \to 0$ the existence of two solutions defines a physical scale $R_*$: when the specific solution exterior to a known source is decomposed in terms of these two modes $R_*$ is the distance at which the two modes are comparable in size. $R_*$ turns out to be the scale that determines the size of scattering cross sections, since the $S$-wave cross section turns out to be of order $\sigma_s \sim \pi R_*^2$. 

For instance, for nuclear-strength interactions -- such as is relevant for pionic atoms -- $R_*$ is comparable to the nuclear radius $R$. By contrast, for electromagnetic scattering $R_* \sim \alpha R$ is much smaller (where $\alpha = e^2/4\pi$ is the fine-structure constant). What is unusual about fermion-dyon scattering is there is only a single $S$-wave mode, and so decomposing a specific solution in terms of this mode does not introduce a scale similar to $R_*$. Instead scattering turns out to be scale invariant and predicts $\sigma_s \sim 2\pi/k^2$ (where $k$ is the relativistic fermion energy). This is the EFT translation of the standard understanding \cite{Rubakov:1981rg, Callan:1982ah} of why dyon-fermion $S$-wave scattering does not scale with the dyon size.

Although the above is a good start, the full story is of course more complicated because the fermion-dyon interaction polarizes the fermion ground state \cite{Goldhaber:1977xw, Ezawa:1983vi, CallanSMatrix, Yamagishi:1982wp, Grossman:1983yf, Kazama:1983rt, Isler:1987xn}. Within the PPEFT approach this can be interpreted as a further renormalization of the effective couplings as the radius of the underlying Gaussian pillbox grows to become larger than the Compton wavelength of any light fermion species. For the particular dyonic system we study here this has been argued to simply turn off the underlying charge-changing dyon-fermion interactions in a calculable way \cite{Polchinski}; converting them into helicity-flipping interactions. We argue that the basic PPEFT scattering rates apply equally well in this case, provided one works with the appropriate long-wavelength effective couplings that contain these interaction effects.

We treat the fermion-dyon scattering within the Born-Oppenheimer approximation with the idea that the fermions represent the fast degrees of freedom and the dyonic response provides the slow degrees of freedom. Within this framework the PPEFT formalism also allows us to discuss how the dyon responds to fermion scattering and in principle allows estimates for the size of effects such as vacuum angle evolution due to fermion scattering. We show how such evolution can occur although we have been unable to find a regime where the dynamics envisaged in \cite{Brennan:2021ewu} takes place in detail. Although we derive some expressions that show how the dyon response to fermion scattering is consistent with the flow of conserved quantities, we have not been able to explicitly derive this effective dyon Hamiltonian by integrating out the fermions.

Our results are presented as follows. \S\ref{sec:DyonReview} reviews the basic construction of the dyon solutions we explore in detail and \S\ref{sec:PPEFTs} sets up the PPEFT description for the scattering of relativistic fermions with the dyon. It turns out that the number of effective operators in the simplest effective description is larger than the number of observable parameters and we argue why this difference can be understood in terms of redundancy of PPEFT interactions. (A similar thing happens for electromagnetic interactions with nuclei in atoms, where it implies the various nuclear moments do not contribute independently to shifts in atomic energy levels to fixed order in an expansion in powers of $Z \alpha$ \cite{Burgess:2020ndx, Zalavari:2020tez}.) \S\ref{sssec:RGcatalysis} then explores the RG evolution of the boundary conditions as the radius of the Gaussian pillbox is varied and shows in more detail why the kinematics of $S$-wave dyon-fermion scattering leads to scale-invariant cross sections even though the same does not happen for small objects without magnetic charge (such as nuclei). This section also discusses how the fermion condensation effects described in \cite{Polchinski} can be incorporated within this framework. Finally \S\ref{sec:SO3Dyon} uses the PPEFT effective couplings to compute fermion polarization and fermion-dyon scattering within the Born-Oppenheimer approximation, after first warming up by computing similar results perturbatively in the fermion-dyon interactions in \S\ref{sec:PerturbativeScattering}. Our conclusions are briefly summarized in \S\ref{sec:Conclusions}.

\section{The Julia-Zee dyon}
\label{sec:DyonReview}

To start we work within a simple model that supports simple dyon solutions: the Georgi-Glashow style $SU(2)$ gauge theory spontaneously broken down to $U(1)$ using an adjoint Higgs field \cite{Georgi:1972cj}. To this we couple a complex doublet of massless fermions\footnote{Because this corresponds to two pseudoreal doublets, it is the minimal anomaly-free fermion content \cite{Witten:1982fp}.} so that we can study fermion-dyon interactions in the dyon's low-energy regime (but where the fermions are nonetheless ultra-relativistic). 

This model is described by the following action,\footnote{The distinction between symmetric differentiation $\overset{\leftrightarrow}{D}_\mu = \overset{\rightarrow}{D}_\mu - \overset{\leftarrow}{D}_\mu$ and the usual one-sided quantity in the Dirac action is a total derivative, but much of the later discussion hinges on being careful with total derivatives and boundary terms.}
\be\label{SU(2) Georgi-Glashow action}
     {S}=-\int \mathrm{d}^4x\left[\,\frac{1}{4}  F^a_{\mu \nu} \,F^{\mu \nu}_a+\frac{1}{2} D_{\mu} \Phi^a\, D^{\mu} \Phi_a + \frac12 \overline{\psi}\,\gamma^{\mu} \overset{\leftrightarrow}{D}_{\mu}\psi  -\frac{\mu^2}{2}\,\Phi^a\,\Phi_a +
     \frac{\lambda}{4}\,\left(\Phi^a\,\Phi_a\right)^2\right] \,,
\ee
with covariant derivatives 
\be
  D_{\mu}\psi=\left( \partial_{\mu} - \frac{i}{2} \, e A^a_{\mu} \tau_a \right)\psi \quad \hbox{and} \quad 
  (D_{\mu}\Phi)_a=\partial_{\mu}\Phi_a + \epsilon_{abc}\,eA^b_{\mu}\,\Phi^c \,,
\ee
and nonabelian field strength $F^a_{\mu \nu} = \partial_{\mu}A^a_{\nu } -  \partial_{\nu}A^a_{\mu} + e\, \epsilon^{a b c} A_{b\mu}\,A_{c\nu}$. The gauge generators acting on the fermion's doublet gauge indices are given in terms of the usual Pauli matrices by $T_a = \frac12 \tau_a$ and so satisfy the $SU(2)$ commutation relation $[T_a , T_b] = i \epsilon_{abc} T_c$. With respect to the gauge field gauging rotations in the $T_3$ direction (say) we see that the members of the fermion doublet have charge $\pm \frac12 \, e$ and the two gauge bosons spanned by $A^1_\mu$ and $A^2_\mu$ have charge $\pm e$. The components of the adjoint Higgs multiplet $\Phi^a$ similarly carry charges zero and $\pm e$ for this symmetry.

The model's parameters $\lambda$, $e$ and $\mu^2$ are all real and positive, with dimensionless couplings $\lambda$ and $e$ chosen small enough to justify the standard semiclassical analysis. In the semiclassical regime the choice $\mu^2 > 0$ ensures the vacuum expectation value $w^a := \langle \Phi^a \rangle$ satisfies $w^2 := w^a w_a = \mu^2/\lambda \neq 0$ so that the gauge symmetry is spontaneously broken down to $U(1)$. The two charged gauge bosons then `eat' the two charged fields in $\Phi^a$ and by doing so acquire nonzero mass $m_g \simeq e w = \beta \mu$, where $\beta^2 := e^2/\lambda$. The remaining spin-one particle is massless and so can be regarded as the `photon' that gauges the unbroken $U(1)$ symmetry. The uneaten physical scalar is also neutral under the unbroken gauge symmetry and has mass $m_s \simeq \sqrt{2\lambda}\,  w = \sqrt2 \, \mu$.

\subsection{Classical dyon}

There are a variety of soliton solutions to the field equations of this model. Of these the so-called Julia-Zee dyon \cite{tHooft:1974kcl, Julia:1975ff, Wu:1975es} is of most interest for the present purposes. This has the form
\be \label{Julia-Zee dyon spherical gauge}
    e{\cA}^a_{i} = {\epsilon^a}_{i b} \,\hat{r}^b\, \left[\frac{1 - \cK(r)}{r} \right] \,, \quad
    e{\cA}^a_{0} = \hat{r}^a\, \frac{\cJ(r)}{ r} \quad \hbox{and} \quad
    e\varphi^a = \hat{r}^a\, \frac{\cH(r)}{ r} \,,
\ee
with vanishing $\psi$. Here $\hat{\bm r}$ denotes the unit vector in the radial direction and the dimensionless functions $\cK(r)$, $\cH(r)$ and $\cJ(r)$ depend on $r$ only through the combination $\mu r$. They satisfy second-order coupled ordinary differential equations given explicitly in \cite{Julia:1975ff} that depend on $e$ and $\lambda$ only through the ratio $e^2/\lambda = \beta^2$. 

The integration constants are chosen to ensure boundedness at the origin --- and so $\cK \to 1$ and $\cJ,\cH \to 0$ as $r \to 0$. They also have a finite-energy falloff to the vacuum solution as $r \to \infty$, implying the asymptotic behaviour
\be \label{Asymptotic behavior of J, K, H}
    \cK(r) \to 0 \,, \quad \cJ(r) \to e(Q - vr  )\quad \hbox{and} \quad
      \cH(r) \to h r \quad \hbox{as } \, r \to \infty \,,
\ee
where $h = ew = \beta\mu$ and only one of $Q$ or $v$ is an independent parameter because one combination of $Q$ and $v$ is a calculable function of $\beta$ and $\mu$. The approach to these asymptotic forms is exponential in $\mu r$, and this sets the `size' of the solution. For instance the function $\cK(r)$ behaves for large $r$ as $\cK(r) \sim  e^{- a r}$ with $a = \sqrt{h^2 - (ev)^2}$, which shows the solutions damp exponentially provided $h > ev$. 

In the special case where $\lambda \rightarrow 0$ and $e \to 0$ with $\beta^2 = e^2/\lambda$ and $\mu$ fixed the functions $\cK(r), \cJ(r)$ and $\cH(r)$ can be solved in closed form \cite{BPS solution}. In this {\it Prasad-Sommerfield} limit the solution that is regular at the origin and for which $\varphi^a \varphi_a$ approaches a constant\footnote{When $\lambda \to 0$ it is no longer necessary to require $\cH /r$ to approach the specific constant $h$ as $r \to \infty$.} at infinity is given explicitly by
\be \label{PrasadSommerfieldLimit}
    \cK(r)=\frac{c r }{\sinh (c r)}, \quad \cJ(r)=\sinh \varpi \, \Bigl[ 1-c r \coth(c r) \Bigr] \quad\text{and} \quad \cH(r)=\cosh \varpi \,\Bigl[ c r \coth(c r)-1 \Bigr],
\ee
with $\varpi$ an arbitrary real constant and the field equations require $c = \beta \mu$. Comparing with the large-$r$ limit of $\cJ(r)$ shows that $v$ and $Q$ are given in terms of these constants by $ev = c \sinh \varpi$ and $eQ = \sinh \varpi$ and so $Q = v/c$. 

At any particular spacetime point the nonzero fields break the $SU(2)$ gauge invariance down to a $U(1)$ subgroup, so just like for the vacuum the dyon preserves a $U(1)$ symmetry and as a result one of the gauge modes -- the photon -- remains precisely masses. But for the dyon the particular embedding of the unbroken $U(1)$ within $SU(2)$ varies from place to place, as can be seen by the mixing of gauge and spatial indices in \pref{Julia-Zee dyon spherical gauge}. This makes it convenient to change gauge to a form for which the massless gauge mode corresponds to the same gauge direction everywhere in spacetime. 

If the asymptotic gauge field is chosen to point along the $T_3$ direction in $SU(2)$ space, the required gauge transformation is 
\be \label{Gauge function for R_-} 
     U({\bm r}) =\frac{1}{\sqrt{2}}\left[ \sqrt{1-\hat{\bmz} \cdot\hat{\bmr}}+  \frac{i{\vec\tau}\cdot(\hat{\bmr}\times \hat{\bmz})}{\sqrt{1-\hat{\bmz} \cdot \hat{\bmr}}}\right],
\ee
where we use arrows to denote vectors in gauge space and bold-face for vectors in physical space and $\hat{\bmz}$ denotes the unit vector in the `3' or $z$ direction. This is a singular gauge transformation inasmuch as it introduces a previously non-existent singularity into the asymptotic vector potential if it is performed everywhere. Because the singularity is a gauge artefact we can remove it by following the procedure outlined in \cite{Wu:1975es} where we define instead gauge-related but nonsingular configurations on each of two hemispheres that enclose the dyon. 

That is, defining the two overlapping regions, $R_+$ and $R_{-}$, by
\be 
    R_+: \left\{ 0 \leq \theta < \frac{\pi}{2}+\delta , \, r>0,\, 0 \leq \phi < 2\pi \right\}\quad \hbox{and} \quad
    R_-: \left\{ \frac{\pi}{2}-\delta < \theta \leq \pi ,\, r>0,\, 0 \leq \phi < 2\pi \right\} ,
\ee
where $(\theta,\phi)$ are the usual spherical-polar angular coordinates, $0 < \delta \leq  \frac{\pi}{2}$ is otherwise arbitrary and we define the gauge field in region $R_-$ using the gauge function $U(\bmr)$ while the gauge field in $R_+$ is additionally transformed using the gauge function $V(\bmr)=e^{ i \phi \,\tau_3}$. Defined in this way, the Julia-Zee dyon field configuration is given by
 \be \label{Julia-Zee dyon abelian gauge}
    e\mathcal{A}^{a\pm }_{i}(\vec{r})= \frac{1}{r} \left[\frac{\pm 1 - \cos \theta }{\sin \theta} \hat{\phi}_i \, \delta^a_3 - \cK(r) \, \zeta^a_{i \pm} \right], \quad 
    e\mathcal{A}^a_{0}(\vec{r})=- \frac{\cJ(r)}{r}\, \delta^a_3 \quad \hbox{and} \quad
    e \Phi ^a(\vec{r})= - \frac{\cH(r)}{r}\, \delta^a_3.
\ee
where
\be\label{zeta}
   {\bm \zeta}_{i\pm} := \frac12 \, \zeta^a_{i \pm} \tau_a = \frac12 \left(i\hat{\theta}_i - \hat{\phi}_i\right)e^{\pm i \phi} \tau_+ - \frac12 \left(i\hat{\theta}_i+ \hat{\phi}_i\right) e^{\mp i \phi}  \tau_- 
\ee
arises as  the gauge transform of $\frac12 \hat{\bm r}\times  {\vec \tau} $. Here the subscripts $\pm$ refer to the regions $R_{\pm}$ and $\tau_{\pm}$ are defined as usual by $\tau_\pm :=\frac12(\tau_1\pm i \tau_2)$. 

The asymptotic large-$r$ form of the solution in this gauge makes its electromagnetic properties explicit, 
\be\label{Asymptotic form of gauge potential in abelian gauge}
    e{\mathcal{A}}^{a \pm}_i  \to \frac{\pm 1 - \cos \theta}{r \sin \theta} \hat{\phi}_i \,\delta^a_3 \qquad \hbox{and} \quad  
     e\mathcal{A}^a_0 \to \left(ev - \frac{e Q}{ r }\right) \delta ^a_3 \qquad \hbox{as} \quad r \rightarrow \infty \,,
\ee
since using this in Gauss' law
\be
    q_{\ssE} = \frac{1}{4\pi}\int_0^{4\pi} \bfE \cdot \hat{\bfr} \,\exd^2 S  =  Q \quad \hbox{and} \quad
    q_{\ssM} = \frac{1}{4\pi} \int_0^{4\pi} \bfB \cdot \hat{\bfr} \,\exd^2S = \frac{1}{ e},
\ee
reveals it to be a dyon that carries magnetic charge $q_\ssM = 1/e$ and electric charge $q_\ssE = Q$. Notice the magnetic charge is consistent with the Dirac quantization condition that requires $2 q_\ssM q$ to be an integer for any electric charge $q$ given that the particle with the smallest electric charge is the fermion, for which $q_\psi = \pm \frac12 e$. The parameter $v$ is also seen to be the electrostatic potential difference between the origin and infinity, which in the general case is not independent of $Q$, with 
\be \label{ClassicalDyonCharge}
   ev =  \mu \; \cZ\left( \beta , eQ \right) \,, 
\ee
for a calculable order-unity function $\cZ$. (For instance in the Prasad-Sommerfield limit \pref{PrasadSommerfieldLimit} we have $\cZ = eQ\beta$.) The monopole limit $Q \to 0$ corresponds to taking $v \to 0$ so $\cZ(\beta,0) = 0$.

The classical dyon mass is similarly given by evaluating the energy of the classical solution:
\be \label{ClassicalDyonMass}
   M \simeq \frac{4\pi \beta \mu}{e^2} \; F\left(\beta, v/\mu \right) \sim \cO\left( \frac{m_g}{\alpha}  \right) \,,
\ee
where $m_g \simeq \beta \mu = e w$ is the gauge boson mass, $\alpha = e^2/4\pi$ is the fine-structure constant and $F$ is an explicitly calculable function that is order unity when $e^2 \sim \lambda$ and $v \sim \mu$. The mass is related to the dyon size $R \sim m_g^{-1}$ by $MR \sim4\pi/e^2  \gg 1$ in the semiclassical regime and so its Compton wavelength is much smaller than its radius.

\subsection{Semiclassical quantization} 

The interactions of a dyon with other degrees of freedom are found by performing a semiclassical expansion of the quantum fields around the classical dyonic background, with
\be \label{SCExp}
    A_{\mu}^a(x)  =  \mathcal{A}_{\mu}^a(x)  + \widehat{A}_{\mu}^a(x) \quad \hbox{and} \quad
    \Phi^a(x)  =  {\varphi}^a(x)  + \widehat{\Phi}^a(x) ,  
\ee
where the fields $\widehat{A}^a_{\mu}(x)$ and $\widehat{\Phi}^a(x)$ join the fermion field ${\psi}(x)$ as quantum operators. 

\subsubsection{Collective coordinates}

A special role is played by those fluctuations that are Goldstone directions corresponding to symmetry transformations that act nontrivially on the background dyon, since these cost little energy and so are important for understanding a dyon's low-energy response to external probes. 

The counting of these zero modes is subtle for dyons \cite{SubtleCounting}, but for the case of the Julia-Zee dyon the physical collective coordinates are the dyon's center-of-mass position, ${\bm y}(t)$, and a charged collective coordinate, $\mfa(t)$. These can be regarded as the special cases where $\widehat A_\mu^a$ and $\widehat \Phi^a$ are obtained by transforming the dyon by a time-dependent spatial translation or a time-dependent gauge rotation in the unbroken $U(1)$ gauge symmetry direction (generated by $\tau_3$ in the gauge \pref{Julia-Zee dyon abelian gauge}),   
\be
    \delta \cA_{\mu}^a(x)  =  y^i \partial_i \cA^a_\mu + \partial_\mu y^i \cA^a_i  + \frac{1}{e}\, \delta^a_3 \, \partial_\mu \mfa   -  \mfa  \, \epsilon^{a3b} \cA^b_\mu \quad\hbox{and} \quad 
   \delta  \varphi^a(x)  = y^i \partial_i \varphi^a    -  \mfa  \, \epsilon^{a3b} \varphi^b   , 
\ee
and the dependence on the background fields ensures these are localized near the dyon's location. 

The semiclassical expansion proceeds by using \pref{SCExp} in the action \pref{SU(2) Georgi-Glashow action} and Taylor expanding in the fluctuation fields. The classical contribution to the action (with $\widehat A_\mu^a = \widehat \Phi^a = \psi = 0$) is order $4\pi/e^2$ in the same way that eq.~\pref{ClassicalDyonMass} implies that $M/m_g$ is order $4\pi/e^2$. The leading dependence on fluctuations is quadratic and describes free quantum fields evolving within the dyonic background. Because the quadratic term is independent of $e$ the 
masses of the quanta destroyed by the fluctuation fields are suppressed compared to the dyon mass by $e^2/4\pi$. It is because successive terms in this expansion are suppressed by still more powers of $e^2/4\pi$ that semiclassical methods are under control in the weakly coupled regime. 

In particular, the kinetic term for the collective coordinate $\mfa$ is found by evaluating the Maxwell kinetic term using the ansatz
\be 
    e\mathcal{A}^{\pm}_i =\frac{\pm1 - \cos \theta}{2r \sin \theta} \hat{\phi}_i \tau_3 - \frac{\cK(r)}{ 2r }\Big[ ( i \hat{\theta}_i - \hat{\phi}_i ) e^{ i \mfa(t)}e^{\pm i \phi} \tau_{+} - ( i \hat{\theta}_i + \hat{\phi}_i ) e^{- i \mfa(t)}e^{\mp i \phi} \tau_{-}  \Big] 
\ee
for which the field strength contains
\be 
    \cF_{0i} \ni \partial_0 \mathcal{A}^{\pm}_i= -\frac{i \dot{\mfa}\,\cK(r)}{ 2er }\Big[ ( i \hat{\theta}_i - \hat{\phi}_i ) e^{ i \mfa}e^{\pm i \phi} \tau_{+} + ( i \hat{\theta}_i + \hat{\phi}_i ) e^{- i \mfa}e^{\mp i \phi} \tau_{-}  \Big] \,, 
\ee
%
and so
\be \label{alphakin}
    -  \text{Tr}(\cF_{0i} \cF^{0i}) = \dot{\mfa}^2 (-i)^2 \left[\frac{\cK(r)}{ er }\right]^2 (i \hat{\theta}_i -\hat{\phi}_i)\cdot (i \hat{\theta}_i +\hat{\phi}_i)\,\frac{\text{Tr}(\tau_+ \tau_-)}{2} =  \left[\frac{\cK(r)}{ er }\right]^2 \dot{\mfa}^2  \,.
\ee
This is localized near the position of the dyon because $\cK(r)$ falls to zero for large $r$. Writing this as $\frac12  \,\cI \dot \mfa^2 \, \delta^3(x)$ implies the constant $\cI$ is 
\be \label{DyonS0}
  \mathcal{I}=\frac{8\pi}{e^2} \int_0^\infty \exd r \; \cK^2(r) \sim \frac{1}{\alpha m_g} \,,
\ee
and so is parametrically large compared to the dyon size $R \sim m_g^{-1}$.

In the absence of fermion interactions the kinetic lagrangian for $\mfa$ is $\frac12 \, \cI \, \dot \mfa^2$ and because $\mfa $ is a periodic variable it behaves like a quantum rigid rotor, whose canonical momentum $\mfp$ is both conserved and quantized. Because $\mfa $ shifts under electromagnetic gauge transforms, with $\delta \mfa = \omega$ when $\delta A_\mu^3 = \frac{1}{e} \partial_\mu \omega$, the conserved rotor momentum is proportional to the rotor's contribution to the dyon-localized electric charge\footnote{We return to a more precise statement of rotor quantization including the effects of a vacuum angle in \S\ref{ssec:RotorBO} below.}
\be \label{rotorcharge}
    \cQ_\ssD = - e \,  \mfp \,,
\ee
and for $\mfa$ identified with $\mfa + 2\pi$ the canonical rotor momentum takes integer values $\mfp | n \rangle = n | n \rangle$ (and so $\cQ_\ssD = - e n$). Unit steps in the rotor levels differ in charge by $\delta \cQ_\ssD = \pm  e$ and differ in energy by  
\be \label{DyonEnergy}
   \delta E_n = E_{n+1}-E_n = \frac{2n+1}{\cI} \sim \frac{e^2 m_g}{4\pi}  \ll m_g  \,.
\ee

These estimates show that dyonic excitations belong in the effective theory for energies below $m_g$ precisely because the characteristic loop-counting factor $e^2/4\pi$ suppresses $\delta E$ relative to $m_g$. But because the steps in rotor energy are proportional to $e^2/4\pi$ changes to rotor energies due to transitions can be negligible at a given order in the semiclassical expansion, so care must be taken about the ordering of the limits $E/m_g \to 0$ and $e^2/4\pi \to 0$ when working at low energies within the semiclassical limit.  

\subsubsection{Fermion $S$-wave modes}
\label{ssec:BulkModes}

Our main focus in this paper is how the bulk fermion interacts with the dyon, so we now focus on the leading dependence of the action on the fermions, which is given by
\be \label{Fermonic action to order e^0}
     S_{\psi}=-\int \mathrm{d}^4x\, \overline{\bm{\psi}}
     \,\gamma^{\mu}\left[\partial_{\mu} - \frac{i}{2} \, e \mathcal{A}^a_{\mu}  \tau_a \right]\bm{\psi}.
\ee
Far from the dyon core it is straightforward to find the fermionic energy eigenstates predicted by this action. Near-source fermion-dyon interactions are trickier to deal with since in principle they depend in detail on the functions $\cK(r)$ and $\cJ(r)$. The next section shows how EFT reasoning allows the small size of the dyon to be exploited to circumvent this issue for fermion states with $E \ll \mu$. Before doing so we first identify the eigenmodes far from the dyon.
 
At low energies the partial waves that dominate in dyon scattering are those that minimize the centrifugal barrier that must be penetrated in order to reach the dyon core. What makes monopole and dyon interactions special is the way that their magnetic charge adds a new contribution to the angular momentum of an incident charged particle, and how this changes how the angular momentum barrier depends on the orbital and total angular momenta \cite{Rubakov:1981rg, Callan:1982ah}. For spinless charged particles the new contribution implies there is always a centrifugal barrier, even with zero orbital angular momentum. Unusually, it is only for spin-half charged particles that the absence of a barrier is possible, and this only occurs for a very specific type of $S$-wave state (on which we now focus). 

In the present instance the conserved angular momentum $\vec{J}$ of an external isodoublet doublet  fermion moving in the presence of the Julia-Zee dyon is
\be \label{Def: Angular momentum}
     \vec{J}=\vec{L}+\vec{S}+\vec{T},
\ee
where $\vec{L}$ and $\vec{S}$ are the usual orbital and spin angular momentum while $\vec{T}$ is the gauge isospin of the particle state. This last term is the new one and it arises because the dyon background breaks the freedom to perform independent gauge and physical rotations but leaves unbroken the diagonal subgroup where gauge and physical rotations are peformed in unison. As a consequence the radial magnetic field sourced by the dyon's magnetic charge contributes to a charged particle's angular momentum. 

For the isodoublet fermions of interest here, $\vec T^2$ and $\vec S^2$ both have eigenvalue $\frac34$ as appropriate for a spin-half contribution to the total angular momentum. The usual rules for combining angular momenta show that the combination $\vec S + \vec T$ can either carry spin zero or spin one. The state with no angular momentum barrier is the one where $\vec{S} + \vec{T}$ combine with orbital angular momentum $\vec L$ to give zero total angular momentum for $\vec J$. Writing the eigenvalues of $\vec L^2 = \ell(\ell + 1)$ and $\vec{J}^2 = j(j+1)$ the rules for combining angular momenta show that $j = 0$ can be obtained by combining the spin and magnetic angular momentum with either of the two orbital angular momenta $\ell = 0, 1$. 

The explicit form for this type of $S$-wave fermion mode for the Julia-Zee dyon was found for the solution \eqref{Julia-Zee dyon spherical gauge} by Jackiw and Rebbi \cite{Jackiw and Rebbi s-wave}, and writing the spin index $i$ and isospin index $a$ as a matrix $M_{ia}$ for each chirality of field the $S$-wave configuration has the structure $M(x) = s(r,t) + i p(r,t) \, \hat{\bmr} \cdot \vec \tau$ and so $\psi_\ssL$ and $\psi_\ssR$ each involve two independent functions. Once transformed to the gauge where the solution has the form \eqref{Julia-Zee dyon abelian gauge} the result becomes
\be \label{s wave fermions}
     \bm{\psi}(x)=\frac{1}{r}\begin{bmatrix}  \psi_+(x) \\ \psi_-(x)  \end{bmatrix} \quad \hbox{where} \quad
     \psi_\pm(x) = \begin{pmatrix}f_\pm(r, t)\, \eta_{\pm}(\theta, \phi)\\[1mm]g_\pm(r, t)\, \eta_{\pm}(\theta, \phi)\end{pmatrix}  \,,
\ee
where square brackets denote gauge isodoublets and round brackets denote 4-component Dirac spinors in a basis for which $\gamma_5 = \hbox{diag}(I, -I)$. The 2-component Weyl spinors $\eta_{\pm}$ are defined to satisfy $\sigma^r \eta_\pm = \pm \eta_\pm$ (see Appendix \ref{App:GammaConventions} for our Dirac matrix conventions) and so are given explicitly by
\be \label{etapmdefs}
    \eta_{+}(\theta, \phi)=\frac{1}{\sqrt{4\pi}}  \left(\begin{array}{c}
  \cos{\frac{\theta}{2}} \, e^{- i \phi} \\  \sin{\frac{\theta}{2}} 
\end{array}\right) \quad \hbox{and} \quad  \eta_{-}(\theta, \phi) = \frac{1}{\sqrt{4\pi}} \left(\begin{array}{c}
-\sin{\frac{\theta}{2}} \\
\cos{\frac{\theta}{2}}\, e^{i \phi}
\end{array}\right)
\ee
in the region $R_-$ and by $\eta'_{\pm}(\theta, \phi)=\eta_{\pm}(\theta, \phi) \, e^{ \pm i \phi}$ in the region $R_+$. 

At large distances from the dyon core only the far-field electromagnetic parts of the monopole fields matter and so  
the $S$-wave sector Dirac equation simplifies to
\be  \label{4d Dirac equation}
  \left( -\partial_t + \gamma^0\gamma^r \partial_r + \frac{i}{2} \, e \mathcal{A}^3_0 \, \tau_3 \right) (r\bm{\psi})  
  = \begin{bmatrix} \left( -\partial_t + \partial_r + \frac{i}{2} \, e \mathcal{A}^3_0  \right) {f_+ \eta_+} \\  \left( -\partial_t - \partial_r + \frac{i}{2} \, e \mathcal{A}^3_0  \right) g_+ \eta_+ \\  \left( -\partial_t - \partial_r - \frac{i}{2} \, e \mathcal{A}^3_0  \right) f_- \eta_- \\  
  \left( -\partial_t + \partial_r - \frac{i}{2} \, e \mathcal{A}^3_0  \right) g_- \eta_- \end{bmatrix}   = 0
\ee
which uses $\gamma^0 \gamma^r = \hbox{diag}(\sigma^r, - \sigma^r)$ (see Appendix \ref{App:GammaConventions}) and $\sigma^r \eta_\pm = \pm \eta_\pm$. In principle we can simultaneously diagonalize $\gamma_5$, $\gamma^0\gamma^r$ and $\tau_3$, each of which has eigenvalues $\pm 1$, leading to eight possible unique sets of quantum numbers, $\{\mfc, \mfh, \mfs\}$, where we denote the eigenvalues of $\gamma_5$, $\gamma^0 \gamma^r$ and $\tau_3$ respectively by $\mfc$ (chirality), $\mfh$ (helicity) and $\mfs$ (for $\tau_3)$. Of these, the $S$-wave condition \pref{s wave fermions} says that the eigenvalues of $\tau_3$ and $\gamma_5 \gamma^0 \gamma^r = \hbox{diag}(\sigma^r, \sigma^r)$ are the same and so $\mfs = \mfc \mfh$. Also eq.~\pref{4d Dirac equation} shows that the direction of motion (radially ingoing or radially outgoing) is correlated with the eigenvalue of $\gamma^0 \gamma^r$ with $\mfh = +1$ corresponding to infalling modes and $\mfh=-1$ pairing with outgoing modes. 

For an $S$-wave fermion passage through the origin inevitably brings a change of radial fermion direction and this must change the sign of $\mfh$. Because $\mfh = \mfc \mfs$ we see why $S$-wave scattering famously must involve a change in  chirality ($\mfc$) or in electric charge ($\mfs$). If it should be true that the microscopic properties of the dyon preserve chirality -- as is indeed the case {\it e.g.}~when these interactions are modeled by solving the Dirac equation in the presence of a fixed dyon background \cite{Goldhaber:1977xw, Blaer:1981ps, Marciano:1983md, Ezawa:1983vi} -- then $\mfc$ cannot change and so only charge changing processes $\mfs' \neq \mfs$ are possible. Famously, the rotor-fermion interactions can significantly modify the fermionic vacuum which can in turn allow more complicated behaviour \cite{Rubakov:1981rg, Callan:1982ah, CallanSMatrix, Yamagishi:1982wp, Grossman:1983yf, Kazama:1983rt, Polchinski, Csaki:2021ozp, Csaki:2022tvb, vanBeest:2023dbu, vanBeest:2023mbs} (more about which later). For this reason we do not assume below that it is $\mfs$ that must change during $S$-wave fermion-dyon scattering.
 
We label the four independent modes by their quantum numbers $\mfs$ and $\mfc$ (from which the $S$-wave condition implies $\mfh = \mfc\mfs$), and then integrate to find the mode functions with frequency $\omega$ once $\cA_0^3$ is evaluated using \pref{Julia-Zee dyon abelian gauge} specialized to the asymptotic form \pref{Asymptotic form of gauge potential in abelian gauge}. This gives the following explicit basis of far-field positive-frequency solutions $u_{\mfs\mfc \omega}(x)$, 
\be \label{Chirality basisu}
    \begin{bmatrix} u_{+ \mfc \omega}(x) \\ 0 \end{bmatrix} \,, \; 
    \begin{bmatrix} 0 \\ u_{ - \mfc\omega}(x) \end{bmatrix} \quad \hbox{where} \quad
    u_{\mfs\mfc\omega}(x) := \frac{1}{r} \, \xi_\mfc \otimes \eta_\mfs(\theta, \phi)   \, e^{-i \omega t}e^{-i (\mfs \omega +\frac12 ev) \,\mfc \,r } \left( \frac{r}{r_0} \right)^{i \mfc\, eQ/2},  
\ee 
with $\sigma_3 \xi_\mfc = \mfc \,\xi_\mfc$ and we write $\gamma_5 = \sigma_3 \otimes I$. The length scale $r_0$ is arbitrary and is introduced on dimensional grounds due to the singular Coulomb phase. Here $\omega$ is bounded from below, but the asymptotic voltage $v$ implies the floor is $\omega \geq - \frac12  \mfs ev$. The oscillatory factors in this expression can be equivalently written $e^{-i (k- \frac12 \mfs  e v) t}e^{-i \mfs  \mfc k r }$ for $k \geq 0$. The negative frequency counterparts similarly are
\be \label{Chirality basisv}
    \begin{bmatrix} v_{+ \mfc \omega}(x) \\ 0 \end{bmatrix} \,, \; 
    \begin{bmatrix} 0 \\ v_{ - \mfc\omega}(x) \end{bmatrix} \quad \hbox{where} \quad
    v_{\mfs\mfc\omega}(x) := \frac{1}{r} \, \xi_\mfc \otimes \eta_\mfs(\theta, \phi)   \, e^{i \omega t}e^{i (\mfs \omega - \frac{1}{2} ev) \,\mfc \,r } \left( \frac{r}{r_0} \right)^{i \mfc\, eQ/2} \,.
\ee 
Frequency $\omega$ is again bounded from below, but in this case the floor is $\omega \geq + \frac12  \mfs ev$ and the oscillatory factors can be written $ e^{i (k + \frac12\mfs e v ) t} e^{i  \mfs \mfc k r }$ for $k \geq 0$. 

Keeping in mind that particle probability flux points in the opposite direction to the momentum for negative-frequency states (see for example \cite{Hansen:1980nc}), the sign of the radial direction of motion for these modes is given by $-\mfs\mfc$ for particles and by $+ \mfs\mfc$ for antiparticles. These modes are normalized so that $i \ol{u}_{\mfs\mfc\omega} \gamma^0 u_{\mfs\mfc\omega} = i \ol{v}_{\mfs\mfc\omega} \gamma^0 v_{\mfs\mfc\omega} = (4\pi r^2)^{-1}$ and as a result $i \ol{u}_{\mfs\mfc\omega} \gamma^r u_{\mfs\mfc\omega} = i \ol{v}_{\mfs\mfc\omega} \gamma^r v_{\mfs\mfc\omega} = -\mfs\mfc /(4\pi r^2) = - \mfh/(4\pi r^2)$. 

\subsubsection{2D formulation}

As has been often noted elsewhere purely $S$-wave dynamics can be usefully rewritten as a Dirac equation in 1+1 dimensions \cite{Rubakov:1981rg, Callan:1982ah}, and we pause here to establish our conventions. In this 1+1 dimensional formulation the only spatial direction corresponds to the radial direction in the 3+1 dimensional formulation, and so runs only along the half-line corresponding to positive $r$, with the dyon interior providing a more complicated background solution only within a distance of order $\mu^{-1}$ about $r = 0$. 

Within the full theory fields are required to be nonsingular at the origin (deep within the dyon) and this can be converted in the 2D formulation into a boundary condition for fields at $r=0$ designed so that reflection at $r = 0$ describes the effect in the 4D theory of passing through the origin. Alternatively we can instead describe $S$-wave scattering within the 2D picture by extending the spatial direction to cover the entire real line and think of, say, $x < 0$ as describing incoming waves and $x > 0$ describing outgoing waves.

In either formulation 4D fermions are described by independent 2D spinors, $\chi$, for each fermion flavour and chirality in the following way\footnote{This is to be compared with the 4D expression \pref{s wave fermions}. These definitions ensure that $\psi_{\mfs}$ and $\chi_{\mfs}$ transform in  the same way under the remaining Lorentz transformation in 2D \textit{i.e.} radial boosts. These are generated by $\gamma^0 \gamma^r$ in 4D and $\Gamma^0\Gamma^1=\Gamma_c$ in 2D, and since $\mfh = - \mfc$ for negatively charged fermions, the order of $f_-(r,t)$ and $g_-(r,t)$ is switched when going to 2D.}
\be \label{2d s-wave fields}
    \chi_{+}(r,t) = \begin{pmatrix}f_+(r,t)\\g_+(r,t)\end{pmatrix} \quad \hbox{and} \quad
    \chi_{-}(r,t) = \begin{pmatrix}g_-(r,t)\\f_-(r,t)\end{pmatrix} = i \Gamma^0 \begin{pmatrix}f_-(r,t)\\g_-(r,t)\end{pmatrix}, 
\ee
with the subscript $\pm$ indicating both the isospin and the $U(1)$ charge of each spinor (see Appendix \ref{App:GammaConventions} for our 2D Dirac-matrix conventions). Comparing to \eqref{4d Dirac equation} shows that their equations of motion can then be compactly written as a 2D Dirac equation in the background potential $\mathcal{A}^3_0(r)$ of the form:
\be 
    \left(-\partial_t +\Gamma^0 \Gamma^1 \partial_1 + \frac{i \mfs}{2} \, e\mathcal{A}^3_0\right)\chi_\mfs=0 \,,
\ee
such as would follow from the bulk action
\be \label{2DDirac}
   S_2 = - \frac12 \sum_{\mfs = \pm} \int \exd^2x \; \ol \chi_\mfs  \,\Gamma^\alpha \overset{\leftrightarrow}{D}_\alpha \chi_\mfs \,,
\ee
where $\ol \chi := i \chi^\dagger \Gamma^0$ and $D_\alpha \chi_\mfs = ( \partial_\alpha -  \tfrac12 i e\mfs  \cA_\alpha^3 ) \chi_\mfs$. 

Notice that \pref{s wave fermions} shows that the eigenvalue of the 2D chirality $\Gamma_c := \Gamma^0 \Gamma^1 = \sigma_3$ in this representation is $\mfc \mfs$ where $\mfc$ is the eigenvalue of 4D chirality and so is the same as $\mfh$ and therefore should correlate with the direction of motion. This correlation in the 2D language is a direct consequence of the Dirac equation because $\Gamma^0(\Gamma^\alpha \partial_\alpha) = - \partial_t + \Gamma_c \partial_1$.

Dimensional reduction of the $S$-wave modes strips away their angular content, but otherwise leaves them much as described in \pref{Chirality basisu} and \pref{Chirality basisv} above. For instance, writing the doublet built from $\chi_+$ and $\chi_-$ by $\bm{\chi}$ (and continuing to denote the isodoublets by square brackets and spinor doublets with round brackets), the positive-frequency basis of solutions $\mfu_{\mfs\mfc \omega}(x)$ corresponding to \pref{Chirality basisu} are
\be \label{Chirality basisu2D}
    \begin{bmatrix} \mfu_{+ \mfc \omega}(x) \\ 0 \end{bmatrix} \,, \; 
    \begin{bmatrix} 0 \\ \mfu_{ - \mfc\omega}(x) \end{bmatrix} \quad \hbox{where} \quad
    \mfu_{\mfs\mfc\omega}(x) :=  \xi_{\mfs \mfc} \, e^{-i \omega t}e^{-i (\mfs \omega +\frac12 ev) \,\mfc \,r } \left( \frac{r}{r_0} \right)^{i \mfc\, eQ/2},  
\ee 
with 
$\Gamma_c\, \xi_{\mfs \mfc} = \sigma_3 \,\xi_{\mfs \mfc} = \mfs \mfc \,\xi_{\mfs \mfc}$. The negative frequency counterparts similarly are
\be \label{Chirality basisvx2D}
    \begin{bmatrix} \mfv_{+ \mfc \omega}(x) \\ 0 \end{bmatrix} \,, \; 
    \begin{bmatrix} 0 \\ \mfv_{ - \mfc\omega}(x) \end{bmatrix} \quad \hbox{where} \quad
    \mfv_{\mfs\mfc\omega}(x) :=  \xi_{\mfs \mfc}  \, e^{i \omega t}e^{i (\mfs \omega - \frac{1}{2} ev) \,\mfc \,r } \left( \frac{r}{r_0} \right)^{i \mfc\, eQ/2} \,. 
\ee 

The frequency $\omega$ is (as above) bounded from below, with floors $\omega \geq - \frac12  \mfs ev$ and $\omega \geq + \frac12  \mfs ev$ for particle and antiparticle respectively. These modes are normalized so that $i \ol{\mfu}_{\mfs\mfc\omega} \Gamma^0 \mfu_{\mfs\mfc\omega} = i \ol{\mfv}_{\mfs\mfc\omega} \Gamma^0 \mfv_{\mfs\mfc\omega} = 1$ and so the 2D and 4D fluxes are related by $i \ol{\mfu}_{\mfs\mfc\omega} \Gamma^r \mfu_{\mfs\mfc\omega} = 4\pi r^2 (i \ol{u}_{\mfs\mfc\omega} \gamma^r u_{\mfs\mfc\omega})$ and similarly for $\mfv$ and $v$.  

\section{Dyonic EFT}
\label{sec:PPEFTs}

The far-field modes given in \pref{Chirality basisu} and \pref{Chirality basisv} suffice to fully describe the state of both an incident and departing fermion (or antifermion) when it is far from the dyon, but scattering calculations relate the size of the departing wave to the initially incident one and in principle this requires solving the full Dirac equation obtained from \pref{Fermonic action to order e^0} including the dyon's interior structure. It is only once this structure is included that the fermionic modes can be required to be nonsingular at the origin, providing the information that ultimately links the incoming and outgoing modes. 

But solving the full Dirac equation including the dyon interior is difficult (see however \cite{Goldhaber:1977xw, Blaer:1981ps, Marciano:1983md, Ezawa:1983vi}) and also likely overkill, at least for incident fermion energy $E$ small compared with the inverse dyon size $\mu$. In this regime the dyon is so small that its physical implications should be capturable by a choice of boundary condition near the origin. But this is where EFT techniques can usefully be applied and to this end we apply here the PPEFT formalism \cite{PPEFT, PPEFT2, PPEFT3} --- a framework explicitly designed for the purpose of constructively deriving such boundary conditions starting from the low-energy effective action for the compact object (for a review see {\it e.g.}~\cite{EFTBook}). This section applies this formalism to determine the form of the boundary conditions required for the Julia-Zee dyon (and its more complicated counterparts), at leading order in the small ratio $E/\mu$.  

\subsection{$S$-wave PPEFT}
\label{ssec:BCs}

The strategy, as always for EFTs, starts by replacing the full dyon with an effective description that captures its low-energy interactions. The required EFT is defined along the dyonic world-line and describes the interactions between low-energy bulk fields and the low-energy dyonic collective coordinates. This begins with an enumeration of all possible lowest-dimension interactions allowed by the assumed symmetries and particle content. 

The dyon-localized fields to be included in the dyonic effective action include the dyonic collective coordinates like the centre-of-mass position, $x^\mu = y^\mu(s)$,  of the dyon's world-line $W$ and the charge-excitation field $\mfa(s)$, where $s$ is an arbitrary parameter along the world-line. The `bulk' fields\footnote{Here `bulk' fields mean fields that are defined everywhere in spacetime and not only along the dyon's world-line.} to be included are those with masses much smaller than $\mu$: the fermion doublet ${\bm \psi}(x)$ and the massless bulk electromagnetic field $\widehat A_\mu(x) =  \widehat A^{\,3}_\mu(x)$, and (in principle) the spacetime metric $g_{\mu\nu}(x)$. 

The symmetries to be imposed are ($i$) world-line reparameterizations of $s$; ($ii$) Poincar\'e invariance in spacetime (or general coordinate invariance if a general metric $g_{\mu\nu}$ is included); ($iii$) invariance under electromagnetic gauge transformations; and ($iv$) any low-energy flavour and/or discrete symmetries. Spacetime symmetries are built in by constructing the action using the pull-backs of spacetime tensors to the world-line,\footnote{Interactions involving the normals to the world-line can also be constructed but do not play any role in what follows.}
\be
  \psi(s) = \psi[y(s)] \,, \quad \widehat A(s) = \dot y^\mu(s) \widehat A_\mu[y(s)] 
  \quad \hbox{and} \quad \gamma(s) = \dot y^\mu(s) \dot y^\nu(s) g_{\mu\nu}[y(s)] \,,
\ee
where over-dots denote differentiation with respect to $s$. One builds from these a reparameterization invariant action in the standard way. 

For these fields the electromagnetic gauge transformations are 
\be \label{dyongauge}
   \delta {\bm \psi}(s) = \frac{i }{2} \,\omega(s) \,  \tau_3{\bm \psi}(s) \,, \quad
   \delta \widehat A(s) = \frac{1}{e} \, \dot \omega(s) \quad \hbox{and} \quad
   \delta \mfa(s) = \omega(s) \,, 
\ee
where $\omega(s)$ is related to the spacetime dependent $SU(2)$ gauge transformation parameters $\omega^a(x)$ by $\omega(s) = \omega^3[y(s)]$. This symmetry requires derivatives of $\mfa$ to appear within a covariant derivative 
\be
  D \mfa := \dot \mfa - e \widehat A \,.
\ee

\subsubsection{Dyon world-line effective action}

For these variables the lowest-dimension interactions involving the rotor field $\mfa$ and the dyon displacement $y^\mu$ located at the dyonic position become 
\be \label{dyonEFT1}
  S_{\rm dyon} =  \int_\ssW \exd s \left\{ \sqrt{-\gamma}\left[ -M- \frac{\cI }{2\gamma} (D\mfa)^2- \frac12 \,\ol{\bm \psi} \, \mfC(\mfa) \, {\bm \psi}  + \cdots \right] + \frac{\vartheta}{2\pi} D \mfa \right\}
\ee
where $M$ is the classical dyon mass while $\cI$ is the `rotor' coefficient \pref{DyonS0} and the ellipses include a variety of other higher-dimension terms whose contributions to physics should be suppressed at low energies and so whose detailed form is not required in what follows.\footnote{Notice we only include here the couplings of the fluctuation fields and not also the dyon interactions giving rise to the background fields themselves. This is why no term like $Q \widehat A$ appears describing the dyon's classical charge, and why no term is required expressing the dyon's magnetic charge \cite{MagCharge, MagCharge1}. The same could also have been done for the metric if we'd expanded about the dyon's gravitational back-reaction, but we do not do so. \label{fn:Qterm}} The most general fermion bilinear consistent with the field content, gauge and spacetime symmetries is
\bea \label{PPEFTList0}
\mfC(\mfa) &:=&  \hat\mfc^s_{1} + i\, \hat\mfc^{ps}_1\gamma_5 + i\, \hat\mfc^{v}_1 \gamma_{\mu} \dot{y}^{\mu} + i\, \hat\mfc^{pv}_1 \gamma_5\gamma_{\mu} \dot{y}^{\mu}   + \Bigl(\hat\mfc^{s}_3 + i \,\hat\mfc^{ps}_3\gamma_5 + i \,\hat\mfc^{v}_3\gamma_{\mu}\dot{y}^{\mu} + i\, \hat\mfc^{pv}_3 \gamma_5\gamma_{\mu}\dot{y}^{\mu}\Bigr) \tau_3 \nn\\
&& \qquad\qquad +\Bigl( \hat\mfc^s_{+} + i\, \hat\mfc^{ps}_+\gamma_5 + i\, \hat\mfc^{v}_+ \gamma_{\mu} \dot{y}^{\mu} + i\, \hat\mfc^{pv}_+ \gamma_5\gamma_{\mu} \dot{y}^{\mu} \Bigr) e^{i\mfa} \tau_+   \\
&& \qquad\qquad\qquad\qquad +  \Bigl(\hat\mfc^{s}_- + i \,\hat\mfc^{ps}_-\gamma_5 + i \,\hat\mfc^{v}_-\gamma_{\mu}\dot{y}^{\mu} + i\, \hat\mfc^{pv}_- \gamma_5\gamma_{\mu}\dot{y}^{\mu}\Bigr) e^{-i\mfa} \tau_- \,,\nn
\eea 
where (as before) $\tau_\pm = \frac12(\tau_1 \pm i \tau_2)$. 

This action is meant to capture the low-energy interactions of the underlying dyon and the quantities $M$, $\mfC$ and so on are found by matching to the properties of the dyon's microscopic description. In particular, this relates the parameter $\vartheta$ to the vacuum angle appearing in the bulk electromagnetic theta-term 
\be
  \cL_\vartheta =  \frac{\vartheta e^2}{4\pi^2} \, \bfE \cdot \bfB \,,
\ee
since both ultimately descended from the underlying theta-term for the microscopic nonabelian $SU(2)$ gauge interactions. (Equivalently, this connection can also be established using invariance of the system under large gauge transformations.) 

To interpret \pref{dyonEFT1} it is convenient to specialize to a dyon that is perturbatively close to being at rest so $\dot{y}^\mu(s) = \delta^\mu_0 + \delta \dot{y}^\mu(s)$, choose the Minkowski metric and choose the parameter $s = t$ to be time in the background dyon's rest frame. In this case $-\gamma = -\eta_{\mu\nu} \dot y^\mu \dot y^\nu = 1 - \dot{\bm y} \cdot \dot {\bm y} + \cdots$ (where ${\bm y}$ is the spatial part of $\delta y^\mu$) and \pref{dyonEFT1} becomes
\be \label{dyonEFT2}
  S_{\rm dyon} \simeq \int_\ssW \exd t \left[\frac{M}{2} \dot {\bm y} \cdot \dot {\bm y} + \frac{\cI}{2}  \Bigl(\dot \mfa - e \widehat A_0 \Bigr)^2 - \frac12  \,\ol{\bm \psi} \, \mfC (\mfa) \, {\bm \psi} + \frac{\vartheta}{2\pi} \Bigl(\dot \mfa - e \widehat A_0 \Bigr) + \cdots \right] 
\ee
where to leading order -- {\it i.e.}~neglecting dyon recoil effects -- the integral is evaluated along the dyon's world-line, which we choose to be located at $\bmr = {\bm 0}$. In the same approximation the matrix $\mfC(\mfa)$ controlling the fermion-dyon couplings becomes
\bea \label{PPEFTList0RF}
\mfC(\mfa) &:=&  \hat\mfc^s_{1} + i\, \hat\mfc^{ps}_1\gamma_5 - i\, \hat\mfc^{v}_1 \gamma^{0} - i\, \hat\mfc^{pv}_1 \gamma_5\gamma^{0}   + \Bigl(\hat\mfc^{s}_3 + i \,\hat\mfc^{ps}_3\gamma_5 - i \,\hat\mfc^{v}_3\gamma^{0} - i\, \hat\mfc^{pv}_3 \gamma_5\gamma^{0} \Bigr) \tau_3 \\
&& \quad  +\Bigl( \hat\mfc^s_{+} + i\, \hat\mfc^{ps}_+\gamma_5 - i\, \hat\mfc^{v}_+ \gamma^{0}  - i\, \hat\mfc^{pv}_+ \gamma_5\gamma^{0}  \Bigr) e^{i\mfa} \tau_+   +  \Bigl(\hat\mfc^{s}_- + i \,\hat\mfc^{ps}_-\gamma_5 - i \,\hat\mfc^{v}_-\gamma^{0} - i\, \hat\mfc^{pv}_- \gamma_5\gamma^{0} \Bigr) e^{-i\mfa} \tau_- \,,\nn
\eea 
which drops ${\bm \gamma} \cdot \dot\bmy$ dyon-recoil terms.

On dimensional grounds all sixteen of the effective couplings\footnote{Since the size of $\mfc$ might naturally be expected to be set by the dyon size $R \sim \mu^{-1}$ this is usually where the suppression by the monopole scale would naively enter into observable scattering. Part of the purpose of this exercise is to understand why this suppression does {\it not} actually arise in fermion-dyon scattering, and how general a phenomenon this is.} $\hat\mfc$ have dimensions (length)${}^2$. All of these effective fermion interactions conserve electric charge, and this dictates the $\mfa$-dependence of $\mfC(\mfa)$. If treated perturbatively these interactions describe fermion scattering from the dyon, possibly associated with $\mfa$ excitation. The perturbative dyon response to fermion scattering can be seen because the canonical momentum for $\mfa$ is also its conserved charge,  
\be \label{dyonmomischarge}
  \mfp := \frac{\delta S_{\rm dyon}}{\delta \dot \mfa} = \cI D \mfa +\frac{\vartheta}{2\pi}= - \frac{\cQ_\ssD}{e} \,, 
\ee
where -- see {\it e.g.}~eq.~\pref{BCGauss} -- $\cQ_\ssD$ is the contribution of dyonic excitations to the fluctuations' electric charge: $\cQ = \cQ_\ssD + \cQ_\ssF$ (where $\cQ_\ssF$ is the electric charge carried by the fermions). The canonical commutation relation $[\mfp(t) \,, \mfa(t)] = -i$ therefore implies the quantities $e^{\pm i \mfa}$ act as raising/lowering operators for $\cQ_\ssD$, since
\be
   \Bigl[ \cQ_\ssD/e \,, e^{\pm i \mfa} \Bigr] = \mp   e^{\pm i \mfa} \,. \nn
\ee
Interactions proportional to $e^{\pm i \mfa}$ therefore describe transitions that raise or lower the dyon charge by $e$, as required by charge conservation for the reactions involving $\tau_\pm$ in \pref{dyonEFT2} (which change the fermion charge from $-\frac12 e$ to $+\frac12 e$ or vice versa). 

Because these fermion-$\mfa$ couplings are not suppressed by powers of small couplings like $e$ they are not intrinsically negligible even at leading loop order, which does not justify expanding the exponentials in powers of $\mfa$. Furthermore the energy exchanged by exciting or de-exciting $\mfa$ is order $\cI^{-1}$ which \pref{DyonEnergy} reveals is suppressed by $\alpha = e^2/(4\pi)$ compared to the characteristic scale $\mu$ and so has little intrinsic cost in the semiclassical limit. Such transitions can be important to fermion scattering, though there is an order-of-limits issue when working both at low fermion energies, $E \ll \mu$, and at leading nontrivial order in the loop expansion, $\alpha = e^2/(4\pi) \ll 1$, because it matters in practice whether or not the fermion energy $E$ is larger or smaller than the dyon excitation scale $\alpha \mu$. As argued below, a natural way to handle this fermion-dyon dynamics in the effective theory -- at least in the regime $\alpha \mu \ll E \ll \mu$ -- is through the Born-Oppenheimer approximation \cite{BornOppenheimer} (in which the fermions play the role of the `fast' degrees of freedom while the large size of $\cI$ makes the dyonic excitations `slow').

The interactions in \pref{PPEFTList0} can also be classified by how they transform under global `flavour' transformations acting on the fermion field. In the present instance the limited field content restricts this to two types of such symmetries: an axial symmetry for which $\delta_\ssA {\bm \psi} = i \omega_\ssA \, \gamma_5 \, {\bm \psi}$ and a fermion-number `baryon' symmetry for which $\delta_\ssB {\bm \psi} = i \omega_\ssB  \, {\bm \psi}$ (notice the difference between this and the gauge transformation \pref{dyongauge}). We denote the corresponding conserved charges by $\cQ_\ssA$ and $\cQ_\ssB$ to distinguish them from the fermionic contribution to the gauge charge $\cQ_\ssF$. The Noether currents for these two symmetries are $J^\mu_\ssB = i \ol \bmpsi \gamma^\mu \bmpsi$ 
and $J^\mu_\ssA = i \ol \bmpsi \gamma^\mu \gamma_5 \bmpsi$ 
and satisfy
\be
  \partial_\mu J^\mu_\ssB = 0 \quad \hbox{and} \quad 
  \partial_\mu J^\mu_\ssA = \frac{e^2}{32\pi^2} \, \epsilon^{\mu\nu\lambda\rho} F^a_{\mu\nu} F^a_{\lambda\rho}= -\frac{e^2}{4\pi^2} \, \bfE^a \cdot \bfB^a \,, 
\ee
which shows that the axial symmetry is anomalous. Table \ref{Classification of interactions for Julia-Zee dyon} identifies which of the various effective interactions of \pref{PPEFTList0} preserves each of these two flavour symmetries. 

\begin{table}[ht!]
\centering
\setlength{\arrayrulewidth}{0.3mm}
\setlength{\tabcolsep}{5pt}
\renewcommand{\arraystretch}{1.4}

\begin{tabular}[ht!]{c|cccc}
& $\Delta \cQ_\ssF =\Delta \cQ_\ssA=0$  & $\Delta \cQ_\ssF =0$, $\Delta \cQ_\ssA \neq 0$  & $\Delta \cQ_\ssA=0$ and $\Delta \cQ_\ssF \neq 0$  & $\Delta \cQ_\ssA \neq 0$ and $\Delta \cQ_\ssF \neq 0$  \\
 \hline
 \hline
 coupling & $\hat\mfc^v_1, \hat\mfc^v_3 , \hat\mfc^{pv}_1, \hat\mfc^{pv}_3$   & $\hat\mfc^s_1, \hat\mfc^s_3, \hat\mfc^{ps}_1, \hat\mfc^{ps}_3$   & $\hat\mfc^v_{\pm}, \hat\mfc^{pv}_{\pm}$ & $\hat\mfc^s_{\pm}, \hat\mfc^{ps}_{\pm}$ \\
\end{tabular}
\caption{\small Classification of interaction terms in the worldline action according to whether they conserve the fermionic charges $\cQ_\ssF$ (electric charge) and $\cQ_\ssA$ (axial charge) as defined in the main text. All interactions conserve $\cQ_\ssB$ (fermion number). Total electric charge (including the contribution from $\mfa$) is also conserved by all interactions.}
\label{Classification of interactions for Julia-Zee dyon}
\end{table}

\subsubsection{Induced boundary conditions}

In practice we wish to excise the dyon from the external world and replace it with a gaussian `pillbox', $\cP_\epsilon$, whose radius is chosen much larger than the dyon's, $\epsilon \gg R \sim m_g^{-1}$, but also much smaller than the distances of interest for the bulk fields used for low-energy probes of the dyon. The idea is to replace the dyon with a set of near-dyon boundary conditions on the surface $r = \epsilon$ of this pillbox, whose detailed form is chosen to reproduce the interactions implied by \pref{dyonEFT1}. It is these boundary conditions that communicate dyonic physics to the external world from which the dyon is excised. 

The boundary conditions can be obtained by a variety of means \cite{PPEFT, PPEFT2, PPEFT3}, such as by integrating the field equations with the dyon-localized interaction written proportional to $\delta^3(x)$ and then regularizing the appearance of divergent bulk fields, like $\widehat A(0)$, with its value on the surface of the pillbox, $\widehat A(\epsilon)$. Applied to \pref{dyonEFT2} this leads to a familar expression: Gauss' law itself, in the form
\be \label{BCGauss}
  \oint \exd^2\Omega \, \Bigl(  r^2 \partial_r \widehat A_0 \Bigr)_{r = \epsilon} = 4\pi \Bigl(  r^2 \partial_r \widehat A_0 \Bigr)_{r = \epsilon} =  \left( \frac{\delta S_{\rm dyon}}{\delta \widehat A_0} \right)_{r=\epsilon} = -e \mfp = -e\left( \cI D \mfa + \frac{\vartheta}{2\pi} \right)\,.
\ee

Comparing this to the bulk far-field Coulomb solution $\widehat A_0 \simeq - \cQ_\ssD/(4\pi r)$ with integration constant $\cQ_\ssD$ given by the dyonic fluctuation's charge, this boundary condition fixes $\cQ_\ssD = - e \mfp$ as used above (and in passing shows how the dyon acquires a $\vartheta$-dependent charge through the Witten effect \cite{Witten:1979ey}). 

We wish to make the same argument for the bulk fermion field. The same steps \cite{PPEFT3} lead to the boundary condition  
\be \label{fermbcgen}
  \frac12 \oint \exd^2\Omega \, \Bigl(  r^2 \gamma^r \bmpsi \Bigr)_{r = \epsilon} = \left( \frac{\delta S_{\rm dyon}}{\delta \ol \bmpsi} \right)_{r=\epsilon} =  - \frac12 \, \Bigl[ \mfC(\mfa) \, \bmpsi  \Bigr]_{r=\epsilon} \,, 
\ee
once projected onto the $S$ wave state. This can formally be recast as the non-differential condition
\bea \label{s-wave boundary condition 1}
  && 0=\Bigl\{\Bigl[ \gamma^r  + \Bigl( \hat{\cC}^s_{1} + i\, \hat{\cC}^{ps}_1\gamma_5 - i\, \hat{\cC}^{v}_1 \gamma^{0} - i\, \hat{\cC}^{pv}_1 \gamma_5\gamma^{0}  \Bigr) + \Bigl(\hat{\cC}^{s}_3 + i \,\hat{\cC}^{ps}_3\gamma_5 - i \,\hat{\cC}^{v}_3\gamma^{0} - i\, \hat{\cC}^{pv}_3 \gamma_5\gamma^{0} \Bigr) \tau_3 \\
&& \quad  +\Bigl( \hat{\cC}^s_{+} + i\, \hat{\cC}^{ps}_+\gamma_5 - i\, \hat{\cC}^{v}_+ \gamma^{0}  - i\, \hat{\cC}^{pv}_+ \gamma_5\gamma^{0}  \Bigr) e^{i\mfa} \tau_+   +  \Bigl(\hat{\cC}^{s}_- + i \,\hat{\cC}^{ps}_-\gamma_5 - i \,\hat{\cC}^{v}_-\gamma^{0} - i\, \hat{\cC}^{pv}_- \gamma_5\gamma^{0} \Bigr) e^{-i\mfa} \tau_-  \Bigr] \bmpsi \Bigr\}_{r=\epsilon} \nonumber
\eea
where the dimensionless coefficients $\hat{\cC}^\ssA_\ssI$ are related to the coefficients $\hat \mfc^\ssA_\ssI$ by
\be \label{dimensionlessCs}
    \hat{\cC}^\ssA_\ssI = \frac{\hat \mfc^\ssA_\ssI}{4 \pi \epsilon^2} \,,
\ee
for all $A = s, ps, v, pv$ and $I = 1,3,+,-$. 

This expression is only `formal' because the projection onto the $S$-wave while regulating by displacing the field to $r = \epsilon$ is more subtle because of the different angular dependence imposed on $\psi_+$ relative to $\psi_-$ by the presence of the magnetic monopole background configuration.  These issues are summarized in Appendix \ref{App:Codimension1Action} and the result is most conveniently expressed in terms of the two-dimensional fields $\chi$ whose bulk dynamics is described by \pref{2DDirac}. In this 2D formulation the fermionic terms in the dyon's point-particle world line action given by \pref{dyonEFT2} and \pref{PPEFTList0} can be written 
\be \label{dyonEFT22D}
     S_{\rm dyon2} =-\frac12 \sum_{\mfs,\mfs'=\pm} \int \exd t\, \overline{\chi}_{\mfs} \Bigl(\mathcal{C}^s_{\mfs\mfs'}+i\mathcal{C}^{ps}_{\mfs\mfs'} \Gamma_c+i\mathcal{C}^{v}_{\mfs\mfs'}\Gamma_{\alpha}\dot{y}^{\alpha}+i\mathcal{C}^{pv}_{\mfs\mfs'} \Gamma^{\alpha}\epsilon_{\alpha \beta}\dot{y}^{\beta}\Bigr)\,\chi_{\mfs'}  \; e^{\frac{i}{2} (\mfs-\mfs')\, \mfa} \,, 
\ee
where the fields are evaluated at $r = \epsilon$. Here (as above) $\Gamma_c := \Gamma^0 \Gamma^1$ has eigenvalues $\mfh = \pm 1$ (and is diagonal in the basis used here) while the 2D Dirac matrices satisfy $\Gamma_c \Gamma^\alpha = \epsilon^{\alpha\beta} \Gamma_\beta$ where our Levi-Civita convention chooses $\epsilon^{01} = +1$. The coefficients $\cC^\ssA_{ij}$ are dimensionless in the same way that the $\hat{\cC}^\ssA_\ssI$ are, and must satisfy $\cC^{\ssA*}_{\mfs\mfs'} = \cC^\ssA_{\mfs'\mfs}$ if the action $S_{\rm dyon2}$ is real (in which case they contain 16 independent real parameters). 

Combining this `boundary' action with the `bulk' action \pref{2DDirac} leads to the following near-dyon boundary condition for the 2D fermions near a static dyon (for which $\dot y^\alpha \simeq \delta^\alpha_0$)
\be \label{1+1d:s-wave boundary condition 1}
    \sum_{\mathfrak{s}'}\Bigl( \delta_{\mathfrak{s} \mathfrak{s}'}\Gamma^1+\mathcal{C}^s_{\mathfrak{s} \mathfrak{s'}}+i \mathcal{C}^{ps}_{\mathfrak{s} \mathfrak{s'}}\Gamma_c - i \mathcal{C}^{v}_{\mathfrak{s} \mathfrak{s'}}\Gamma^{0}+i \mathcal{C}^{pv}_{\mathfrak{s} \mathfrak{s'}}\Gamma^{1} \Bigr)\,e^{\frac{i}{2}(\mathfrak{s}-\mathfrak{s}')\mfa(t)}{\chi}_{\mathfrak{s}'}(\epsilon,t)=0,
\ee
for all $t$. Equivalently
\be \label{Matrix2DBC}
    \Bigl[\Gamma^1+ O_{\mathcal{B}}(\mfa)\Bigr] \left[ \begin{matrix}
        \chi_{+} \\ \chi_{-}   \end{matrix} \right]_{r=\epsilon} =0,
\ee
where 
\bea \label{OBdef}
    O_{\mathcal{B}}(\mfa) &= &\begin{pmatrix}
        \mathcal{C}^{s}_{++} &  \mathcal{C}^{s}_{+-} e^{i \mfa}\\\mathcal{C}^{s}_{-+} e^{-i \mfa}& \mathcal{C}^{s}_{--}
    \end{pmatrix}+i \begin{pmatrix}
        \mathcal{C}^{ps}_{++} &  \mathcal{C}^{ps}_{+-} e^{i \mfa}\\\mathcal{C}^{ps}_{-+} e^{-i \mfa}& \mathcal{C}^{ps}_{--}
    \end{pmatrix}\Gamma_c\nonumber\\ 
    &&\qquad\qquad\qquad - i\begin{pmatrix}
        \mathcal{C}^{v}_{++} &  \mathcal{C}^{v}_{+-} e^{i \mfa}\\\mathcal{C}^{v}_{-+} e^{-i \mfa}& \mathcal{C}^{v}_{--}
    \end{pmatrix}\Gamma^{0} +i\begin{pmatrix}
        \mathcal{C}^{pv}_{++} &  \mathcal{C}^{pv}_{+-} e^{i \mfa}\\\mathcal{C}^{pv}_{-+} e^{-i \mfa}& \mathcal{C}^{pv}_{--}
    \end{pmatrix}\Gamma^{1} \,.
\eea

Some intuition about the physical meaning of this boundary condition can be obtained by considering what it implies for the flux of fermionic currents through the surface at $r = \epsilon$. Consider, for example, a fermion current of the form $j^\alpha = i \overline{\bm{\chi}} \Gamma^\alpha M \bm{\chi}$ where $M$ is some matrix in spin and isospin space. When the action \pref{2DDirac} is real the boundary condition allows $\Gamma^1 \bmchi(\epsilon,t)$ to be replaced by terms involving the effective couplings $\cC^\ssA_{ij}$ and this can be used to learn something about the flux $j^1(\epsilon,t)$. As applied to the currents for fermion number, $j^{\alpha}_\ssB =i \overline{\bm{\chi}}\Gamma^{\alpha} \bm{\chi}$, electric charge, $j^{\alpha}_{\ssF} = \tfrac12 i e\overline{\bm{\chi}}\Gamma^{\alpha}\, \tau_3 \bm{\chi}$ and axial symmetry\footnote{$j^{\alpha}_\ssA$ is chosen in the 2D theory to match the $4$D axial current $J^\mu_\ssA = i \ol \bmpsi \gamma^\mu \gamma_5 \bmpsi$ 
for $S$-wave states, up to factors of $4\pi r^2$.} $j^{\alpha}_\ssA = i \overline{\bm{\chi}} \Gamma^{\alpha} \Gamma_c \tau_3 \bm{\chi}$ (for real boundary action) the boundary conditions imply\footnote{Divergences associated with evaluating these fermion bilinears can be regulated in a way that preserves this relation, as we show in appendix \ref{App: Currents}.} $j^r_\ssB(\epsilon, t)=0$,  
\be \label{boundarycurrents}
     j^r_\ssF(\epsilon, t) = -\tfrac14 i e \,\overline{\bm{\chi}}(\epsilon,t)\Bigl[ \tau_3, O_{\mathcal{B}}(\mfa) \Bigr] \bm{\chi}(\epsilon, t) \quad \hbox{and} \quad  j^r_\ssA(\epsilon, t) = \tfrac12 i \, \overline{\bm{\chi}}(\epsilon,t) \Bigl\{\Gamma_c \tau_3, O_{\mathcal{B}}(\mfa) \Bigr\}\bm{\chi}(\epsilon, t) \,,
\ee
where $O_\cB$ is defined by \pref{OBdef}. These show that none of the dyon-fermion interactions transfer fermion number to or from the dyon while the terms in the boundary action involving $\tau_\pm$ contribute nonzero flux of electrical current at $r = \epsilon$, as required by charge conservation when exciting or de-exciting the dyonic field $\mfa$.  

Several things about the boundary condition \pref{Matrix2DBC} are noteworthy (for more details  and a discussion of how things look for a non-dyonic source see Appendix \ref{App:PPEFT3} and \cite{PPEFT3}).
\begin{itemize}
\item {\it Linearity:} This boundary condition is linear in ${\bm \psi}$ (or $\bmchi$) because the action \pref{dyonEFT2} is quadratic in ${\bm \psi}$ (or \pref{dyonEFT22D} is quadratic in $\bmchi$). Furthermore, the action is quadratic because this is the lowest-dimension interaction consistent with the field content and symmetries and it is the lowest-dimension interactions that dominate at low energies within an effective theory. It is through arguments like this that PPEFTs explain the ubiquity of linear boundary conditions for many systems as being a generic consequence of the low-energy limit.
\item {\it Algebraic:} This boundary condition does not involve derivatives -- unlike \pref{BCGauss}, for instance -- because the fermion bulk action is linear in derivatives of the fields. The first term in \pref{s-wave boundary condition 1} comes from integrating by parts in the bulk action, and when the same exercise is performed for bosonic fields one instead finds first-derivative Robin-style boundary conditions because the bulk field equations are second order in derivatives. We discuss some novel implications of non-derivative boundary conditions at some length below.
\end{itemize}

\subsubsection{Implications for mode functions}

We next work out how these boundary conditions influence the bulk fermion mode functions. This is complicated by the fact that the boundary condition \pref{s-wave boundary condition 1}  
explicitly involves the collective coordinate $\mfa$, and this is important because it shows how the dynamics of $\mfa$ feeds back onto the fermionic field as it interacts with the dyon. As emphasized in \cite{Polchinski} it is the nonperturbative nature of this fermion-dyon interaction that drives much of the unusual dynamics of the dyonic effective theory, generating effects that are correlated between the ${\bm \psi}$ and $\mfa$ sectors. In practice this complicates things by making the fermion boundary condition \pref{s-wave boundary condition 1} field-dependent. For the remainder of this section we regard the field $\mfa$ to be a specified classical function while we explore the implications of \pref{Matrix2DBC}  for the fermion field. \S\ref{ssec:AlphaPartyLine} returns to justify how to interpret the resulting $\mfa$-dependence of our results using the Born-Oppenheimer approximation.

Eqs.~\pref{Chirality basisu} and \pref{Chirality basisv} provide the general solution for the ${\bm \psi}$ mode functions and so form a complete basis of solutions to the bulk field equations. Energy eigenstates in the presence of the dyon are found by asking these also to satisfy the near-dyon boundary condition \pref{Matrix2DBC}, since this captures the implications of the dyon for its surroundings. Subtleties to do with the angle-dependence of the $S$-wave mode functions can be avoided if the boundary conditions are phrased in terms of the two-dimensional fields ${\bm \chi}$. For instance, if a general positive-frequency solution is given as a linear combination of the solutions \pref{Chirality basisu2D}, with coefficients $\bm{\kappa}_{\mfs\mfc}$, then a near-source boundary condition of the form \eqref{Matrix2DBC} can be written
\be \label{system of boundary eqs2D}
     {\cB}(\mfa)  \begin{bmatrix}
     \bm{\kappa}_{++}\, \mfu_{+ + \omega}(\epsilon,t) + \bm{\kappa}_{+-} \,\mfu_{+- \omega}(\epsilon,t) \\[1mm]\bm{\kappa}_{-+}\,\mfu_{-+ \omega}(\epsilon,t) + \bm{\kappa}_{--}\, \mfu_{- - \omega}(\epsilon,t)
     \end{bmatrix}=0 
\ee
where $\cB(\mfa)\coloneqq\Gamma^0[\Gamma^1+O_{\cB}(\mfa)]$ and so
{\small{
\be \label{BofaDef}
   \mathcal{B}({\mfa})= \begin{pmatrix}
        1+i(\mathcal{C}^{pv}_{++}+\mathcal{C}^{v}_{++})&-(\mathcal{C}^{ps}_{++}+i \mathcal{C}^{s}_{++})& i (\mathcal{C}^{pv}_{+-}+ \mathcal{C}^{v}_{+-})e^{i \mfa}& - (\mathcal{C}^{ps}_{+-}+i \mathcal{C}^{s}_{+-})e^{i \mfa}\\[2mm]\mathcal{C}^{ps}_{++}-i \mathcal{C}^{s}_{++}&  -1-i(\mathcal{C}^{pv}_{++} -\mathcal{C}^{v}_{++}) &  (\mathcal{C}^{ps}_{+-}-i \mathcal{C}^{s}_{+-})e^{i \mfa}& - i(\mathcal{C}^{pv}_{+-}- \mathcal{C}^{v}_{+-})e^{i \mfa}\\[2mm] i(\mathcal{C}^{pv}_{-+}+ \mathcal{C}^{v}_{-+})e^{-i \mfa} & -(\mathcal{C}^{ps}_{-+}+i \mathcal{C}^{s}_{-+})e^{-i \mfa} & 1+ i (\mathcal{C}^{pv}_{--}+\mathcal{C}^{v}_{--})& -(\mathcal{C}^{ps}_{--} + i \mathcal{C}^{s}_{--})\\[2mm] (\mathcal{C}^{ps}_{-+}-i \mathcal{C}^{s}_{-+})e^{-i \mfa} & - i(\mathcal{C}^{pv}_{-+}- \mathcal{C}^{v}_{-+})e^{-i \mfa} & \mathcal{C}^{ps}_{--}- i \mathcal{C}^{s}_{--}& -1 -i (\mathcal{C}^{pv}_{--}- \mathcal{C}^{v}_{--})
    \end{pmatrix} \,.
\ee}} 
Notice that when the $\cC^\ssA_{ij}$ couplings are hermitian -- {\it i.e.}~when the dyon-fermion action \pref{dyonEFT22D} is real -- the matrix $\cB(\mfa)$ satisfies the useful identity 
\be\label{BadjRel}
    \mathcal{B}^{\dagger}= 2 \begin{pmatrix}
        \Gamma_c &0\\0&\Gamma_c
    \end{pmatrix} -\mathcal{B}
\ee
for all $\mfa$, where $\Gamma_c = \sigma_3$ is the diagonal Pauli matrix. The most general matrix satisfying this condition has 16 real parameters, corresponding to the 16 real effective couplings contained in the $\cC^\ssA_{ij}$. 

The physical implications of this boundary condition depend sensitively on the rank of the matrix $\mathcal{B}$. For instance, if all couplings $\mathcal{C}^\ssA_{ij}$ were independent of one another $\mathcal{B}$ would generically have rank 4, in which case the only solution to \pref{system of boundary eqs2D} is ${\bm \chi}(\epsilon,t) = 0$. If $\mathcal{B}$ instead has rank $4-n$ for $n = 0,1,2,3$ then there would be $n$ linearly independent nonzero values for ${\bm \chi}(\epsilon,t)$ allowed by \pref{system of boundary eqs2D}. 

Now comes a key observation: when the boundary action $S_{\rm dyon2}$ is real -- and so the constants $\cC^\ssA_{ij}$ are hermitian -- then the rank of $\mathcal{B}$ cannot be smaller than two. This is because both of the following quantities are $\mfa$-independent and always nonzero:
\bea \label{non-zero determinants}
    \left|\begin{array}{c c}
        \mathcal{B}_{22} & \mathcal{B}_{24}\\ \mathcal{B}_{42}&  \mathcal{B}_{44}
    \end{array}\right| &=& 1+i (\mathcal{C}^{pv}_{--}-\mathcal{C}^{v}_{--}+\mathcal{C}^{pv}_{++}-\mathcal{C}^{v}_{++})-(\mathcal{C}^{pv}_{++}-\mathcal{C}^{v}_{++}) (\mathcal{C}^{pv}_{--}-\mathcal{C}^{v}_{--})+\left|\mathcal{C}^{pv}_{+-}-\mathcal{C}^{v}_{+-}\right|^2, \nonumber\\  
    \left|\begin{array}{c c}
        \mathcal{B}_{11} & \mathcal{B}_{13}\\ \mathcal{B}_{31}&  \mathcal{B}_{33}
    \end{array}\right| &=& 1+i (\mathcal{C}^{pv}_{--}+\mathcal{C}^{v}_{--}+\mathcal{C}^{pv}_{++}+\mathcal{C}^{v}_{++})-(\mathcal{C}^{pv}_{++}+\mathcal{C}^{v}_{++}) (\mathcal{C}^{pv}_{--}+\mathcal{C}^{v}_{--})+\left|\mathcal{C}^{pv}_{+-}+\mathcal{C}^{v}_{+-}\right|^2 . \nn\\
    &&
\eea
To see why these cannot vanish, consider the first case. Notice that its imaginary part can only vanish if $\mathcal{C}^{pv}_{--}-\mathcal{C}^{v}_{--}=-(\mathcal{C}^{pv}_{++}-\mathcal{C}^{v}_{++})$, but if this is true then the real part becomes $1+(\mathcal{C}^{pv}_{++}-\mathcal{C}^{v}_{++})^2+\left|\mathcal{C}^{pv}_{+-}-\mathcal{C}^{v}_{+-}\right|^2\ge 1$. A similar argument goes through also for the second case. 

As a consequence it is always possible to solve for two of the $\bm \kappa$'s in terms of the other two. For instance, solving for $\bm{\kappa}_{++}$ and $\bm{\kappa}_{--}$ gives
\be \label{1+1d,ML:kappa++sol}
\bm{\kappa}_{+ +} = -\bm{\kappa}_{+ -} e^{2 i (\omega + \frac{1}{2} e v) \epsilon}\,\left(\frac{\epsilon}{r_0}\right)^{- i e Q}\frac{\left|\begin{array}{ll}
\hat{\mathcal{B}}_{12}& \hat{\mathcal{B}}_{13} \\
\hat{\mathcal{B}}_{32}\, & \hat{\mathcal{B}}_{33}
\end{array}\right|}{\left|\begin{array}{ll}
\hat{\mathcal{B}}_{11} & \hat{\mathcal{B}}_{13} \\
\hat{\mathcal{B}}_{31} & \hat{\mathcal{B}}_{33}
\end{array}\right|}-\bm{\kappa}_{-+} e^{  2 i \omega\epsilon}e^{i \mfa} \frac{\left|\begin{array}{ll}
 \hat{\mathcal{B}}_{14}  & \hat{\mathcal{B}}_{13} \\
\hat{\mathcal{B}}_{34} & \hat{\mathcal{B}}_{33}
\end{array}\right|}{\left|\begin{array}{ll}
\hat{\mathcal{B}}_{11} & \hat{\mathcal{B}}_{13} \\
\hat{\mathcal{B}}_{31} & \hat{\mathcal{B}}_{33}
\end{array}\right|}, 
\ee
and
\be \label{1+1d,ML:kappa--sol}
     \bm{\kappa}_{--}= -\bm{\kappa}_{+ -} e^{  2 i \omega\epsilon}e^{-i \mfa}\frac{\left|\begin{array}{ll}
\hat{\mathcal{B}}_{11}& \hat{\mathcal{B}}_{12} \\
\hat{\mathcal{B}}_{31}\, & \hat{\mathcal{B}}_{32}
\end{array}\right|}{\left|\begin{array}{ll}
\hat{\mathcal{B}}_{11} & \hat{\mathcal{B}}_{13} \\
\hat{\mathcal{B}}_{31} & \hat{\mathcal{B}}_{33}
\end{array}\right|}-\bm{\kappa}_{-+} e^{ 2 i (\omega- \frac{1}{2}e v)\epsilon}\left(\frac{\epsilon}{r_0}\right)^{ i e Q}  \frac{\left|\begin{array}{ll}
 \hat{\mathcal{B}}_{11}  & \hat{\mathcal{B}}_{14} \\
\hat{\mathcal{B}}_{31} & \hat{\mathcal{B}}_{34}
\end{array}\right|}{\left|\begin{array}{ll}
\hat{\mathcal{B}}_{11} & \hat{\mathcal{B}}_{13} \\
\hat{\mathcal{B}}_{31} & \hat{\mathcal{B}}_{33}
\end{array}\right|} \,,
\ee
which defines $\hat{\mathcal{B}}_{ij} := \mathcal{B}_{ij}(\mfa=0)$ so that all $\mfa$-dependence is explicit. In later sections it is sometimes useful instead to solve for $\bm \kappa_{+-}$ and $\bm \kappa_{-+}$, which instead gives 
\be \label{1+1d,ML:kappa+-sol}
      \bm{\kappa}_{+ -}= -\bm{\kappa}_{ ++} e^{-2 i (\omega + \frac{1}{2}e v) \epsilon}\,\left(\frac{\epsilon}{r_0}\right)^{ i eQ}\frac{\left|\begin{array}{ll}
\hat{\mathcal{B}}_{2 1}& \hat{\mathcal{B}}_{24} \\
\hat{\mathcal{B}}_{41}\, & \hat{\mathcal{B}}_{44}
\end{array}\right|}{\left|\begin{array}{ll}
\hat{\mathcal{B}}_{22} & \hat{\mathcal{B}}_{24} \\
\hat{\mathcal{B}}_{42} & \hat{\mathcal{B}}_{44}
\end{array}\right|}-\bm{\kappa}_{--} e^{ - 2 i \omega\epsilon} e^{i \mfa}\frac{\left|\begin{array}{ll}
\hat{\mathcal{B}}_{23}  & \hat{\mathcal{B}}_{24} \\
\hat{\mathcal{B}}_{43} & \hat{\mathcal{B}}_{44}
\end{array}\right|}{\left|\begin{array}{ll}
\hat{\mathcal{B}}_{22} & \hat{\mathcal{B}}_{24} \\
\hat{\mathcal{B}}_{42} & \hat{\mathcal{B}}_{44}
\end{array}\right|}, 
\ee
and
\be \label{1+1d,ML:kappa-+sol}
      \bm{\kappa}_{- +}= -\bm{\kappa}_{++} e^{-  2i  \omega\epsilon}e^{-i \mfa}\frac{\left|\begin{array}{ll}
\hat{\mathcal{B}}_{22}& \hat{\mathcal{B}}_{21} \\
\hat{\mathcal{B}}_{42}\, & \hat{\mathcal{B}}_{41}
\end{array}\right|}{\left|\begin{array}{ll}
\hat{\mathcal{B}}_{22} & \hat{\mathcal{B}}_{24} \\
\hat{\mathcal{B}}_{42} & \hat{\mathcal{B}}_{44}
\end{array}\right|}-\bm{\kappa}_{- -}e^{ -2 i (\omega-\frac{1}{2}{ev})\epsilon}\left(\frac{\epsilon}{r_0}\right)^{ -i e Q} \frac{\left|\begin{array}{ll}
 \hat{\mathcal{B}}_{22}  & \hat{\mathcal{B}}_{23} \\
\hat{\mathcal{B}}_{42} & \hat{\mathcal{B}}_{43}
\end{array}\right|}{\left|\begin{array}{ll}
\hat{\mathcal{B}}_{22} & \hat{\mathcal{B}}_{24} \\
\hat{\mathcal{B}}_{42} & \hat{\mathcal{B}}_{44}
\end{array}\right|}.
\ee 

If $\mathcal{B}$ is rank two then this is all that can be learned because the other two equations in \pref{system of boundary eqs2D} are not independent. When $\mathcal{B}$ has rank two the boundary condition has precisely enough information to determine `out' states from the `in' states (with no extra constraints) as is required for scattering problems, so we henceforth assume the effective couplings $\mathcal{C}^\ssA_{ij}$ satisfy the conditions required to ensure rank$(\mathcal{B})=2$. This should automatically be the case when the microscopic physics of the source allows out states to be inferred from arbitrary in states -- such as for fermion scattering in the classical dyon background of the full nonabelian theory -- since the effective couplings obtained by matching must give a consistent description.\footnote{We have examples of effective couplings for which rank$(\cB) = 3$, but defer an exploration of their microscopic physical significance to future work.} As is shown in Appendix \ref{App:T amplitudes} the requirement that $\cB$ be rank two imposes a total of 8 conditions on its coefficients (four of which amount to unitarity conditions) and so removes half of the 16 real parameters that could have been encoded in $\cB$ for general choices of hermitian $\cC^\ssA_{ij}$'s.

\subsection{Scattering states}
\label{ssec:ScatteringState}

We next construct the explicit single-particle scattering states appropriate for fermion dyon scattering, assuming the rank of the matrix $\cB$ is two. This allows us to identify which combinations of the effective couplings actually appear in scattering processes.  

To this end we construct a basis of energy eigenmodes that either correspond to a single type of particle moving towards the dyon ($in$ state) or a single type of particle moving away from the dyon ($out$ state). These correspond to choosing specific couplings $\bm{\kappa}_{\mfs\mfc}$ to vanish, as described in detail below. Because the direction of motion of an $S$-wave state correlates with the value of the quantum number $\mfh = \mfs \mfc$, it suffices to label in and out states using just $\mfs$ and momentum $k$ (or frequency $\omega$), leading to positive (negative) frequency modes $\mfu^\ins_{\mfs, k}$ and $\mfv^\ins_{\mfs, k}$ or $\mfu^\outs_{\mfs, k}$ and $\mfv^\outs_{\mfs, k}$.

\subsubsection{In modes}

We define the positive-frequency in-states to be those modes for which there is only one component with incoming momentum (heading towards the dyon). Labelling these by the sign of the incoming particle's electric charge $\mfs$, they are given explicitly by 
\be \label{uinfirst}
     {{\mfu}}^{\rm in}_{+, k}  =  \begin{bmatrix*}[l]\begin{pmatrix}e^{-i k r}\\e^{i k (r- 2\epsilon)} \left(\frac{\epsilon}{r}\right)^{i e Q} \mathcal{T}^{++}_{\ins}\end{pmatrix}\\[4mm]\begin{pmatrix}0\\ e^{i (k- e v)r}e^{- i (2 k  - e v)\epsilon}\,e^{- i \mfa}\cT^{-+}_{\ins}\end{pmatrix}\end{bmatrix*}{ e^{- i (k-\frac{e v }{2}) t }\left(\frac{r}{r_0}\right)^{i e Q/2}}
\ee
and
\be
     {{\mfu}}^{\rm in}_{-, k}  =  \begin{bmatrix*}[l]\begin{pmatrix}0\\e^{i (k+ e v) r} e^{- i (2k+ e v) \epsilon}\,e^{i  \mfa}\cT^{+-}_{\ins}\end{pmatrix}\\[4mm]\begin{pmatrix} e^{-i k r}\\e^{i  k (r-2\epsilon)}\,\left(\frac{r}{\epsilon}\right)^{ i  e Q}\,\mathcal{T}^{--}_{\ins}\end{pmatrix}\end{bmatrix*}{ e^{- i (k+\frac{e v}{2}) t }\left(\frac{r}{r_0}\right)^{-ie Q/2} \,}\,.
\ee 
where 
$k = \omega + \frac12 \mfs ev \geq 0$. (These can be obtained from \pref{1+1d,ML:kappa+-sol} and \pref{1+1d,ML:kappa-+sol} by choosing $\bm{\kappa}_{++}=1, \bm{\kappa}_{--}=0$ for $\mfu^{\ins}_{+, k}$ and $\bm{\kappa}_{--}=1, \bm{\kappa}_{++}=0$ for $\mfu^{\ins}_{-, k}$.) The negative-frequency in-modes are similarly defined by
\be 
     {{\mfv}}^{\rm in}_{+, k} = \begin{bmatrix*}[l]\begin{pmatrix}e^{i k r}\\e^{-i k (r- 2\epsilon)} \left(\frac{\epsilon}{r}\right)^{i e Q} \mathcal{T}^{++}_{\ins}\end{pmatrix}\\[4mm]\begin{pmatrix} 0\\ e^{-i (k+ e v)r}e^{ i(2 k+ e v)\epsilon}\,e^{- i \mfa}\cT^{-+}_{\ins}\end{pmatrix}\end{bmatrix*}{ e^{ i (k+\frac{ e v}{2}) t }\left(\frac{r}{r_0}\right)^{i e Q/2}} 
\ee
and
\be \label{vinlast}
   {{\mfv}}^{\rm in}_{-,  k} = \begin{bmatrix*}[l]\begin{pmatrix}0\\e^{-i (k- e v) r} e^{ i (2 k - e v) \epsilon} \,e^{ i \mfa}\cT^{+-}_{\ins}\end{pmatrix}\\[4mm]\begin{pmatrix} e^{i k r}
    \\e^{-i k (r-2\epsilon)}\,\left(\frac{r}{\epsilon}\right)^{ i e Q}\mathcal{T}^{--}_{\ins}\end{pmatrix}\end{bmatrix*}{ e^{ i (k - \frac{e v}{2}) t }\left(\frac{r}{r_0}\right)^{-i e Q/2}},
\ee
where $k = \omega - \frac12 \mfs ev \geq 0$. In both of these expressions $r_0$ is an arbitrary length that contributes only to the overall Coulomb phase.

The coefficients $\mathcal{T}^{++}_{\ins}$, $\mathcal{T}^{+-}_{\ins}$, $\mathcal{T}^{-+}_{\ins}$ and $\mathcal{T}^{--}_{\ins}$ appearing in these expressions are $\mfa$-independent quantities given as explicit functions of the fermion-dyon couplings by 
\be 
      \mathcal{T}^{++}_{\rm in}=-\frac{\left|\begin{array}{ll}
\hat{\mathcal{B}}_{21}& \hat{\mathcal{B}}_{24} \\
\hat{\mathcal{B}}_{41}\, & \hat{\mathcal{B}}_{44}
\end{array}\right|}{\left|\begin{array}{ll}
\hat{\mathcal{B}}_{22} & \hat{\mathcal{B}}_{24} \\
\hat{\mathcal{B}}_{42} & \hat{\mathcal{B}}_{44}
\end{array}\right|}, \;
     \cT^{+-}_{\rm in}= -\frac{\left|\begin{array}{ll}
\hat{\mathcal{B}}_{23}& \hat{\mathcal{B}}_{24} \\
\hat{\mathcal{B}}_{43}\, & \hat{\mathcal{B}}_{44}
\end{array}\right|}{\left|\begin{array}{ll}
\hat{\mathcal{B}}_{22} & \hat{\mathcal{B}}_{24} \\
\hat{\mathcal{B}}_{42} & \hat{\mathcal{B}}_{44}
\end{array}\right|},  \;
     \cT^{-+}_{\rm in}= - \frac{\left|\begin{array}{ll}
 \hat{\mathcal{B}}_{22}  & \hat{\mathcal{B}}_{21} \\
\hat{\mathcal{B}}_{42} & \hat{\mathcal{B}}_{41}
\end{array}\right|}{\left|\begin{array}{ll}
\hat{\mathcal{B}}_{22} & \hat{\mathcal{B}}_{24} \\
\hat{\mathcal{B}}_{42} & \hat{\mathcal{B}}_{44}
\end{array}\right|}, \;
     \mathcal{T}^{- -}_{\rm in}=- \frac{\left|\begin{array}{ll}
 \hat{\mathcal{B}}_{22}  & \hat{\mathcal{B}}_{23} \\
\hat{\mathcal{B}}_{42} & \hat{\mathcal{B}}_{43}
\end{array}\right|}{\left|\begin{array}{ll}
\hat{\mathcal{B}}_{22} & \hat{\mathcal{B}}_{24} \\
\hat{\mathcal{B}}_{42} & \hat{\mathcal{B}}_{44}
\end{array}\right|} \,.
    \ee
 Written directly in terms of  the boundary couplings, these become
\bea \label{TinvsC}
    \mathcal{T}^{++}_{\ins}&=&\frac{(\mathcal{C}^s_{++}+i \mathcal{C}^{ps}_{++})(i-\mathcal{C}^{pv}_{--}+ \mathcal{C}^{v}_{--})+(\mathcal{C}^s_{-+}+i \mathcal{C}^{ps}_{-+})(\mathcal{C}^{pv}_{+-}- \mathcal{C}^{v}_{+-})}{-|\mathcal{C}^{pv}_{+-}-\mathcal{C}^{v}_{+-}|^2+(-i +\mathcal{C}^{pv}_{++}-\mathcal{C}^{v}_{++})(\mathcal{C}^{pv}_{--}-\mathcal{C}^{v}_{--}-i)}, \nonumber\\ 
     \mathcal{T}^{--}_{\ins}&=&\frac{(\mathcal{C}^s_{--}+i \mathcal{C}^{ps}_{--})(i-\mathcal{C}^{pv}_{++}+ \mathcal{C}^{v}_{++})+(\mathcal{C}^s_{+-}+i \mathcal{C}^{ps}_{+-})(\mathcal{C}^{pv}_{-+}- \mathcal{C}^{v}_{-+})}{-|\mathcal{C}^{pv}_{+-}-\mathcal{C}^{v}_{+-}|^2+(-i +\mathcal{C}^{pv}_{++}-\mathcal{C}^{v}_{++})(\mathcal{C}^{pv}_{--}-\mathcal{C}^{v}_{--}-i)}, \\ 
     \cT^{-+}_{\ins}&=&\frac{(\mathcal{C}^s_{-+}+i \mathcal{C}^{ps}_{-+})(i-\mathcal{C}^{pv}_{++}+ \mathcal{C}^{v}_{++})+(\mathcal{C}^s_{++}+i \mathcal{C}^{ps}_{++})(\mathcal{C}^{pv}_{-+}- \mathcal{C}^{v}_{-+})}{-|\mathcal{C}^{pv}_{+-}-\mathcal{C}^{v}_{+-}|^2+(-i +\mathcal{C}^{pv}_{++}-\mathcal{C}^{v}_{++})(\mathcal{C}^{pv}_{--}-\mathcal{C}^{v}_{--}-i)}, \nonumber\\ 
    \cT^{+-}_{\ins}&=&\frac{(\mathcal{C}^s_{+-}+i \mathcal{C}^{ps}_{+-})(i-\mathcal{C}^{pv}_{--}+ \mathcal{C}^{v}_{--})+(\mathcal{C}^s_{--}+i \mathcal{C}^{ps}_{--})(\mathcal{C}^{pv}_{+-}- \mathcal{C}^{v}_{+-})}{-|\mathcal{C}^{pv}_{+-}-\mathcal{C}^{v}_{+-}|^2+(-i +\mathcal{C}^{pv}_{++}-\mathcal{C}^{v}_{++})(\mathcal{C}^{pv}_{--}-\mathcal{C}^{v}_{--}-i)}.\nn
\eea

When the boundary action $S_{\rm dyon2}$ is real (and so the $\cC^\ssA_{ij}$ are hermitian) these satisfy the following unitarity conditions
\be \label{unitarityT1}
    |\mathcal{T}^{+ -}_{\ins}|^2 =  |\mathcal{T}^{-+}_{\ins}|^2=1- |\mathcal{T}^{++}_{\ins}|^2=1- |\mathcal{T}^{--}_{\ins}|^2 \,,
\ee
as well as
  \be \label{unitarityT}
  \mathcal{T}^{--}_{\ins}\cT^{-+\,*}_{\ins}+\cT^{+-}_{\ins}\mathcal{T}^{++\,*}_{\ins}=0,
\ee
as identities (see Appendix \ref{App:T amplitudes} for a derivation). These relations imply that the \textit{in} amplitudes only carry 4 real parameters' worth of information; they can always be written
\be  \label{Tinparam1}
    \mathcal{T}^{++}_{\ins}=\rho \,e^{i\theta_{++}} \quad\text{and} \quad\mathcal{T}^{--}_{\ins}=\rho\, e^{i\theta_{--}},
\ee
and  
\be \label{Tinparam2}
    {\mathcal{T}}^{+-}_{\ins}=\sqrt{1-\rho^2}\, e^{i\theta_{+-}}\quad  \text{and}  \quad {\mathcal{T}}^{-+}_{\ins}=-\sqrt{1-\rho^2}  \,e^{i(\theta_{++}+ \theta_{--}-\theta_{+-})},
\ee
where $\rho$, $\theta_{++}$, $\theta_{+-}$ and $\theta_{--}$ are the four independent real parameters.  
 
\subsubsection{Out modes}

A set of outgoing modes can be constructed in precisely the same way, with the positive frequency modes given by
\be \label{uoutfirst}
     {{\mfu}}^{\outs}_{+, k}=\begin{bmatrix*}[l]\begin{pmatrix}e^{-i k (r- 2\epsilon)}\left(\frac{r}{\epsilon}\right)^{ i e Q} \mathcal{T}^{++}_{\outs}\\e^{i k r}\end{pmatrix}\\[4mm] \begin{pmatrix}e^{-i (k-e v)r}e^{i (2 k - e v)\epsilon}\,e^{-i\mfa}\cT^{-+}_{\outs}\\0\end{pmatrix}\end{bmatrix*}{ e^{ -i (k -\frac{e v}{2}) t }\left(\frac{r}{r_0}\right)^{-i e Q/2} }, 
\ee
and 
\be
  {{\mfu}}^{\outs}_{-, k}=\begin{bmatrix*}[l]\begin{pmatrix}e^{-i (k+ e v)r}e^{i (2 k + e v) \epsilon}\,e^{ i \mfa} \cT^{+-}_{\outs}\\0\end{pmatrix}\\[4mm] \begin{pmatrix}e^{-i k (r- 2\epsilon)}\left(\frac{\epsilon}{r}\right)^{i e Q}\,\mathcal{T}^{--}_{\outs}\\e^{i k r}\end{pmatrix}\end{bmatrix*}{ e^{ -i (k+\frac{e v}{2}) t }\left(\frac{r}{r_0}\right)^{i e Q/2}},
\ee
 with 
 $k = \omega + \frac12 \mfs ev \geq 0$ (these can be obtained from \pref{1+1d,ML:kappa++sol} and \pref{1+1d,ML:kappa--sol} by choosing $\bm{\kappa}_{+-}=1$, $ \bm{\kappa}_{-+}=0$ for $\mfu^{\outs}_{+, k}$ and $\bm{\kappa}_{-+}=1, \bm{\kappa}_{+-}=0$ for $\mfu^{\outs}_{-, k}$). The negative-frequency out-modes are
\be 
 {{\mfv}}^{\outs}_{+, k}=\begin{bmatrix*}[l]\begin{pmatrix}e^{i k(r- 2\epsilon)}\left(\frac{r}{\epsilon}\right)^{ i e Q}\, \mathcal{T}^{++}_{\outs}\\e^{-i k r}\end{pmatrix}\\[4mm] \begin{pmatrix}e^{i (k+ e v)r}e^{- i (2 k + e v)\epsilon}\,e^{-i \mfa}\cT^{-+}_{\outs}\\ 0\end{pmatrix}\end{bmatrix*}{ e^{i (k+\frac{e  v }{2}) t }\left(\frac{r}{r_0}\right)^{-i e Q/2} },
\ee
and
\be \label{voutlast}
    {{\mfv}}^{\outs}_{-, k}=\begin{bmatrix*}[l]\begin{pmatrix}e^{i (k- e v)r}e^{-i(2 k - e v) \epsilon}\, e^{ i \mfa}\cT^{+-}_{\outs}\\0\end{pmatrix}\\[4mm]\begin{pmatrix}e^{i k (r- 2\epsilon)}\left(\frac{\epsilon}{r}\right)^{i e Q}\mathcal{T}^{--}_{\outs}\\e^{-i k r}\\\end{pmatrix}\end{bmatrix*}{ e^{ i (k- \frac{ e v}{2}) t }\,\left(\frac{r}{r_0}\right)^{i e Q/2}},
\ee
for which $k = \omega - \frac12 \mfs ev \geq 0$. 

The $\mfa$-independent coefficients $\mathcal{T}^{++}_{\outs}, \mathcal{T}^{+-}_{\outs}, \mathcal{T}^{-+}_{\outs}, \mathcal{T}^{--}_{\outs}$ appearing in these expressions are defined in terms of the dyon-fermion couplings by
\be 
     \mathcal{T}^{++}_{\outs}= -\frac{\left|\begin{array}{ll}
\hat{\mathcal{B}}_{12}& \hat{\mathcal{B}}_{13} \\
\hat{\mathcal{B}}_{32}\, & \hat{\mathcal{B}}_{33}
\end{array}\right|}{\left|\begin{array}{ll}
\hat{\mathcal{B}}_{11} & \hat{\mathcal{B}}_{13} \\
\hat{\mathcal{B}}_{31} & \hat{\mathcal{B}}_{33}
\end{array}\right|}, \;
\cT^{- +}_{\outs}= -\frac{\left|\begin{array}{ll}
\hat{\mathcal{B}}_{11}& \hat{\mathcal{B}}_{12} \\
\hat{\mathcal{B}}_{31}\, & \hat{\mathcal{B}}_{32}
\end{array}\right|}{\left|\begin{array}{ll}
\hat{\mathcal{B}}_{11} & \hat{\mathcal{B}}_{13} \\
\hat{\mathcal{B}}_{31} & \hat{\mathcal{B}}_{33}  
\end{array}\right|}, \;
\cT^{+-}_{\outs}= - \frac{\left|\begin{array}{ll}
 \hat{\mathcal{B}}_{14}  & \hat{\mathcal{B}}_{13} \\
\hat{\mathcal{B}}_{34} & \hat{\mathcal{B}}_{33}
\end{array}\right|}{\left|\begin{array}{ll}
\hat{\mathcal{B}}_{11} & \hat{\mathcal{B}}_{13} \\
\hat{\mathcal{B}}_{31} & \hat{\mathcal{B}}_{33}
\end{array}\right|}, \;
\mathcal{T}^{- -}_{\outs}= - \frac{\left|\begin{array}{ll}
 \hat{\mathcal{B}}_{11}  & \hat{\mathcal{B}}_{14} \\
\hat{\mathcal{B}}_{31} & \hat{\mathcal{B}}_{34}
\end{array}\right|}{\left|\begin{array}{ll}
\hat{\mathcal{B}}_{11} & \hat{\mathcal{B}}_{13} \\
\hat{\mathcal{B}}_{31} & \hat{\mathcal{B}}_{33}
\end{array}\right|} ,
    \ee
and so
\bea \label{ToutvsC}
    \mathcal{T}^{++}_{\outs}&=&\frac{i(\mathcal{C}^{ps}_{++}+i \mathcal{C}^{s}_{++})(-i+\mathcal{C}^{pv}_{--}+ \mathcal{C}^{v}_{--})+(-i \mathcal{C}^{ps}_{-+}+ \mathcal{C}^{s}_{-+})(\mathcal{C}^{pv}_{+-}+ \mathcal{C}^{v}_{+-})}{|\mathcal{C}^{pv}_{+-}+\mathcal{C}^{v}_{+-}|^2-(-i +\mathcal{C}^{pv}_{--}+\mathcal{C}^{v}_{--})(\mathcal{C}^{pv}_{++}+\mathcal{C}^{v}_{++}-i)}, \nonumber\\ 
     \mathcal{T}^{--}_{\outs}&=&\frac{i(\mathcal{C}^{ps}_{--}+i \mathcal{C}^{s}_{--})(-i+\mathcal{C}^{pv}_{++}+ \mathcal{C}^{v}_{++})+(-i \mathcal{C}^{ps}_{+-}+ \mathcal{C}^{s}_{+-})(\mathcal{C}^{pv}_{-+}+ \mathcal{C}^{v}_{-+})}{|\mathcal{C}^{pv}_{+-}+\mathcal{C}^{v}_{+-}|^2-(-i +\mathcal{C}^{pv}_{--}+\mathcal{C}^{v}_{--})(\mathcal{C}^{pv}_{++}+\mathcal{C}^{v}_{++}-i)}, \\ 
     \cT^{-+}_{\outs}&=&\frac{i(\mathcal{C}^{ps}_{-+}+i \mathcal{C}^{s}_{-+})(-i+\mathcal{C}^{pv}_{++}+ \mathcal{C}^{v}_{++})+(-i \mathcal{C}^{ps}_{++}+ \mathcal{C}^{s}_{++})(\mathcal{C}^{pv}_{-+}+ \mathcal{C}^{v}_{-+})}{|\mathcal{C}^{pv}_{+-}+\mathcal{C}^{v}_{+-}|^2-(-i +\mathcal{C}^{pv}_{--}+\mathcal{C}^{v}_{--})(\mathcal{C}^{pv}_{++}+\mathcal{C}^{v}_{++}-i)}, \nonumber\\ 
    \cT^{+-}_{\outs}&=&\frac{i(\mathcal{C}^{ps}_{+-}+i \mathcal{C}^{s}_{+-})(-i+\mathcal{C}^{pv}_{--}+ \mathcal{C}^{v}_{--})+(-i \mathcal{C}^{ps}_{--}+ \mathcal{C}^{s}_{--})(\mathcal{C}^{pv}_{+-}+ \mathcal{C}^{v}_{+-})}{|\mathcal{C}^{pv}_{+-}+\mathcal{C}^{v}_{+-}|^2-(-i +\mathcal{C}^{pv}_{--}+\mathcal{C}^{v}_{--})(\mathcal{C}^{pv}_{++}+\mathcal{C}^{v}_{++}-i)}. \nn
\eea
   
When the boundary action $S_{\rm dyon2}$ is real (so the $\cC^\ssA_{ij}$'s are hermitian) these satisfy
\be \label{unitarityTout1}
    |\mathcal{T}^{+ -}_{\outs}|^2 =  |\mathcal{T}^{-+}_{\outs}|^2=1- |\mathcal{T}^{++}_{\outs}|^2=1- |\mathcal{T}^{--}_{\outs}|^2 \,.
\ee
and
\be \label{unitarityTout2}
\mathcal{T}^{--}_{\outs}\cT^{-+\,*}_{\outs}+\cT^{+-}_{\outs}\mathcal{T}^{++\,*}_{\outs}=0 ,
\ee
in the same way as found earlier for $\cT_{\ins}$. There is of course a good reason why $\cT_{\outs}$ and $\cT_{\ins}$ satisfy similar conditions: they are not independent of one another. Since both the \textit{in} and \textit{out} bases are complete, the $\mathcal{T}_{\outs}$ amplitudes can be expressed in terms of the $\mathcal{T}_{\ins}$ amplitudes. This leads to the relations (see Appendix \ref{App:T amplitudes})
    \be 
        \mathcal{T}^{++}_{\outs}=(\mathcal{T}^{++}_{\ins})^*, \quad \mathcal{T}^{--}_{\outs}=(\mathcal{T}^{--}_{\ins})^*, \quad \mathcal{T}^{+-}_{\outs}=(\mathcal{T}^{-+}_{\ins})^* \quad  \text{and}\quad\mathcal{T}^{-+}_{\outs}=(\mathcal{T}^{+-}_{\ins})^* \,,
    \ee
and so the \textit{out} amplitudes also depend only on the four parameters given in \pref{Tinparam1} and \pref{Tinparam2}.

The implications of the $\cT_{\ins}$ and $\cT_{\outs}$ amplitudes for fermion-dyon scattering problems are explored in some detail in \S\ref{sec:PerturbativeScattering} and \S\ref{sec:SO3Dyon} below and suffice it to say that they carry  all of the information about the dyon that is relevant to describing transitions amongst the $S$-wave fermion modes described above, at least at leading order at low energies. See \S\ref{ssec:Matching} for a discussion of the values predicted for these quantities by specific microscopic choices for the underlying dyon solution. 

But the above discussion introduces a minor puzzle: why are there only four free real parameters within, say, the $\cT_{\ins}$'s when there are 16 possible  real parameters in the hermitian effective couplings $\cC_{ij}^\ssA$ appearing in $S_{\rm dyon2}$ in {\it e.g.}~eqs.~\pref{dyonEFT2} and \pref{PPEFTList0RF}? We argue in \S\ref{ssec:RedundantInteractions} that all but four of the effective couplings $\cC_{ij}^\ssA$ are redundant (in the precise EFT sense \cite{EFTBook}), but before doing so the next section first develops a required tool. Along the way it also resolves a dangling technical issue: how to handle systematically the $\mfa$-dependence that is embedded in boundary conditions like \pref{Matrix2DBC} or \pref{system of boundary eqs2D}.

\subsection{Dyonic response}
\label{ssec:AlphaPartyLine}

Up to this point we have treated the matrix $\cB$ as if it were specified completely once the effective couplings $\cC^\ssA_{ij}$ are, but the appearance of the field $\mfa$ within $\cB$ makes this not quite true. Having $\cB$ be a function of $\mfa$ complicates its use -- as in \pref{system of boundary eqs2D} -- to find fermionic mode functions. This section provides two complementary ways to handle this field-dependence: perturbation theory and the Born-Oppenheimer approximation.

\subsubsection{Fermion-dyon perturbation theory}
\label{sssec:Fermion-DyonPtbnThy}

The first approach starts with the observation that not {\it every} term in the matrix $\cB$ depends on $\mfa$. If the coefficients $\cC^\ssA_{\mfs\mfs'}$ with $\mfs \neq \mfs'$ were for some reason much smaller than the others then they could be ignored when formulating the near-dyon fermion boundary conditions with the effects of $\mfa$-dependent terms then included perturbatively. 

For instance, suppose $S_{\rm dyon2}$ of \pref{dyonEFT22D} could be written $S_{\rm dyon2}^{(0)} + S_{\rm dyon2}^{\rm int}$ with\footnote{We choose the unperturbed boundary terms here fairly arbitrarily and all that is important for our arguments is that the dominant piece be $\mfa$-independent.}
\be \label{dyonEFT22Dfree}
     S_{\rm dyon2}^{(0)} :=-\frac12 \sum_{\mfs=\pm} \int \exd t\, i\overline{\chi}_{\mfs}   \Gamma_c \,\chi_{\mfs}   \,,
\ee
and
\be \label{dyonEFT22Dpert}
     S_{\rm dyon2}^{\rm int} =-\frac12 \sum_{\mfs,\mfs'=\pm} \int \exd t\, \overline{\chi}_{\mfs} \Bigl(\delta\mathcal{C}^s_{\mfs\mfs'}+i \delta\mathcal{C}^{ps}_{\mfs\mfs'} \Gamma_c-i \delta\mathcal{C}^{v}_{\mfs\mfs'}\Gamma^{0}+i\delta \mathcal{C}^{pv}_{\mfs\mfs'} \Gamma^{1}\Bigr)\,\chi_{\mfs'}  \; e^{\frac{i}{2} (\mfs-\mfs')\, \mfa} \,, 
\ee
where  the coefficients $\delta \cC^\ssA_{ij}$ are regarded as being perturbatively small\footnote{Although, as noted above, the $\cC^\ssA_{ij}$'s are not generically suppressed by the loop-counting parameter $\alpha = e^2/(4\pi)$ it is possible that in some circumstances some of them are suppressed by another small quantity.} and we again specialize to the case of an approximately static dyon (for which $\dot y^\alpha \simeq \delta^\alpha_0$). In this case only $S_{\rm dyon2}^{(0)}$ need play a role in determining the boundary conditions of the free modes appearing in the field expansions within the interaction picture.

With this choice the boundary condition matrix of \pref{OBdef} becomes $O_{\mathcal{B}}^{(0)}  =  i I \otimes \Gamma_c$ and so the matrix appearing in \pref{BofaDef} becomes
\be  \label{B0def}
   \mathcal{B}^{(0)}= \Gamma^0\left[\Gamma^1 + O_\cB^{(0)}\right] = \begin{pmatrix}
        1 &-1 & 0 & 0 \\[2mm] 1 &  -1  &  0 & 0 \\[2mm] 0 & 0 & 1 & -1 \\[2mm] 0 & 0 & 1 & -1   \end{pmatrix} \,. 
\ee
which clearly has rank two. The boundary condition satisfied by the free mode functions at $r=\epsilon$ therefore is 
\be \label{Ptbn0BC}
  \Bigl( \Gamma^1+ i \Gamma_c \Bigr) \bm{\chi}(\epsilon, t)=0 \,.
\ee

The continuum normalized free \textit{in} and \textit{out} mode functions that satisfy \pref{Ptbn0BC} then are
{\scriptsize
\be \label{Ptuin0}
     {\mfu}^{\ins 0}_{+,  k }=\begin{bmatrix*}[l]\begin{pmatrix}e^{-i k r}\\e^{i k (r- 2\epsilon)} \left(\frac{\epsilon}{r}\right)^{i e Q} \end{pmatrix}\\[5mm]\hspace{5mm}\begin{pmatrix}\hspace{5mm}0\hspace{5mm}\\ \hspace{5mm}0\hspace{5mm}\end{pmatrix}\end{bmatrix*}{ e^{- i (k - \frac{ e v}{2}) t }\left(\frac{r}{r_0}\right)^{i e Q/2}}, \quad
    {\mfu}^{\ins 0}_{-, k}=\begin{bmatrix*}[l]\hspace{5mm}\begin{pmatrix}\hspace{5mm}0\hspace{5mm}\\ \hspace{5mm}0\hspace{5mm}\end{pmatrix}\\[5mm]\begin{pmatrix} e^{-i k r}\\e^{i k (r-2\epsilon)}\,\left(\frac{r}{\epsilon}\right)^{ i e Q}\end{pmatrix}\end{bmatrix*}{ e^{- i (k+ \frac{ e v}{2}) t }\left(\frac{r}{r_0}\right)^{-ie Q/2} \,},
\ee
\be \label{Ptvin0}
    {\mfv}^{\ins 0}_{+, k}=\begin{bmatrix*}[l]\begin{pmatrix}e^{i k  r}\\e^{-i k (r- 2\epsilon)} \left(\frac{\epsilon}{r}\right)^{i e Q} \end{pmatrix}\\[5mm]\hspace{5mm}\begin{pmatrix}\hspace{5mm}0\hspace{5mm}\\ \hspace{5mm}0\hspace{5mm}\end{pmatrix}\end{bmatrix*}{ e^{ i (k + \frac{ e v}{2}) t }\left(\frac{r}{r_0}\right)^{i e Q/2}} , \quad
    {\mfv}^{\ins 0}_{-, k }=\begin{bmatrix*}[l]\hspace{5mm}\begin{pmatrix}\hspace{5mm}0\hspace{5mm}\\ \hspace{5mm}0\hspace{5mm}\end{pmatrix}\\[5mm]\begin{pmatrix} e^{i k r}
    \\e^{-i k (r-2\epsilon)}\,\left(\frac{r}{\epsilon}\right)^{ i e Q}\end{pmatrix}\end{bmatrix*}{ e^{ i (k - \frac{e v }{2}) t }\left(\frac{r}{r_0}\right)^{-i e Q/2}} ,
\ee 
}
(where $k \geq 0$) and
{\scriptsize 
\be \label{Ptuout0}
    {\mfu}^{\outs 0}_{+, k}=\begin{bmatrix*}[l]\begin{pmatrix}e^{-i k(r- 2\epsilon)}\left(\frac{r}{\epsilon}\right)^{ i e Q} \\e^{i k r}\end{pmatrix}\\[5mm]\hspace{5mm}\begin{pmatrix}\hspace{5mm}0\hspace{5mm}\\ \hspace{5mm}0\hspace{5mm}\end{pmatrix}\end{bmatrix*}{ e^{ -i (k- \frac{ e v}{2}) t }\left(\frac{r}{r_0}\right)^{-ie  Q/2} }, \quad
    {\mfu}^{\outs 0}_{-, k}=\begin{bmatrix*}[l]\hspace{5mm}\begin{pmatrix}\hspace{5mm}0\hspace{5mm}\\ \hspace{5mm}0\hspace{5mm}\end{pmatrix}\\[5mm] \begin{pmatrix}e^{-i k(r- 2\epsilon)}\left(\frac{\epsilon}{r}\right)^{i e Q}\\e^{i k r}\end{pmatrix}\end{bmatrix*}{ e^{ -i (k +\frac{ e v}{2}) t }\left(\frac{r}{r_0}\right)^{i e Q/2}}, 
\ee
\be\label{Ptvout0}
 {\mfv}^{\outs 0}_{+,  k}=\begin{bmatrix*}[l]\begin{pmatrix}e^{i k(r- 2\epsilon)}\left(\frac{r}{\epsilon}\right)^{ i e Q}\\e^{-i k r}\end{pmatrix}\\[5mm] \hspace{5mm}\begin{pmatrix}\hspace{5mm}0\hspace{5mm}\\ \hspace{5mm}0\hspace{5mm}\end{pmatrix}\end{bmatrix*}{ e^{i (k+\frac{ e v}{2}) t }\left(\frac{r}{r_0}\right)^{-i e Q/2} },\quad
  {\mfv}^{\outs 0}_{-, k}=\begin{bmatrix*}[l]\hspace{5mm}\begin{pmatrix}\hspace{5mm}0\hspace{5mm}\\ \hspace{5mm}0\hspace{5mm}\end{pmatrix}\\[5mm]\begin{pmatrix}e^{i k (r- 2\epsilon)}\left(\frac{\epsilon}{r}\right)^{i e Q}\\e^{-i kr}\\\end{pmatrix}\end{bmatrix*}{ e^{ i (k - \frac{ e v}{2}) t }\,\left(\frac{r}{r_0}\right)^{i e Q/2}},
\ee
}
and so the unperturbed modes satisfy $\cT_{\ins}^{++} = \cT_\ins^{--} = 1$ and $\cT_{\ins}^{+-} = \cT_{\ins}^{-+} = 0$. These modes are normalized so that
\be 
     \int^{\infty}_{\epsilon} \mathrm{d}r \,({\mfu}^{\ins 0}_{\mathfrak{s}, k})^{\dagger}\,{\mfu}^{\ins 0}_{\mathfrak{s}', k'} =  \int^{\infty}_{\epsilon} \mathrm{d}r \,({\mfv}^{\ins 0}_{\mathfrak{s}, k})^{\dagger}\,{\mfv}^{\ins 0}_{\mathfrak{s}', k'} = 2\pi \delta(k - k')\,\delta_{\mathfrak{s} \mathfrak{s}'}, 
\ee
and
\be
   \int^{\infty}_{\epsilon} \mathrm{d}r \,({\mfu}^{\ins 0}_{\mathfrak{s}, k})^{\dagger}\,{\mfv}^{\ins 0}_{\mathfrak{s}', k'}=  2\pi \delta(k + k')\,\delta_{\mathfrak{s} \mathfrak{s'}},
\ee
and similarly for the \textit{out} modes. 

From here on perturbation theory proceeds in the standard way by moving to the interaction picture and expanding the fermion field operator using these mode functions,
\bea 
    \bm{\chi}(x) &=& \sum_{\mfs=\pm}\int^{\infty}_{0} \frac{\mathrm{d}k}{\sqrt{2\pi}}\Bigl[ \mfu^{\ins 0}_{\mfs,  k }(x)\, {a}^{\ins}_{\mfs, k}+\mfv^{\ins 0}_{\mfs,  k}(x)\, (\overline{a}^{\ins}_{\mfs, k})^{\star}\Bigr] \nn\\
    &=& \sum_{\mfs=\pm}\int^{\infty}_{0} \frac{\mathrm{d}k}{\sqrt{2\pi}}\Bigl[ \mfu^{\outs 0}_{\mfs,  k }(x)\, {a}^{\outs}_{\mfs, k}+\mfv^{\outs 0}_{\mfs,  k}(x)\, (\overline{a}^{\outs}_{\mfs, k})^{\star}\Bigr] ,
\eea
where the creation/annihilation operators satisfy the usual algebra $\{a^{\ins}_{\mathfrak{s}, k}, \,(a^{\ins}_{\mathfrak{s}', k'})^{\star}\} = \delta(k - k')\,\delta_{\mathfrak{s} \mathfrak{s}'}$ and $\{\overline{a}^{\ins}_{\mathfrak{s}, k}, \,(\overline{a}^{\ins}_{\mathfrak{s}', k'})^{\star}\}=  \delta(k - k')\,\delta_{\mathfrak{s} \mathfrak{s}'}$ appropriate for fermions (and similarly for the \textit{out} operators). Because both the \textit{in} and \textit{out} modes are complete, each can be expanded in terms of the other, allowing the \textit{in} and \textit{out} operators to be related by
\be 
     a^{\outs}_{\mfs, k}= e^{-2i k \epsilon}\left(\frac{\epsilon}{r_0}\right)^{ i \mfs e Q}a^{\ins}_{\mfs, k }\quad \text{and} \quad (\ol{a}^{\outs}_{\mfs, k})^{\star}= e^{2 i k\epsilon}\left(\frac{\epsilon}{r_0}\right)^{ i \mfs e Q }\, (\ol{a}^{\ins}_{\mfs, k})^{\star}, 
\ee
and their adjoints. The above Bogoliubov relations depend on the arbitrary scale $r_0$. Physical predictions should not depend on $r_0$ and in \S \ref{sec:PerturbativeScattering} -- which uses the perturbative framework set up here -- we rephase the $out$ operators to absorb the $r_0$ dependence. 

The perturbative formalism is useful for some kinds of questions -- such as the discussion of redundancy in \S\ref{ssec:RedundantInteractions} --  and is used in \S\ref{sec:PerturbativeScattering} to compute some scattering processes (which can be compared with calculations performed in \S\ref{sec:SO3Dyon} using the alternative approach we describe next). But it is not guaranteed that specific microscopic realizations of the dyon must yield effective couplings for which such a perturbative approach is a good approximation. This is true in particular for massless fermion scattering in the classical nonabelian dyon background because in this case conservation of chirality implies the vanishing of all of the $\mfa$-independent boundary couplings $\cC^\ssA_{\mfs\mfs'}$ with $\mfs'=\mfs$.
 
\subsubsection{Born-Oppenheimer evolution of $\mfa$}
\label{sssec:BOIntro}

A more ambitious approach to computing the fermion-$\mfa$ interactions takes advantage of the fact that the mass $\cI  \sim (\alpha m_g)^{-1}$ appearing in the $\mfa$ kinetic term is much larger than the generic dyon scale $m_g^{-1}$ in the semiclassical limit (for which $\alpha = e^2/4\pi \ll 1$). This means that the energy associated with $\mfa$ excitation is order $\cI^{-1} \sim \alpha m_g$ and so is negligible -- even within the low energy theory -- for fermion momenta in the range $\alpha m_g \ll k \ll m_g$. 

Furthermore, the equations of motion say that $\dot\mfa \propto \mfp/\cI$ and so the timescale for $\mfa$ to respond to order-unity changes in $\mfp$ are order $\tau \sim \cI$. By contrast the timescale for a relativistic fermionic wave packet of size $L$ to interact with a much smaller dyon is order $L$ (which cannot be smaller than $1/k$ for fermions with momentum $k$). So the rotor response is slow compared to the fermion interaction time for fermion momenta in the range $\alpha m_g \ll L^{-1} \lsim k$.

In the regime $\alpha m_g \ll k$ the `rotor' field response to fermions is slow and costs little energy in much the same way that the position of heavy atomic nuclei respond to much lighter and faster electrons in everyday solids. This electron/nucleus analogy suggests using the Born-Oppenheimer approximation \cite{BornOppenheimer} to describe the interactions of relativistic fermions with the slower dyonic excitation $\mfa$. In this approximation one first solves for the motion of the fast degrees of freedom (the light fermions) with the slow degrees of freedom (the nuclear positions $\bfR$ or the field $\mfa$) imagined to be fixed classical variables. The idea is that the slow variables behave classically because they do not have time to respond at all when hit by the fast ones. Once the fast evolution is computed one calculates an effective Hamiltonian describing the dynamics of the slow degrees of freedom obtained by averaging over the fast degrees of freedom. This Hamiltonian is then used to solve for how the slower variables evolve in response to its interactions with the much faster system.

As applied to dyon-fermion interactions the first part of the Born-Oppenheimer procedure asks for a solution to fermion evolution with the field $\mfa$ regarded as a fixed classical background. This is precisely what is done in \S\ref{ssec:ScatteringState} above when finding 
the fermion scattering states in the presence of an $\mfa$-dependent boundary condition like \pref{system of boundary eqs2D}. These are used in \S\ref{sec:SO3Dyon} to calculate 
scattering rates also for fixed dyon configurations. We return to the issue of how to determine 
the dyon reponse in \S\ref{ssec:RotorBO} below, but first close this section by gathering together several loose ends and conceptual issues to do with using the PPEFT framework in the monopole/dyon setting. 

\subsection{Redundant interactions}
\label{ssec:RedundantInteractions}

In this section we return to the puzzle alluded to at the end of \S\ref{ssec:ScatteringState} above: why are the physical effects of the 16 real parameters in the couplings $\cC^\ssA_{ij}$ in the action $S_{\rm dyon2}$ all encoded in amplitudes $\cT_{\ins}$ (or $\cT_{\outs}$) that involve only 4 parameters. Why are there not more scattering options available given the number of possible effective interactions? Part of the reason for this reduction is the rank-2 condition we assume for $\cB(\mfa)$, since -- as shown in Appendix \ref{Appssec:Rank2} -- this imposes 8 real conditions on the components of  $\cB(\mfa)$ (and so also on the $\cC^\ssA_{ij}$'s). But the fact that the remaining 8 independent effective couplings only produce a 4-parameter number of scattering outcomes strongly suggests that some of the remaining effective couplings are actually redundant (in the precise EFT sense reviewed for example in \cite{EFTBook}). 

One way in which an effective operator can be seen to be redundant is if it can be removed using a field redefinition. When working in perturbation theory there is a very simple diagnostic for when such a field redefinition exists: one asks whether the operator vanishes when it is evaluated at the solution to the leading order field equations. In the present instance the fermion field equation at the position $r = \epsilon$ is actually just the boundary condition itself, and this suggests a redundancy test that can be performed, at least if the dyon-fermion interactions are treated perturbatively (as they are in \S\ref{sssec:Fermion-DyonPtbnThy} when perturbing the dyon effective couplings around the action $S^{(0)}_{\rm dyon2}$).  

When perturbing in this way we first ask what conditions the $\delta \cC^\ssA_{ij}$'s must satisfy to ensure that the perturbed boundary condition remains rank two. Perturbing the rank-$2$ conditions like \eqref{Rank 2 conditions} of the appendix to linear order in the $\delta \cC$s shows that $\cB$ remains rank two only if the coefficients $\delta \cC^{pv}_{\mfs\mfs'}$ and $\delta \cC^{ps}_{\mfs\mfs'}$ all vanish. For the remaining couplings the redundancy test then asks how many of the effective couplings in \pref{dyonEFT22Dpert} survive when simplified using the lowest-order fermion boundary condition $(\Gamma^1 + i \Gamma_c ) \bmchi (\epsilon) = 0$ that follows from the unperturbed dyon-fermion action \pref{dyonEFT22Dfree}. Notice that because $\Gamma_c = \Gamma^0 \Gamma^1$ this is equivalent to the condition $i\Gamma^0 \bmchi(\epsilon) =  \bmchi(\epsilon)$ and because $\Gamma^0$ is antihermitian this implies $\bmchi^\dagger(i \Gamma^0) = \bmchi^\dagger$ when evaluated at $r = \epsilon$, and so also $\overline{\bm{\chi}} = \bm{\chi}^{\dagger}$ there. These conditions allow us to rewrite  
\be 
     \overline{\chi}_\mfs (\epsilon) \chi_{\mfs'}(\epsilon)= \chi^{\dagger}_{\mathfrak{s}}(\epsilon)i\Gamma^0\chi_{\mathfrak{s}'}(\epsilon)= \chi^{\dagger}_{\mathfrak{s}}(\epsilon)\chi_{\mathfrak{s}'}(\epsilon)= \overline\chi_{\mathfrak{s}}(\epsilon)i\Gamma^0\chi_{\mathfrak{s}'}(\epsilon),
\ee
where the first equality uses the definition $\ol \bmchi := i \bmchi^\dagger \Gamma^0$. This shows that within perturbative calculations the couplings $\delta \mathcal{C}^s_{\mathfrak{s} \mathfrak{s}'} \ol \chi_\mfs \chi_{\mfs'} - i \delta \cC^v_{\mfs\mfs'} \ol \chi_\mfs \Gamma^0 \chi_{\mfs'}$ appearing in expression \pref{dyonEFT22Dpert} for $S_{\rm dyon2}^{\rm int}$ should only contribute to physical predictions through the combination $\delta \mathcal{C}^s_{\mathfrak{s} \mathfrak{s}'}  -  \delta \cC^v_{\mfs\mfs'}$.  

This is consistent with the more general expressions like \pref{TinvsC} for the amplitudes $\cT_\ins$ and $\cT_\outs$ within the Born-Oppenheimer framework, since for rank-2 perturbations about the unperturbed couplings $\cC^{ps}_{++}=\cC^{ps}_{--} = 1$ these give
\bea \label{TinvsCpert}
   && \mathcal{T}^{++}_{\ins} \simeq 1   + i(\delta \cC^{v}_{++}  -\delta \cC^{s}_{++})   , \quad 
     \mathcal{T}^{--}_{\ins} \simeq  1   + i(\delta \cC^{v}_{--}  -\delta \cC^{s}_{--})  , \nn\\
   &&  \cT^{+-}_{\ins} \simeq    i(\delta \cC^{v}_{+-}  -\delta \cC^{s}_{+-})  , \quad \hbox{and}   \quad
     \cT^{-+}_{\ins} \simeq   i(\delta \cC^{v}_{-+}  -\delta \cC^{s}_{-+}) ,
\eea
showing at linear order that $\delta \cC^s_{\mfs\mfs'}$ and $\delta \cC^v_{\mfs\mfs'}$ really do only appear in the combination $\delta \cC^s_{\mfs\mfs'} - \delta \cC^v_{\mfs\mfs'}$.  

These arguments suggest that within the perturbative framework one combination of $\delta \cC^{s}_{\mfs\mfs'}$ and $\delta \cC^{v}_{\mfs\mfs'}$ is redundant in the sense that it can be removed by performing a field redefinition at the position of the dyon worldline. Indeed, writing 
\be \label{s+vands-v}
 \ol \bmchi  \cC^s  \bmchi - i \ol \bmchi \cC^v  \Gamma^0 \bmchi = \tfrac12\ol \bmchi  (\cC^s + \cC^v)  (1 -i \Gamma^0) \bmchi + \tfrac12 \ol \bmchi (\cC^s - \cC^v)(1 + i \Gamma^0) \bmchi \,,
\ee
we seek a redefinition that removes the first term. But under a small variation $\bmchi \to \bmchi + \delta \bmchi$ at $r = \epsilon$ the variation of the bulk action and the unperturbed boundary action \pref{dyonEFT22Dfree} becomes
\be
  \delta S(r = \epsilon) =  -\tfrac12 \ol \bmchi \Bigl( -\Gamma^1 + i \Gamma_c \Bigr) \delta \bmchi - \tfrac12 \delta \ol \bmchi \Bigl(  \Gamma^1 + i \Gamma_c \Bigr) \bmchi \,,
\ee 
which for $\delta \bmchi = iA \, \Gamma_c \bmchi$ (with hermitian $A$ in flavour space) gives
\be
   \delta S(r = \epsilon) = - \tfrac12  i\, \ol \bmchi A\Bigl[ (-\Gamma^1 + i \Gamma_c) \Gamma_c + \Gamma_c(\Gamma^1 + i \Gamma_c) \Bigr] \bmchi =  \, \ol \bmchi A\Bigl(1 - i \Gamma^0 \Bigr) \bmchi \,,
\ee
showing that $A$ can be chosen to remove the $\cC^s + \cC^v$ term in \pref{s+vands-v}, as claimed.

A very similar reduction in the number of independent couplings also happens for more prosaic applications of the PPEFT framework to fermions. When used to describe the influence of a non-pointlike nucleus on electronic energy levels within an atom there turn out to be more effective interactions describing various types of nuclear effective couplings on the nuclear worldline than there are independent nuclear contributions to electronic energy levels. In that case the redundancy of many of the apparent nuclear moments at low energies ensures that nuclear uncertainties enter into atomic calculations in fewer ways than one would naively expect \cite{Burgess:2020ndx, Zalavari:2020tez}  

\section{RG methods and catalysis}
\label{sssec:RGcatalysis}

We now turn to the issue of $\epsilon$-dependence. The mode functions and scattering amplitudes found above depend in detail on the boundary conditions imposed at the surface $\partial P$ of the gaussian pillbox that surrounds each source. But how can it be that physical quantities depend on an arbitrary radius $r = \epsilon$ that defines the boundary of this pillbox? 

This question is precisely what PPEFT methods are good for: they show how effective couplings like the $\cC^\ssA_{ij}$'s appearing in the dyon's effective action must depend implicitly on ({\it i.e.}~run with) $\epsilon$ in order to ensure that physical predictions are $\epsilon$-independent. A side benefit of this discussion is that it shows how divergences that arise (even at the classical level) in the values of the fields as $\epsilon \to 0$ get renormalized into the effective couplings of the dyon action, along the lines described in \cite{Goldberger:2001tn}. In this language the physical scales of the UV physics -- such as $\mu$ in the action \pref{SU(2) Georgi-Glashow action} -- are described within the EFT as RG-invariant scales associated with this running. 

In this section we set up how this works for $S$-wave scattering from the dyon and in particular ask why this running makes scattering for monopoles so different from scattering from other small massive objects like nuclei in atoms (for which the corresponding issues are described in \cite{PPEFT3} and briefly summarized in Appendix \ref{App:PPEFT3}). We discuss in turn two ways the running of effective couplings of fermions to dyons differ from their couplings to more run-of-the-mill compact objects: ($i$) dimensional scaling changes associated with the kinematics of the fermion $S$-wave (which lies at the root of why dyon-fermion scattering is insensitive to dyon size), and ($ii$) the large-scale effects to do with the nontrivial fermionic interacting vacuum (what is sometimes called the fermion `condensate').  

\subsection{Scaling for compact objects}
\label{ssec:ScalingCompact}

The most important difference between fermion-dyon and fermion-nucleus scattering is the size of the interaction rates to which one is led. For fermion-nucleus scattering the RG invariant scale that sets the size of scattering cross sections depends on the nucleus' small radius.\footnote{More precisely, the RG scales expressing the implications of the nuclear strong interactions of pionic atoms, say, are set by the nuclear radius while the nucleus' electromagnetic effects relevant for electron-nuclear interactions are suppressed relative to this by powers of the fine-structure constant \cite{Burgess:2020ndx, Zalavari:2020tez}.} But for fermion-dyon scattering the $S$-wave cross sections are {\it not} suppressed by the small dyonic size $ R \sim m_g^{-1}$ even at energies $E \ll m_g$. Although the mechanism for this (the special kinematics of $S$-wave scattering) has been well-understood for quite some time \cite{Rubakov:1981rg, Callan:1982ah} our discussion here embeds this understanding into the wider PPEFT framework and shows how it can be understood using the same standard EFT reasoning that also applies to nuclei.  

\subsection*{Scaling for nuclei} 

For nuclei the bulk fermion fields satisfy the Dirac equation away from the nucleus (in the `bulk') 
\be \label{DiracBeom}
  (\Dsl + m) \psi  = 0  \,,
\ee
subject to nucleus-dependent boundary conditions at the surface $r = \epsilon$ of some small gaussian pillbox described by an effective action similar to the one described above. The main difference from the dyon case is that the absence of a magnetic monopole implies the total spin $j$ is half-integral rather than integral (see Appendix \ref{App:PPEFT3}). 

For instance, for $r > \epsilon$ the parity-even solutions are
\be \label{bulk+}
 \psi =  \left( \begin{array}{c}  \psi_\ssL \\  \psi_\ssR  \end{array}  \right)  =  \left( \begin{array}{c}  f(r) \,U^+(\theta,\phi) +i g(r) \,U^-(\theta,\phi) \\  f(r) \,U^+(\theta,\phi) -i g(r) \,U^-(\theta,\phi) \end{array}  \right)  \,,
\ee
and a similar expression exists for parity-odd ones. Here $U^\pm$ are appropriate spinor harmonics and the radial functions $f(r)$ and $g(r)$ with mode frequency $\omega$ solve the radial equations 
\be \label{fgpluseqstxt}
  f' = \left( m + \omega  \right) g \quad \hbox{and} \quad
  g' + \frac{2g}{r} = \left( m - \omega   \right) f \,,
\ee
where primes denote differentiation with respect to $r$. The general solution to these coupled first-order equations is a linear combination of two linearly independent solutions
\be \label{fgvsAC}
 f(r) = C_1 f_{1}(r) + C_2 f_{2}(r) \qquad \hbox{and} \qquad g(r) = C_1 g_{1}(r) + C_2 g_{2}(r) \,,
\ee
where $C_1$ and $C_2$ are the two expected integration constants.

Physical predictions ({\it e.g.}~for atomic energy levels or scattering cross sections) can be expressed in terms of $C_2/C_1$ and this ratio is in turn determined in terms of nuclear properties through the near-nucleus boundary conditions, which give the ratio of the functions $f$ and $g$ at $r = \epsilon$ in terms of the effective couplings $\hat\mfc_i$ for fermion bilinears of the form $\int_\ssW \exd t \; \hat \mfc_i \, \ol\psi M_i \psi$ in the nucleus' world-line action (where $M_i$ are a basis of Dirac matrices). More precisely, the boundary condition at the surface of a gaussian pillbox centered on the nucleus is -- compare with eqs.~\pref{dyonEFT2} and \pref{PPEFTList0RF}:
\be  \label{cscvtofg}
  \hat \cC  = \left( \frac{  g }{f} \right)_{r=\epsilon}  \,.
\ee
Here $\hat \cC$ is a linear combination of dimensionless quantities $\hat \cC_i = \hat \mfc_i/(4\pi\epsilon^2)$ -- compare with \pref{dimensionlessCs} (see \cite{PPEFT3} for details). 
Eq.~\pref{cscvtofg} provides the solution to how the properties of the source influence the bulk solutions for $\psi$ in the source's vicinity: the couplings $\hat \cC_i$ determine $(g/f)_{r=\epsilon}$ through the near-source boundary condition and this in turn fixes the ratios of integration constants $C_2/C_1$ that control the size of observable phenomena.

But it is still a potential puzzle why physical predictions can depend on the relatively arbitrary radius, $r=\epsilon$, of the Gaussian pillbox, which is just a scale that regularizes the boundary conditions that need not be directly related to the underlying physical scales such as the physical size of the actual compact object within the pillbox (which we denote by $R$). The precise value of $\epsilon$ must therefore drop out of predictions for observables. In detail, this happens because any explicit $\epsilon$-dependence arising in calculations of an observable cancels an implicit $\epsilon$-dependence buried within the `bare' quantities $\hat \cC_i$. Physical predictions remain $\epsilon$-independent if quantities like $\hat \cC_i(\epsilon)$ are chosen to be $\epsilon$-dependent in a way that ensures that ratios like $C_2/C_1$ remain fixed as $\epsilon$ is varied. 

This gives us another way to interpret eq.~\pref{cscvtofg}: rather than reading it as fixing $f/g$ at $r = \epsilon$ given fixed values of the $\hat \cC_i$ we can instead read it as telling us how the $\hat \cC_i(\epsilon)$ must depend on $\epsilon$ in order to ensure that $C_2/C_1$ remains $\epsilon$-independent. Within this rereading physical observables depend only on the {\it trajectory} $(\epsilon, \hat \cC_i(\epsilon))$ rather than depending on $\epsilon$ and $\hat \cC_i(\epsilon)$ separately. Changes of $\epsilon$ with $C_2/C_1$ fixed (and so also with physical observables fixed) can be regarded as defining a renormalization-group (RG) flow of the $\hat \cC_i(\epsilon)$'s, and physical observables must be invariant with respect to this flow. 

It turns out that this kind of RG flow defines a natural RG-invariant length scale, and it is this length scale that both appears in physical predictions (such as for scattering cross sections) and is determined in terms of physical scales like $R$ when matching effective couplings to the full UV theory of the source. To see how this scale arises in terms of radial mode functions like $f(r)$ and $g(r)$ it is convenient to choose $\epsilon$ such that $R \ll \epsilon \ll a$ where $a$ is a characteristic length scale of the bulk theory far from the source (for instance, for nuclei in atoms we might have $R$ of order the nuclear size and $a$ of order the Bohr radius). Radial mode functions are often well-approximated by power laws in this regime, with 

\be \label{fgvsACsmallr}
 f(r) = C_1 \left( \frac{r}{a} \right)^{\mfz -1} + C_2 \left( \frac{r}{a} \right)^{-\mfz-1} \qquad \hbox{and} \qquad g(r) = \widetilde C_1 \left( \frac{r}{a} \right)^{\mfz-1} + \widetilde C_2 \left( \frac{r}{a} \right)^{-\mfz-1} \,,
\ee
for some power\footnote{Within atoms, for instance, the relevant power is $\mfz = [(j + \tfrac12)^2 - (Z\alpha)^2]^{1/2}$.} $\mfz$ with $\widetilde C_i \propto C_i$ in a way that depends on the relative small-$r$ asymptotic behaviour of $f_i(r)$ and $g_i(r)$. One of these solutions dominates for sufficiently small $r$ while the other wins when $r$ is sufficiently big. The  precise crossover radius $R_{\star}$ between these two regimes depends only on the value of the constant $C_2/C_1$, and so provides a convenient RG-invariant characterization of the coupling evolution, and one typically finds {\it e.g.}~scattering cross sections with bulk fields of size $\sigma \sim \pi R_\star^2$ for scattering of long-wavelength modes ($k R_\star \ll 1$) despite couplings like $\hat \cC_i$ being dimensionless --  see \cite{PPEFT3}.

\subsubsection*{$S$-wave scaling for dyons and catalysis}

With the above story in mind we can now use the same EFT language to see why $S$-wave scattering from dyons is so different from scattering from other compact objects. The key issue is not whether couplings like $\hat \cC_i$ are dimensionless or not (they are dimensionless for both nuclei and dyons). The key issue is the size to be expected for RG-invariant scales like $R_\star$. 

The main issue is the difference between eqs.~\pref{fgpluseqstxt} for nuclei and \pref{4d Dirac equation} for dyons. For nuclei \pref{fgpluseqstxt} has two linearly independent solutions and so admits two different power-law type asymptotic forms like \pref{fgvsACsmallr} the transition between which defines the RG-invariant scale $R_\star$. But for dyons the $S$-wave condition removes one of these solutions leaving just a single first-order equation \pref{fgvsACsmallr} for each choice of external quantum numbers. This means there is never a transition between two asymptotic power-law regimes; there is only one power-law for each type of mode. As a result there is no RG-invariant scale $R_\star$ on which measureable things like scattering cross sections can depend.  

For $S$-wave scattering from dyons the situation is similar to what would have happened for nucleons if for some reason we were required to choose $C_2 = 0$. In this case using the asymptotic form \pref{fgvsACsmallr} in \pref{cscvtofg} would imply that $\hat \cC$ is $\epsilon$-independent. The $\epsilon$-independence of physical quantities for $S$-wave dyons similarly requires quantities like $\cT_\ins$ or $\cT_\outs$ to be $\epsilon$-independent (as we see below explicitly), and this asks all of the dimensionless rank-two couplings $\cC_i$ to themselves directly be $\epsilon$-independent.  This makes $S$-wave dyon scattering from massless fermions scale invariant and so its size is mainly set by the projection of any incoming wave onto the $S$-wave state, leading to cross sections that vary as $\sigma \propto \pi/k^2$ when $k R \ll 1$ rather than $\sigma \propto \pi R^2$ (as we verify explicitly below). This is true for any $S$-wave process regardless of whether or not the reaction in question violates a flavour symmetry (like baryon number). 

Arguments like these relying on the uniqueness of the $S$-wave kinematics are standard ones \cite{Rubakov:1981rg, Callan:1982ah} for explaining the large size of catalysis cross sections. What the above arguments do is provide them with an EFT veneer that shows why they do not undermine the usual notions of decoupling (once these are carefully formulated). 

\subsection*{Matching}
\label{ssec:Matching}

To this point the effective couplings $\cC_i$ and amplitudes $\cT_\ins$ have been treated as arbitrary parameters. But they really should be regarded as being functions of microscopic parameters in any specific theory and so take definite values once a given microscopic dyon construction is chosen. Physical predictions for things like fermion-dyon scattering cross sections within specific microscopic theories are then obtained by substituting the appropriate 
values for the $\cC_i$'s (or $\cT_\ins$) into the general expressions for {\it e.g.}~scattering cross sections given in \S\ref{sec:PerturbativeScattering} and \S\ref{sec:SO3Dyon} below. 

For simple calculations of semiclassical scattering of massless fermions moving in a classical dyon field chirality is conserved by the Dirac equation. As discussed below eq.~\pref{4d Dirac equation} the change of direction of radial motion required by $S$-wave scattering implies $\mfh = \mfc \mfs$ must change sign during scattering and so conservation of $\mfc$ implies $\mfs$ must change. This means that fermion charge is always exchanged with the dyon and so $\cT_\ins^{++} = \cT_\ins^{--} = 0$ for this type of semiclassical scattering. This then implies -- {\it c.f.}~the unitarity conditions \pref{unitarityT1} -- ${\cT}^{-+}_{\ins}=e^{-i \delta}, {\cT}^{+-}_{\ins}=e^{i \delta'}$ for some phases $\delta, \delta'$. This is the situation  that applies to the majority of microscopic fermion scattering calculations (where typically $\delta'=\delta$) performed in the presence of a classical dyon \cite{Goldhaber:1977xw, Blaer:1981ps, Marciano:1983md, Ezawa:1983vi}.
 
Alternatively Kazama {\it et.al.}~\cite{Kazama:1976fm} compute scattering processes where the fermion moving within the dyonic background has an anomalous magnetic moment. In this case chirality is not conserved and chirality-changing processes dominate, corresponding to the case $\mathcal{T}^{++}_{\ins}=\mathcal{T}^{--}_{\ins}=i \,\text{sign}(\kappa)$ (where $\kappa$ is a parameter of their model) and $\mathcal{T}^{+-}_{\ins}=\mathcal{T}^{-+}_{\ins}=0$. In both chirality-preserving and chirality-breaking cases our expressions for cross sections and currents found in later sections agree with theirs once restricted to these choices.

\subsection{Interaction effects}
\label{ssec:InteractionEffects}

In practice the motion of a free fermion within a fixed dyonic background does not provide a good approximation to fermion-dyon scattering. The free-fermion-moving-in-a-fixed-background approximation breaks down because charge-changing fermion interactions with the rotor field $\mfa$ described by the amplitudes $\cT_\ins^{+-}$ and $\cT_\ins^{-+}$ significantly distort the ground state within the fermionic sector (more about this in \S\ref{ssec:FermionBO} below) and this distortion cannot be neglected \cite{Rubakov:1981rg, Callan:1982ah, CallanSMatrix, Yamagishi:1982wp, Grossman:1983yf, Kazama:1983rt, Polchinski}. The radial extent of the fermionic vacuum polarization can extend outside the dyon to distances of order the fermion Compton wavelength and so can be much larger than the underlying classical dyon solution itself.

The back-reaction of this kind of dynamics can appreciably alter the RG flow of couplings like the $\cC_i$'s once $\epsilon$ is large enough to include a significant component of fermionic polarization within the gaussian pillbox. In this case it is the new values for $\cC_i(\epsilon)$ that are relevant when computing quantities like $\cT_\ins$ and the above conclusion that $\cT_\ins$'s are $\epsilon$-independent changes. The results of this type of evolution are studied in \cite{Rubakov:1981rg, Callan:1982ah, CallanSMatrix, Polchinski}, and for general models the full interpretation of the resulting physics remains incomplete (see for example \cite{vanBeest:2023dbu, vanBeest:2023mbs} for recent discussions). In the specific model considered here, however, the upshot is fairly simple: the Coulomb energies associated with the fermionic vacuum distortions turn out to convert the UV semiclassical prediction $\cT_\ins^{++} = \cT_\ins^{--} = 0$ for $\epsilon$ of order the monopole scale into the new prediction $\mathcal{T}^{+-}_{\ins}=\mathcal{T}^{-+}_{\ins}=0$ for $\epsilon$ large enough to include the fermionic vacuum distortions \cite{Polchinski}. This conversion is intuitive inasmuch as the underlying distortion of the fermionic vacuum is driven by nonzero $\cT_\ins^{+-}$ and $\cT_\ins^{-+}$ (as we see in \S\ref{sec:SO3Dyon} below)

More generally the coefficients appropriate to any other particular microscopic dyon construction can in principle be obtained in a similar way by matching the EFT to the microscopic theory of interest, once this is known. When doing this matching we typically choose $\epsilon$ such that $R \ll \epsilon \ll a$. For the simplest applications $R \sim m_g^{-1}$ is of order the dyon size, but for applications including fermion condensation in the dyon field $R$ is instead of order the fermion's Compton wavelength $R \sim m_\psi^{-1}$. In either case taking $R \ll \epsilon \ll a$ remains justified provided the low-energy focus is on sufficiently large $a$. 

\section{Perturbative calculations}
\label{sec:PerturbativeScattering}

We next turn to calculating some simple dyon-fermion processes as functions of the general effective couplings described above. We do so first in this section treating the fermion-dyon interactions perturbatively, and then in the next section repeat the process within the Born-Oppenheimer approximation (which works to all orders in the fermion-dyon interactions when the rotor field $\mfa$ is treated classically).

We start by computing several reaction rates treating the dyon-fermion interactions perturbatively, within the framework described in \S\ref{sssec:Fermion-DyonPtbnThy} above. This choice is just made for convenience to explore some perturbative consquences of the fermion-dyon interactions, and might not be the dominant reaction for particular microscopic descriptions of the underlying dyon. We return to the more broadly applicable general case without this perturbative approximation in \S\ref{sec:SO3Dyon} below. 

The dyon-localized part of the hamiltonian governing dyon-fermion interactions obtained from the lagrangian \pref{dyonEFT2} (and including the electrostatic background and fluctuation field) is
\be \label{EffDyonH}
  H_{\rm dyon} =  e \mfp \, \widehat A_0(r=\epsilon) + \frac{1}{2\cI} \left( \mfp - \frac{\vartheta}{2\pi} \right)^2 +  \frac12  \Bigl[ \ol{\bm \psi} \, \mfC (\mfa) \, {\bm \psi}  \Bigr]_{r = \epsilon}
\ee
where $\mfp$ is the canonical momentum for $\mfa$ given in \pref{dyonmomischarge}. In this section we drop the Coulomb fluctuation $\widehat A_0$ and perturb the fermion boundary action about a simple $\mfa$-independent boundary action 
along the lines described in \S\ref{sssec:Fermion-DyonPtbnThy}. Specializing to the $S$-wave and switching to 2D fields, we split the bulk and boundary-localized hamiltonian into unperturbed and perturbed parts, $H_0 + H_{\rm int}$. The unperturbed hamiltonian is\footnote{We follow standard practice here and keep the rotor kinetic term despite it being order $(e^2/4\pi) \mu$ in magnitude and so nominally being a higher-loop size.}  
\bea \label{EffDyonH0}
H_0= \frac{1}{2\cI} \left( \mfp - \frac{\vartheta}{2\pi} \right)^2  +\frac{i}{2}\sum_{\mfs=\pm}\left(\ol{{\chi}}_{\mfs}\Gamma_c \chi_{\mfs}\right)_{r=\epsilon}+\frac{1}{2}\sum_{\mfs=\pm}\int^{\infty}_{\epsilon} \exd r \,\ol{{\chi}}_{\mfs}\left[\Gamma^1\overset{\leftrightarrow}{\partial}_1- i \mfs \,\Gamma^0\left(e v - \frac{e Q}{r}\right) \right]\chi_{\mfs},
\eea
in which the first term describes the free rotor dynamics and the second term gives the boundary conditions at $r = \epsilon$ satisfied by the fermions, whose bulk dynamics in the presence of the background dyon charge is given by the last term. 

The unperturbed boundary term $\frac{i}{2}\left(\ol{\bm \chi}\Gamma_c \bm{\chi}\right)_{r=\epsilon}$ is chosen such that ($i$) it does not involve the rotor field $\mfa$; ($ii$) the corresponding boundary matrix $\cB^{(0)}$  has rank $2$ and ($iii$) the resulting modes describe fermion reflection from the dyon with no phase change at $r=\epsilon$: that is, $\cT^{++}=\cT^{--}=1$ to zeroeth order in perturbation theory. This choice implies the interaction picture field $\bm{\chi}$ satisfies the boundary condition \eqref{Ptbn0BC} and can be expanded in terms of either the modes $\mfu^{\rm in 0}_{\mfs, k}, \mfv^{\rm in 0}_{\mfs, k}$ or $\mfu^{\rm out 0}_{\mfs, k}, \mfv^{\rm out 0}_{\mfs, k}$, given in equations \eqref{Ptuin0}-\eqref{Ptvout0}. 

The interaction hamiltonian in the interaction picture is then given by 
\be \label{HintFieldBasis}
     H_{\rm int} =\frac12 \sum_{\mfs,\mfs'=\pm} \overline{\chi}_{\mfs} \Bigl(\delta\mathcal{C}^s_{\mfs\mfs'}+i \delta\mathcal{C}^{ps}_{\mfs\mfs'} \Gamma_c-i \delta\mathcal{C}^{v}_{\mfs\mfs'}\Gamma^{0}+i\delta \mathcal{C}^{pv}_{\mfs\mfs'} \Gamma^{1}\Bigr)\,\chi_{\mfs'}  \;  e^{\frac{i}{2} (\mfs-\mfs')\, \mfa_\ssI}  \,,
\ee
where the fermion fields are evaluated at $r=\epsilon$ and $\mfa_\ssI$ is the interaction picture rotor field  
\be
   \mfa_\ssI(t) :=  \; e^{\frac{i}{2\mathcal{I}} \Pi^2 t} \mfa\, e^{-\frac{i}{2\mathcal{I}} \Pi^2 t} \,,
\ee
with $\Pi := \mfp - (\vartheta/2\pi)$. As discussed in \S\ref{ssec:RedundantInteractions} this perturbed problem also has rank two only when $\delta \cC^{pv}_{\mfs\mfs'} = \delta \cC^{ps}_{\mfs\mfs'} = 0$, which we henceforth assume (though this is not crucial for this perturbative discussion). In terms of creation and annihilation operators eq.~\pref{HintFieldBasis} can be written in the normal-ordered form
\bea\label{Hint in terms of in0 modes}
     H_{\rm int}&=& E_0 + \sum_{\mathfrak{s}, \mathfrak{s}'=\pm}\int^{\infty}_0\frac{\mathrm{d}k\,\mathrm{d}k'}{2\pi}\,(\delta\mathcal{C}^s_{\mathfrak{s} \mathfrak{s}'}-\delta\mathcal{C}^v_{\mathfrak{s} \mathfrak{s}'})    \Big[({a}^{\ins}_{\mathfrak{s}, k})^{\star} {a}^{\ins}_{\mathfrak{s}', k'} e^{i (k-k')t} +({a}^{\ins}_{\mathfrak{s}, k})^{\star} (\overline{{a}}^{\ins}_{\mathfrak{s}', k'})^{\star} e^{i (k+k')t} \nn\\
     && \qquad\qquad + \overline{{a}}^{\ins}_{\mathfrak{s}, k} {{a}}^{\ins}_{\mathfrak{s}', k'} e^{-i (k+k')t}  - (\overline{{a}}^{\ins}_{\mathfrak{s}', k'})^{\star} \overline{{a}}^{\ins}_{\mathfrak{s}, k} e^{-i (k-k') t}  \Big] \; e^{\frac{i}{2\mathcal{I}} \Pi^2 t}e^{\frac{i}{2}(\mathfrak{s}-\mathfrak{s}') (\mfa- e v t)}e^{-\frac{i}{2\mathcal{I}} \Pi^2 t}  
\eea
where we absorb an $r_0$-dependent phase using an appropriate constant shift of $\mfa$. We also drop factors of $e^{i(k\pm k')\epsilon}$ because our EFT framework requires we choose $\epsilon$ to be smaller than the characteristic bulk length scales of interest, as described in \S \ref{sssec:RGcatalysis}. In particular this requires us to restrict our attention to the regime where $k \epsilon$, $k' \epsilon$ and  $e v \epsilon$ are all much smaller than unity. 

In \eqref{Hint in terms of in0 modes} $E_0$ denotes the  vacuum expectation value  of the interacting hamiltonian $H_{\rm int}$, given by
\be\label{Hint vev}
E_0 := \bra{0}H_{\rm int}\ket{0}=\frac{e Q}{4\pi \epsilon}(\delta\mathcal{C}^{s}_{--}-\delta\mathcal{C}^{v}_{--}-\delta\mathcal{C}^{s}_{++}+\delta\mathcal{C}^{v}_{++}),
\ee
as shown in Appendix \ref{App: Currents}. Although this diverges as $\epsilon \to 0$ it can be absorbed into the counterterm describing the mass of the dyon. Notice these expressions depend on the boundary couplings only through the combination $\delta \cC^s- \delta \cC^v$, as argued must be the case in \S\ref{ssec:RedundantInteractions}. 

The remainder of this section uses the above setup to calculate some scattering observables within this perturbative framework.

\subsection{Processes involving dyon charge eigenstates}

We start by recording the amplitudes for single-fermion scattering processes assuming the dyon is chosen to be in a charge -- {\it i.e.}~momentum -- eigenstate at both the initial and final times and working to first order in perturbation theory. By focusing here on single-particle scattering we avoid the complications of multiparticle state-definitions in a monopole background described in \cite{Csaki:2021ozp, Csaki:2022tvb} and the twist operators described in \cite{vanBeest:2023dbu, vanBeest:2023mbs}. 

\subsubsection{Pair production}

Among the charge-changing processes mediated at leading order by \pref{Hint in terms of in0 modes} is the production or absorption of particle-antiparticle pairs carrying net charge. The amplitude obtained at leading order from \pref{Hint in terms of in0 modes} for the production of a pair of positive charge -- {\it i.e.}~for $\ket{0}\ket{{\Pi}}\rightarrow ({a}^{\outs}_{+, k})^{\star}(\overline{a}^{\outs}_{-, k'})^{\star}\ket{0}\ket{{\Pi}'}$ -- is  
\bea \label{f+PP-Pi}
  &&\cA \Bigl[ \Pi \to \Pi' +  f_+(k) + \bar f_-(k') \Bigr]  = - \delta_{{\Pi}', {\Pi}+1}  \delta\left(\omega_{+, k}+\ol{\omega}_{-, k'}+ \frac{\Pi'^2 -\Pi^2}{2\mathcal{I}}\right)  i \left(\delta\mathcal{C}^{s}_{+-}-\delta\mathcal{C}^{v}_{+-}\right), 
\eea
where $\omega_{\mfs, k} = k- \frac12 \mfs e v$ is the energy of a particle with charge $\mfs e/2$ and momentum $k$ and $\overline{\omega}_{\mfs,k}= k + \frac12 \mfs e v$ is the energy of an antiparticle with charge $-\mfs{e}/2$ and momentum $k$.

The amplitude for producing a negatively charged pair -- {\it i.e.}~for $\ket{0}\ket{{\Pi}}\rightarrow ({a}^{\outs}_{-, k})^{\star}(\overline{a}^{\outs}_{+, k'})^{\star}\ket{0}\ket{{\Pi}'}$ -- similarly is
\bea
 &&\cA \Bigl[\Pi \to \Pi' +  f_-(k) + \bar f_+(k')  \Bigr] = - \delta_{{\Pi}', {\Pi}-1}    \delta\left(\omega_{-,k}+\ol{\omega}_{+, k'}+ \frac{\Pi'^2-\Pi^2}{2\mathcal{I}}\right)  i \left(\delta\mathcal{C}^{s}_{-+}-\delta\mathcal{C}^{v}_{-+}\right). 
\eea

These show that the rotor level can only change by one unit, as required for it to absorb or emit the charge lost or gained by the fermions. The fermion similarly gains or loses the energy required by this transition. These processes cause rotor states with nonzero $\Pi$ to decay towards $\Pi = 0$ by emitting fermion pairs. It is energetically possible to emit positively charged pairs when $k+k'=e v -\frac{1+2\Pi}{2\cI}\ge 0$ and so is only possible when $\Pi \leq \cI e v-\frac{1}{2}$. It is similarly possible to emit negatively charged pairs if $k+k'=-e v -\frac{1-2\Pi}{2\cI}\ge 0$ and this is only possible when $\Pi \geq \cI e v+\frac{1}{2}$.  No pair production occurs for electrically neutral particle-antiparticle pairs since energy conservation implies this can only happen if $k=k'=0$. 

Writing $\cA = \cM \, \delta(E_f - E_i)$ and using these amplitudes in Fermi's golden rule gives the following differential decay rate 
\be
 \mathrm{d}\Gamma  = 2\pi |\cM|^2 \, \delta(E_f - E_i) \, \frac{\exd k}{2\pi} \, \frac{\exd k'}{2\pi} \,.
\ee
Using this and performing the final-state momentum integrals, the integrated rate for emitting positively charged fermions is nonzero for initial dyon momenta satisfying $\Pi< \cI e v - \frac{1}{2} \sim (4\pi/e)(v/\mu)$, and is given by 
\be
\Gamma\Bigl[ \Pi \to (\Pi+1) +  f_+(k) + \bar f_-(k') \Bigr] =\frac{1}{2\pi}\left( e v - \frac{2\Pi +1}{2\cI}\right)|\delta \cC^s_{+-}-\delta \cC^v_{+-}|^2 \,.
\ee
The integrated rate for emitting negatively charged pairs is similarly nonzero when $\Pi>\cI e v + \frac{1}{2}$ with
\be
\Gamma\Bigl[ \Pi \to (\Pi-1) +  f_-(k) + \bar f_+(k') \Bigr]  =\frac{1}{2\pi}\left(-e v + \frac{2\Pi -1}{2\cI}\right)|\delta \cC^s_{-+}-\delta \cC^v_{-+}|^2 \,.
\ee

\subsubsection{Scattering cross sections}

The Hamiltonian \pref{Hint in terms of in0 modes} also describes scattering processes. The amplitude for the charge-changing process $({a}^{\ins}_{+, k})^{\star}\ket{0}\ket{{\Pi}}\rightarrow ({a}^{\outs}_{-, k'})^{\star}\ket{0}\ket{{\Pi}'}$ is given by
\be\label{Pt:+to-}
    \cA\Bigl[ f_+(k) + \Pi \to \Pi' + f_-(k') \Bigr]  = - \delta_{{\Pi}', {\Pi}-1}  \delta\left(\omega_{-, k'} - \omega_{+, k}+\frac{{\Pi'}^2-\Pi^2}{2\mathcal{I}}\right) \,i \left(\delta\mathcal{C}^{s}_{-+}-\delta\mathcal{C}^{v}_{-+}\right) ,
\ee
where (as before) $\omega_{\mfs, k} = k- \mfs \frac{e v }{2}$. Similarly the amplitude for $({a}^{\ins}_{-, k})^{\star}\ket{0}\ket{{\Pi}}\rightarrow ({a}^{\outs}_{+, k'})^{\star}\ket{0}\ket{{\Pi}'}$ is
\be\label{Pt:-to+}
    \cA\Bigl[ f_-(k) + \Pi \to \Pi' + f_+(k') \Bigr] = - \delta_{{\Pi}', {\Pi}+1}  \delta\left(\omega_{+, k'} - \omega_{-, k}+\frac{{\Pi'}^2-\Pi^2}{2\mathcal{I}}\right) \,i \left(\delta\mathcal{C}^{s}_{+-}-\delta\mathcal{C}^{v}_{+-}\right). 
\ee
The analogous amplitudes for charge-changing antiparticle transitions can be inferred from those above by crossing symmetry. These reactions can proceed so long as the initial fermion energy satisfies $\omega_i \ge \frac{1}{2\cI}( \mfs' 2\Pi+1) -\mfs' \frac{e v}{2}$,  where $\mfs'$ is the charge of the final particle, since $\omega_f \ge - \mfs' \frac{e v}{2}$ for single particle 
scattering.  

The effective interactions in \pref{Hint in terms of in0 modes}  also describe scattering processes that do not exchange charge or energy with the dyon (and so necessarily flip the $4$D chirality).
The amplitude for an incoming positively charged particle to scatter to an outgoing particle of the same charge is
\be\label{Pt:+to+}
 \cA\Bigl[ f_+(k) + \Pi \to \Pi' +  f_+(k')  \Bigr]  =-\delta_{{\Pi}', {\Pi}} \,\delta\left(\omega_{+, k'}-\omega_{+, k}  \right) i\Bigl[\,\int^{\infty}_{-\infty} \exd t\, \bra{0}H_{\rm int}\ket{0}+ \left(\delta\mathcal{C}^{s}_{++}-\delta\mathcal{C}^{v}_{++}\right)\Bigr], 
\ee
where we omit the zeroth order term in the perturbative expansion of the $S$-matrix. The above expression includes a contribution from the vacuum survival amplitude, given in terms of 
the vacuum expectation value of $H_{\rm int}$. Although we formally include this in amplitudes such as  \eqref{Pt:+to+} and \eqref{Pt:-to-} below, we are primarily interested in {\it inclusive} scattering processes in which the number of pairs produced by the dyon  is unmeasured, as explained in more detail in \S \ref{sec:SO3Dyon}. In that case, the vacuum survival amplitude, which describes a process in which no pairs are produced, drops out of physical predictions such as cross section results. The corresponding amplitude for a negatively charged incoming particle is
\be\label{Pt:-to-}
  \cA\Bigl[ f_-(k) + \Pi \to \Pi' +  f_-(k') \Bigr]  = -\delta_{{\Pi}', {\Pi}} \delta\left(\omega_{-, k'}-\omega_{-, k} \right) i \Bigl[\,\int^{\infty}_{-\infty} \exd t\, \bra{0}H_{\rm int}\ket{0}+ \left(\delta\mathcal{C}^{s}_{--}-\delta\mathcal{C}^{v}_{--}\right)\Bigr]. 
\ee
The amplitudes for transitions between antiparticles of the same charge can be calculated similarly. 

Fermion-dyon scattering reactions are most usefully described in terms of 4D cross sections rather than 2D scattering rates, so we pause to make the connection to these explicit. Because incoming initial states in 4D are plane waves far from the dyon, they are not prepared in the $S$-wave. Their scattering rates are therefore the product of the 2D $S$-wave scattering rate times the probability, $p_s$, of finding the incoming plane-wave in the $S$-wave. The 4D cross section then is obtained by dividing by the incoming 4D particle flux $\mfF_i$. 

Combining these factors leads to the following factorized expression for the 4D single-particle $S$-wave differential cross section $\exd \sigma_s$ computed using the above amplitudes with $\cA = \cM \, \delta(E_f - E_i)$: 
\be\label{Cross section definition}
  \exd \sigma_s \Bigl[ f_{\mfs\mfc}(k) + \Pi \to \Pi' + f_{\mfs'\mfc'}(k') \Bigr] = \frac{p_{s}}{\mfF_i} \; \exd  \Gamma_2  = \frac{\pi}{k^2} \, \delta_{\mfs, \mfc} \delta_{\mfs', -\mfc'} \, |\cM|^2\, \delta(E_f-E_i) \, \exd k' \,,
\ee
where the factors $p_s$ and $\mfF_i$ are computed explicitly in Appendix \ref{App:Scattering states, new}, where we also show how the $2$D rates are calculated. The factor $\delta_{\mfs, \mfc}$ ensures the result is nonzero only when $\mfs = \mfc$ corresponding to the observation made below eq.~\pref{Chirality basisv} that $\mfs\mfc = +1$ for any incoming $S$-wave fermion. Similarly $\mfs' \mfc' = -1$ for any outgoing fermion.\footnote{We restore the chirality label $\mfc'$ on the final $S$-wave state to make all the quantum numbers explicit in $\exd \sigma_s$} Notice the proportionality to $1/k^2$ ensures the cross section scales with energy as does the unitarity bound (so long as the fermion mass is negligible). 

For instance, using the amplitudes for fermion scattering given 
above we find the charge-changing cross section to be 
\be
  \exd\sigma_s\Bigl[ f_{++}(k) + \Pi \rightarrow \Pi' + f_{-+}( k') \Bigr]  = \frac{\pi}{k^2} \, \delta_{\Pi',\Pi-1} \delta\left(k'-k+ e v+\frac{1-2\Pi}{2\cI}\right) |\delta \cC^s_{-+}-\delta \cC^v_{-+}|^2 \,{\exd k'} \,.
\ee
Integrating over the final fermion momentum and marginalizing over the unmeasured final dyon momentum then gives the total charge-changing cross section
\be
\sigma_s\Bigl[ f_{++}(k) + \Pi \rightarrow (\Pi-1) + f_{-+}( k') \Bigr]  = \Theta\left(k- e v+\frac{2\Pi-1}{2\cI} \right) \,\frac{\pi}{k^2} \,| \, \delta \cC^s_{-+}-\delta \cC^v_{-+}|^2 \,.
\ee
In the same way for the charge-changing $f_{-\,-}(k) + \Pi \rightarrow (\Pi+1) + f_{+-}(k')$ transition, we get  
\bea
\sigma_s\Bigl[ f_{- \, -}(k) + \Pi \rightarrow (\Pi+1) + f_{+-}( k') \Bigr] = \Theta\left(k+ e v-\frac{2\Pi +1}{2\cI} \right)\frac{\pi}{k^2}|\delta \cC^s_{+-}-\delta \cC^v_{+-}|^2.
\eea

Similar expressions can be found for the cross section for processes where the $4$D chirality of the fermions changes but not their charge (such as the amplitudes \pref{Pt:+to+} and \pref{Pt:-to-}) but we do not provide them explicitly here because they depend more sensitively on our initial choice of unperturbed boundary conditions. Expressions for these processes are instead derived in \S\ref{sec:SO3Dyon} below using the Born-Oppenheimer approximation. 

\subsection{Transitions between dyon field eigenstates}

For the purposes of comparing with Born-Oppenheimer results in \S\ref{sec:SO3Dyon} it is worth computing the same processes as above but this time choosing the initial and final rotor states to be eigenstates of $\mfa$ rather than $\mfp$ or $\Pi$. Strictly speaking, the interaction picture operator 
\be 
  \mfa_\ssI(t)=\mfa + \frac{\Pi\, t}{\cI} \,,
\ee
is not conserved and so its eigenstates need not agree at different times. However because $\cI^{-1} \sim \alpha \mu \ll \mu$ in the semiclassical limit, it is a good approximation to neglect the time-dependence of $\mfa$ and so amplitudes for transitions between $\mfa$ eigenstates simplify considerably, because $\mfa$ is approximately conserved.

\subsubsection{Pair production}

For instance, for static $\mfa$ the amplitude to produce a particle-antiparticle pair of positive charge is found by taking the matrix element of \pref{HintFieldBasis}, leading to
\be
  \cA\Bigl[ \mfa \to \mfa' + f_+(k) + \bar f_-(k') \Bigr] 
  \simeq - \delta_{{\mfa}' {\mfa}}  \delta\left(\omega_{+, k}+\ol{\omega}_{-, k'}\right)  e^{i \mfa}\,i \left(\delta\mathcal{C}^{s}_{+-}-\delta\mathcal{C}^{v}_{+-}\right).
\ee
The corresponding amplitude for producing a negatively charged pair vanishes in this approximation because the energy conservation condition now implies $k+k'+ e v =0$, which is never satisfied for $k,k'\ge 0$, $e v >0$. As we shall see, the above amplitudes exactly match the ones listed in  \S\ref{sec:SO3Dyon} below, once these are evaluated in the perturbative regime.

The resulting total rate for producing fermion pairs (marginalized over their unmeasured quantum numbers and the dyon final state) then is
\be \label{PairProdPertRate}
\Gamma\Bigl[ \mfa \to \mfa' + f_+(k) + \bar f_-(k') \Bigr] \simeq \frac{e v}{2\pi} \Bigl|\delta \cC^s_{+-}-\delta \cC^v_{+-} \Bigr|^2 .
\ee

\subsubsection{Scattering}

The $S$-wave amplitudes for charge-changing fermion scattering in the same static-$\mfa$ limit are
\be
    \cA\Bigl[ f_{+}(k) + \mfa \to \mfa' + f_{-}(k') \Bigr] = - \delta_{\mfa' \mfa}  \delta\left(\omega_{-, k'} - \omega_{+, k}\right)  e^{-i \mfa}\,i \left(\delta\mathcal{C}^{s}_{-+}-\delta\mathcal{C}^{v}_{-+}\right),
\ee
and
\be
    \cA\Bigl[ f_{-}(k) + \mfa \to \mfa' + f_{+}(k') \Bigr]  = - \delta_{{\mfa}' \mfa}  \delta\left(\omega_{+, k'} - \omega_{-, k}\right)  e^{i \mfa}\,i \left(\delta\mathcal{C}^{s}_{+-}-\delta\mathcal{C}^{v}_{+-}\right).
\ee
The corresponding total total cross section for charge-exchange processes in which the dyon remains in an $\mfa$ eigenstate then is
\be
\sigma_s\Bigl[f_{\mfs\mfs}(k) + \mfa \to \mfa + f_{-\mfs  \mfs}(k')  \Bigr]   = \Theta\left(k- \mfs e v\right) \, \frac{\pi}{k^2} \, \Bigl|\delta \cC^s_{-\mfs \mfs}-\delta \cC^v_{-\mfs \mfs} \Bigr|^2 \,.
\ee

Notice that both pair production and scattering involve only a fairly simple condition on $k$ (if any) as opposed to the fairly complicated restrictions on $\Pi$ that arose when the dyon was prepared in a charge eigenstate. This relative simplicity arises because the $\mfa$ eigenstate always has an overlap with $\Pi$ eigenstates for which the processes are energetically allowed. The $\mfa$-eigenstate expressions are easier to compare with the Born-Oppenheimer, and agree within their common domain of validity with the cross sections found in \S\ref{sec:SO3Dyon}.

\section{Born-Oppenheimer approximation}
\label{sec:SO3Dyon}

We next compute the pair-production and scattering implied by the full mode functions described within the Born-Oppenheimer approximation of \S\ref{ssec:ScatteringState}, for which the fermions are quantized with the bosonic field $\mfa$ initially treated as a classical field. We start in \S\ref{ssec:FermionBO} by examining how the fast degrees of freedom (the fermions) evolve in the presence of a static classical rotor field $\mfa$, then continue in \S\ref{ssec:RotorBO} with some observations about the rotor's response.  

For these purposes recall that dyon interactions with $S$-wave fermions are described by the hamiltonian $H = H_{\rm dyon} + H_{\rm bulk}$ that is the sum of the dyon-localized terms of \pref{EffDyonH}, reproduced for convenience here,\footnote{We follow standard practice here and keep the rotor kinetic term despite it being order $(e^2/4\pi) m_g$ in magnitude and so nominally being higher-loop in size. It can make sense to do so to the extent that all of the rotor's responses arise at this same order (and are included).}  
\be \label{EffDyonH2}
  H_{\rm dyon} = \frac{1}{2\cI} \left[ \mfp(t) - \frac{\vartheta}{2\pi} \right]^2  + e \mfp \, \widehat A_0(r=\epsilon,t) +  \frac12  \Bigl[ \ol{\bm \chi} \, O_\cB(\mfa) \, {\bm \chi}  \Bigr]_{r = \epsilon,t}
\ee
where $O_\cB(\mfa)$ is given in terms of the couplings $\cC_{\mfs\mfs'}^\ssA$ by \pref{OBdef} and the bulk 2D hamiltonian is
\bea
H_{\rm bulk} =    H_\ssC +  \frac{1}{2}\sum_{\mfs=\pm}\int^{\infty}_{\epsilon} \exd r \,\ol{{\chi}}_{\mfs}\left[\Gamma^1\overset{\leftrightarrow}{\partial}_1- i \mfs \,\Gamma^0\left(e v - \frac{e Q}{r} +e \widehat A_0 \right) \right]\chi_{\mfs} \,,
\eea
with $H_\ssC$ denoting the part of the Maxwell action depending on the Coulomb field $\widehat A_0$. We follow previous work and perturb in the Coulomb interactions involving $\widehat A_0$, whose contributions to the energy are suppressed by powers of $e^2$. We differ from the warm-up calculations of \S\ref{sec:PerturbativeScattering} by {\it not} splitting $O_\cB(\mfa)$ into an unperturbed and perturbed piece; instead treating the interaction with $\mfa$ using Born-Oppenheimer methods (as motivated in \S\ref{sssec:BOIntro}). In practice this means that we treat $\mfa$ as a fixed classical field when determining the  
fermionic response and then return to ask how this slower rotor field evolves in response to interactions with the faster fermions.  

\subsection{Fermion evolution}
\label{ssec:FermionBO}

We start by calculating the Bogoliubov coefficients that relate $in$ and $out$ modes, neglecting the Coulomb back-reaction of the distortions of the fermion ground state. Subsequent subsections then consider some of the implications of this fermionic distortion such as to pair-production rates and scattering cross sections. 

\subsubsection{Bogoliubov relations}
\label{ssec:Bogoliubov}

The starting point expands the $S$-wave fermion field in terms of the $in$ and $out$ bases for the fermion modes in the presence of a dyon background described in \S\ref{ssec:ScatteringState}:
\bea\label{The s-wave Dirac field}
   \bm{\chi}(x) &=&  \sum_{\mfs=\pm}\int^{\infty}_{0} \frac{\exd k}{\sqrt{2\pi}}\,\Bigl[ \mfu^\ins_{\mfs, k}(x) \,  a^\ins_{\mfs, k } + \mfv^\ins_{\mfs, k}(x) \,( \ol{a}^\ins_{\mfs, k} )^\star\Bigr] \nn\\
   &=&   \sum_{\mfs=\pm}\int^{\infty}_{0} \frac{\exd k}{\sqrt{2\pi}}\,\Bigl[ \mfu^\outs_{\mfs, k}(x) \,  a^\outs_{\mfs, k } + \mfv^\outs_{\mfs, k}(x) \,( \ol{a}^\outs_{\mfs, k} )^\star\Bigr]
\eea
with mode functions defined in eqs.~\pref{uinfirst} through \pref{vinlast} and \pref{uoutfirst} through \pref{voutlast} and particle and antiparticle creation and annihilation operators satisfying
\bea \label{anticom1}
    \Bigl\{{ a}^\ins_{\mfs, k}, ({ a}^\ins_{\mfs', k'})^\star \Bigr\} &=&  \Bigl\{\ol{{ a}}^\ins_{\mfs, k}, (\ol{{ a}}^\ins_{\mfs', k'})^\star \Bigr\} =   \delta(k-k') \,\delta_{\mfs \mfs'},  \nn\\
\hbox{and} \quad
    \Bigl\{{ a}^\outs_{\mfs, k}, ({ a}^\outs_{\mfs', k'})^\star \Bigr\} &=&  \Bigl\{\ol{{ a}}^\outs_{\mfs, k}, (\ol{{ a}}^\outs_{\mfs', k'})^\star \Bigr\} =  \delta(k-k') \,\delta_{\mfs \mfs'},  
\eea
with all other anticommutators vanishing. 

Since both the $in$ and $out$ basis are complete, each can be expanded in terms of the other. For each $k \geq 0$ we have
\bea
     \mfu^\ins_{-,k}(x) &=&   \mathcal{T}^{--}_{\ins} e^{- 2 i k \epsilon} \left(\frac{\epsilon}{r_0} \right)^{-i eQ}\,\mfu^\outs_{-, k}(x)+\cT^{+-}_{\ins} e^{-  i (2 k + e v)\epsilon}\, e^{i\mfa} \, \mfu^\outs_{+, k + e v}(x)  \nonumber \\ 
     \mfv^\ins_{+, k}(x) &=& \mathcal{T}^{++}_{\ins} e^{ 2 i k \epsilon}\left( \frac{\epsilon}{r_0} \right)^{i eQ}\,\mfv^\outs_{+, k}(x)+\cT^{-+}_{\ins} e^{  i (2 k + e v) \epsilon}\, e^{-i\mfa}\,\mfv^\outs_{-, k+ e v}(x) \nonumber\\
     \mfu^\ins_{+, k}(x) &=& \mathcal{T}^{++}_{\ins} e^{ -2 i k \epsilon}\left( \frac{\epsilon}{r_0} \right)^{i eQ} \,\mfu^\outs_{+, k}(x) \\
   && \qquad + \cT^{-+}_{\ins} e^{ - i (2k -e v) \epsilon  }\, e^{-i\mfa} \Bigl[ \Theta\left(k-
   ev\right) \mfu^\outs_{-, k - e v}(x) + \Theta\left(-k+ev\right) \mfv^\outs_{-,  -k +  e v}(x) \Bigr] \nonumber\\ 
     \mfv^\ins_{-, k}(x) &=& \mathcal{T}^{--}_{\ins} e^{ 2 i k \epsilon}\left( \frac{\epsilon}{r_0} \right)^{-i eQ}\, \mfv^\outs_{-, k}(x) \nonumber\\
     &&\qquad +\cT^{+-}_{\ins} e^{  i (2 k - e v)\epsilon  }\, e^{i\mfa} \Bigl[\Theta\left(k -ev\right) \mfv^\outs_{+, k - e v}(x) + \Theta\left(-k+ ev\right) \mfu^\outs_{+, -k + e v}(x)\Bigr],\nonumber
\eea
and similarly
\bea
     \mfu^\outs_{-, k}(x) &=&   \mathcal{T}^{--}_{\outs} e^{ 2 i k \epsilon} \left( \frac{\epsilon}{r_0} \right)^{i eQ}\,\mfu^\ins_{-, k}(x)+\cT^{+-}_{\outs} e^{ i (2 k + e v) \epsilon}\, e^{i\mfa}\,\mfu^\ins_{+, k+ e v}(x) \nn\\
     \mfv^\outs_{+, k}(x) &=& \mathcal{T}^{++}_{\outs} e^{ -2 i k\epsilon}\left( \frac{\epsilon}{r_0} \right)^{-i eQ}\,\mfv^\ins_{+,  k}(x) + \cT^{-+}_{\outs} e^{-  i (2 k + e v) \epsilon}\, e^{-i\mfa}\,\mfv^\ins_{-, k + e v}(x) \nn\\
     \mfu^\outs_{+, k}(x) &=& \mathcal{T}^{++}_{\outs} e^{ 2 i k \epsilon}\left(\frac{\epsilon}{r_0} \right)^{-i eQ}\,\mfu^\ins_{+, k}(x) \\
     && \qquad +\cT^{-+}_{\outs} e^{  i (2 k - e v) \epsilon  }\, e^{-i\mfa} \Bigl[ \Theta\left(k - ev\right) \mfu^\ins_{-, k- e v}(x) + \Theta\left(-k+ev\right) \mfv^\ins_{-, -k+ e v}(x)\Bigr] \nn\\
      \mfv^\outs_{-, k}(x) &=& \mathcal{T}^{--}_{\outs} e^{- 2 i k \epsilon}\left(\frac{\epsilon}{r_0} \right)^{i eQ}\,\mfv^\ins_{-, k}(x) \nn\\
      &&\qquad +\cT^{+-}_{\outs} e^{-  i (2 k - e v)\epsilon  }\, e^{i\mfa} \Bigl[\Theta\left(k-ev\right)\mfv^\ins_{+, k- e v}(x) + \Theta\left(-k+ev\right) \mfu^\ins_{+,  -k+ e v}(x)\Bigr].\nn
\eea
These lead to the following Bogoliubov relations between the $in$ and $out$ operators for each $k \geq 0$ (see appendix \ref{App:Scattering states, new} for derivation) 
\bea \label{Bogoutfromin}
  &&  a^\outs_{-, k} =  \mathcal{T}^{- -}_{\ins}  \; a^\ins_{-, k}+\cT^{- +}_{\ins} \, e^{-i\mfa} \;  a^\ins_{+, k+ e v} 
  \,, \qquad (\ol{a}^\outs_{+, k})^\star = \mathcal{T}^{+ +}_{\ins}  \;  (\ol{a}^\ins_{+, k})^\star + \cT^{+ -}_{\ins}  \, e^{i\mfa }\; (\ol{a}^\ins_{-, k+ e v})^\star \nn\\
   && \qquad a^\outs_{+, k} = \mathcal{T}^{+ +}_{\ins}   a^\ins_{+, k}  + \cT^{+ -}_{\ins} \, e^{i\mfa } \Bigl[ \Theta\left(k - ev\right) a^\ins_{-, k - e v} + \Theta\left(-k + ev\right) (\ol{a}^\ins_{-, -k + e v})^\star \Bigr] \\
  && \quad  (\ol{a}^\outs_{-, k})^\star = \mathcal{T}^{- -}_{\ins}  (\ol{a}^\ins_{-, k})^\star  +\cT^{- +}_{\ins}  \, e^{-i\mfa }\Bigl[ \Theta\left(k-ev \right) (\ol{a}^\ins_{+, k- e v})^\star +\Theta\left(-k+ev\right) a^\ins_{+, -k+ e v} \Bigr] ,\nn
\eea
as well as their inverses
\bea \label{Boginfromout}
   && a^\ins_{-, k} = \mathcal{T}^{- -}_{\outs} \; a^\outs_{-, k}+\cT^{- +}_{\outs} \, e^{-i\mfa } \;  a^\outs_{+, k+ e v} \,, \qquad
    (\ol{a}^\ins_{+, k})^\star = \mathcal{T}^{+ +}_{\outs}  \;  (\ol{a}^\outs_{+, k})^\star + \cT^{+ -}_{\outs}  \, e^{i\mfa }\; (\ol{a}^\outs_{-, k+ e v})^\star \nn\\
  && \qquad  a^\ins_{+, k} = \mathcal{T}^{+ +}_{\outs}   a^\outs_{+, k}  + \cT^{+ -}_{\outs} \, e^{i\mfa } \Bigl[ \Theta\left(k - ev\right) a^\outs_{-, k - e v} + \Theta\left(-k + ev\right) (\ol{a}^\outs_{-, -k + e v})^\star \Bigr] \\
  &&  (\ol{a}^\ins_{-, k})^\star = \mathcal{T}^{- -}_{\outs}  (\ol{a}^\outs_{-, k})^\star  +\cT^{- +}_{\outs} \, e^{-i\mfa }\Bigl[ \Theta\left(k-ev \right) (\ol{a}^\outs_{+, k- e v})^\star +\Theta\left(-k+ev\right) a^\outs_{+, -k+ e v} \Bigr] .\nn
\eea
These are consistent with the anticommutation relations \pref{anticom1} by virtue of the unitarity identities \pref{unitarityT1} and \pref{unitarityT} -- and their counterparts \pref{unitarityTout1} and \pref{unitarityTout2} -- satsified by the $\cT$'s. As in \S \ref{sec:PerturbativeScattering}, we drop powers of $k\epsilon$, $k' \epsilon$ and $ev\epsilon$ and shift $\mfa$ and rephase the out-state creation and-annihilation operators to remove an $r_0$-dependent phase.

\subsubsection{Pair production}
\label{ssec:SO3DyonPP}

The mixing of creation and annihilation operators in the Bogoliubov transformations \pref{Bogoutfromin} and \pref{Boginfromout} shows that the system is unstable to pair production if $\cT_{\ins/\outs}^{+-}$ is nonzero. 

To compute the pair-production rate define the $in$ and $out$ vacua $\ket{0_{\ins}}$ and $\ket{0_{\outs}}$ as usual: 
\be
  \bm{ a}^\ins_{\mfs, k} \ket{0_{\ins}} = \ol{\bm{ a}}^\ins_{\mfs, k} \ket{0_{\ins}} = 0
  \quad \hbox{and} \quad 
  \bm{ a}^\outs_{\mfs, k} \ket{0_{\outs}} = \ol{\bm{ a}}^\outs_{\mfs, k} \ket{0_{\outs}} = 0 \,.
\ee
We switch here for convenience to discretely normalized momentum states, for which we denote the creation and annihilation operators using bold-faced fonts, as in ($\bm{a}, \ol{\bm a}$). (See appendix \ref{App:Scattering states, new} for relation between discrete and continuum normalized states.) 

Writing the total vacuum as a tensor product over momenta, $\ket{0}= \underset{k>0}{\prod}\ket{0^{\, k}}$, and using the above Bogoliubov transformations implies for each $k \geq 0$ we have
\be \label{BogInvsOutVac}
     \ket{{0}^{\,{k}}_{\ins}}= \left\{ \Theta\left({k}-  e v\right)+\Theta\left(-{k} + e v\right) \left[ \mathcal{T}^{--}_{\ins} +\cT^{+ -}_{\ins}e^{i \mfa} (\bm{a}^{\outs}_{+, {k}})^{\star}(\bm{\overline{a}}^{\outs}_{-, {-k+ e v}})^{\star}\right] \right\} \ket{{0}^{\,{k}}_{\outs}}
\ee
and
\be \label{BogOutvsInVac}
    \ket{{0}^{\,{k}}_{\outs}}= \left\{ \Theta\left({k} - e  v\right)+\Theta\left(-{k} + e v\right) \left[ \mathcal{T}^{--\,*}_{\ins} +\cT^{-+\, *}_{\ins}e^{i \mfa} (\bm{a}^{\ins}_{+, {k}})^{\star}(\bm{\overline{a}}^{\ins}_{-, {-k+e  v}})^{\star}\right] \right\} \ket{{0}^{\,{k}}_{\ins}} \,,
\ee
where as usual $\Theta(x)$ denotes the Heaviside step function. See appendix \ref{App: In and Out vacuum} for derivation of the above expressions.

If ${\mathcal{T}}^{+-} = 0$ then \pref{unitarityT1} shows that ${\cT}^{-+}$ also vanishes and $|\cT^{--}| = |\cT^{++}| = 1$. In this case \pref{BogInvsOutVac} and \pref{BogOutvsInVac} imply the $in$ and $out$ vacua are the same state. But when ${\mathcal{T}}^{+-} \neq 0$ for some momenta the $in$ vacuum contains occupied $out$ particles. This shows that when charge-changing effective dyon-fermion interactions are present the dyon spontaneously emits a fermion with charge $+ \frac12 e$ and the antiparticle of the charge $- \frac12 e$ state, ensuring the net charge emission of $\frac12 e + \frac12 e = e$. 

Total electric charge is conserved because this emission is accompanied by a transition between rotor levels for $\mfa$ that removes one unit $+e$ of charge from the dyon, as can be seen\footnote{This is made very explicit in \S\ref{sec:PerturbativeScattering} for those who need convincing.} from the $\mfa$-dependence of \pref{EmitAmp}. The sign of the charge removed is dictated by the overall sign of the dyon charge $Q$ which we've chosen to be positive: $Q > 0$. These pairs are produced through the Schwinger effect \cite{Schwinger2D, Blaer:1981ps} by the external voltage $v$ of the dyon between $r \to \infty$ and $r= 0$, and act to discharge the dyon's net charge. 

The energetics of the process is somewhat obscured within the Born-Oppenheimer approximation because the rotor is treated as a classical field, which if regarded as an eigenstate of $\mfa$ is not an energy eigenstate. When calculated 
perturbatively in \S\ref{sec:PerturbativeScattering} we do find that fermion emission extracts an energy given by the spacing between rotor steps, which is of order $\cI^{-1} \sim \alpha m_g$. But in a semiclassical approximation this energy transfer is the same size (relative to $m_g$) as loop corrections, which to this order we neglect, and this is why \pref{EmitAmp} allows pair-production for the entire momentum range $k \in (0, ev)$. Implicit in this treatment is the assumption that the fermion energies of interest are much higher than the rotor gap; one manifestation of the noncommuting order of limits discussed below eq.~\pref{DyonEnergy}.

The amplitude for this emission (for fixed $\mfa$) in a specific momentum mode is 
\be \label{EmitAmp}
\bra{0^{\, k}_{\outs}} \,\bm{\ol{a}}^\outs_{-,-k + e v} \, \bm{a}^\outs_{+,k } \ket{0^{\, k}_\ins} =\Theta(-k + e v )  \, \cT^{+-}_{\ins} \, e^{i\mfa} \,,
\ee
which has squared modulus $\cP_k = |\cT^{+-}_\ins|^2 = |\cT^{+-}_\outs|^2$ when $0 < k < ev$.  The vacuum-survival amplitude for this specific mode is similarly
\be\label{BogInvsOutOverlap}
   \braket{0^{\, k}_{\outs}|{0}^{\,{k}}_{\ins}}  = \Theta\left({k}-  e v\right) + \Theta\left(-{k} + e v\right)  \mathcal{T}^{--}_{\ins}  \,,
\ee
which squares to unity when $k > ev$ but has squared modulus $| \mathcal{T}^{--}_{\ins} |^2 = 1 - | \mathcal{T}^{+-}_{\ins}|^2 = 1 - \cP_k$ when $0 < k < ev$. The likelihood of producing zero or one pairs sums to unity because for fermions these are the only possible options.

The {\it exclusive} probability for the full vacuum to produce exactly one pair in a specific mode $k\in (0, e v)$ is given by combining the above result for all $k$, giving 
\be 
 {P}^{}_{\text{pair}}(k) =|\bra{0_{\outs}} \bm{\ol{a}}^\outs_{-,-k + e v} \, \bm{a}^\outs_{+,k } \ket{0_{\ins}} |^2= \cP_k \overset{q< ev}{\prod_{q \neq k}} (1 - \cP_q) = |\cT^{+-}_\ins(k) |^2 \overset{q< ev}{\prod_{q \neq k}}| \mathcal{T}^{--}_{\ins} (q)|^2.
\ee
The total probability factorizes because the likelihood of pair-production in each mode is independent of what happens for the other modes. More useful is the {\it inclusive} probability for pair production of a specific mode with the other modes unmeasured (and so marginalized over). The above expressions show this is given by 
\be
  p_{\text{pair}}(k)= \cP_k = |\cT^{+-}_{\ins}|^2
  \quad \hbox{and so} \quad
  p_{\rm no\,pair}(k) = 1-\cP_k =  |\mathcal{T}^{--}_{\ins}|^2 \,.
\ee

We can calculate the average number of particles produced by making use of the Bogoliubov relations in the form
\bea 
 \bra{0_{\ins}} N_\outs \ket{0_{\ins}}&=& 
\sum^{\infty}_{k=0}\bra{0_{\ins}} (\bm{a}^{\outs}_{+ k})^{\star}\,\bm{a}^{\outs}_{+ k} \ket{0_{\ins}}\nn \\ &=& \sum^{e v}_{k=0} |\cT^{+-}_{\ins}|^2 \bra{0_{\ins}}  \overline{\bm{a}}^{\ins}_{-, -k + e v}(\overline{\bm{a}}^{\ins}_{-, -k + e v})^{\star} \ket{0_{\ins}}=\frac{e v }{\Delta k}|\cT^{+-}_{\ins}|^2, 
\eea
where $\Delta k = \pi/L$ is the spacing between momentum states (when these are discretely normalized).\footnote{We discretize momenta by putting the system in a box $-L <r< L$ with near-dyon boundary conditions imposed at $r = \epsilon\approx 0$. In these conventions the relevant length of the system is $2L$, and the density of states is $2\pi/(2L) = \pi/L$} The number of particles per unit length is obtained by dividing by $2 L$, and this has a sensible continuum limit as  $L \to \infty$, with $\exd\langle N \rangle /\exd x = ev|\cT_\ins^{+-}|^2/(2\pi)$. Since each produced particle moves to larger $r$ at the speed of light this means that these particles emerge at infinity with a rate $\Gamma_\infty = ev|\cT_\ins^{+-}|^2/(2\pi)$.  A similar counting also applies to the number of produced antiparticles. Because each pair contains one positively charged particle and one positively charged antiparticle the number of produced {\it pairs} is also
\be \label{PairRate}
  \Gamma_{\rm pair} = \frac{ev}{2\pi} | \cT_\ins^{+-}|^2 \,.
\ee
This result is independently computed below using the expectation values for the fermionic currents.

Eq.~\pref{PairRate} can also be compared with \pref{PairProdPertRate} (when restricted to the perturbative domain). To this end we must expand the coefficients $\cT_\ins$ perturbatively to the same order in the $\delta \cC$s used in \S\ref{sec:PerturbativeScattering}. To linear order the $\cT_{\ins}$ amplitudes are given by
 \be\label{Pt:Tin}
    \cT^{\mfs \mfs}_{\ins}=1+i (\delta\cC^{v}_{\mfs \mfs}-\delta \cC^{s}_{\mfs \mfs})\quad \text{and} \quad \cT^{+-}_{\ins}=-\cT^{-+\,*}_{\ins}= i (\delta \cC^{v}_{+-}-\delta \cC^{s}_{+-}), 
\ee
once the rank-2 conditions $\delta \cC^{ps}_{\mfs\mfs'} = \delta \cC^{pv}_{\mfs\mfs'} = 0$ are used. Using these in \pref{PairRate} then agrees with \pref{PairProdPertRate}.

 \subsubsection{Vacuum currents}
 \label{sssec:VacCur}

An alternative characterization of dyonic pair production that lends itself to the continuum limit is the integrated contribution of the produced pairs to current flow in the fermion sector. To display these currents we evaluate the expectation value of the various conserved currents in the $in$ vacuum. Since these expectation values in general diverge we regulate them by point-splitting the two fermion fields involved by a distance $\varepsilon$, writing 
\be
   \overline{\bm{\chi}}(r, t) M \bm{\chi}(r, t) \to   \overline{\bm{\chi}}(r+\varepsilon/2, t) M \bm{\chi}(r-\varepsilon/2, t) \,, 
\ee
with $\varepsilon \to 0$ taken at the end, after renormalizing. We quote here expressions for the regularized currents -- see appendix  \ref{App: Currents} for details of the matrix-element calculations.

The regularized components of the fermion number current\footnote{We do not go through the exercise of renormalizing the fermion number current here, since we only use $ \bra{0_{\ins}}{j}^1_\ssB(x)\ket{0_{\ins}}$ to evaluate the conservation equations for  $j^{\alpha}_\ssF, j^{\alpha}_\ssA$. } $j_\ssB^\alpha =i \ol\bmchi \Gamma^\alpha \bmchi$ are given by 
\be
\bra{0_{\ins}}{j}^0_\ssB(x)\ket{0_{\ins}} = \bra{0_{\outs}}{j}^0_\ssB(x)\ket{0_{\outs}}=0 \quad
\hbox{and} \quad \bra{0_{\ins}}{j}^1_\ssB(x)\ket{0_{\ins}} = \bra{0_{\outs}}{j}^1_\ssB(x)\ket{0_{\outs}}= \frac{2 i }{\pi \varepsilon}, 
\ee
while those of the fermionic electromagnetic current $j^\alpha_\ssF = \frac12 i e \ol\bmchi \Gamma^\alpha \tau_3 \bmchi$ are  
\bea \label{2DEMFlux}
  \bra{0_{\ins}}{j}^0_\ssF(r,t)\ket{0_{\ins}} &=& \bra{0_{\outs}}{j}^0_\ssF(r,t)\ket{0_{\outs}}=\frac{e^2 v}{2\pi}|\mathcal{T}^{+-}_{\ins}|^2-\frac{e^2 Q}{2\pi r} \nn\\
  \hbox{and} \quad
   \bra{0_{\ins}}{j}^1_\ssF(r,t)\ket{0_{\ins}} &=& -\bra{0_{\outs}}{j}^1_\ssF(r,t)\ket{0_{\outs}}=  \frac{e^2 v }{2 \pi}|\mathcal{T}^{+-}_{\ins}|^2 \,.
\eea
We define the axial current by $j^\alpha_\ssA = i \ol\bmchi \Gamma^\alpha \Gamma_\ssA \bmchi$ with $\Gamma_\ssA\coloneqq \Gamma_c \, \tau_3$ rather than $\Gamma_c$ because this agrees with the 4D axial current $i\ol\bmpsi \gamma^\mu \gamma_5 \bmpsi$ for $S$-wave states (up to the usual 2D normalization factor of $4 \pi r^2$). Its vacuum matrix elements are
\bea
   \bra{0_{\ins}}{j}^0_\ssA(r,t)\ket{0_{\ins}} &=& -\bra{0_{\outs}}{j}^0_\ssA(r,t)\ket{0_{\outs}}=-\frac{e  v }{\pi}|\mathcal{T}^{+-}_{\ins}|^2 \nn\\
   \hbox{and} \quad \bra{0_{\ins}}{j}^1_\ssA(r,t)\ket{0_{\ins}} &=& \bra{0_{\outs}}{j}^1_\ssA(r,t)\ket{0_{\outs}}=-\frac{ e v }{\pi}|\mathcal{T}^{+-}_{\ins}|^2+\frac{ e Q}{\pi r} \,.
\eea

Recalling that the 2D modes are normalized so that the 2D flux $j^1$ gives the integrated radial flux $4\pi r^2 J^r$ for the corresponding 4D $S$-wave current, we see that at spatial infinity there is a nonzero flux of both electric and axial charge:
\be
   \bra{0_{\ins}}{j}^1_\ssF(\infty,t)\ket{0_{\ins}} = e \mfF \quad \hbox{and} \quad
    \bra{0_{\ins}}{j}^1_\ssA(\infty,t)\ket{0_{\ins}} = -2 \mfF \quad \hbox{where}\quad
    \mfF = \frac{ev}{2\pi} |\cT_\ins^{+-}|^2 \,.
\ee
This has a simple interpretation as a flux of pair-produced particles, since each such pair carries electric charge $e$ and axial charge\footnote{See appendix \ref{App:Scattering states, new} for discussion of asymptotic charges of $in, out$ states.} $-2$ (and no net fermion number), with integrated particle flux (or rate with which particle pairs appear at infinity) given by $\Gamma_{\rm pair} = \mfF$, in agreement with \pref{PairRate}. Our result for the flux of the axial current is consistent with  \cite{Blaer:1981ps}, when the $\cT$ amplitudes are chosen to match theirs\footnote{Note that the axial current in \cite{Blaer:1981ps} has a relative minus sign compared to our definition.} \textit{i.e.} when $|\cT^{+-}|=1$. 

These expressions are also consistent with (anomalous) current conservation. As shown in appendix \ref{App: Currents}, the above currents satisfy
\bea
    \partial_{\alpha}j^{\alpha}_\ssB(x) &=&  \frac{e Q}{2 r^2}\underset{\varepsilon \rightarrow 0}{\lim}  \Bigl[ \varepsilon\,\overline{\bm{\chi}}(r+\varepsilon/2, t)\Gamma^0\, \tau_3\bm{\chi}(r-\varepsilon/2, t)\Bigr]=- \frac{iQ}{r^2} \underset{\varepsilon \rightarrow 0}{\lim}\Bigl[ \varepsilon j^{0}_\ssF(x) \Bigr] = 0, \nonumber\\
     \partial_{\alpha}j^{\alpha}_\ssF(x) &=& \frac{e^2 Q}{4 r^2}\underset{\varepsilon \rightarrow 0}{\lim}  \Bigl[ \varepsilon\,\overline{\bm{\chi}}(r+\varepsilon/2, t)\Gamma^0\bm{\chi}(r-\varepsilon/2, t) \Bigr] =- \frac{i e^2 Q}{4 r^2} \underset{\varepsilon \rightarrow 0}{\lim} \Bigl[ \varepsilon j^{0}_\ssB(x) \Bigr] = 0  , \nonumber\\ 
    \partial_{\alpha}j^{\alpha}_\ssA(x) &=& -\frac{e Q}{2 r^2} \underset{\varepsilon \rightarrow 0}{\lim}  \Bigl[ \varepsilon\,\overline{\bm{\chi}}(r+\varepsilon/2, t)\Gamma^1\bm{\chi}(r-\varepsilon/2, t) \Bigr] = \frac{ie Q}{2 r^2}\underset{\varepsilon \rightarrow 0}{\lim}\Bigl[ \varepsilon j^{1}_\ssB(x) \Bigr] = - \frac{eQ}{\pi r^2} \,.
\eea
These agree with the standard 2D anomaly expressions, which in the present instance tell us that $j^\alpha_\ssB$ and $j^\alpha_\ssF$ are anomaly free and give
\be
 \partial_{\alpha} j^{\alpha}_\ssA =\left[\frac{e}{2}-\left(-\frac{e}{2}\right)\right]\frac{1}{2\pi}\epsilon^{\alpha \beta} \cF_{\alpha \beta} =  \frac{  e}{ \pi}\Bigl[ \partial_0 \mathcal{A}_r -\partial_r \mathcal{A}_0 \Bigr] = \frac{1}{ \pi}\partial_r\left (   \frac{ e Q}{r} -ev\right)=-\frac{e Q}{ \pi r^2}, 
\ee
when evaluated with the background dyonic Coulomb field.

It is also noteworthy that \pref{2DEMFlux} implies that particle production significantly polarizes the fermionic ground state, inducing a charge density with the opposite sign to the dyon charge that (for massless fermions) falls off only as a power law as one moves away from the dyon. As has been remarked elsewhere \cite{Preskill:1984gd, Ezawa:1983vi, CallanSMatrix, Yamagishi:1982wp, Grossman:1983yf, Kazama:1983rt}, such charging of the fermion ground state puffs up the dyon into a much bigger object than was the underlying classical dyon configuration.

\subsubsection{Dyon-fermion scattering}
\label{ssec:SO3DyonScat}

The Bogoliubov relations in the 2D EFT also allow us to calculate the cross section for any $S$-wave fermion-dyon scattering process, by evaluating amplitudes of the form $\langle A_\outs | B_\ins \rangle$.  The Bogoliubov transformation provides a succinct listing of the options for $\bra{A_\outs}$ that give nonzero amplitudes given a single-particle $\ket{B_\ins}$ that can be seen by expanding the incoming state $(a^{\ins}_{\mathfrak{s}, k})^{\star}$ in terms of $out$ operators -- see \pref{Boginfromout}. The presence of pair-production means that any such process can be accompanied by some number of spontaneously produced pairs.

For instance the $4$D chirality-changing amplitude 
for $f_{-}(k) \to f_{-}(k')$ accompanied by the emission of $n$ pairs is given by
\bea \label{- to -}
&&\bra{0_{\outs}}\overline{a}^{\outs}_{-, -q_n+ e v}{a}^{\outs}_{+, q_n}...\overline{a}^{\outs}_{-, -q_1+ e v}{a}^{\outs}_{+, q_1}\,{a}^{\outs}_{-, k'}({a}^{\ins}_{-, k})^{\star}\ket{0_{\ins}}  \\
&& \qquad  =  \delta({k- k'})\mathcal{T}^{--}_{\ins}\bra{0_{\outs}}\overline{a}^{\outs}_{-, -q_n+ e v}{a}^{\outs}_{+, q_n}...\overline{a}^{\outs}_{-, -q_1+ e v}{a}^{\outs}_{+, q_1}\ket{0_{\ins}}, \nn
\eea
where $n = 0,1,2,\cdots$. The  amplitude for 
charge-changing processes like $f_{\pm}(k) \to f_{\mp}(k')$ (accompanied by $n$ spontaneously produced pairs) is similarly 
\bea \label{- to +}
    &&\bra{0_{\outs}}\overline{a}^{\outs}_{-, -q_n+ e v}{a}^{\outs}_{+, q_n}...\overline{a}^{\outs}_{-, -q_1+ e v}{a}^{\outs}_{+, q_1}\,{a}^{\outs}_{-, k'}({a}^{\ins}_{+, k})^{\star}\ket{0_{\ins}} \\
    &&\qquad  =  \delta( k'+ e v -k)e^{-i \mfa}\cT^{-+}_{\ins}   \bra{0_{\outs}}\overline{a}^{\outs}_{-, -q_n+ e v}{a}^{\outs}_{+, q_n}...\overline{a}^{\outs}_{-, -q_1+ e v}{a}^{\outs}_{+, q_1}\ket{0_{\ins}}, \nn
\eea
and
\bea \label{+ to -}
    &&\bra{0_{\outs}}\overline{a}^{\outs}_{-, -q_n+ e v}{a}^{\outs}_{+, q_n}...\overline{a}^{\outs}_{-, -q_1+ e v}{a}^{\outs}_{+, q_1}\,{a}^{\outs}_{+, k'}({a}^{\ins}_{-, k})^{\star}\ket{0_{\ins}} \\
    &&\qquad  =  \delta({k'- e v- k})e^{i \mfa}\cT^{+-}_{\ins}  \bra{0_{\outs}}\overline{a}^{\outs}_{-, -q_n+ e v}{a}^{\outs}_{+, q_n}...\overline{a}^{\outs}_{-, -q_1+ e v}{a}^{\outs}_{+, q_1}\ket{0_{\ins}} \,.\nn
\eea
In particular the different voltages seen by the two charge states imply the reaction $f_+(k) \to f_-(k')$ (plus pair production) vanishes unless $k > ev$ for want of fermion final states with the required energy. 

The remaining reaction obtained with an initial incoming positively charged fermion is slightly more complicated. On one hand the $4$D chirality-flipping  
process $f_+(k) \to f_+(k')$ (plus the production of $n$ pairs) proceeds much as above. However when $k < ev$ the part of the Bogoliubov transformation relating $({a}^{\ins}_{+, k})^{\star}$ to $\ol{a}^{\outs}_{-, -k+ e v}$ also contributes to give an amplitude for spontaneously emitting $n+1$ pairs from the vacuum (with one of the pairs corresponding to a particle and antiparticle of momentum $k'$ and $- k+ e v$, respectively), leading to:
\bea\label{+ to +1}
 &&\bra{0_{\outs}}\overline{a}^{\outs}_{-, -q_n+ e v}{a}^{\outs}_{+, q_n}...\overline{a}^{\outs}_{-, -q_1+ e v}{a}^{\outs}_{+, q_1}\,{a}^{\outs}_{+, k'}({a}^{\ins}_{+, k})^{\star}\ket{0_{\ins}} \nn\\  
 &&\qquad = \delta(k- k')\mathcal{T}^{++}_{\ins}
 \bra{0_{\outs}}\overline{a}^{\outs}_{-, -q_n+ e v}{a}^{\outs}_{+, q_n}...\overline{a}^{\outs}_{-, -q_1+ e v}{a}^{\outs}_{+, q_1}\ket{0_{\ins}} \\ 
 && \qquad \quad+\Theta\left(-k + e v\right) e^{-i \mfa}\cT^{-+}_{\ins}\bra{0_{\outs}}\overline{a}^{\outs}_{-, -q_n+ e v}{a}^{\outs}_{+, q_n}...\overline{a}^{\outs}_{-, -q_1+ e v}{a}^{\outs}_{+, q_1}{a}^{\outs}_{+, k'}\ol{a}^{\outs}_{-, -k+ e v}\ket{0_{\ins}} .\nn
\eea
In 
equations \eqref{- to -}-\eqref{+ to +1}, the momenta $k', q_1, \cdots q_n$ are all distinct and satisfy $e v> q_1, \cdots q_n>0$ as well as $k, k'>0$. Similar formulae can be derived for the amplitudes with a single antiparticle in the initial state. 

The next step is to evaluate the pair production amplitudes appearing in \eqref{- to -}-\eqref{+ to +1} by using the 
$out$ particle content of the $in$ vacuum, as in appendix \ref{App: In and Out vacuum}. 
When counting pairs it is more convenient to switch to a discrete normalization for momentum eigenstates, as we now do, in which case the
maximum number of $out$ pairs in the $in$ vacuum, $N$, is a finite but large number\footnote{$N$ is defined as the maximum value of $\bra{0_{\ins}}N_{\rm out}\ket{0_{\ins}}= {e v }/{\Delta k}\,|\cT^{+-}_{\ins}|^2$ and so is given by $N = { e v}/{\Delta k}$. In the continuum limit, $N$ goes to infinity as the spacing $\Delta k$ between states vanishes (see Appendix \ref{App:Scattering states, new} for details).}.
The 
discrete normalization analogue of the 
amplitude for emitting $n$ pairs appearing in \pref{- to -} to \pref{+ to -} and the first line of the right-hand side of \eqref{+ to +1} evaluates to
\be\label{A pair}
\bm \cA^n_{\rm pair}=\bra{0_{\outs}}\overline{\bm a}^{\outs}_{-, -q_n+ e v}\bm{a}^{\outs}_{+, q_n}...\overline{\bm a}^{\outs}_{-, -q_1+ e v}{\bm a}^{\outs}_{+, q_1}\ket{0_{\ins}}=\left(\cT_{\ins}^{--}\right)^{N- n}\left(\cT_{\ins}^{+-}\right)^{ n} e^{i n \mfa},
\ee
for $n\le N$, and vanishes otherwise.\footnote{We denote transition amplitudes between discretely normalized states by $\bm \cA$ and their continuum normalization counterparts by $\cA$.} For $k < ev$ the  amplitude for emitting $n+1$ pairs encountered in the second line of \pref{+ to +1} similarly corresponds to
\bea\label{A pair2}
\bm \cA^{n+1}_{\rm pair}(k, k') &=& \bra{0_{\outs}}\overline{\bm a}^{\outs}_{-, -q_n+ e v}\bm{a}^{\outs}_{+, q_n}...\overline{\bm a}^{\outs}_{-, -q_1+ e v}\bm{a}^{\outs}_{+, q_1}\bm{a}^{\outs}_{+, k'}\ol{\bm a}^{\outs}_{-, -k+ e v}\ket{0_{\ins}} \nn\\
&=& -\delta_{k k'} \left(\cT_{\ins}^{--}\right)^{N- (n+1)}\left(\cT_{\ins}^{+-}\right)^{ n+1} e^{i (n+1) \mfa},
\eea
when $n<N$ and vanishes otherwise.  The two lines on the right-hand side of \eqref{+ to +1} then combine to become
\bea\label{+ to +1.1}
 &&\bra{0_{\outs}}\overline{\bm a}^{\outs}_{-, -q_n+ e v}\bm{a}^{\outs}_{+, q_n}...\overline{\bm a}^{\outs}_{-, -q_1+ e v}\bm{a}^{\outs}_{+, q_1}\,\bm{a}^{\outs}_{+, k'}(\bm{a}^{\ins}_{+, k})^{\star}\ket{0_{\ins}} \nn\\  
 &&\qquad = \delta_{k k'}\Bigl[\mathcal{T}^{++}_{\ins}\mathcal{T}^{--}_{\ins}-\cT^{-+}_{\ins}\cT^{+-}_{\ins}\Bigr]\left(\cT_{\ins}^{--}\right)^{N- (n+1 )}\left(\cT_{\ins}^{+-}\right)^{ n} e^{i n \mfa} ,
\eea
for $k< e v$  and $n<N$, when evaluated using discretely normalized states. The corresponding continuum normalization amplitudes have a very similar form.

Notice that the amplitudes \eqref{- to -}-\eqref{+ to -} factorize into a product of a single-particle transition amplitude, $\cA_{sc}$, from a particle with quantum numbers $\mathfrak{s}, k$ to one with quantum numbers $\mathfrak{s}', k'$, times  a product of pair-production (or vacuum survival) amplitudes for all the modes. The amplitude \eqref{+ to +1} factorizes in the same way for initial momenta $k> ev$, while for $k< e v$ the product over pair production (or vacuum survival) amplitudes runs over all but the $k$-th mode\footnote{For $k<e v$, the amplitude \eqref{+ to +1} factorizes differently than \eqref{- to -}-\eqref{+ to -}
because it describes two processes in which the number of produced pairs is not the same. 
We define  $\cA_{sc}$ for this process as in \eqref{ScatAmp05}  so that the single-particle amplitude captures the relevant contribution of \eqref{+ to +1} to inclusive observables.
}. 
The amplitudes factorize in this way because scattering and pair production for different modes are statistically independent. Inspection of the above formulae shows that the single-particle transition amplitudes $\mathcal{A}_{sc} [ f_{\mfs}(k) \to f_{\mathfrak{s}'}(k')]$ are given by
\be \label{ScatAmp01}
   \mathcal{A}_{sc} [ f_{-}(k) \to f_{-}(k') ] = \mathcal{T}^{--}_{\ins} \; \delta({k- k'}),  
\ee 
\be
   \mathcal{A}_{sc} [ f_{+}(k) \to f_{-}(k') ] = e^{-i \mfa}\cT^{-+}_{\ins}  \; \delta( k'+ e v -k) , 
\ee
\be \label{ScatAmp04}
 \mathcal{A}_{sc} [ f_{-}(k) \to f_{+}(k') ] = e^{i \mfa}\cT^{+-}_{\ins} \; \delta({k'- e v- k})\,,
\ee
and 
\be \label{ScatAmp05}
   \mathcal{A}_{sc} [ f_{+}(k) \to f_{+}(k') ]  =\Bigl[ \Theta(k- e v)\mathcal{T}^{++}_{\ins}+\Theta\left(-k + e v\right)\left[\cT^{++}_{\ins}\cT^{--}_{\ins}-\cT^{+-}_{\ins}\cT^{-+}_{\ins}\right]\Bigr] \, \delta(k- k') \,.
\ee
Notice that the unitarity constraints on the $\cT_{\ins}$ amplitudes given in \eqref{unitarityT1} and \eqref{unitarityT} imply that the expression $\cT^{++}_{\ins}\cT^{--}_{\ins}-\cT^{+-}_{\ins}\cT^{-+}_{\ins}$ simplifies to a phase. 

Low-energy scattering rates can now be computed much as in \S \ref{sec:PerturbativeScattering} by projecting any incoming plane wave onto the $S$-wave. 4D cross sections can then be computed by dividing by the appropriate incident particle flux. For instance, factoring out the energy-conserving delta function from amplitudes \pref{ScatAmp01}-\pref{ScatAmp04} as $\cA_{sc} = \cM \, \delta(E_f - E_i)$, the $4$D cross section for scattering with no pair production is
\be
  \exd \sigma^s_{\rm exclusive} \Bigl[ f_{\mfs\mfc}(k) \to  f_{\mfs' \mfc'}(k') \Bigr] = \frac{p_{s}}{\mfF_i} \; \exd  \Gamma_{\rm exclusive}  = \frac{\pi}{k^2} \, \delta_{\mfs, \mfc} \delta_{\mfs', -\mfc'} \, |\cM|^2\;|\braket{0_{\outs}|0_{\ins}}|^2 \, \delta(E_f-E_i) \, \exd k' \,,
\ee
where the probability for the plane wave to be found in an $S$-wave, $p_{s}$ and the initial particle flux $\mfF_i$ are defined as in \S \ref{sec:PerturbativeScattering} and are calculated in Appendix \ref{App:Scattering states, new} along with the 2D exclusive differential interaction rate $\exd\Gamma_{\rm exclusive}$. Similarly, the cross section for scattering with no additional pair production for the final amplitude \pref{ScatAmp05} is
\bea
  \exd \sigma^s_{\rm exclusive} \Bigl[ f_{+ +}(k) \to  f_{+ -}(k') \Bigr]  &=& \frac{\pi}{k^2}  \, |\cM|^2\;|\braket{0_{\outs}|0_{\ins}}|^2\, \delta(E_f-E_i) \, \exd k' \nn \\ && \qquad \qquad \quad \times\Bigl[\Theta(k- e v)+\Theta(-k+ e v)\left|\cT^{--}_{\ins}\right|^{-2}\Bigr] \,,
\eea
where $\cM$ is defined through $\cA_{sc} = \cM \, \delta(E_f - E_i)$ and the factor of $\left|\cT^{--}_{\ins}\right|^{-2}$ is cancelled by similar factors in the overlap $|\braket{0_{\outs}|0_{\ins}}|^2$, making the cross section finite even in the $\cT^{--}_{\ins} \to 0$ limit.

Of more practical interest are {\it inclusive} cross sections for which the number of associated pair productions is unmeasured and so marginalized over. Appendix \ref{App: In and Out vacuum} -- see the discussion below eq.~\pref{UnitaritySum} -- explicitly performs the marginalization over the number of produced pairs and shows that the resulting cross section 
can be expressed purely in terms of the single-particle scattering amplitudes $\cA_{sc}$, given in \eqref{ScatAmp01}-\eqref{ScatAmp05}. The ability to do so is a consequence of unitarity, and is also the reason the parameters $n$ and $N$ drop out of our final results. The inclusive 4D cross section becomes
\be\label{cross section}
 \exd \sigma_{s} \Bigl[ f_{\mfs\mfc}(k) \to  f_{\mfs' \mfc'}(k') \Bigr] = \frac{p_{s}}{\mfF_i} \; \exd  \Gamma_{\rm inclusive}  = \frac{\pi}{k^2} \, \delta_{\mfs, \mfc}\delta_{\mfs', -\mfc'} \, |\cM|^2 \, \delta(E_f-E_i) \, \exd k' \, \,,
\ee
where the 2D inclusive differential interaction rate $\exd\Gamma_{\rm inclusive}$ is defined in Appendix \ref{App:Scattering states, new}.

Combining results and integrating over the final momentum $k'$ leads to the total inclusive $S$-wave cross sections. For charge-exchange processes with $\mathfrak{s}=\mathfrak{c}$ we have
\be
 \sigma_s[ f_{\pm} \to f_{\mp} ]  = \frac{\pi}{k^2} \Bigl| \mathcal{T}^{+-}_{\ins} \Bigr|^2 \Theta(k - \mfs e v). 
\ee
Similarly for processes in which the $4$D chirality changes, but the charge doesn't the cross section when $\mathfrak{s}=\mathfrak{c}=-$ is
\be
   \sigma_s[ f_{-} \to f_{-} ] = \frac{\pi}{k^2} \Bigl| \mathcal{T}^{--}_{\ins} \Bigr|^2,
\ee
and 
\be
    \sigma_s[ f_{+} \to f_{+} ]  = \frac{\pi}{k^2}\Big[\Theta(k- e v)\Bigl|\mathcal{T}^{++}_{\ins}\Bigr|^2+\Theta\left(-k + e v\right)\Big],
\ee
when $\mathfrak{s}=\mathfrak{c}=+$.  These agree with the corresponding perturbative expressions when their domains of validity overlap, and are consistent with the cross section results given in \cite{Kazama:1976fm}, when we restrict to their choices:  $|\cT^{++}_{\ins}|=|\cT^{--}_{\ins}|=1$, $\cT^{+-}_{\ins}=\cT^{-+}_{\ins}=0$. 
 
These cross sections display catalysis inasmuch as they are independent of the scale $R \sim m_g^{-1}$ of the underlying classical dyon \cite{Preskill:1984gd, Rubakov:1988aq, Schnir}. They also do not depend directly on the dyon magnetic or electric charge (though there is a large Coulomb phase that drops out of the cross section that does see the dyon's electric charge). Their size scales like the unitarity bound $\sigma \propto 1/k^2$ whose origin comes purely from the projection of the incoming plane wave onto the $S$-wave that dominates at low energies. Finally the rates are directly controlled by the size of the effective couplings hidden within the $\mathcal{T}_{\ins}$ amplitudes, rather than through the more microscopic scales associated with the RG-invariants that would normally be needed once couplings are renormalized to remove the spurious $\epsilon$-dependence. In the present instance -- and just for the special kinematics of the monopole $S$-wave -- the existence of only a single solution to the radial equation implies that the dimensionless magnitudes $|\cT_\ins|$ themselves are already RG-invariant. 

\subsection{Effective dyon dynamics}
\label{ssec:RotorBO}

Having described fermion behaviour in the approximation where the `slow' degree of freedom $\mfa$ is a fixed background we here return to the question of how $\mfa$ responds to fermion scattering on longer time-scales. The leading dynamics of $\mfa$ is governed by the hamiltonian \pref{EffDyonH2}
\be \label{EffDyonH3}
  H_{\rm dyon} = \frac{1}{2\cI} \left[ \mfp(t) - \frac{\vartheta}{2\pi} \right]^2  +  \frac12  \Bigl[ \ol{\bm \chi} \, O_\cB(\mfa) \, {\bm \chi}  \Bigr]_{r = \epsilon,t}
\ee
where $\mfp$ is the conjugate momentum for $\mfa$ given in \pref{dyonmomischarge}, repeated here for convenience:
\be \label{mfaEOM}
  \mfp := 
  \frac{\vartheta}{2\pi}+ \cI \dot{\mfa} \,. 
\ee

\subsubsection{Charge conservation and rotor evolution}

In the Heisenberg picture the momentum $\mfp$ satisfies the equation of motion
\be \label{p EOM}
e \,\dot{\mfp}(t)=\frac{i}{2}\left(\ol{\bm{\chi}}\, \left[O_{\cB}(\mfa(t)), \frac{e}{2}\tau_3\right]\,\bm{\chi}\right)_{r=\epsilon}=j^1_\ssF(\epsilon, t) \,,
\ee
where $O_\cB(\mfa)$ is given in \pref{OBdef} and the last equality equating the result to the fermionic electromagnetic current flux is obtained by using the boundary condition \pref{Matrix2DBC} {\it c.f.}~eq.~\pref{boundarycurrents}: 
\be
\left[\Gamma^1+ O_{\cB}(\mfa(t))\right]\bm{\chi}(\epsilon, t)=0.
\ee
Eq.~\pref{p EOM} expresses conservation of total electric charge in the sense that it equates the change in the dyon charge $-e \mfp(t)$ to the radial flux of fermion electric charge $j^1_\ssF(\epsilon,t)$ evaluated near the dyon. 

Equation \eqref{p EOM} integrates to give
\be
\mfp(t)=\mfp(t_0) +\frac{1}{e}\int^t_{t_0} \exd t'j^1_\ssF(\epsilon, t'),
\ee
and this result can be used in \pref{mfaEOM} to evolve $\mfa$, leading to 
\bea
\mfa(t)&=&\mfa(t_0)+\frac{1}{\cI}\int^t_{t_0} \exd t'\left[\mfp(t_0)-\frac{\vartheta}{2\pi} +\frac{1}{e}\int^{t'}_{t_0} \exd t^{''}j^1_\ssF(\epsilon, t'')\right]\nn \\ 
&=&\mfa(t_0)+\frac{t-t_0}{\cI} \left[ \mfp(t_0) - \frac{\vartheta}{2\pi} \right] +\frac{1}{e \cI}\int^t_{t_0} \exd t'\int^{t'}_{t_0} \exd t^{''}j^1_\ssF(\epsilon, t'').
\eea
These expressions show how the rotor operator acquires a component that acts within the fermionic part of the Hilbert space for times $t > t_0$. 

Now comes the main point. We wish to use the above expressions to determine 
how the slow variable $\mfa$ evolves in response to its interactions with the relativistic fermion field. If we use how nuclei are handled within the Born-Oppenheimer approximation applied to atoms, the answer seems simple. For atoms one first computes the electronic state $|\Psi(\bfR) \rangle$ as a function of fixed classical nuclear positions, $\bfR$, and then determines the nuclear positions by minimizing the nuclear energy $V(\bfR) = \langle \Psi(\bfR) | H | \Psi (\bfR) \rangle$ given these atomic states. In the present instance the first step corresponds to computing the fermion state $\ket{\psi_i; \mfa}$ as a function of a classical initial value for $\mfa$. Then we compute 
an interaction Hamiltonian that captures the correct dynamics within the fermionic state $\ket{\psi_i; \mfa}$ and use it to find 
how the field $\mfa$ evolves.

Suppose we assume that the rotor and fermion sectors are initially unrelated to one another at $t = t_0$, at which point $\mfa(t_0) = \mfa_0$. The above reasoning suggests these slow-moving rotor degrees of freedom see only an average over the fast variables and so \pref{p EOM} is approximately given by
\be \label{p EOM BO}
  \dot{\mfp}_{\rm eff}(t) \simeq \frac{1}{e} \bra{\mfa_0; \psi_i}j^1_\ssF(\epsilon, t)\ket{\psi_i; \mfa_0} \,,
\ee
which integrates to
\be \label{mfpeffvst}
\mfp_{\rm eff}(t)=\mfp(t_0) +\frac{1}{e}\int^t_{t_0} \exd t'\bra{\mfa_0; \psi_i}j^1_\ssF(\epsilon, t')\ket{\psi_i; \mfa_0},
\ee
as well as
\be \label{mfaeffvst}
\mfa_{\rm eff}(t) = \mfa_0+\frac{t-t_0}{\cI}\left[ \mfp(t_0) - \frac{\vartheta}{2\pi} \right]  +\frac{1}{e \cI}\int^t_{t_0} \exd t'\int^{t'}_{t_0} \exd t^{''}\bra{ \mfa_0; \psi_i}j^1_\ssF(\epsilon, t'')\ket{\psi_i; \mfa_0}.
\ee

For instance, for fermions initially prepared in the vacuum state $\ket{\psi_i; \mfa_0}=\ket{0_{\ins}}$ eq.~\pref{2DEMFlux} gives a time-independent current expectation value
\be \label{p EOM BOvac}
  \dot{\mfp}_{\rm eff}(t) \simeq \frac{1}{e} \bra{0_\ins}j^1_\ssF(\epsilon, t)\ket{0_\ins} =\frac{e v }{2 \pi}|\cT^{+-}_{\ins}|^2 \,,
\ee
and so the time-evolution of $\mfa$ and its conjugate momentum $\mfp$ are approximately given by
\be \label{mfpeffvstvac}
\mfp_{\rm eff}(t)=\mfp(t_0)+\frac{e v (t-t_0)}{2 \pi}|\cT^{+-}_{\ins}|^2,
\ee
and
\be
\mfa_{\rm eff}(t) = \mfa(t_0)+\frac{t-t_0}{\cI}\left[ \mfp(t_0) - \frac{\vartheta}{2\pi} \right]  +  \frac{e v (t-t_0)^2}{4 \pi \cI}|\cT^{+-}_{\ins}|^2 \simeq \mfa_0 \,,
\ee
where the final approximate equality drops $\alpha$-supressed terms involving $\cI^{-1} \sim \alpha m_g$ (which also assumes $t - t_0$ is not too large). 

\subsubsection{Effective rotor hamiltonian}

One can ask: is there an effective rotor hamiltonian whose equations of motion have the same form as eqs.~\pref{mfaEOM} and \pref{p EOM}? Strictly speaking, such a hamiltonian need not exist because the rotor is an open quantum system once the fermion degrees of freedom are ignored. For such systems an effective hamiltonian only exists to the extent that there is a mean-field description for which fluctuations in the ignored degrees of freedom -- the `environment' -- can be neglected relative to the mean evolution (for a review of these issues, including a more precise statement of the mean-field criterion -- see \cite{EFTBook}).   

But even if an effective rotor hamiltonian exists, its description is likely not as simple as generating a potential $V(\mfa)$ for $\mfa$, which would be the analogy expected based on the Born-Oppenheimer description of the energetics of nuclear positions within an atom. In particular any such hamiltonian must produce nonzero $\dot \mfp$ that is independent of $\mfa$. But while it is true that Hamilton's equation $\dot \mfp = -\partial H/\partial \mfa$ seems to imply that nonzero $\dot \mfp$ requires $H$ to depend on $\mfa$ -- such as by adding a potential $V(\mfa)$ to $H$ -- there is also no choice for a potential satisfying $V(\mfa + 2\pi) = V(\mfa)$ that can produce a nonzero $\dot \mfp$ that is independent of $\mfa$. 

To explore what a successful choice for a rotor Hamiltonian would look like consider as a starting point the rotor dynamics implied by the Lagrangian of \pref{dyonEFT2}, repeated here (including the coupling to $\widehat A_0$)
\be \label{rotorEFT3}
 L_{\rm dyon} \ni \frac{\vartheta_{\rm eff}(t)}{2\pi} \Bigl(\dot \mfa - e \widehat A_0 \Bigr)  +   \frac{\cI}{2}  \Bigl(\dot \mfa - e \widehat A_0 \Bigr)^2 +  \cdots \,,
\ee
which writes the terms in order of dominance at low-energies. For later purposes we temporarily entertain the possibility that $\vartheta_{\rm eff} = \vartheta_{\rm eff}(t)$ is a specified function of time. 

Consider first keeping just the leading term,
\be \label{rotorEFT0}
  S_{\rm rotor} = \int_\ssW \exd t \; \frac{\vartheta_{\rm eff}(t)}{2\pi} \Bigl(\dot \mfa - e \widehat A_0 \Bigr)    \,,
\ee
in which case the canonical momentum and Hamiltonian are
\be
  \mfp := \frac{\delta S_{\rm rotor}}{\delta \dot\mfa} = \frac{\vartheta_{\rm eff}(t)}{2\pi} \quad \hbox{and} \quad
  H = \mfp \, \dot \mfa - L _{\rm rotor}= \frac{e \vartheta_{\rm eff}(t)}{2\pi} \, \widehat A_0   \,,
\ee
where the momentum equation can be regarded as a constraint. The Hamiltonian reveals the sole energy associated with this interaction to be the Coulomb energy due to the dyon acquiring an additional induced charge $-e\vartheta_{\rm eff}/2\pi$ (as expected from the Witten effect \cite{Witten:1979ey}). The evolution equation for $\mfp$ is then
\be
  \dot\mfp = \frac{\dot \vartheta_{\rm eff}}{2\pi}.
\ee

Interestingly, this equation does agree with \pref{p EOM BO} provided we identify
\be \label{thetamatch}
  \vartheta_{\rm eff}(t) \simeq \vartheta_0 + \frac{2\pi}{e} \int_{t_0}^t \exd t' \; \bra{\mfa_0; \psi_i}j^1_\ssF(\epsilon, t')\ket{\psi_i; \mfa_0} \,.
\ee
In the special case where $\ket{\psi_i; \mfa_0} = \ket{0_\ins}$ this becomes
\be \label{thetamatchvac}
  \vartheta_{\rm eff}(t) \simeq \vartheta_0 + ev |\cT^{+-}_{\ins}|^2 (t-t_0) \,.
\ee
The idea that fermion scattering might cause vacuum angle evolution was earlier discussed in \cite{Brennan:2021ewu}.

Extending the above to include the subdominant kinetic term for $\mfa$ appearing in \pref{rotorEFT3} -- and again entertaining the possibility that $\vartheta_{\rm eff} = \vartheta_{\rm eff}(t)$ is a function of time -- instead leads to the canonical momentum and Hamiltonian
\be \label{rotorHvsthetaeff}
  \mfp := \frac{\delta S}{\delta \dot\mfa} = \frac{\vartheta_{\rm eff}(t)}{2\pi} + \cI (\dot\mfa - e \widehat A_0) \quad \hbox{and} \quad
  H = \mfp \, \dot \mfa - L = e \mfp \, \widehat A_0 + \frac{1}{2\cI} \left(\mfp - \frac{\vartheta_{\rm eff}}{2\pi} \right)^2\,,
\ee
which shows how the previous momentum constraint $\mfp = \vartheta_{\rm eff}/(2\pi)$ emerges  as the momentum choice that minimizes the energy,  for fixed $e \widehat A_0$. 
To the extent that the rotor evolution minimizes its energy one expects
\be
  \dot\mfp \simeq \frac{\dot \vartheta_{\rm eff}}{2\pi}  \quad\hbox{and} \quad 
  \dot \mfa =  \frac{1}{\cI} \left( \mfp - \frac{\vartheta_{\rm eff}}{2\pi} \right) \simeq0 \,,
\ee 
which  agree with eqs.~\pref{mfaEOM} and \pref{p EOM BOvac} in the limit that $\cI^{-1}$ can be neglected (so $\dot \mfa \simeq 0$), when $\vartheta_{\rm eff}(t)$ is given by \pref{thetamatchvac}.  Although the rotor charge $\mfp$ evolves as it follows $\vartheta_{\rm eff}$ the minimized value of the energy remains unchanged.

The above picture apparently relies on $\mfa$ and $\mfp$ approximately behaving as slow classical variables so that $\mfp$ can evolve continuously with time as $\vartheta_{\rm eff}$ does. This is indeed a good approximation for the fermion energies $\alpha m_g \ll E \ll m_g$ for which the Born-Oppenheimer approximation applies. In a fuller quantum treatment the initial value $\mfa_0$ appearing in \pref{mfaeffvst} should be regarded as an operator satisfying $[\mfa_0 \,, \mfp_0] = i$, and commutes with the fermion degrees of freedom evaluated at $t_0$. The requirement that $\mfa_0$ also be a periodic variable with $\mfa_0 \sim \mfa_0 + 2\pi$ implies that its canonical momentum $\mfp_0$ is quantized with eigenvalues $\lambda_n = n$ where $n$ is an arbitrary integer. Evaluating the hamiltonian of \pref{rotorHvsthetaeff} as a function of $\mfp_{\rm eff}(t)$ given in \pref{mfpeffvstvac} with $\vartheta_{\rm eff}(t)$ given by \pref{thetamatchvac} then shows -- in the absence of $\widehat A_0$ -- that
\be
   H =  \frac{1}{2\cI} \left[\mfp_{\rm eff}(t) - \frac{ev}{2\pi} |\cT^{+-}_{\ins}|^2(t-t_0) -\frac{\vartheta_0}{2\pi}\right]^2 = \frac{1}{2\cI} \left(\mfp_{0} - \frac{\vartheta_0}{2\pi} \right)^2
\ee
and so has energy eigenvalues
\be
   E_n = \frac{1}{2\cI} \left( n - \frac{\vartheta_0}{2\pi} \right)^2 \,,
\ee
that are independent of time and quantized with step size of order $\cI^{-1} \sim \alpha m_g$ even as $\mfp_{\rm eff}(t)$ varies continuously.  

\subsection{Vacuum-angle evolution}

The upshot of the previous section is that -- somewhat surprisingly -- an effective Hamiltonian can exist that captures the rotor's evolution equations \pref{mfaEOM} and \pref{p EOM} if tracing out the fermion were to produce an effective contribution to the effective lagrangian of the form $\Delta L = \frac{1}{2\pi}\vartheta_{\rm eff}(t) D \mfa$ with $\vartheta_{\rm eff}(t)$ satisfying the matching condition \pref{thetamatch}. How might such an effective interaction actually be generated when explicitly integrating out the bulk fermion?\footnote{This issue is also discussed in \cite{Brennan:2021ewu}, though in a way that invokes a bulk coupling of $\mfa$ to fermions.}

An effective lagrangian of the form \pref{rotorEFT3} with a time-dependent $\vartheta_{\rm eff}$ might arise if the fermion-dyon interaction involves operators that contain one extra derivative relative to those in \pref{dyonEFT2}, such as
\be \label{DeltaLexpnew}
  \Delta L =  \cO(\ol\bmchi, \bmchi) \, D \mfa \,,
\ee
where $\cO$ is an operator built from the bulk fermion field and $D\mfa = \dot \mfa - e \widehat A_0$ 
as before. Taking the expectation value of this in the $in$ vacuum shows that $\langle \cO \rangle$ plays the role of $\vartheta_{\rm eff}/(2\pi)$ and this can be time-dependent (and calculable) if $\langle \cO \rangle$ is. In principle the operator we seek should satisfy
\be \label{Omatchcond}
  \partial_t \cO \simeq \frac{1}{e} j^1_\ssF(\epsilon, t') = - \frac{1}{2}j^0_\ssA(\epsilon, t')  \,.
\ee
in order to ensure that \pref{thetamatch} is true. 

This last condition suggests a guess for what the operator $\cO$ should be. In 1+1 dimensions our two Dirac fermions $\chi_\mfs$ can be bosonized into two real scalars $\phi_\mfs$, where the map between bosons and fermions implies -- see also \pref{boundarycurrents} 
\be \label{boundarycurrentsbosonized}
     j^\alpha_\ssA(\epsilon, t) \propto \partial^\alpha \phi_+ - \partial^\alpha \phi_- \,.
\ee
Comparing this to \pref{Omatchcond} suggests that the operator $\cO$ we seek has a simple expression in terms of the bosonized field: $\cO \propto \phi_+ - \phi_-$. If so then \pref{DeltaLexpnew} represents a dyon-localized kinetic mixing between $\phi_\mfs$ and $\mfa$. 

We have not yet found a convincing derivation of why $\Delta L$ of the form \pref{DeltaLexpnew} is generated once the fermions are integrated out, or why it arises with the right coefficients. But the above discussion reinforces earlier work \cite{Callan:1982ah, CallanSMatrix, Preskill:1984gd, Brennan:2021ewu} that suggests that the bosonized formulation of the scattered fermions might be more useful for understanding how the dyonic excitation $\mfa$ evolves.

\section{Conclusions}
\label{sec:Conclusions}

This paper identifies an EFT description of fermion-dyon interactions and uses this to compute several simple reaction rates in the presence of a dyon. This is done by adapting a framework -- point-particle effective field theories (PPEFTs) -- that are designed to describe how small compact sources interact with surrounding relativistic `bulk' fields. 

As typical for an EFT approach the starting point is an effective action -- in this case a world-line action that describes how bulk fields couple to one another and to dyon-localized degrees of freedom (such as collective coordinates). Such an action is organized as usual as a succession of operators of higher and higher dimension, with the lowest-dimension operators dominating at low energies. For the dyon case the important terms in this action governing the interactions with light fermions are given in \pref{dyonEFT1} and are not that remarkable since they resemble the kinds of terms that also arise in more prosaic examples (like those describing nuclei within atoms \cite{PPEFT3}).

The main novelty in PPEFT methods is an algorithm that maps this effective action onto a set of boundary conditions for the bulk fields as they approach the source. These boundary conditions depend explicitly on the couplings appearing in the effective action and this is ultimately how the news about the nature of the source reaches the observables of low-energy bulk physics. For fermions interacting with dyons this leads to the boundary conditions \pref{fermbcgen} or \pref{s-wave boundary condition 1}, which are also not that different from the kinds of things that appear for fermions interacting with other compact sources. 

The novelty of dyons and monopoles enters once one projects the bulk fields onto the $S$-wave mode that dominates the interactions at low energies since the dyon's magnetic charge gives this mode a distinctive angular dependence that is not shared by $S$-waves for sources without magnetic charges. The projection of the low-energy physics onto this $S$-wave allows the full 4D bulk physics to be described by an effective 2D bulk action that couples to a boundary situated outside of the underlying dyon field at $r = \epsilon$, with the effects of dyonic structure communicated through the boundary conditions at $r = \epsilon$ following from the PPEFT action. For dyons the $S$-wave projection become scale invariant in a way that excludes the emergence of a microscopic length scale $R_*$ on which low-energy observables can depend, leading to potentially enormous interaction cross sections that are not suppressed at all by powers of the physical size of the underlying source (unlike for nuclei in atoms). 

We show how the resulting dynamics of the bulk fermion coupled to the dyon's `rotor' mode captures the well-known fermion-dyon physics, but in a way that is very generally characterized by only a few parameters -- {\it e.g.}~the $\cT_\ins^{\mfs\mfs'}$ of \pref{TinvsC} -- whose values can be obtained by matching with microscopic physics for specific dyon configurations. Our general expressions for {\it e.g.}~scattering as a function of the $\cT_\ins^{\mfs\mfs'}$ reduce to those of the literature once these matched values are used. These expressions also capture how scattering includes the nonperturbative effects of fermion-rotor interactions as studied in \cite{Polchinski}, which can be regarded as providing a more complicated matching prescription to the microscopic physics that changes the values found for the $\cT_\ins^{\mfs\mfs'}$ but not the expressions for how observables depend on these parameters.

We also explore how the fermion scattering causes the dyonic excitions to evolve and identify an effective hamiltonian that captures the dynamics required by the model's conservation laws at low energy. Although we do not yet have a microscopic derivation of this Hamiltonian we explore several preliminary options.

\section*{Acknowledgements}
We thank Peter Hayman, Markus Rummel and Laszlo Zalavari for helpful conversations. This research was partially supported by funds from the Natural Sciences and Engineering Research Council (NSERC) of Canada. Research at the Perimeter Institute is supported in part by the Government of Canada through NSERC and by the Province of Ontario through MRI.  

\changelocaltocdepth{1}
\begin{appendix}

\section{Gamma-matrix conventions}
\label{App:GammaConventions}

This appendix summarizes our Dirac matrix conventions in both four and two dimensions.

\subsection{4D Dirac matrices}

We follow Weinberg's spinor and metric conventions, so our metric has signature $(-,+,+,+)$ and a Weyl representation for the gamma matrices in four dimensions is given by
\be 
    \gamma^0 = -i\,\left(\begin{array}{ccc}
0 & & I \\
I && 0
\end{array}\right),\quad \gamma^j = \left(\begin{array}{ccc}
0 & & -i \sigma_j \\
i  \sigma_j && 0
\end{array}\right),
\ee
where  $I$ is the $2$ by $2$ unit matrix and $\sigma_j$ are Pauli matrices (acting in spin space, as opposed to the Pauli matrices $\tau_a$ acting on gauge doublet indices). These satisfy the Clifford algebra $\left\{\gamma^{\mu}, \gamma^{\nu}\right\}=2 \eta^{\mu \nu}$ where $\eta^{\mu \nu}$ is the inverse Minkowski metric, and the representation is chosen to diagonalize 
\be 
   \gamma^5=-i \gamma^0 \gamma^1 \gamma^2 \gamma^3 = \left(\begin{array}{ccc}
I & & 0 \\
0 && -I
\end{array}\right) \,.
\ee

Dirac conjugation in these conventions is given by $\overline{\psi}=i \psi^{\dagger}\gamma^0$. In these conventions the left- and right-handed chirality projectors are $\gamma_\ssL := \frac12(1 + \gamma_5)$ and $\gamma_\ssR := \frac12(1 - \gamma_5)$.

\subsubsection*{Spherical coordinates}

The gamma matrices adapted to spherical coordinates are given in terms of the unit coordinate vectors 
\bea 
    \hat {\bm r} := \hat r_i \, \bfe_i &=& \sin \theta\cos{\phi}\, \bfe_x + \sin \theta \sin \phi \, \bfe_y + \cos \theta \, \bfe_z,\nonumber\\
    \hat {\bm \theta} := \hat{\theta}_i \, \bfe_i &=& \cos \theta\cos{\phi}\,\bfe_x + \cos \theta \sin \phi \,\bfe_y - \sin \theta \,\bfe_z,\nonumber\\
 \hat {\bm \phi} := \hat{\phi}_i \, \bfe_i &=& -\sin {\phi}\,\bfe_x + \cos \phi\, \bfe_y .
\eea
by 
$\gamma^r := \gamma^i \,\hat r_i$, $\gamma^\theta := \gamma^i \,\hat \theta_i$ and $\gamma^\phi := \gamma^i \, \hat \phi_i/\sin \theta$, 
and so
\be 
\gamma^{r} 
=\left(\begin{array}{ccc}
0 & & -i\sigma^r \\
i\sigma^r && 0
\end{array}\right) \,, \quad
\gamma^{\theta}
 = \left(\begin{array}{ccc}
0 & & -i\sigma^{\theta} \\
i\sigma^{\theta} && 0
\end{array}\right) \quad \hbox{and} \quad
\gamma^{\phi}
  = \frac{1}{\sin \theta}\,\left(\begin{array}{ccc}
0 && -i \sigma^{\phi} \\
i\sigma^{\phi} && 0
\end{array}\right).
\ee
where the Pauli matrices in spherical coordinates are defined by 
\be 
    \sigma^r = \left(\begin{array}{cc}
\cos \theta &  e^{-i \phi}\sin \theta \\
e^{i \phi}\sin \theta & -\cos \theta 
\end{array}\right) , \; \sigma^{\theta}= \left(\begin{array}{cc}
-\sin \theta &  e^{-i \phi}\cos \theta \\
e^{i \phi}\cos \theta & \sin \theta 
\end{array}\right)  \; \hbox{and} \;\; \sigma^{\phi} = \left(\begin{array}{cc}
0 &  - i e^{-i \phi} \\
i e^{i \phi} & 0
\end{array}\right).
\ee

\subsection{2D Dirac matrices}

In $D = 1+1$ spacetime dimensions we label coordinates with $x^\alpha = \{ t, r \}$ for $\alpha = 0,1$ and use the following representation of gamma matrices 
\be 
    \Gamma^0 = -i \sigma_1, \quad \Gamma^1= \sigma_2  \,.
\ee
These satisfy the algebra $\{\Gamma^{\alpha}, \Gamma^{\beta}\}=2 \eta^{\alpha \beta}$, where $\eta^{\alpha \beta} = \hbox{diag}(-1,1)$ is the inverse Minkowski metric in $1+1$ dimensions. The chiral matrix in 1+1 dimensions we then define to be
\be 
    \Gamma_c := \Gamma^0 \Gamma^1= \sigma_3 \,,
\ee
which has eigenvalues $\pm 1$ (and is diagonal in the basis used here). Notice that these definitions imply $\Gamma_c \Gamma^\alpha = \epsilon^{\alpha\beta} \Gamma_\beta$ where our Levi-Civita convention chooses $\epsilon^{01} = +1$. Dirac conjugation is again given by $\overline{\chi}=i \chi^{\dagger}\Gamma^0$ and in these conventions the left- and right-handed helicity projectors are $\Gamma_+ := \frac12(1 + \Gamma_c)$ and $\Gamma_- := \frac12(1 - \Gamma_c)$.

\section{PPEFTs and relativistic fermions}
\label{App:PPEFT3}

This Appendix summarizes the results of \cite{PPEFT3} describing how PPEFTs couple to relativistic fermions. Formally we wish to couple a Dirac fermion to a point-like source located at the origin
\be \label{DiracBaction}
 S = - \int \exd^4 x \Bigl[ \psibar (\Dsl + m) \psi +   \psibar N \psi   \, \delta^3(x) \Bigr]   \,,
\ee
where $N$ is a Dirac matrix. For definiteness in this Appendix we follow \cite{PPEFT3} and consider the rotational and parity invariant case $N = \hat\mfc_s + i \hat\mfc_v \gamma^0$ where $\hat\mfc_s$ and $\hat\mfc_v$ are coupling constants (but more general possibilities are entertained in the main text). The $\psi$ equation of motion including the coupling to the source is
\be \label{DiracBeom2}
  (\Dsl + m) \psi +  N \psi   \, \delta^3(x) = 0  \,.
\ee

Formally we'd like to trade the delta function for a near-source boundary condition, and following usual practice this would be obtained by integrating \pref{DiracBeom2} over a small Gaussian pillbox, $P$, of radius $\epsilon$ centred on the source. This gives, in the limit $\epsilon \to 0$ of vanishingly small pillbox, the result
\be \label{App:PillboxIntegrated}
   \int_{\partial P} \exd^2 x   \; n_\mu \gamma^\mu \psi =  \int \exd^2\Omega \, \epsilon^2  \;  \gamma^r \,\psi \simeq-N \psi(0) \,,
\ee
where the solid-angle measure is $\exd^2\Omega = \sin\theta \, \exd \theta \exd \phi$ and $n_\mu$ is an outward-pointing unit normal to the pillbox so in polar coordinates $n_\mu \exd x^\mu = \exd r$ and $\psi(0)$ denotes the value of the field at the position of the source. The final approximate equality drops the $m \psi$ term, as is appropriate for a sufficiently small pillbox since this vanishes as $\epsilon \to 0$ provided $\psi$ is sufficiently smooth near the origin. Our conventions on gamma-matrices in polar coordinates are given in Appendix \ref{App:GammaConventions}.

\subsection{Boundary action}

The problem with the formal argument is that bulk fields like $\psi$ are typically {\it not} smooth as $r \to 0$, which both complicates the neglect of the $m \psi$ term when integrating \pref{DiracBeom2} over the pillbox and makes $\psi(0)$ undefined. The PPEFT way for dealing with both of these issues is essentially to regulate the source action by replacing it by a boundary action on the boundary of the pillbox $\partial P$ at $r = \epsilon$. For configurations that are spherically symmetric\footnote{Non-spherically symmetric configurations can also be handled by decomposing into spherical harmonics and treating each harmonic separately on $\partial P$.} very near the source this is particularly simple to do by replacing the world-line action by its value integrated over $\partial P$:
\be
   \int_{r=0} \exd t \; \ol \psi N \psi \to \frac{1}{4\pi \epsilon^2} \int_{\partial P} \exd^2\Omega \,\exd t \; \epsilon^2\, \ol \psi N \psi \,.
\ee
This procedure makes the replacement $N\psi(0) \to (4\pi \epsilon^2)^{-1} \int \exd^2\Omega\, \epsilon^2 N(\epsilon) \psi(\epsilon)$ on the right-hand side of \pref{App:PillboxIntegrated}, leading (in the limit where $\epsilon$ is much smaller than all other scales of interest) to the regulated boundary condition
\be \label{diracBC0}
  \int_{r = \epsilon} \exd^2 \Omega \;  \left[ \gamma^r + \frac{N}{4\pi\epsilon^2 } \right] \psi  
  = \int_{r = \epsilon} \exd^2 \Omega \;  \left[ \gamma^r + \frac{1}{4\pi\epsilon^2 }\Bigl( \hat \mfc_s + i \hat \mfc_{v} \gamma^0 \Bigr) \right] \psi  = 0 \,.
\ee
Notice this boundary condition is trivially satisfied if we'd tried to make the same derivation using a pillbox that does {\it not} contain the source position. This is because in this case the $N \psi$ term is no longer present and so the boundary condition states $\int \exd^2\Omega \, \gamma^r \psi = 0$. Since $\psi$ varies very slowly in a small enough region not containing the source, it can be taken to be approximately constant across the pillbox and so the integral over all directions of $\gamma^r$ gives zero trivially without restricting $\psi$. 

Returning to the case where the pillbox does enclose the source, the boundary condition \pref{diracBC0} can be written as $\int \exd^2 \Omega \; B_\epsilon \, \psi(\epsilon) = 0$ where
\be
 B_\epsilon :=  \gamma^r  + \frac{N}{4\pi \epsilon^2}  
   =   \gamma^r  + \hat \cC_s  + i\hat \cC_v \gamma^0  
   =  \left( \begin{array}{cc}
 \hat \cC_s & \hat \cC_v  -i \sigma^r    \\
  {}  \hat \cC_v +i  \sigma^r  & \hat \cC_s \end{array}\right) \,,
\ee
where $\hat \cC_s = \hat \mfc_s/(4\pi\epsilon^2)$ and $\hat \cC_v = \hat \mfc_{v}/(4\pi\epsilon^2)$ are now dimensionless effective couplings. The subscript $\epsilon$ on $B_\epsilon$ is meant to emphasize that the constants $\hat \cC_a$ (and in general also the original couplings $\hat\mfc_i$ themselves) also must carry an implicit $\epsilon$-dependence if physical quantities are to remain unchanged as $\epsilon$ is varied (more about which below). In terms of the left- and right-handed parts of $\psi$ the boundary condition becomes
\be \label{2partbcv0}
 - \hat \cC_s \int_\epsilon \exd^2 \Omega \,   \psi^\pm_\ssL= \int_\epsilon \exd^2\Omega \,  \Bigl(   \hat \cC_v   -i  \sigma^r  \Bigr) \psi^\pm_\ssR \quad \hbox{and} \quad
     - \int_\epsilon \exd^2\Omega \,  \Bigl(  \hat \cC_v   +i  \sigma^r \Bigr)\, \psi^\pm_\ssL = \hat \cC_s  \int_\epsilon \exd^2\Omega \,   \psi^\pm_\ssR  \,. 
\ee

To see what these boundary conditions imply, imagine solving the bulk equation $(\Dsl + m)\psi = 0$ for $r > \epsilon$ and decomposing the result into rotation and parity eigenstates. The parity-even solutions are
\be \label{bulk+2}
 \psi^+ =  \left( \begin{array}{c}  \psi^+_\ssL \\  \psi^+_\ssR  \end{array}  \right)  =  \left( \begin{array}{c}  f_+(r) \,U^+(\theta,\phi) +i g_+(r) \,U^-(\theta,\phi) \\  f_+(r) \,U^+(\theta,\phi) -i g_+(r) \,U^-(\theta,\phi) \end{array}  \right)  \,,
\ee
while the parity-odd ones are
\be \label{bulk-}
 \psi^- =  \left( \begin{array}{c}  \psi^-_\ssL \\  \psi^-_\ssR  \end{array}  \right)  =  \left( \begin{array}{c}  f_-(r)\, U^-(\theta,\phi) +i g_-(r)\, U^+(\theta,\phi) \\  f_-(r)\, U^-(\theta,\phi) -i g_-(r)\, U^+(\theta,\phi) \end{array}  \right)  \,,
\ee
where $U^\pm$ are the spinor harmonics that combine the particle's spin-half with orbital angular momenta $\ell = j \mp \frac12$ to give total angular momentum $j = \frac12, \, \frac32, \cdots$. The radial functions $f_\pm(r)$ and $g_\pm(r)$ with mode frequency $\omega$ solve the radial equations 
\be \label{fgpluseqstxtApp}
  f_+' = \left( m + \omega  \right) g_+ \quad \hbox{and} \quad
  g_+' + \frac{2g_+}{r} = \left( m - \omega   \right) f_+ \,,
\ee
together with 
\be \label{fgminuseqstxt}
  g_-' = \left( m - \omega  \right) f_- \quad \hbox{and} \quad
  f_-' + \frac{2f_-}{r} = \left( m + \omega   \right) g_- \,,
\ee
where primes denote differentiation with respect to $r$. The boundary conditions \pref{2partbcv0} fix the ratio of the functions $f$ and $g$ at $r = \epsilon$ once the angular integrations are performed, giving
\be  \label{cscvtofgApp}
  \hat \cC_s +   \hat \cC_v    = \left( \frac{  g_+ }{f_+} \right)_{r=\epsilon} \qquad  \hbox{and} \qquad
  \hat \cC_s -   \hat \cC_v    = \left( \frac{ f_-}{g_-}  \right)_{r=\epsilon} \,.
\ee

\subsection{RG evolution}
\label{sec:RG}

Eq.~\pref{cscvtofgApp} provides the solution to how the properties of the source influence the bulk solutions for $\psi$ in the source's vicinity. Given the general solution, 
\be \label{fgvsAC2}
 f_\pm(r) = C^\pm_1 f_{1\pm}(r) + C^\pm_2 f_{2\pm}(r) \qquad \hbox{and} \qquad g_\pm(r) = C^\pm_1 g_{1\pm}(r) + C^\pm_2 g_{2\pm}(r) \,,
\ee
to the radial part of the Dirac field equation we see that \pref{cscvtofgApp} show that the couplings $\hat \cC_s$ and $\hat \cC_v$ determine the ratios of integration constants $C^+_2/C^+_1$ and $C^-_2/C^-_1$ that specify $(g_\pm/f_\pm)_{r=\epsilon}$. Energy levels for states of either parity and scattering amplitudes are then determined by the values of $C^\pm_2/C^\pm_1$. 

But it is still a potential puzzle why physical predictions can depend on the radius, $r=\epsilon$, of the Gaussian pillbox which is not a physical scale (arising just as a way to regularize the boundary conditions). The precise value of $\epsilon$ must therefore drop out of predictions for observables (unlike the physical size, $R$, of the underlying source, say). In detail, this happens because any explicit $\epsilon$-dependence arising in a calculation of an observable cancels an implicit $\epsilon$-dependence buried within the `bare' quantities $\hat \mfc_s$ and $\hat \mfc_v$. Physical predictions remain $\epsilon$-independent if $\hat \mfc_s(\epsilon)$ and $\hat \mfc_v(\epsilon)$ are chosen to ensure the ratios $C^\pm_2/C^\pm_1$ are held fixed as $\epsilon$ is varied. 

This gives us another way to interpret eq.~\pref{cscvtofgApp}. Rather than reading \pref{cscvtofgApp} as fixing $f_\pm/g_\pm$ at a specific radius given known values of $\epsilon$, $\hat \mfc_s$ and $\hat \mfc_v$ we can instead read the equations
\begin{equation} \label{csvmatch}
 \hat \mfc_s(\epsilon) = \left[  \frac{g_+(\epsilon)}{f_+(\epsilon)} + \frac{f_-(\epsilon)}{g_-(\epsilon)} \right] 2\pi \epsilon^2 \quad \hbox{and} \quad
 \hat \mfc_{v}(\epsilon) = \left[  \frac{g_+(\epsilon)}{f_+(\epsilon)} - \frac{f_-(\epsilon)}{g_-(\epsilon)} \right] 2\pi \epsilon^2 \,,
\end{equation}
as telling us how $\hat \mfc_s(\epsilon)$ and $\hat \mfc_{v}(\epsilon)$ must depend on $\epsilon$ in order to ensure that $C^\pm_2/C^\pm_1$ remains $\epsilon$-independent. Since we choose $\epsilon$ much smaller than the typical scale of the external problem (such as the Bohr radius, for applications to atoms), it suffices to use the leading small-$r$ form of the solutions $f_\pm$ and $g_\pm$ when using \pref{csvmatch}. In this regime solutions are usually well described by power laws, with \pref{fgvsAC2} reducing to
\be \label{fgvsACsmallrApp}
 f_\pm(r) = C^\pm_1 \left( \frac{r}{a} \right)^{\mfz-1} + C^\pm_2 \left( \frac{r}{a} \right)^{-\mfz-1} \qquad \hbox{and} \qquad g_\pm(r) = \widetilde{C}^\pm_1  \left( \frac{r}{a} \right)^{\mfz-1} + \widetilde{C}^\pm_2 \left( \frac{r}{a} \right)^{-\mfz-1} \,,
\ee
with $\widetilde{C}^{\pm}_i \propto C^{\pm}_i$ in a way that depends on the relative small-$r$ asymptotic behaviour of $f_i(r)$ and $g_i(r)$. For such solutions the choice of $C^\pm_2/C^\pm_1$ controls the precise radius at which one of these solutions dominates the other one, and as a result the RG evolution of the couplings implied by \pref{csvmatch} in this regime describes the cross-over between these two types of evolution.

\section{Regularized codimension-1 boundary action}
\label{App:Codimension1Action}

This Appendix addresses the question of how to regularize the fermion-dyon interactions defined  on the dyon world-line, which appear in equations such as \eqref{dyonEFT2}.

As argued in the main text, the lowest-dimension interactions between the fermion doublet $\bmpsi(x)$ and the dyon collective coordinates $y^{\mu}(t), \mfa(t)$ are given by 
{
\small
\bea \label{fermion-dyon worldline interactions}
S^{\rm int}_{\rm dyon}&=&-\frac{1}{2}\int \exd t \,\ol{\bm \psi}\, \mfC(\mfa) \,\bm \psi\nn \\ &=&  -\frac{1}{2}\int \exd t\,\ol{\bm \psi}\Bigl[\hat\mfc^s_{1} + i\, \hat\mfc^{ps}_1\gamma_5 - i\, \hat\mfc^{v}_1 \gamma^{0} - i\, \hat\mfc^{pv}_1 \gamma_5\gamma^{0}   + \Bigl(\hat\mfc^{s}_3 + i \,\hat\mfc^{ps}_3\gamma_5 - i \,\hat\mfc^{v}_3\gamma^{0} - i\, \hat\mfc^{pv}_3 \gamma_5\gamma^{0} \Bigr) \tau_3  \\
&& \quad  +\Bigl( \hat\mfc^s_{+} + i\, \hat\mfc^{ps}_+\gamma_5 - i\, \hat\mfc^{v}_+ \gamma^{0}  - i\, \hat\mfc^{pv}_+ \gamma_5\gamma^{0}  \Bigr) e^{i\mfa} \tau_+   +  \Bigl(\hat\mfc^{s}_- + i \,\hat\mfc^{ps}_-\gamma_5 - i \,\hat\mfc^{v}_-\gamma^{0} - i\, \hat\mfc^{pv}_- \gamma_5\gamma^{0} \Bigr) e^{-i\mfa} \tau_- \Bigr]\, \bm \psi \nn 
\eea
}
\noindent where $\bmpsi$ is evaluated at $\bm{r}=0$ and we specialize to the dyon rest frame and neglect dyon recoil effects so that $\dot{y}^{\mu}\approx \delta^{\mu}_0$.
We can  regulate the operators appearing in $S^{\rm int}_{\rm dyon}$ by replacing them with appropriate interaction terms, defined on the boundary of a Gaussian pillbox at $r=\epsilon$. Appendix \ref{App:PPEFT3} shows that for fermion field configurations which are spherically symmetric near the source,  the regularization procedure amounts to replacing fermion bilinears such as $\ol{\bm{\psi}} M \bmpsi$, where $M$ is a matrix in spin and isospin space, with  their average  over the  pillbox boundary $ ({4\pi \epsilon^2})^{-1}\int \exd^2\Omega\, \epsilon^2\,\ol{\bm{\psi}}(\epsilon) {M} \bmpsi(\epsilon)$. This is a valid prescription for operators in \eqref{fermion-dyon worldline interactions} describing $\psi_+, \psi_-$ self-interactions (mediated by $\hat \mfc_1, \hat \mfc_3$). For operators that couple $\psi_+$ and $\psi_-$ we must use another prescription however,  since the angular dependence of the two components of the doublet is different, even when restricting to the same partial wave.  As a result, regularized fermion bilinears such as $\ol{\bm{\psi}}(\epsilon)\, \hat \mfc^s_+ \tau_+\bmpsi(\epsilon)$ are generally not rotation-invariant and so can vanish after integration over the boundary of the pillbox. This can be  seen explicitly by specializing to $S$-wave states for which we get \textit{e.g.}:
\be
\frac{1}{4\pi \epsilon^2}\int \exd^2\Omega\, \epsilon^2\,\ol{\bm{\psi}}(\epsilon)\, \hat \mfc^s_+ \tau_+\bmpsi(\epsilon)=\frac{f^*_+(\epsilon,t)\, g_- (\epsilon,t)+g^*_+ (\epsilon,t)\,f_-(\epsilon,t)}{4\pi \epsilon^2}\hat \mfc^s_+\int {\exd^2\Omega}\,\eta^{\dagger}_+ \, \eta_-=0,
\ee
in the gauge where the Julia-Zee solution has the form \eqref{Julia-Zee dyon abelian gauge} and $S$-wave fermions are given by \eqref{s wave fermions}. That the  form of angular momentum eigenstates depends on their electric charge can be traced back to  the expression for the angular momentum operator $\vec{J}$, which includes a contribution from the gauge isospin $\vec{T}$, as in \eqref{Def: Angular momentum}. 

To couple the two components of the doublet at the $r=\epsilon$ boundary, it suffices to introduce additional matrices $P_{\pm}$, which turn the angular dependence of $\psi_+$ into that of $\psi_-$ and vice versa. For $S$-wave states and in the `abelian' gauge of \eqref{Julia-Zee dyon abelian gauge}, these matrices act in spin space and are defined through:
\be \label{Def: P_+, P_- 1}
P_+\, \frac{1}{r} \begin{pmatrix}
    f_+(r, t) \,\eta_+\\ g_+(r, t) \, \eta_+
\end{pmatrix}= \frac{1}{r} \begin{pmatrix}
    f_+(r, t) \, \eta_-\\ g_+(r, t) \, \eta_-
\end{pmatrix}, \quad \text{and} \quad P_-\, \frac{1}{r} \begin{pmatrix}
    f_-(r, t) \,\eta_-\\ g_-(r, t) \,\eta_-
\end{pmatrix}= \frac{1}{r} \begin{pmatrix}
    f_-(r, t) \,\eta_+\\ g_-(r, t) \,\eta_+
\end{pmatrix}.
\ee
Since these equations do not uniqely determine $P_{\pm}$, we further choose:
\be \label{Def: P_+, P_- 2}
P_-\, \frac{1}{r} \begin{pmatrix}
    f_+(r, t) \,\eta_+\\ g_+(r, t) \, \eta_+
\end{pmatrix}=  \begin{pmatrix}
   0\\ 0
\end{pmatrix}, \quad \text{and} \quad P_+\, \frac{1}{r} \begin{pmatrix}
    f_-(r, t) \,\eta_-\\ g_-(r, t) \,\eta_-
\end{pmatrix}= \begin{pmatrix}
    0 \\ 0
\end{pmatrix},
\ee
which leads us to the following expressions for $P_{\pm}$ in the $R_-$ patch
\be \label{P+-}
P_+=-\frac{i e^{ i \phi}}{2}\begin{pmatrix}
    i \sigma_{\theta}+\sigma_{\phi} && 0\\ 0 &&   i \sigma_{\theta}+\sigma_{\phi}
\end{pmatrix}, \quad \text{and} \quad P_-=-\frac{i e^{- i \phi}}{2}\begin{pmatrix}
    i \sigma_{\theta}-\sigma_{\phi} && 0\\ 0 &&   i \sigma_{\theta}-\sigma_{\phi}
\end{pmatrix},
\ee
as well as $P'_{\pm}\coloneqq e^{\mp 2 i \phi}P_{\pm}$ in the $R_+$ patch. It follows that for $S$-wave states, the world-line interactions in \eqref{fermion-dyon worldline interactions} can be regulated using the following boundary action:
{
\small
\bea \label{fermion-dyon boundary interactions}
I^{\rm int}_{\rm dyon}&=& -\frac{1}{2}\underset{r=\epsilon}{\int} \exd t\,\exd^2 \Omega\, \epsilon^2 \,\ol{\bm \psi}\Bigl[\hat{\cC}^s_{1} + i\, \hat{\cC}^{ps}_1\gamma_5 - i\, \hat{\cC}^{v}_1 \gamma^{0} - i\, \hat{\cC}^{pv}_1 \gamma_5\gamma^{0}   + \Bigl(\hat{\cC}^{s}_3 + i \,\hat{\cC}^{ps}_3\gamma_5 - i \,\hat{\cC}^{v}_3\gamma^{0} - i\, \hat{\cC}^{pv}_3 \gamma_5\gamma^{0} \Bigr) \tau_3  \\
&& \quad  +\Bigl( \hat{\cC}^s_{+} + i\, \hat{\cC}^{ps}_+\gamma_5 - i\, \hat{\cC}^{v}_+ \gamma^{0}  - i\, \hat{\cC}^{pv}_+ \gamma_5\gamma^{0}  \Bigr) e^{i\mfa} P_-\,\tau_+   +  \Bigl(\hat{\cC}^{s}_- + i \,\hat{\cC}^{ps}_-\gamma_5 - i \,\hat{\cC}^{v}_-\gamma^{0} - i\, \hat{\cC}^{pv}_- \gamma_5\gamma^{0} \Bigr) e^{-i\mfa} P_+\,\tau_- \Bigr]\, \bm \psi \nn
\eea
}
where we introduce the boundary couplings $\hat{\cC}^A_I=\frac{1}{4\pi \epsilon^2}\,\hat \mfc^A_I$. The boundary condition satisfied by $S$-wave fermions is then given by
{\small
\bea 
  && 0=\Bigl\{\Bigl[ \gamma^r  + \Bigl( \hat{\cC}^s_{1} + i\, \hat{\cC}^{ps}_1\gamma_5 - i\, \hat{\cC}^{v}_1 \gamma^{0} - i\, \hat{\cC}^{pv}_1 \gamma_5\gamma^{0}  \Bigr) + \Bigl(\hat{\cC}^{s}_3 + i \,\hat{\cC}^{ps}_3\gamma_5 - i \,\hat{\cC}^{v}_3\gamma^{0} - i\, \hat{\cC}^{pv}_3 \gamma_5\gamma^{0} \Bigr) \tau_3 \\
&& \quad  +\Bigl( \hat{\cC}^s_{+} + i\, \hat{\cC}^{ps}_+\gamma_5 - i\, \hat{\cC}^{v}_+ \gamma^{0}  - i\, \hat{\cC}^{pv}_+ \gamma_5\gamma^{0}  \Bigr) e^{i\mfa} P_-\,\tau_+   +  \Bigl(\hat{\cC}^{s}_- + i \,\hat{\cC}^{ps}_-\gamma_5 - i \,\hat{\cC}^{v}_-\gamma^{0} - i\, \hat{\cC}^{pv}_- \gamma_5\gamma^{0} \Bigr) e^{-i\mfa}P_+\, \tau_-  \Bigr] \bmpsi \Bigr\}_{r=\epsilon}, \nonumber
\eea
}
\noindent instead of \eqref{s-wave boundary condition 1} and is equivalent to the $2$D boundary condition \eqref{1+1d:s-wave boundary condition 1} when the $4$D and $2$D boundary couplings are related as follows:
\be
\cC^s_{\mfs \mfs}=\hat{\cC}^s_1 + \mfs \hat{\cC}^s_3,\quad  \cC^{ps}_{\mfs \mfs}=\mfs \hat{\cC}^{ps}_1 +  \hat{\cC}^{ps}_3, \quad \cC^v_{\mfs \mfs}=\hat{\cC}^v_1 + \mfs \hat{\cC}^v_3,\quad  \cC^{pv}_{\mfs \mfs}=-(\mfs \hat{\cC}^{pv}_1 +  \hat{\cC}^{pv}_3)
\ee
as well as
\be
\cC^s_{+-}=-\hat{\cC}^v_{+},\quad  \cC^{ps}_{+ -}= i\hat{\cC}^{pv}_{+}, \quad \cC^v_{+-}=-\hat{\cC}^s_+,\quad  \cC^{pv}_{+-}= i \hat{\cC}^{ps}_+
\ee
and
\be
\cC^s_{-+}=-\hat{\cC}^v_{-},\quad  \cC^{ps}_{-+}= -i\hat{\cC}^{pv}_{-}, \quad \cC^v_{-+}=-\hat{\cC}^s_-,\quad  \cC^{pv}_{-+}=- i \hat{\cC}^{ps}_-.
\ee
Notice that there is a mismatch between $2$D and $4$D chirality, which explains why some (pseudo)scalar and (pseudo)vector $4$D couplings correspond to (pseudo)vector and (pseudo)scalar $2$D couplings respectively,  as well as the presence of additional minus signs in the matching of pseudoscalar and pseudovector couplings.

The $S$-wave boundary action could have equivalently been formulated in the original spherical gauge of \eqref{Julia-Zee dyon spherical gauge}, in which we get: 

{
\small
\bea
I^{\rm int}_{\rm dyon}&=& -\frac{1}{2}\underset{r=\epsilon}{\int} \exd t\,\exd^2 \Omega\, \epsilon^2 \,\ol{\bm \psi}\Bigl[\hat{\cC}^s_{1} + i\, \hat{\cC}^{ps}_1\gamma_5 - i\, \hat{\cC}^{v}_1 \gamma^{0} - i\, \hat{\cC}^{pv}_1 \gamma_5\gamma^{0}   +\Bigl(\hat{\cC}^{s}_3 + i \,\hat{\cC}^{ps}_3\gamma_5 - i \,\hat{\cC}^{v}_3\gamma^{0} - i\, \hat{\cC}^{pv}_3 \gamma_5\gamma^{0} \Bigr)(- \tau_r)  \\
&&  -\Bigl( \hat{\cC}^s_{+} + i\, \hat{\cC}^{ps}_+\gamma_5 - i\, \hat{\cC}^{v}_+ \gamma^{0}  - i\, \hat{\cC}^{pv}_+ \gamma_5\gamma^{0}  \Bigr) e^{i\mfa} P_-\,\tau^{-r}_+   -  \Bigl(\hat{\cC}^{s}_- + i \,\hat{\cC}^{ps}_-\gamma_5 - i \,\hat{\cC}^{v}_-\gamma^{0} - i\, \hat{\cC}^{pv}_- \gamma_5\gamma^{0} \Bigr) e^{-i\mfa} P_+\,\tau^{-r}_- \Bigr]\, \bm \psi, \nn 
\eea
}
where  $P_{\pm}$ are given by \eqref{P+-} and $\tau^{-r}_{\pm}\coloneqq\frac{1}{2}e^{\pm i \phi}\left(\tau_{\theta}\mp i \tau_{\phi}\right)$ act as raising and lowering operators on eigenstates of $-\tau_{r}$, that is
\be
\tau^{-r}_{+}\eta_+(\theta, \phi)=\eta_-(\theta, \phi), \quad \tau^{-r}_{-}\eta_-(\theta, \phi)=\eta_+(\theta, \phi),\quad \text{and} \quad \tau^{-r}_{-}\eta_+(\theta, \phi)=\tau^{-r}_{+}\eta_-(\theta, \phi)=0,
\ee
where $\eta_{\pm}(\theta, \phi)$ are vectors in isospin space given by \eqref{etapmdefs}, which satisfy $(-\tau_r) \eta_{\pm}(\theta, \phi)=\mp \eta_{\pm}(\theta, \phi)$.

\section{Properties of the amplitudes $\mathcal{T}_{\ins}, \mathcal{T}_{\outs}$}
\label{App:T amplitudes}

This appendix derives and summarizes several useful properties of the amplitudes $\mathcal{T}_{\ins}$ and $\mathcal{T}_{\outs}$ that appear in the construction of the $in$ and $out$ states. We assume when doing so that the dyon-fermion action is real (and so the couplings $\cC^\ssA_{ij}$ are hermitian). 

\subsection{Definition of the amplitudes}

The main text shows that the $\mathcal{T}_{\ins}, \mathcal{T}_{\outs}$ amplitudes are given in terms of the boundary matrix $\hat{\mathcal{B}}$ by
 \be  \label{TinvsB}
      \mathcal{T}^{++}_{\ins}=-\frac{\left|\begin{array}{ll}
\hat{\mathcal{B}}_{21}& \hat{\mathcal{B}}_{24} \\
\hat{\mathcal{B}}_{41}\, & \hat{\mathcal{B}}_{44}
\end{array}\right|}{\left|\begin{array}{ll}
\hat{\mathcal{B}}_{22} & \hat{\mathcal{B}}_{24} \\
\hat{\mathcal{B}}_{42} & \hat{\mathcal{B}}_{44}
\end{array}\right|}, 
     \cT^{+ -}_{\ins}= -\frac{\left|\begin{array}{ll}
\hat{\mathcal{B}}_{23}& \hat{\mathcal{B}}_{24} \\
\hat{\mathcal{B}}_{43}\, & \hat{\mathcal{B}}_{44}
\end{array}\right|}{\left|\begin{array}{ll}
\hat{\mathcal{B}}_{22} & \hat{\mathcal{B}}_{24} \\
\hat{\mathcal{B}}_{42} & \hat{\mathcal{B}}_{44}
\end{array}\right|},  
     \cT^{-+}_{\ins}= - \frac{\left|\begin{array}{ll}
 \hat{\mathcal{B}}_{22}  & \hat{\mathcal{B}}_{21} \\
\hat{\mathcal{B}}_{42} & \hat{\mathcal{B}}_{41}
\end{array}\right|}{\left|\begin{array}{ll}
\hat{\mathcal{B}}_{22} & \hat{\mathcal{B}}_{24} \\
\hat{\mathcal{B}}_{42} & \hat{\mathcal{B}}_{44}
\end{array}\right|}, 
     \mathcal{T}^{- -}_{\ins}=- \frac{\left|\begin{array}{ll}
 \hat{\mathcal{B}}_{22}  & \hat{\mathcal{B}}_{23} \\
\hat{\mathcal{B}}_{42} & \hat{\mathcal{B}}_{43}
\end{array}\right|}{\left|\begin{array}{ll}
\hat{\mathcal{B}}_{22} & \hat{\mathcal{B}}_{24} \\
\hat{\mathcal{B}}_{42} & \hat{\mathcal{B}}_{44}
\end{array}\right|},
    \ee
and 
 \be \label{ToutvsB}
     \mathcal{T}^{++}_{\outs}= -\frac{\left|\begin{array}{ll}
\hat{\mathcal{B}}_{12}& \hat{\mathcal{B}}_{13} \\
\hat{\mathcal{B}}_{32}\, & \hat{\mathcal{B}}_{33}
\end{array}\right|}{\left|\begin{array}{ll}
\hat{\mathcal{B}}_{11} & \hat{\mathcal{B}}_{13} \\
\hat{\mathcal{B}}_{31} & \hat{\mathcal{B}}_{33}
\end{array}\right|}, \cT^{- +}_{\outs}= -\frac{\left|\begin{array}{ll}
\hat{\mathcal{B}}_{11}& \hat{\mathcal{B}}_{12} \\
\hat{\mathcal{B}}_{31}\, & \hat{\mathcal{B}}_{32}
\end{array}\right|}{\left|\begin{array}{ll}
\hat{\mathcal{B}}_{11} & \hat{\mathcal{B}}_{13} \\
\hat{\mathcal{B}}_{31} & \hat{\mathcal{B}}_{33}
\end{array}\right|}, \cT^{+-}_{\outs}= - \frac{\left|\begin{array}{ll}
 \hat{\mathcal{B}}_{14}  & \hat{\mathcal{B}}_{13} \\
\hat{\mathcal{B}}_{34} & \hat{\mathcal{B}}_{33}
\end{array}\right|}{\left|\begin{array}{ll}
\hat{\mathcal{B}}_{11} & \hat{\mathcal{B}}_{13} \\
\hat{\mathcal{B}}_{31} & \hat{\mathcal{B}}_{33}
\end{array}\right|}, \mathcal{T}^{- -}_{\outs}= - \frac{\left|\begin{array}{ll}
 \hat{\mathcal{B}}_{11}  & \hat{\mathcal{B}}_{14} \\
\hat{\mathcal{B}}_{31} & \hat{\mathcal{B}}_{34}
\end{array}\right|}{\left|\begin{array}{ll}
\hat{\mathcal{B}}_{11} & \hat{\mathcal{B}}_{13} \\
\hat{\mathcal{B}}_{31} & \hat{\mathcal{B}}_{33}
\end{array}\right|}.
    \ee
Written directly in terms of  the boundary couplings, these become
\bea \label{AppTinvsC}
    \mathcal{T}^{++}_{\ins}&=&\frac{(\mathcal{C}^s_{++}+i \mathcal{C}^{ps}_{++})(i-\mathcal{C}^{pv}_{--}+ \mathcal{C}^{v}_{--})+(\mathcal{C}^s_{-+}+i \mathcal{C}^{ps}_{-+})(\mathcal{C}^{pv}_{+-}- \mathcal{C}^{v}_{+-})}{-|\mathcal{C}^{pv}_{+-}-\mathcal{C}^{v}_{+-}|^2+(-i +\mathcal{C}^{pv}_{++}-\mathcal{C}^{v}_{++})(\mathcal{C}^{pv}_{--}-\mathcal{C}^{v}_{--}-i)}, \nonumber\\ 
     \mathcal{T}^{--}_{\ins}&=&\frac{(\mathcal{C}^s_{--}+i \mathcal{C}^{ps}_{--})(i-\mathcal{C}^{pv}_{++}+ \mathcal{C}^{v}_{++})+(\mathcal{C}^s_{+-}+i \mathcal{C}^{ps}_{+-})(\mathcal{C}^{pv}_{-+}- \mathcal{C}^{v}_{-+})}{-|\mathcal{C}^{pv}_{+-}-\mathcal{C}^{v}_{+-}|^2+(-i +\mathcal{C}^{pv}_{++}-\mathcal{C}^{v}_{++})(\mathcal{C}^{pv}_{--}-\mathcal{C}^{v}_{--}-i)}, \\ 
     \cT^{-+}_{\ins}&=&\frac{(\mathcal{C}^s_{-+}+i \mathcal{C}^{ps}_{-+})(i-\mathcal{C}^{pv}_{++}+ \mathcal{C}^{v}_{++})+(\mathcal{C}^s_{++}+i \mathcal{C}^{ps}_{++})(\mathcal{C}^{pv}_{-+}- \mathcal{C}^{v}_{-+})}{-|\mathcal{C}^{pv}_{+-}-\mathcal{C}^{v}_{+-}|^2+(-i +\mathcal{C}^{pv}_{++}-\mathcal{C}^{v}_{++})(\mathcal{C}^{pv}_{--}-\mathcal{C}^{v}_{--}-i)}, \nonumber\\ 
    \cT^{+-}_{\ins}&=&\frac{(\mathcal{C}^s_{+-}+i \mathcal{C}^{ps}_{+-})(i-\mathcal{C}^{pv}_{--}+ \mathcal{C}^{v}_{--})+(\mathcal{C}^s_{--}+i \mathcal{C}^{ps}_{--})(\mathcal{C}^{pv}_{+-}- \mathcal{C}^{v}_{+-})}{-|\mathcal{C}^{pv}_{+-}-\mathcal{C}^{v}_{+-}|^2+(-i +\mathcal{C}^{pv}_{++}-\mathcal{C}^{v}_{++})(\mathcal{C}^{pv}_{--}-\mathcal{C}^{v}_{--}-i)}, \nn
\eea
and
\bea \label{AppToutvsC}
    \mathcal{T}^{++}_{\outs}&=&\frac{i(\mathcal{C}^{ps}_{++}+i \mathcal{C}^{s}_{++})(-i+\mathcal{C}^{pv}_{--}+ \mathcal{C}^{v}_{--})+(-i \mathcal{C}^{ps}_{-+}+ \mathcal{C}^{s}_{-+})(\mathcal{C}^{pv}_{+-}+ \mathcal{C}^{v}_{+-})}{|\mathcal{C}^{pv}_{+-}+\mathcal{C}^{v}_{+-}|^2-(-i +\mathcal{C}^{pv}_{--}+\mathcal{C}^{v}_{--})(\mathcal{C}^{pv}_{++}+\mathcal{C}^{v}_{++}-i)}, \nonumber\\ 
     \mathcal{T}^{--}_{\outs}&=&\frac{i(\mathcal{C}^{ps}_{--}+i \mathcal{C}^{s}_{--})(-i+\mathcal{C}^{pv}_{++}+ \mathcal{C}^{v}_{++})+(-i \mathcal{C}^{ps}_{+-}+ \mathcal{C}^{s}_{+-})(\mathcal{C}^{pv}_{-+}+ \mathcal{C}^{v}_{-+})}{|\mathcal{C}^{pv}_{+-}+\mathcal{C}^{v}_{+-}|^2-(-i +\mathcal{C}^{pv}_{--}+\mathcal{C}^{v}_{--})(\mathcal{C}^{pv}_{++}+\mathcal{C}^{v}_{++}-i)}, \\ 
     \cT^{-+}_{\outs}&=&\frac{i(\mathcal{C}^{ps}_{-+}+i \mathcal{C}^{s}_{-+})(-i+\mathcal{C}^{pv}_{++}+ \mathcal{C}^{v}_{++})+(-i \mathcal{C}^{ps}_{++}+ \mathcal{C}^{s}_{++})(\mathcal{C}^{pv}_{-+}+ \mathcal{C}^{v}_{-+})}{|\mathcal{C}^{pv}_{+-}+\mathcal{C}^{v}_{+-}|^2-(-i +\mathcal{C}^{pv}_{--}+\mathcal{C}^{v}_{--})(\mathcal{C}^{pv}_{++}+\mathcal{C}^{v}_{++}-i)}, \nonumber\\ 
     \cT^{+-}_{\outs}&=&\frac{i(\mathcal{C}^{ps}_{+-}+i \mathcal{C}^{s}_{+-})(-i+\mathcal{C}^{pv}_{--}+ \mathcal{C}^{v}_{--})+(-i \mathcal{C}^{ps}_{--}+ \mathcal{C}^{s}_{--})(\mathcal{C}^{pv}_{+-}+ \mathcal{C}^{v}_{+-})}{|\mathcal{C}^{pv}_{+-}+\mathcal{C}^{v}_{+-}|^2-(-i +\mathcal{C}^{pv}_{--}+\mathcal{C}^{v}_{--})(\mathcal{C}^{pv}_{++}+\mathcal{C}^{v}_{++}-i)}. \nn
\eea

\subsection{Rank-2 conditions}
\label{Appssec:Rank2}
The above definitions assume the denominators do not vanish. This is satisfied  when the $\cC^\ssA_{ij}$ are all hermitian, as shown in equation \eqref{non-zero determinants} and the text directly after it.

\subsubsection*{Conditions for rank-two boundary conditions}

We next ask what is required to ensure that $\text{rank}(\mathcal{B})=2$. When this is true one pair of linearly independent columns of $\mathcal{B}$ can be written as a linear combination of the other two linearly independent columns. When the $\cC^\ssA_{ij}$ are hermitian we've seen the linearly independent pairs consist of the first and third columns of $\mathcal{B}$ and the second and fourth columns of $\cB$. Consequently there exist nonzero coefficients $A_1$, $A_3$, $B_1$ and $B_3$ such that
\bea\label{Rank 2 condition: c1,3 in terms of c2,4}
    \begin{pmatrix}
        \mathcal{B}_{11} & \mathcal{B}_{21} & \mathcal{B}_{31} & \mathcal{B}_{41} 
    \end{pmatrix} &=& A_1  \begin{pmatrix}
        \mathcal{B}_{12} & \mathcal{B}_{22} & \mathcal{B}_{32} & \mathcal{B}_{42} 
    \end{pmatrix}+ B_1  \begin{pmatrix}
        \mathcal{B}_{14} & \mathcal{B}_{24} & \mathcal{B}_{34} & \mathcal{B}_{44} 
    \end{pmatrix}, \nonumber\\ 
     \begin{pmatrix}
        \mathcal{B}_{13} & \mathcal{B}_{23} & \mathcal{B}_{33} & \mathcal{B}_{43} 
    \end{pmatrix}&=& A_3  \begin{pmatrix}
        \mathcal{B}_{12} & \mathcal{B}_{22} & \mathcal{B}_{32} & \mathcal{B}_{42} 
    \end{pmatrix}+ B_3  \begin{pmatrix}
        \mathcal{B}_{14} & \mathcal{B}_{24} & \mathcal{B}_{34} & \mathcal{B}_{44} 
    \end{pmatrix},
\eea
The required coefficients are given by
\be
    A_1=\frac{\left|\begin{array}{c c}
        \mathcal{B}_{21} & \mathcal{B}_{24}\\ \mathcal{B}_{41}&  \mathcal{B}_{44}  \end{array}\right|}{\left|\begin{array}{c c}
        \mathcal{B}_{22} & \mathcal{B}_{24}\\ \mathcal{B}_{42}&  \mathcal{B}_{44} \end{array}\right|}, 
    \quad  B_1=\frac{\left|\begin{array}{c c} \mathcal{B}_{22} & \mathcal{B}_{21}\\ \mathcal{B}_{42}&  \mathcal{B}_{41}
    \end{array}\right|}{\left|\begin{array}{c c}
        \mathcal{B}_{22} & \mathcal{B}_{24}\\ \mathcal{B}_{42}&  \mathcal{B}_{44}    \end{array}\right|},
        \quad  A_3=\frac{\left|\begin{array}{c c}
        \mathcal{B}_{23} & \mathcal{B}_{24}\\ \mathcal{B}_{43}&  \mathcal{B}_{44}  \end{array}\right|}{\left|\begin{array}{c c}
        \mathcal{B}_{22} & \mathcal{B}_{24}\\ \mathcal{B}_{42}&  \mathcal{B}_{44}    \end{array}\right|},
        \quad  B_3=\frac{\left|\begin{array}{c c}
        \mathcal{B}_{22} & \mathcal{B}_{23}\\ \mathcal{B}_{42}&  \mathcal{B}_{43}
    \end{array}\right|}{\left|\begin{array}{c c}
        \mathcal{B}_{22} & \mathcal{B}_{24}\\ \mathcal{B}_{42}&  \mathcal{B}_{44}
    \end{array}\right|} .
\ee

Comparing these solutions to \pref{TinvsB} implies 
\be
  A_1 =-\mathcal{T}^{++}_{\ins} ,\quad  B_1= -\cT^{-+}_{\ins} e^{-i \mfa},\quad  A_3=-\cT^{+-}_{\ins} e^{i \mfa},
  \quad  B_3=-\mathcal{T}^{--}_{\ins} \,,
\ee
and using these to trade $A_{1,3}, B_{1,3}$ for the $\cT_{\ins}$'s in equation \eqref{Rank 2 condition: c1,3 in terms of c2,4} gives four tautologies as well as the following four complex conditions that can be regarded as conditions required if the matrix $\cB$ is to have rank 2: 
\bea\label{Rank 2 conditions}
    \mathcal{E}_{1}&:=&\hat{\mathcal{B}}_{13} + \cT^{+-}_{\ins}   \hat{\mathcal{B}}_{12}+ \mathcal{T}^{--}_{\ins}   \hat{\mathcal{B}}_{14}=0,\nonumber\\    
    \mathcal{E}_{2}&:=&\hat{\mathcal{B}}_{11} + \mathcal{T}^{++}_{\ins}   \hat{\mathcal{B}}_{12}+ \cT^{-+}_{\ins}   \hat{\mathcal{B}}_{14}=0,\\
    \mathcal{E}_{3}&:=&\hat{\mathcal{B}}_{33} + \cT^{+-}_{\ins}   \hat{\mathcal{B}}_{32}+ \mathcal{T}^{--}_{\ins}   \hat{\mathcal{B}}_{34}=0, \nonumber\\ 
    \mathcal{E}_{4}&:=&\hat{\mathcal{B}}_{31} + \mathcal{T}^{++}_{\ins}   \hat{\mathcal{B}}_{32}+ \cT^{-+}_{\ins}   \hat{\mathcal{B}}_{34}=0. \nn
\eea

These four complex equations amount to eight real conditions. Four of these state
\be
      \Re(\mathcal{E}_2)= 0,\quad \Re(\mathcal{E}_3)= 0,\quad  \text{and}\quad  \mathcal{E}_1+\mathcal{E}^{*}_4=0, 
\ee
and,  once expressions \pref{TinvsB} are used, can be written purely in terms of the $\mathcal{T}_{\ins}$  which must therefore satisfy
\bea\label{Conditions on Tin amplitudes}
     \Re(\mathcal{E}_2) &=& -(|\mathcal{T}^{++}_{\ins}|^2+|\cT^{-+}_{\ins}|^2-1) =0,\nonumber\\ 
     \Re(\mathcal{E}_3) &=& -(|\mathcal{T}^{--}_{\ins}|^2+|\cT^{+-}_{\ins}|^2-1)=0, \\ 
       \mathcal{E}_1+\mathcal{E}^{*}_4&=&- 2(\mathcal{T}^{--}_{\ins}\cT^{-+\,*}_{\ins}+\cT^{+-}_{\ins}\mathcal{T}^{++\,*}_{\ins})=0,\nonumber
\eea
as used in the main text. In  deriving the above expressions we use the identity \pref{BadjRel}, which follows when the boundary coupling constants are hermitian.  The relations \pref{Conditions on Tin amplitudes} simply express the unitarity of the $S$ matrix since the first two of these equations impose that the fermion number density of the $in$ modes is conserved during scattering processes, while the third  imposes that the off-diagonal matrix elements of $S^{\dagger} S$ vanish.

The remaining four rank $(\mathcal{B})=2$ conditions within \pref{Rank 2 conditions} are
\be
    \mathcal{E}_1=0, \quad \Im(\mathcal{E}_2)=0, \quad \text{and} \quad \Im(\mathcal{E}_3)=0.
\ee
These depend on more than just the $\cT_{\ins}$ amplitudes but can be expressed in terms of both $\cT_{\ins}$ and $\cT_{\outs}$. To see why the rank-two conditions \eqref{Rank 2 conditions} can be written entirely in terms of $\mathcal{T}_{\ins}, \mathcal{T}_{\outs}$ take the following linear combinations 
\bea
    -\left|\begin{array}{ll} \hat{\mathcal{B}}_{11} & \hat{\mathcal{B}}_{13} \\ \hat{\mathcal{B}}_{31} & \hat{\mathcal{B}}_{33}
\end{array}\right|^{-1}\left(\hat{\mathcal{B}}_{33}\,\mathcal{E}_1-\hat{\mathcal{B}}_{13}\,\mathcal{E}_3\right)&=& \mathcal{T}^{--}_{\ins}\cT^{+-}_{\outs}+\cT^{+-}_{\ins}\mathcal{T}^{++}_{\outs}=0, \nonumber\\ \left|\begin{array}{ll} \hat{\mathcal{B}}_{11} & \hat{\mathcal{B}}_{13} \\ \hat{\mathcal{B}}_{31} & \hat{\mathcal{B}}_{33}
\end{array}\right|^{-1}\left(\hat{\mathcal{B}}_{31}\,\mathcal{E}_1-\hat{\mathcal{B}}_{11}\,\mathcal{E}_3\right)&=& \mathcal{T}^{--}_{\ins}\mathcal{T}^{--}_{\outs}+\cT^{+-}_{\ins}\cT^{-+}_{\outs}-1=0, \\\left|\begin{array}{ll} \hat{\mathcal{B}}_{11} & \hat{\mathcal{B}}_{13} \\ \hat{\mathcal{B}}_{31} & \hat{\mathcal{B}}_{33}
\end{array}\right|^{-1}\left( \hat{\mathcal{B}}_{31}\,\mathcal{E}_2-\hat{\mathcal{B}}_{11}\,\mathcal{E}_4 \right) &=& \cT^{-+}_{\ins}\mathcal{T}^{--}_{\outs}+\mathcal{T}^{++}_{\ins}\cT^{-+}_{\outs}=0, \nonumber\\  -\left|\begin{array}{ll} \hat{\mathcal{B}}_{11} & \hat{\mathcal{B}}_{13} \\ \hat{\mathcal{B}}_{31} & \hat{\mathcal{B}}_{33}
\end{array}\right|^{-1}\left(\hat{\mathcal{B}}_{33}\,\mathcal{E}_2-\hat{\mathcal{B}}_{13}\,\mathcal{E}_4\right)&=& \cT^{-+}_{\ins}\cT^{+-}_{\outs}+\mathcal{T}^{++}_{\ins}\mathcal{T}^{++}_{\outs}-1=0.\nonumber
\eea
where we again use $\left|\begin{array}{ll} \hat{\mathcal{B}}_{11} & \hat{\mathcal{B}}_{13} \\ \hat{\mathcal{B}}_{31} & \hat{\mathcal{B}}_{33}
\end{array}\right|\neq 0$. These can be solved to give the $\mathcal{T}_{\outs}$ amplitudes in terms of $\mathcal{T}_{\ins}$ amplitudes and vice versa, giving
\bea\label{Tout amplitudes in terms of Tin amplitudes}
    &&\cT^{+-}_{\outs}=\frac{\cT^{+-}_{\ins} }{\mathcal{T}^{-+}_{\ins}\mathcal{T}^{+-}_{\ins}-\mathcal{T}^{--}_{\ins}\mathcal{T}^{++}_{\ins}}, \qquad\, \mathcal{T}^{--}_{\outs}=-\frac{\mathcal{T}^{++}_{\ins}}{\mathcal{T}^{-+}_{\ins}\mathcal{T}^{+-}_{\ins}-\mathcal{T}^{--}_{\ins}\mathcal{T}^{++}_{\ins}}\nonumber\\
    &&\mathcal{T}^{++}_{\outs}=-\frac{\mathcal{T}^{--}_{\ins}}{\mathcal{T}^{-+}_{\ins}\mathcal{T}^{+-}_{\ins}-\mathcal{T}^{--}_{\ins}\mathcal{T}^{++}_{\ins}}, \qquad \,\cT^{-+}_{\outs}=\frac{\cT^{-+}_{\ins}}{\mathcal{T}^{-+}_{\ins}\mathcal{T}^{+-}_{\ins}-\mathcal{T}^{--}_{\ins}\mathcal{T}^{++}_{\ins}}
\eea
and
\bea\label{Tin amplitudes in terms of Tout amplitudes}
    &&\cT^{+-}_{\ins}=\frac{\cT^{+-}_{\outs}}{\mathcal{T}^{-+}_{\outs}\mathcal{T}^{+-}_{\outs}-\mathcal{T}^{--}_{\outs}\mathcal{T}^{++}_{\outs}}, \qquad \,\mathcal{T}^{--}_{\ins}=-\frac{\mathcal{T}^{++}_{\outs}}{\mathcal{T}^{-+}_{\outs}\mathcal{T}^{+-}_{\outs}-\mathcal{T}^{--}_{\outs}\mathcal{T}^{++}_{\outs}}\nonumber\\
    &&\mathcal{T}^{++}_{\ins}=-\frac{\mathcal{T}^{--}_{\outs}}{\mathcal{T}^{-+}_{\outs}\mathcal{T}^{+-}_{\outs}-\mathcal{T}^{--}_{\outs}\mathcal{T}^{++}_{\outs}},\qquad \,\cT^{-+}_{\ins}=\frac{\cT^{-+}_{\outs}}{\mathcal{T}^{-+}_{\outs}\mathcal{T}^{+-}_{\outs}-\mathcal{T}^{--}_{\outs}\mathcal{T}^{++}_{\outs}}.
\eea

\subsubsection*{Modulus and phase of $\mathcal{T}_{\ins}$ and $\cT_{\outs}$}

The constraint conditions \eqref{Conditions on Tin amplitudes} imply that the four complex $\cT_{\ins}$ amplitudes satisfy four real constraints and so actually only contain four independent real parameters. 
To see why notice that the first two conditions in \eqref{Conditions on Tin amplitudes} can be used to write the general amplitudes as 
 \be 
     \mathcal{T}^{++}_{\ins}=\rho_{+} \,e^{i \theta_{++}}, \quad   \cT^{-+}_{\ins}=\sqrt{1-\rho_{+}^2} \,e^{i \theta_{-+}},\quad  \mathcal{T}^{--}_{\ins}=\rho_{-}\, e^{i \theta_{--}}, \quad  \text{and} \quad \cT^{+-}_{\ins}=\sqrt{1-\rho_{-}^2} \,e^{i \theta_{+-}},
 \ee
for the six real parameters $\rho_{\pm} \ge 0$ and $\theta_{++}, \theta_{+-}, \theta_{-+},\theta_{--}$. Plugging these expressions into the third condition in \eqref{Conditions on Tin amplitudes} then gives
\be 
   \rho_- \sqrt{1-\rho^2_+} \,e^{i(\theta_{--}-\theta_{-+})}=\rho_+ \sqrt{1-\rho^2_-} \,e^{-i(\theta_{++}-\theta_{+-}+\pi)},
\ee
and so taking the modulus squared of each side of the equation shows $\rho_+=\rho_-$. Additionally, the four phases are not independent because they satisfy 
\be 
    \theta_{++}+\theta_{--}-(\theta_{-+}+\theta_{+-})+ \pi =2\pi n,
\ee
where $n$ is an integer. Consequently one of the phases can be eliminated in favor of the other three. We see that the most general form of the $\mathcal{T}_{\ins}$ amplitudes for a hermitian theory with a rank 2 boundary matrix is then
 \be \label{AppMostGeneralTin}
     \mathcal{T}^{++}_{\ins}=\rho \,e^{i \theta_{++}}, \quad   \mathcal{T}^{+-}_{\ins}=\sqrt{1-\rho^2} \,e^{i \theta_{+-}},\quad  \mathcal{T}^{--}_{\ins}=\rho\, e^{i \theta_{--}} \quad  \text{and} \quad \mathcal{T}^{-+}_{\ins}=-\sqrt{1-\rho^2} \,e^{i (\theta_{++}+ \theta_{--}-\theta_{+-})}
   .
 \ee
 Notice that these imply $|\cT_\ins^{++}|^2 = |\cT_\ins^{--}|^2$ and $|\cT_\ins^{+-}|^2 = |\cT_\ins^{-+}|^2$ and that the denominator appearing in \pref{Tout amplitudes in terms of Tin amplitudes} is a pure phase:
 \be \label{AppDenomTs}
  \cT_{\ins}^{-+} \cT_{\ins}^{+-} - \cT_{\ins}^{++} \cT_{\ins}^{--} =- e^{i (\theta_{++}+ \theta_{--})}   \,.
\ee

\subsubsection*{Implications for $\cT_{\outs}$}

Plugging the most general form \pref{AppMostGeneralTin} for $\mathcal{T}_{\ins}$ into \eqref{Tout amplitudes in terms of Tin amplitudes} and using \pref{AppDenomTs} gives
\be \label{AppMostGeneralTout}
    \mathcal{T}^{++}_{\outs}= \rho \,e^{-i \theta_{++}}=(\mathcal{T}^{++}_{\ins})^*, \quad \mathcal{T}^{--}_{\outs}= \rho \,e^{-i \theta_{--}}=(\mathcal{T}^{--}_{\ins})^*,
\ee
and
\be 
    \mathcal{T}^{+-}_{\outs}=- \sqrt{1-\rho^2}\, e^{i (\theta_{+-}-\theta_{++}-\theta_{--})}=(\mathcal{T}^{-+}_{\ins})^*, \quad  \mathcal{T}^{-+}_{\outs}= \sqrt{1-\rho^2}\, e^{-i \theta_{+-}}=(\mathcal{T}^{+-}_{\ins})^*.
\ee
It is clear that the unitarity conditions satisfied by $\cT_{\ins}$ are also satisfied by the $\mathcal{T}_{\outs}$ amplitudes:
\bea \label{Constraints on Tout amplitudes}
&& |\mathcal{T}^{++}_{\outs}|^2 = |\mathcal{T}^{--}_{\outs}|^2 = 1 -|\mathcal{T}^{+-}_{\outs}|^2= 1 - |\mathcal{T}^{-+}_{\outs}|^2,\nonumber\\
&&\qquad \mathcal{T}^{--}_{\outs}\cT^{-+\,*}_{\outs}+\cT^{+-}_{\outs}\mathcal{T}^{++\,*}_{\outs}=0.
\eea

\section{Scattering states}
\label{App:Scattering states, new}

In this appendix, we list some useful properties of the $S$-wave $in$ and $out$ states defined in the main text and derive the Bogoliubov relations given in \S \ref{sec:SO3Dyon}. Additionally, we calculate the probability  for a plane-wave state to be found in an $S$-wave, $p_s$, and derive the cross-section formulas in \S \ref{sec:PerturbativeScattering} and \S \ref{sec:SO3Dyon}.

\subsection{Properties of \textit{in} and \textit{out} states}
\subsection*{Orthogonality and normalization relations}

The unitarity constraints on $in$ amplitudes \eqref{unitarityT1}-\eqref{unitarityT} as well as the analogous $out$ amplitude constraints \eqref{unitarityTout1}-\eqref{unitarityTout2} can be used to show that the $in$ and $out$  modes satisfy the orthogonality and normalization relations
\be
   \int^{\infty}_{\epsilon} \mathrm{d}r \,(\mfu^{\text{d}}_{\mathfrak{s}, k})^{\dagger}\,\mfu^{\text{d}}_{\mathfrak{s}', k'}=  \int^{\infty}_{\epsilon} \mathrm{d}r \,(\mfv^{\text{d}}_{\mathfrak{s}, k})^{\dagger}\,\mfv^{\text{d}}_{\mathfrak{s}', k'}= 2\pi \delta(k - k')\,\delta_{\mfs \mfs'},
\ee
and
\be
\int^{\infty}_{\epsilon} \mathrm{d}r \,(\mfu^{\text{d}}_{\mathfrak{s}, k})^{\dagger}\,\mfv^{\text{d}}_{\mathfrak{s}', k'}=2\pi \delta(k + k')\,\delta_{\mathfrak{s} \mathfrak{s'}}.
\ee
where the label d indicates the direction of motion \textit{i.e.} ${\text{d}}=\{ \ins, \outs\}$. These relations can be used to show that, when the $S$-wave fermion field is expanded in terms of $in$, $out$ modes  as 
\bea
   \bm{\chi}(x) &=&  \sum_{\mfs=\pm}\int^{\infty}_{0} \frac{\exd k}{\sqrt{2\pi}}\,\Bigl[ \mfu^{\text{d}}_{\mfs, k}(x) \,  a^{\text{d}}_{\mfs, k } + \mfv^{\text{d}}_{\mfs, k}(x) \,( \ol{a}^{\text{d}}_{\mfs, k} )^\star\Bigr],
\eea
the particle and antiparticle creation and annihilation operator  anticommutation relations are given by
\be
    \Bigl\{{ a}^{\text{d}}_{\mfs, k}, ({ a}^{\text{d}}_{\mfs', k'})^\star \Bigr\} =  \Bigl\{\ol{{ a}}^{\text{d}}_{\mfs, k}, (\ol{{ a}}^{\text{d}}_{\mfs', k'})^\star \Bigr\} =   \delta(k-k') \,\delta_{\mfs \mfs'},  
\ee
with all other anticommutators vanishing.

For some applications, it is preferable to consider discretely normalized states. We define the discretely normalized $in$, $out$ modes as
\be
    \bm{u}^{\text{d}}_{\mathfrak{s}, k}(x)\coloneqq\, \frac{1}{\sqrt{2L}} \mfu^{\text{d}}_{\mathfrak{s}, k}(x),\quad  \bm{v}^{\text{d}}_{\mathfrak{s}, k}(x)\coloneqq  \, \frac{1}{\sqrt{2L}}\mfv^{\text{d}}_{\mathfrak{s},  k}(x), 
\ee
since the orthogonality and normalization relations for these modes become
\bea
   && \int^{L+\epsilon}_{\epsilon} \mathrm{d}r \,(\bm{u}^{\text{d}}_{\mathfrak{s}, k})^{\dagger}\,\bm{u}^{\text{d}}_{\mathfrak{s}', k'}=\int^{L+\epsilon}_{\epsilon} \mathrm{d}r \,(\bm{v}^{\text{d}}_{\mathfrak{s}, k})^{\dagger}\,\bm{v}^{\text{d}}_{\mathfrak{s}', k'}=\delta_{k k'}\,\delta_{\mfs \mfs'},
\eea
and
\be
 \int^{L+\epsilon}_{\epsilon} \mathrm{d}r \,(\bm{u}^{\text{d}}_{\mathfrak{s}, k})^{\dagger}\,\bm{v}^{\text{d}}_{\mathfrak{s}', k'}=\delta_{k , -k'}\,\delta_{\mfs \mfs'},
\ee
in the large $L$ limit. The $S$-wave fermion field can  be expanded in terms of the discretely normalized bases:
\bea
   \bm{\chi}(x) &=&  \sum_{\mfs=\pm}\sum^{\infty}_{k\ge 0}\,\Bigl[ \bm{u}^{\text{d}}_{\mfs, k}(x) \,  \bm{a}^{\text{d}}_{\mfs, k } + \bm{v}^{\text{d}}_{\mfs, k}(x) \,( \ol{\bm{a}}^{\text{d}}_{\mfs, k} )^\star\Bigr],
\eea
where the discrete normalization particle and antiparticle creation and annihilation operators  satisfy 
\be
    \Bigl\{\bm{a}^{\text{d}}_{\mfs, k}, (\bm{a}^{\text{d}}_{\mfs', k'})^\star \Bigr\} =  \Bigl\{\ol{\bm{a}}^{\text{d}}_{\mfs, k}, (\ol{\bm{a}}^{\text{d}}_{\mfs', k'})^\star \Bigr\} =  \delta_{k k'} \,\delta_{\mfs \mfs'},  
\ee
with all other anticommutators vanishing.

\subsection*{Bogoliubov relations}

The $in$ creation and annihilation operators can be expanded in terms of the corresponding $out$ operators and vice versa, as in the Bogoliubov relations \eqref{Bogoutfromin} and \eqref{Boginfromout}. To see this, note that the $in$ operators satisfy 
\be
a^{\ins}_{\mfs, k}=\frac{1}{\sqrt{8\pi}}\int^{\infty}_{-\infty} \exd t \, (\mfu^{\ins}_{\mfs , k}(r, t))^{\dagger} \bm{\chi}(r,t) , \quad \text{and} \quad (\ol{a}^{\ins}_{\mfs, k})^{\star}=\frac{1}{\sqrt{8\pi}}\int^{\infty}_{-\infty} \exd t \, (\mfv^{\ins}_{\mfs , k}(r, t))^{\dagger} \bm{\chi}(r,t),
\ee
which can be shown by expanding $\bm \chi(x)$ in terms of the $in$ basis. Since $\bmchi(x)$ can equivalently be expanded in terms of $out$ states, the above equations imply that \textit{e.g.} $a^{\ins}_{+, k}$ is given by
\bea \label{New Bog form}
a^{\ins}_{+, k}&=&\frac{1}{\sqrt{8\pi}}\sum_{\mfs'=\pm}\int^{\infty}_{-\infty} \exd t \, \int^{\infty}_{0} \frac{\exd k'}{\sqrt{2\pi}} (\mfu^{\ins}_{+ , k}(r, t))^{\dagger}\,\left( \mfu^\outs_{\mfs', k'}(x) \,  a^\outs_{\mfs', k' } + \mfv^\outs_{\mfs', k'}(x) \,( \ol{a}^\outs_{\mfs', k'} )^\star\right)\nn \\&=&\int^{\infty}_{0}{\exd k'}\,\Big( \cT^{++}_{\outs}e^{2i k\epsilon}\left(\frac{\epsilon}{r_0}\right)^{-i e Q}\, \delta(\omega_{+, k}-\omega_{+, k'})\,a^{\outs}_{+, k'}\nn \\ && \qquad \qquad+ \,\cT^{+-}_{\outs}e^{i (2k - e v)\epsilon}e^{i \mfa}\, \left[\delta(\omega_{+, k}-\omega_{-, k'}) \,a^{\outs}_{-, k'}+\delta(\omega_{+, k}+\ol{\omega}_{-, k'})\,(\ol{a}^{\outs}_{-, k'})^{\star} \right]\Big),
\eea
where $\omega_{\mfs, k}= k - \mfs e v/2$ ($\ol{\omega}_{\mfs, k}= k + \mfs e v/2$) is the energy of a particle (antiparticle) with quantum numbers $\mfs, k$. The above equation reduces to one of the Bogoliubov relations, once the integral over momentum is evaluated:
\bea \label{Old Bog form}
a^{\ins}_{+, k}&=& \cT^{++}_{\outs}\,a^{\outs}_{+, k}+ \cT^{+-}_{\outs}e^{i \mfa}\, \left[\Theta(k- e v) \,a^{\outs}_{-, k- e v}+\Theta(-k+ e v)\,(\ol{a}^{\outs}_{-, -k+ e v})^{\star}\right],
\eea
where we drop powers of $k \epsilon $ and $e v \epsilon$, shift $\mfa$ and rephase the $out$ operators to absorb the $r_0$-dependent phase, as in the main text.  The remaining Bogoliubov relations can be derived in a similar way.

It is sometimes convenient to write the Bogoliubov relations as in the last line of \eqref{New Bog form} since this shows that they  are consistent with energy conservation, which is enforced through the  delta functions that appear after the time integral is performed.  Specifically, this form is useful for evaluating scattering amplitudes such as $\bra{0_{\outs}}a^{\outs}_{-, k'}\,(a^{\ins}_{+, k})^{\star}\ket{0_{\ins}}$, which can be written as:
\bea
\bra{0_{\outs}}a^{\outs}_{-, k'}\,(a^{\ins}_{+, k})^{\star}\ket{0_{\ins}}&=&\cT^{+-\,*}_{\outs}e^{-i \mfa} \int^{\infty}_0 \exd p \,\delta(\omega_{+, k}-\omega_{-, p})\bra{0_{\outs}}a^{\outs}_{-, k'}\,(a^{\outs}_{-, p})^{\star}\ket{0_{\ins}} \nn \\ &=& \cT^{+-\,*}_{\outs}\braket{0_{\outs}|0_{\ins}} e^{-i \mfa} \Theta(k- e v)\delta(\omega_{+, k}-\omega_{-, k'}),
\eea
where in the last line we use the `momentum-conserving' delta function $\delta(p-k')$ implicit in the overlap $\bra{0_{\outs}}a^{\outs}_{-, k'}\,(a^{\outs}_{-, p})^{\star}\ket{0_{\ins}}$ to perform the integral over $p$. The surviving delta function is the original energy-conserving one from \eqref{New Bog form} which is regularized by a factor of the duration of the interaction, $T$, when probabilities are calculated. 

The above argument shows that the Heisenberg picture transition probabilities (which can be obtained from the amplitudes listed in \S \ref{sec:SO3Dyon}) depend on $T$ in the same way as their interaction picture counterparts, and so rates and cross sections can be defined  in much the same way in both pictures. This can also be seen in a simpler way, by re-evaluating the amplitude $\bra{0_{\outs}}a^{\outs}_{-, k'}\,(a^{\ins}_{+, k})^{\star}\ket{0_{\ins}}$ using the  `standard' form of the relevant Bogoliubov relation, \eqref{Old Bog form}, in the following way
\bea \label{Old Bog ampl}
\bra{0_{\outs}}a^{\outs}_{-, k'}\,(a^{\ins}_{+, k})^{\star}\ket{0_{\ins}}&=&\cT^{+-\,*}_{\outs}e^{-i \mfa} \,\Theta(k -e v)\bra{0_{\outs}}a^{\outs}_{-, k'}\,(a^{\outs}_{-, k- e v})^{\star}\ket{0_{\ins}} \nn \\ &=& \cT^{+-\,*}_{\outs}\braket{0_{\outs}|0_{\ins}} e^{-i \mfa} \,\Theta(k -e v)\delta(k-k'-ev ) \nn \\&=&  \cT^{+-\,*}_{\outs}\braket{0_{\outs}|0_{\ins}} e^{-i \mfa} \,\Theta(k -e v)\delta(\omega_{+, k}-\omega_{-, k'}) ,
\eea
where the delta function in the second line  comes from the overlap $\bra{0_{\outs}}a^{\outs}_{-, k'}\,(a^{\outs}_{-, k- e v})^{\star}\ket{0_{\ins}} $ and is rewritten in the third line by performing a change of variables from momentum to energy. Since the final expression in \eqref{Old Bog ampl} involves an energy-conserving delta function, we can regulate it with a factor of $T$ per the usual procedure.

\subsection*{Expectation values of energy and electric, axial charge }

The $in$ and $out$ states are eigenstates of the $S$-wave fermionic hamiltonian (with $\widehat{A}_0=0$):
\be
H_\ssF= \frac{1}{2}\Bigl[\ol{\bm{\chi}}\,O_{\cB}(\mfa)\, \bm{\chi}\Bigr]_{r=\epsilon, t}+\frac{1}{2}\sum_{\mfs=\pm}\int^{\infty}_{\epsilon} \exd r \,\ol{{\chi}}_{\mfs}\left[\Gamma^1\overset{\leftrightarrow}{\partial}_1- i \mfs \,\Gamma^0\left(e v - \frac{e Q}{r}\right) \right]\chi_{\mfs},
\ee
when $\mfa$ is treated as a classical variable within the Born-Oppenheimer approximation. This can be seen from the expansion of $H_\ssF$ in terms of $in$ or $out$ states (see appendix \ref{App: Currents} for details of how calculations involving fermion bilinears are done)
\bea
H_\ssF&&=E^{\text{d}}_{0, \ssF}+\sum_{\mfs =\pm}\int^{\infty}_0 \exd k\,\left[\left(k - \frac{1}{2}\mfs e v \right)(a^{\text{d}}_{\mfs, k})^{\star}\,a^{\text{d}}_{\mfs, k}+\left(k + \frac{1}{2} \mfs e v \right)(\ol{a}^{\text{d}}_{\mfs, k})^{\star}\,\ol{a}^{\text{d}}_{\mfs, k}\right],
\eea
which implies that, relative to the vacuum, the single particle and antiparticle states have energies
\be
\omega_{\mfs, k}= k- \frac{1}{2}\mfs e v \quad \text{and}\quad \ol{\omega}_{\mfs, k}= k+ \frac{1}{2}\mfs e v,
\ee
where $\omega_{\mfs, k}$ ($\ol{\omega}_{\mfs, k}$) is the energy of the particle state $(a^{\text{d}}_{\mfs, k})^{\star}\ket{0_{\text{d}}}$ (antiparticle state $(\ol{a}^{\text{d}}_{\mfs, k})^{\star}\ket{0_{\text{d}}}$) relative to the $\ket{0_{\text{d}}}$ vacuum. The energies of the $in$ and $out$ vacuum are given by $E^{\text{d}}_{0, \ssF}\coloneqq \bra{0_{\text{d}}}H_\ssF\ket{0_{\text{d}}}$. 

Generally, the $in$ and $out$ states are not eigenstates of the fermion electric charge $\cQ_\ssF$ or of the axial charge operator $\cQ_\ssA$. To see this, we expand $\cQ_\ssF, \cQ_\ssA$ in terms of the $in$  basis:
{\scriptsize
\bea
\cQ_\ssF=&& Q^{\ins}_{0, \ssF}+|\mathcal{T}^{++}_{\ins}|^2  \int^{\infty}_0 {\mathrm{d}k }\,\sum_{\mathfrak{s}=\pm}\frac{1}{2}\mathfrak{s} e\,\Big(({a}^{\ins}_{\mathfrak{s}, k})^{\star}\,{a}^{\ins}_{\mathfrak{s}, k} -(\overline{{a}}^{\ins}_{\mathfrak{s}, k})^{\star}\,\overline{{a}}^{\ins}_{\mathfrak{s}, k} \Big)\nonumber\\ &&+  |\mathcal{T}^{+-}_{\ins}|^2\,\sum_{\mathfrak{s}=\pm}\, \mathfrak{s} e\int^{\infty}_0 \frac{\mathrm{d}k }{2\pi}\, PV\Bigg[\int^{\infty}_0 \mathrm{d}k'\Bigg(({a}^{\ins}_{\mathfrak{s}, k})^{\star}\,{a}^{\ins}_{\mathfrak{s}, k'}\, \frac{ i e^{i (k-k')(\epsilon+ t)}}{k-k' } +({a}^{\ins}_{\mathfrak{s}, k})^{\star} \,(\overline{{a}}^{\ins}_{\mathfrak{s}, k'})^{\star}\,\frac{ i e^{i (k+k')(\epsilon+ t)}}{k+k' }
\nonumber\\&& \hspace{55mm} +\overline{{a}}^{\ins}_{\mathfrak{s}, k}
{a}^{\ins}_{\mathfrak{s}, k'}\,\frac{ i e^{-i (k+k')(\epsilon+ t)}}{-k-k'}-(\overline{{a}}^{\ins}_{\mathfrak{s}, k'})^{\star}\,\overline{{a}}^{\ins}_{\mathfrak{s}, k}\,\frac{ i e^{-i (k-k')(\epsilon+ t)}}{-k+k'}   \Bigg)\Bigg]\nonumber\\ &&+e\,\cT^{+-}_{\ins}\mathcal{T}^{++\,*}_{\ins} e^{i \mfa}\int^{\infty}_0 \frac{\mathrm{d}k \,\mathrm{d}k'}{2\pi}\underset{\beta\rightarrow 0+}{\lim}\,\Bigg[({a}^{\ins}_{+, k})^{\star}\,{a}^{\ins}_{-, k'}\,\frac{i e^{i (k-k'- e v)t}e^{i (k-k')\epsilon} }{k'-k+ e v+i\beta }\nonumber\\ &&\hspace{30mm} +({a}^{\ins}_{+, k})^{\star}\,(\overline{{a}}^{\ins}_{-, k'})^{\star}\,\frac{i e^{i (k+k'- e v)t}e^{i (k+k')\epsilon} }{-k'-k+ e v+i \beta }+\overline{a}^{\ins}_{+, k}\,{a}^{\ins}_{-, k'}\,\frac{i e^{i (-k-k'- e v)t}e^{-i (k+k')\epsilon} }{k'+k+ e v+i\beta }\nonumber\\&&\quad \hspace{30mm}-(\overline{a}^{\ins}_{-, k'})^{\star}\,\overline{a}^{\ins}_{+, k}\frac{i e^{i (-k+k'- e v)t}e^{-i (k-k')\epsilon} }{-k'+k+ e v+i\beta } \Bigg]\nonumber\\&&-e\,\cT^{+-\,*}_{\ins}\mathcal{T}^{++}_{\ins} e^{-i \mfa}\int^{\infty}_0 \frac{\mathrm{d}k \,\mathrm{d}k'}{2\pi}\underset{\beta \rightarrow 0+}{\lim}\Bigg[({a}^{\ins}_{-, k})^{\star}\,{a}^{\ins}_{+, k'}\,\frac{i e^{-i (k'-k- e v)t}e^{i (k-k')\epsilon} }{-k'+k+ e v-i\beta }\nonumber\\ &&\hspace{30mm}+({a}^{\ins}_{-, k})^{\star}\,(\overline{a}^{\ins}_{+, k'})^{\star}\,\frac{i e^{-i (-k'-k- e v)t}e^{i (k+k')\epsilon} }{k'+k+ e v-i\beta }+\overline{a}^{\ins}_{-, k}\,{a}^{\ins}_{+, k'}\frac{i e^{-i (k'+k- e v)t}e^{-i (k+k')\epsilon} }{-k'-k+ e v-i\beta }\nonumber\\&&\quad \hspace{30mm}-(\overline{a}^{\ins}_{+, k'})^{\star}\,\overline{a}^{\ins}_{-, k}\,\frac{i e^{-i (-k'+k- e v)t}e^{-i (k-k')\epsilon} }{k'-k+ e v-i\beta}\Bigg], 
\eea
}
and
{\scriptsize
\bea
\cQ_\ssA=&&Q^{\ins}_{0, \ssA}+|\mathcal{T}^{+-}_{\ins}|^2  \int^{\infty}_0 {\mathrm{d}k }\,\sum_{\mathfrak{s}=\pm}\mathfrak{s}\,\Big(({a}^{\ins}_{\mathfrak{s}, k})^{\star}\,{a}^{\ins}_{\mathfrak{s}, k} -(\overline{{a}}^{\ins}_{\mathfrak{s}, k})^{\star}\,\overline{{a}}^{\ins}_{\mathfrak{s}, k} \Big)\nonumber\\ &&+ \,2 |\mathcal{T}^{++}_{\ins}|^2\,\sum_{\mathfrak{s}=\pm}\, \mathfrak{s} \int^{\infty}_0 \frac{\mathrm{d}k }{2\pi}\,PV\Bigg[\int^{\infty}_0  \mathrm{d}k'\Bigg(({a}^{\ins}_{\mathfrak{s}, k})^{\star}\,{a}^{\ins}_{\mathfrak{s}, k'}\, \frac{ i e^{i (k-k')(\epsilon+ t)}}{k-k' } +({a}^{\ins}_{\mathfrak{s}, k})^{\star} \,(\overline{{a}}^{\ins}_{\mathfrak{s}, k'})^{\star}\,\frac{ i e^{i (k+k')(\epsilon+ t)}}{k+k' }
\nonumber\\&&\hspace{30mm}+\overline{{a}}^{\ins}_{\mathfrak{s}, k}
{a}^{\ins}_{\mathfrak{s}, k'}\,\frac{ i e^{-i (k+k')(\epsilon+ t)}}{-k-k'}-(\overline{{a}}^{\ins}_{\mathfrak{s}, k'})^{\star}\,\overline{{a}}^{\ins}_{\mathfrak{s}, k}\,\frac{ i e^{-i (k-k')(\epsilon+ t)}}{-k+k'}   \Bigg)\Bigg]\nonumber\\&&-2\,\cT^{+-}_{\ins}\mathcal{T}^{++\,*}_{\ins} e^{i \mfa}\int^{\infty}_0 \frac{\mathrm{d}k \,\mathrm{d}k'}{2\pi}\underset{\beta \rightarrow 0+}{\lim}\Bigg[({a}^{\ins}_{+, k})^{\star}\,{a}^{\ins}_{-, k'}\,\frac{i e^{i (k-k'- e v)t}e^{i (k-k')\epsilon} }{k'-k+ e v+i\beta }\nonumber\\ &&\hspace{30mm} +({a}^{\ins}_{+, k})^{\star}\,(\overline{{a}}^{\ins}_{-, k'})^{\star}\,\frac{i e^{i (k+k'- e v)t}e^{i (k+k')\epsilon} }{-k'-k+ e v+i\beta }+\overline{a}^{\ins}_{+, k}\,{a}^{\ins}_{-, k'}\,\frac{i e^{i (-k-k'- e v)t}e^{-i (k+k')\epsilon} }{k'+k+ e v+i\beta }\nonumber\\&& \quad\hspace{30mm}-(\overline{a}^{\ins}_{-, k'})^{\star}\,\overline{a}^{\ins}_{+, k}\,\frac{i e^{i (-k+k'- e v)t}e^{-i (k-k')\epsilon} }{-k'+k+ e v+i\beta } \Bigg)\nonumber\\&&+2\,\cT^{+-\,*}_{\ins}\mathcal{T}^{++}_{\ins} e^{-i \mfa}\int^{\infty}_0 \frac{\mathrm{d}k \,\mathrm{d}k'}{2\pi}\underset{\beta \rightarrow 0+}{\lim}\Bigg[({a}^{\ins}_{-, k})^{\star}\,{a}^{\ins}_{+, k'}\,\frac{i e^{-i (k'-k- e v)t}e^{i (k-k')\epsilon} }{-k'+k+ e v-i\beta }\nonumber\\ &&\hspace{30mm}+({a}^{\ins}_{-, k})^{\star}\,(\overline{a}^{\ins}_{+, k'})^{\star}\,\frac{i e^{-i (-k'-k- e v)t}e^{i (k+k')\epsilon} }{k'+k+ e v-i\beta }+\overline{a}^{\ins}_{-, k}\,{a}^{\ins}_{+, k'}\,\frac{i e^{-i (k'+k- e v)t}e^{-i (k+k')\epsilon} }{-k'-k+ e v-i\beta }\nonumber\\&& \quad \hspace{30mm}-(\overline{a}^{\ins}_{+, k'})^{\star}\,\overline{a}^{\ins}_{-, k}\,\frac{i e^{-i (-k'+k- e v)t}e^{-i (k-k')\epsilon} }{k'-k+ e v-i\beta}\Bigg],
\eea
}

\noindent where $Q^{\ins}_{0, \ssF}\coloneqq\bra{0_{\ins}}\cQ_\ssF\ket{0_{\ins}}$, $Q^{\ins}_{0, \ssA}\coloneqq\bra{0_{\ins}}\cQ_\ssA\ket{0_{\ins}}$ , $PV$ refers to the Cauchy principal value of an integral and we  shift $\mfa$ to absorb an $r_0$-dependent phase, as in the main text. The above expansions show that the $in$ states are eigenstates of $\cQ_\ssF$ only if $\cT^{+-}_{\ins}=\cT^{-+}_{\ins}=0$, \textit{i.e.} when only $4$D chirality-changing processes are possible at the boundary, and are eigenstates of $\cQ_\ssA$  in the special case where $\cT^{++}_{\ins}=\cT^{--}_{\ins}=0$ and the boundary action only allows for charge-exchange processes. The expectation values of $\cQ_\ssF, \cQ_\ssA$ in the single particle $in$ states are
\bea\label{expect. value of Q, p in}
&&\bra{0_{\ins}}a^{\ins}_{\mfs, k} \, \cQ_\ssF\,(a^{\ins}_{\mfs, k})^{\star}\ket{0_{\ins}}=\left(Q^{\ins}_{0, \ssF}+  \frac{1}{2}\mfs e|\mathcal{T}^{++}_{\ins}|^2\right)\bra{0_{\ins}}a^{\ins}_{\mfs,  k} \,(a^{\ins}_{\mfs, k})^{\star}\ket{0_{\ins}},\nonumber\\ && \bra{0_{\ins}}a^{\ins}_{\mfs, k} \, \cQ_\ssA\,(a^{\ins}_{\mfs, k})^{\star}\ket{0_{\ins}}=\left(Q^{\ins}_{0, \ssA}+ \mfs |\mathcal{T}^{+-}_{\ins}|^2\right)\bra{0_{\ins}}a^{\ins}_{\mfs,  k} \,(a^{\ins}_{\mfs, k})^{\star}\ket{0_{\ins}},
\eea
and are given by
\bea\label{expect. value of Q, ap in}
&&\bra{0_{\ins}}\ol{a}^{\ins}_{\mfs, k} \, \cQ_\ssF\,(\ol{a}^{\ins}_{\mfs, k})^{\star}\ket{0_{\ins}}=\left(Q^{\ins}_{0, \ssF}-  \frac{1}{2}\mfs e|\mathcal{T}^{++}_{\ins}|^2\right)\bra{0_{\ins}}\ol{a}^{\ins}_{\mfs,  k} \,(\ol{a}^{\ins}_{\mfs, k})^{\star}\ket{0_{\ins}},\nonumber\\ && \bra{0_{\ins}}\ol{a}^{\ins}_{\mfs, k} \, \cQ_\ssA\,(\ol{a}^{\ins}_{\mfs, k})^{\star}\ket{0_{\ins}}=\left(Q^{\ins}_{0, \ssA}- \mfs |\mathcal{T}^{+-}_{\ins}|^2\right)\bra{0_{\ins}}\ol{a}^{\ins}_{\mfs,  k} \,(\ol{a}^{\ins}_{\mfs, k})^{\star}\ket{0_{\ins}},
\eea
for $in$ states with a single antiparticle. $Out$ states similarly have a definite electric charge only if $\cT^{+-}_{\outs}=\cT^{-+}_{\outs}=0$ and a definite axial charge only when $\cT^{++}_{\outs}=\cT^{--}_{\outs}=0$. For more general choices of the $\cT_{\outs}$ amplitudes, the expectation values of the electric and axial charge in single particle or antiparticle $out$ states are
\bea\label{expect. value of Q, p out}
&&\bra{0_{\outs}}a^{\outs}_{\mfs, k} \, \cQ_\ssF\,(a^{\outs}_{\mfs, k})^{\star}\ket{0_{\outs}}=\left(Q^{\outs}_{0, \ssF}+  \frac{1}{2}\mfs e|\mathcal{T}^{++}_{\outs}|^2\right)\bra{0_{\outs}}a^{\outs}_{\mfs,  k} \,(a^{\outs}_{\mfs, k})^{\star}\ket{0_{\outs}},\nonumber\\ && \bra{0_{\outs}}a^{\outs}_{\mfs, k} \, \cQ_\ssA\,(a^{\outs}_{\mfs, k})^{\star}\ket{0_{\outs}}=\left(Q^{\outs}_{0, \ssA}- \mfs |\mathcal{T}^{+-}_{\outs}|^2\right)\bra{0_{\outs}}a^{\outs}_{\mfs,  k} \,(a^{\outs}_{\mfs, k})^{\star}\ket{0_{\outs}},
\eea
and
\bea\label{expect. value of Q, ap out}
&&\bra{0_{\outs}}\ol{a}^{\outs}_{\mfs, k} \, \cQ_\ssF\,(\ol{a}^{\outs}_{\mfs, k})^{\star}\ket{0_{\outs}}=\left(Q^{\outs}_{0, \ssF}-  \frac{1}{2}\mfs e|\mathcal{T}^{++}_{\outs}|^2\right)\bra{0_{\outs}}\ol{a}^{\outs}_{\mfs,  k} \,(\ol{a}^{\outs}_{\mfs, k})^{\star}\ket{0_{\outs}},\nonumber\\ && \bra{0_{\outs}}\ol{a}^{\outs}_{\mfs, k} \, \cQ_\ssA\,(\ol{a}^{\outs}_{\mfs, k})^{\star}\ket{0_{\outs}}=\left(Q^{\outs}_{0, \ssA}+ \mfs |\mathcal{T}^{+-}_{\outs}|^2\right)\bra{0_{\outs}}\ol{a}^{\outs}_{\mfs,  k} \,(\ol{a}^{\outs}_{\mfs, k})^{\star}\ket{0_{\outs}}.
\eea
Equations \eqref{expect. value of Q, p in}-\eqref{expect. value of Q, ap out} show that a measurement of the electric charge in a single-particle state $(a^{\text{d}}_{\mfs, k})^{\star}\ket{0_{\text{d}}}$ or $(\ol{a}^{\text{d}}_{\mfs, k})^{\star}\ket{0_{\text{d}}}$ (relative to the vacuum) does not generally yield the result one would naively expect, naimely $\frac{1}{2}\mfs e $ for particles and $-\frac{1}{2}\mfs e $ for antiparticles. This is because the $in$ and $out$ states agree with the usual notion of particles only in the asymptotic past and future, respectively. More precisely, the state 
$(a^{\ins}_{\mfs, k})^{\star}\ket{0_{\ins}}$ ($(\ol{a}^{\ins}_{\mfs, k})^{\star}\ket{0_{\ins}}$) describes a particle (antiparticle) that approaches the dyon with momentum $k$, charge $\frac{1}{2}\mfs e$ ($-\frac{1}{2}\mfs e$) and $4$D chirality $\mfs$ (-$\mfs$) in the asymptotic past and then scatters either to a particle (antiparticle) with different electric charge and momentum or to one with different $4$D chirality. The $out$  states $(a^{\outs}_{\mfs, k})^{\star}\ket{0_{\outs}}$ ($(\ol{a}^{\outs}_{\mfs, k})^{\star}\ket{0_{\outs}}$) similarly describe particles (antiparticles) of momentum $k$, charge $\frac{1}{2}\mfs e$ ($-\frac{1}{2}\mfs e$) and $4$D chirality $-\mfs$ ($\mfs$) in the asymptotic future and have a more complicated description at early times. This can be shown explicitly by writing single-particle states as the infinitesimally narrow limit of a wave-packet peaked around a given 
momentum, and evaluating the fermionic current expectation values in these states in the asymptotic past (for $in$ states) or asymptotic future ($out$ states).

\subsection{$S$-wave projection of plane-wave states}

For scattering problems, the initial states of interest correspond to $4$D plane waves far from the dyon and so are not prepared in the $S$-wave.
We now show how these plane wave states can be projected onto the $S$-wave sector and use this to compute the cross sections for single-particle $S$-wave scattering given in the main text.

To start, we note that the Dirac equation simplifies considerably at large distances from the dyon, since the background Julia-Zee potential becomes constant \textit{i.e.}  $ e \mathcal{A}^3_{\mu}(x)\underset{r \rightarrow \infty}{\sim}e v \,\delta^0_{\mu}$. The full fermion field can then be expanded in terms of particle and antiparticle mode functions $U_{\mathfrak{s}, \mathfrak{c}, \vec{k}}(x), V_{\mathfrak{s}, \mathfrak{c}, \vec{k}}(x)$
whose asymptotic form is equivalent to that of plane-wave spinors, up to a phase\footnote{Note that this phase has to be defined such that $U_{\mathfrak{s}, \mathfrak{c}, \vec{k}}(x), \,V_{\mathfrak{s}, \mathfrak{c}, \vec{k}}(x)$ are sections \textit{i.e.} it should be defined differently in the $R_+$ and $R_-$ region.}. The particle mode functions then satisfy
\be
 U_{\mathfrak{s}, \mathfrak{c}, \vec{k}}(x)\underset{r \rightarrow \infty}{\sim}{\cN} \,e^{-i (k- \mathfrak{s} \,e v/2)t}e^{i \vec{k}\cdot \vec{x}}e^{i \Phi^{\mathfrak{s}, \mathfrak{c}}_\ssU(x)}\xi_{\mathfrak{s}}\otimes \nu^\ssU_{\mathfrak{c}},
\ee
where $\tau_3 \,\xi_{\mathfrak{s}}= \mathfrak{s}\,\xi_{\mathfrak{s}}$, $ \nu^\ssU_{+}=\begin{pmatrix}
    0&1&0&0
\end{pmatrix}^T$, $ \nu^\ssU_{-}=\begin{pmatrix}
    0&0&1&0
\end{pmatrix}^T$ and $\cN$ is a normalization factor. Similarly, the antiparticle mode functions are asymptotically given by
\be
V_{\mathfrak{s}, \mathfrak{c}, \vec{k}}(x)\underset{r \rightarrow \infty}{\sim}{\cN}\, e^{i (k+ \mathfrak{s} \,e v/2)t}e^{-i \vec{k}\cdot \vec{x}}e^{i  \Phi^{\mathfrak{s}, \mathfrak{c}}_\ssV(x)}
 \xi_{\mathfrak{s}}\otimes \nu^\ssV_{\mathfrak{c}},
\ee
where $ \nu^\ssV_{+}=\begin{pmatrix}
    1&0&0&0
\end{pmatrix}^T$, $ \nu^\ssV_{-}=\begin{pmatrix}
    0&0&0&1
\end{pmatrix}^T$.  The functions $ \Phi^{\mathfrak{s}, \mathfrak{c}}_{\ssU, \ssV}(x)$ are phases whose radial dependence is $(r/r_0)^{i \mathfrak{s} e Q/2}$ \footnote{This choice ensures the asymptotic radial dependence of the incoming spherical waves
in $U, V$ and in partial wave solutions match.
}, and whose angular dependence we discuss shortly. We choose the normalization factor, $\mathcal{N}$, 
such that the mode functions satisfy the following orthogonality and normalization relations
\bea
   \int^{\infty}_{-\infty} \mathrm{d}^3x \,U^{\dagger}_{\mathfrak{s}, \mathfrak{c}, \vec{k}}\,U_{\mathfrak{s}', \mathfrak{c}', \vec{k}'}=  \int^{\infty}_{-\infty} \mathrm{d}^3x \,V^{\dagger}_{\mathfrak{s}, \mathfrak{c}, \vec{k}}\, V_{\mathfrak{s}', \mathfrak{c}', \vec{k}'}=(2\pi)^3\delta(\vec{k}-\vec{k}')\delta_{\mathfrak{s} \mathfrak{s}'}\delta_{\mathfrak{c} \mathfrak{c}'},
\eea
as well as
\be
  \int^{\infty}_{-\infty} \mathrm{d}^3x \,U^{\dagger}_{\mathfrak{s}, \mathfrak{c}, \vec{k}}\,V_{\mathfrak{s}', \mathfrak{c}', \vec{k}'}=0,
\ee

The full fermion field can be expanded in terms of the $U_{\mathfrak{s}, \mathfrak{c}, \vec{k}}(x), \,V_{\mathfrak{s}, \mathfrak{c}, \vec{k}}(x)$ basis as follows:
\be
  \bm{\psi}(x)= \sum_{\mathfrak{s}, \mathfrak{c}=\pm}\int^{\infty}_{-\infty} \frac{\mathrm{d}^3 k}{(2\pi)^{3/2}}\left(a_{\mathfrak{s}, \mathfrak{c}, \vec{k}} \,U_{\mathfrak{s}, \mathfrak{c}, \vec{k}}(x)+(\overline{a}_{\mathfrak{s}, \mathfrak{c}, \vec{k}})^{\star} \,V_{\mathfrak{s}, \mathfrak{c}, \vec{k}}(x)\right),
\ee
where the plane-wave particle and antiparticle creation and annihilation operators satisfy 
\be
\{a_{\mathfrak{s}, \mathfrak{c}, \vec{k}}, (a_{\mathfrak{s}', \mathfrak{c}', \vec{k}'})^{\star}\}= \{\overline{a}_{\mathfrak{s}, \mathfrak{c}, \vec{k}}, (\overline{a}_{\mathfrak{s}', \mathfrak{c}', \vec{k}'})^{\star}\}=\delta(\vec{k}- \vec{k}')\delta_{\mfs \mfs'}\delta_{\mathfrak{c}, \mathfrak{c}'},
\ee
with all other anticommutators being zero. We can also expand $\bm \psi(x)$  in terms of partial wave states,
\be
 \bm{\psi}(x)=\sum_{\mathfrak{s}=\pm}\int^{\infty}_0 \frac{\mathrm{d} k}{\sqrt{2\pi}}\left({a}^{\ins}_{\mathfrak{s}, k} \,u^{\ins}_{\mathfrak{s}, k}(x)+(\overline{{a}}^{\ins}_{\mathfrak{s}, k})^{\star} \,v^{\ins}_{\mathfrak{s}, k}(x)\right)+ \sum_{j>0}\bm{\psi}_j(x)
\ee
where $u^{\ins}_{\mathfrak{s} k}(x), v^{\ins}_{\mathfrak{s} k}(x) $ are the $4$D equivalents of the $2$D $in$ modes\footnote{The $4$D $in$, $out$ modes are given by \eqref{s wave fermions} where $f_{\pm}(r, t), g_{\pm}(r, t)$ are components of the corresponding $2$D basis, as in \eqref{2d s-wave fields}.} and  $\bm{\psi}_j(x)$ is the $j$-th fermion partial wave. Since both the plane-wave and partial wave bases are complete, we can expand the plane-wave particle creation operators $(a_{\mathfrak{s}, \mathfrak{c}, \vec{k}})^{\star}$ in terms of partial wave creation operators 
\be
   ({a}_{\mathfrak{s}, \mathfrak{c}, \vec{k}})^{\star}= \sum_{\mathfrak{s}'=\pm}\int^{\infty}_0 {\mathrm{d} k'}\,\mathcal{D}^{\mathfrak{s}, \mathfrak{c}, \vec{k}}_{\mathfrak{s}', k'}\, ({a}^{\ins}_{\mathfrak{s}', k'})^{\star}+\sum_{j>0, \kappa_j}\mathcal{D}^{\mathfrak{s}, \mathfrak{c}, \vec{k}}_{j, \kappa_j}\, ({a}_{j, \kappa_j})^{\star},
\ee
where $ ({a}_{j, \kappa_j})^{\star}$ is a creation operator for a state in the $j$-th partial wave with quantum numbers $\kappa_j$. To project the plane-wave state $(a_{\mathfrak{s}, \mathfrak{c}, \vec{k}})^{\star}\ket{0}$ onto the $S$-wave sector we simply need to determine the coefficients  $\mathcal{D}^{\mathfrak{s}, \mathfrak{c}, \vec{k}}_{\mathfrak{s}', k'}$, which can be done by expanding the corresponding plane-wave mode function in terms of  $S$-wave mode functions,
\bea\label{expansion of U in partial waves}
 {U}_{\mathfrak{s}, \mathfrak{c}, \vec{k}}(x)&=&(2\pi)^{3/2}\bra{0}\bm{\psi}(x)\,({a}_{\mathfrak{s}, \mathfrak{c}, \vec{k}})^{\star}\ket{0}\nn \\&=&\, 2\pi \sum_{\mathfrak{s}'=\pm}\int^{\infty}_0 {\mathrm{d}k'}\,\mathcal{D}^{\mathfrak{s}, \mathfrak{c}, \vec{k}}_{\mathfrak{s}', k'} \,u^{\ins}_{\mathfrak{s}', k'}(x)+\sum_{j>0, \kappa_j}(2\pi)^{3/2}\bra{0}\bm{\psi}(x)\,\mathcal{D}^{\mathfrak{s}, \mathfrak{c}, \vec{k}}_{j, \kappa_j} ({a}_{j, \kappa_j})^{\star}\ket{0},
\eea
where we use the fact that the $S$-wave annihilation operators also annihilate the full vacuum, $\ket{0}$. Antiparticle states can be projected onto the lowest partial wave in much the same way.

\subsection*{The coefficients  $\mathcal{D}^{\mathfrak{s}, \mathfrak{c}, k \hat{ z}}_{\mathfrak{s}', k'}$}

 For concretness, we now focus on particle plane-wave states which approach the dyon along the $\hat{\bm z}$ axis. As is usually done in scattering problems, we will determine the coefficients $\mathcal{D}^{\mathfrak{s}, \mathfrak{c}, k \hat{ z}}_{\mathfrak{s}', k'}$ by matching terms containing incoming spherical waves on the left side of \eqref{expansion of U in partial waves} to terms with  $j=0$ incoming spherical waves
on the right side of this equation, at large distances from the dyon. To do this, we first isolate the incoming terms in $ U_{\mathfrak{s}, \mathfrak{c}, {k}\hat{z}}$ and find their asymptotic form. Using the plane wave expansion and the asymptotic form of spherical  Bessel functions, we get:
\be
    \left(U_{\mathfrak{s}, \mathfrak{c}, {k} \hat{z}}(x)\right)_{\text{inc.}}\underset{r \rightarrow \infty}{\sim}-\cN\frac{e^{- i kr}}{2 i k r} \,e^{-i (k- \mathfrak{s} \,e v/2)t}e^{i \Phi^{\mathfrak{s}, \mathfrak{c}}_\ssU(x)}\xi_{\mathfrak{s}}\otimes \nu^\ssU_{\mathfrak{c}}\, \overset{\infty}{\underset{\ell=0}{\sum}} (2 \ell + 1 )(-1)^{\ell} P_{\ell} (\cos \theta),
\ee
where the subscript `inc.' refers to only those terms in $U_{\mathfrak{s}, \mathfrak{c}, {k} \hat{z}}(x)$ which include spherical waves that are incident on the dyon. This expression can be simplified further by noticing that $(-1)^{\ell}=P_{\ell}(-1)$ and using the Legendre polynomial completeness relation $\overset{\infty}{\underset{\ell=0}{\sum}} (2 \ell + 1 )P_{\ell} (-1)P_{\ell} (x)= 2\delta(1+x)$ 
\be
   \left( U_{\mathfrak{s}, \mathfrak{c}, {k} \hat{z}}(x)\right)_{\text{inc.}}\underset{r \rightarrow \infty}{\sim}-\cN\frac{e^{-i k r}}{i k r} \,e^{-i (k- \mathfrak{s} \,e v/2)t}\left({r}/r_0\right)^{i \mathfrak{s} e Q/2}e^{i \varphi^{\mathfrak{s}, \mathfrak{c}}_\ssU(\theta)}\xi_{\mathfrak{s}}\otimes \nu^\ssU_{\mathfrak{c}}\delta(1+\cos \theta),
\ee
where we now specialize to the $R_-$ region and rewrite $e^{i \Phi^{\mfs, \mfc}_\ssU(x)}$ as $\left(r/r_0\right)^{i \mfs e Q/2}e^{i \varphi^{\mfs, \mfc}_\ssU(\theta)}$ in this region\footnote{We choose the $\phi$-dependence this way since the plane-wave travels along the $+\hat{\bm z}$ axis and so is incident on the dyon in the $R_-$ region.}

We can expand the top and bottom two components of each Dirac spinor in $U_{\mathfrak{s}, \mathfrak{c}, {k} \hat{z}}(x)$ in terms of eigensections of the total angular momentum and its third component ($J^2$, $J_z$)
\bea\label{Partial wave expansion}
       \left(U_{\mathfrak{s}, \mathfrak{c}, {k} \hat{z}}(x)\right)_{\text{inc.}}\underset{r \rightarrow \infty}{\sim}&&-\cN\frac{e^{- i kr}}{i k r} \,e^{-i (k- \mathfrak{s} \,e v/2)t}\left({r}/r_0\right)^{i \mathfrak{s} e Q/2}\xi_{\mathfrak{s}}\nonumber\\ &&\quad \otimes \Bigg[\begin{pmatrix}
           A^{\mfc}_{0, \mathfrak{s}}{\eta}_{\mathfrak{s}}(\theta, \phi)\\
           B^{\mfc}_{0, \mathfrak{s}} {\eta}_{\mathfrak{s}}(\theta, \phi)
       \end{pmatrix}+ \overset{\infty}{\underset{j>0}{\sum}}\sum^j_{j_z=-j}\sum^2_{i=1}\begin{pmatrix}
           A^{\mfc,(i)}_{j, j_z, \mathfrak{s}} \eta^{(i)}_{j,j_z,\mathfrak{s}}(\theta, \phi)\\[1mm]
          B^{\mfc,(i)}_{j, j_z, \mathfrak{s}} \eta^{(i)}_{j,j_z,\mathfrak{s}}(\theta, \phi)
       \end{pmatrix}\Bigg],
\eea
where ${\eta}_{\mathfrak{s}}(\theta,\phi)$ are defined as in equation \eqref{etapmdefs} and the sentence below it and the higher partial wave angular momentum eigensections $\eta^{(1)}_{j,j_z,\mathfrak{s}}(\theta, \phi)$,  $\eta^{(2)}_{j,j_z,\mathfrak{s}}(\theta, \phi)$ are given by the eigensections $\phi^{(1)}_{j,j_z}(\theta, \phi)$,  $\phi^{(2)}_{j,j_z}(\theta, \phi)$ from \cite{Kazama:1976fm} respectively, with an implied `monopole strength' of $q=\frac{\mathfrak{s}}{2}$. We find the coefficients of interest $ A^{\mfc}_{0, \mathfrak{s}},B^{\mfc}_{0, \mathfrak{s}}$ by multiplying the above equation with $\xi^{\dagger}_{\mathfrak{s}}\otimes\begin{pmatrix}{\eta}^{\dagger}_{\mathfrak{s}}(\theta, \phi) & 0\end{pmatrix}$, $\xi^{\dagger}_{\mathfrak{s}}\otimes\begin{pmatrix}0 &{\eta}^{\dagger}_{\mathfrak{s}}(\theta, \phi) \end{pmatrix}$ and performing the angular integrals. This gives
 \be\label{A-}
  A^{\mfc}_{0, \mathfrak{s}}=\delta_{\mathfrak{c},+}\int \mathrm{d}^2\Omega\,\eta^{\dagger}_{\mathfrak{s}}(\theta, \phi)\begin{pmatrix}
      0\\e^{i \varphi^{\mathfrak{s} \mathfrak{c}}_\ssU(\theta)}
  \end{pmatrix}\delta(1+\cos \theta)=\delta_{\mathfrak{c},+}\delta_{\mathfrak{s},+}e^{i \varphi^{\mathfrak{s} \mathfrak{c}}_\ssU(\pi)}\sqrt{\pi},
 \ee
and
\be\label{B-}
  B^{\mfc}_{0, \mathfrak{s}}=\delta_{\mathfrak{c},-}\int \mathrm{d}^2\Omega\,\eta^{\dagger}_{\mathfrak{s}}(\theta, \phi)\begin{pmatrix}
      e^{i \varphi^{\mathfrak{s} \mathfrak{c}}_\ssU(\theta)}\\0
  \end{pmatrix}\delta(1+\cos \theta)=-\delta_{\mathfrak{c},-}\delta_{\mathfrak{s},-}e^{i \varphi^{\mathfrak{s} \mathfrak{c}}_\ssU(\pi)}\sqrt{\pi},
 \ee
with  ${\eta}_{\mathfrak{s}}(\theta,\phi)$ given by \eqref{etapmdefs} and where we use the fact that ${\eta}_{\mathfrak{s}} (\theta, \phi), {\eta}^{(1)}_{j, j_z,\mathfrak{s}} (\theta, \phi), {\eta}^{(2)}_{j, j_z,\mathfrak{s}} (\theta, \phi)$ are orthonormal. Using \eqref{A-}, \eqref{B-} in equation \eqref{Partial wave expansion} shows that
\be
\left({U}_{\mathfrak{s}, \mathfrak{c}, k \hat{z}}(x)\right)_{\text{inc.}}\underset{r \rightarrow \infty}{\sim}{\cN}
  \,\frac{i\sqrt{ \pi }}{ k}e^{i \varphi^{\mathfrak{s}, \mathfrak{c}}_\ssU(\pi)}\left(\delta_{\mathfrak{c}, +}\delta_{\mathfrak{s}, +}\,u ^{\ins}_{+,k}(x)-\delta_{\mathfrak{c}, -}\delta_{\mathfrak{s}, -}\,u^{\ins}_{-,k }(x)\right)_{\text{inc.}}+(F(x))_{\text{inc.}},
\ee
and so also that
\be
{U}_{\mathfrak{s}, \mathfrak{c}, k \hat{z}}(x)\underset{r \rightarrow \infty}{\sim}{\cN}
  \,\frac{i\sqrt{ \pi }}{ k}e^{i \varphi^{\mathfrak{s}, \mathfrak{c}}_\ssU(\pi)}\left(\delta_{\mathfrak{c}, +}\delta_{\mathfrak{s}, +}\,u^{\ins}_{+,k}(x)-\delta_{\mathfrak{c}, -}\delta_{\mathfrak{s}, -}\,u^{\ins}_{-,k }(x)\right)+F(x),
\ee
where $F(x)$ has no projection onto the $S$-wave. Note that the $S$-wave projection vanishes unless $\mathfrak{c} \mathfrak{s}=+1$, as expected since $S$-wave states satisfy $\mathfrak{h}=\mathfrak{c} \mathfrak{s}$ and we match  $\mathfrak{h}=+1$ spherical waves on each side of the above equation. The coefficients $\mathcal{D}^{\mathfrak{s}, \mathfrak{c}, k \hat{z}}_{\mathfrak{s}', k'}$ are  then:
\be
\mathcal{D}^{\mathfrak{s}, \mathfrak{c}, k \hat{z}}_{\mathfrak{s}', k'}=\frac{i {\cN}}{\sqrt{4\pi } k}e^{i \varphi^{\mathfrak{s}, \mathfrak{c}}_\ssU(\pi)} \delta (k-k')\delta_{\mfs \mfs'}\left(\delta_{\mathfrak{c}, +}\delta_{\mathfrak{s}, +}-\delta_{\mathfrak{c}, -}\delta_{\mathfrak{s}, -}\right).
\ee

\subsection*{$S$-wave scattering cross sections}

We can now calculate the probability for a particle in a plane-wave state to be found in the $S$-wave, which we denote  $p_s$ in the main text. Since 
the probability of being in a  specific momentum eigenstate tends to zero in the continuum limit, we compute $p_s$ for a state described by a phase-space distribution function $\rho(\vec{k})$ which is normalized so that $\exd {\rm n}= \rho(\vec{k}) \,\cV \,\exd^3k /(2\pi)^3$ is the number of momentum states in a volume $\exd^3 k$ around momentum $\vec{k}= k\, \hat{\bm{z}}$. For such an initial state (with electric charge $\mfs$ and $4$D chirality $\mfc$), $p_s$ is given by
\bea
p_s&=&\left[\sum_{\mfs'=\pm}\int^{\infty}_0 \left|\bra{0_{\ins}}a^{\ins}_{\mfs', k'}(a_{\mfs, \mfc, k\hat{z}})^{\star}\ket{0}\right|^2\,\exd k'\right]\,\rho(\vec{k})\,\exd^3 k\nn \\ &=&\left[\sum_{\mfs'=\pm}\int^{\infty}_0 \left|\bra{0_{\ins}}a^{\ins}_{\mfs', k'}(a^{\ins}_{\mfs, k})^{\star}\ket{0_{\ins}}\right|^2\,\exd k'\right]\,\delta_{\mfs, \mfc}\,\cN^2/(4\pi k^2) \,\rho(\vec{k})\,\exd^3 k,
\eea
which evaluates to
\be
p_s =\delta_{\mfs, \mfc} \,\frac{\cN^2}{4\pi k^2}\frac{L}{\pi}\rho(\vec{k})\,\exd^3 k.
\ee
In the above, the factor of ${L}/{\pi}$ is introduced to regulate the momentum-conserving delta function  and is cancelled in the final cross section result by a similar factor in the $2$D scattering rate. The differential  $S$-wave  scattering rates $\exd \Gamma_2$ and $\exd \Gamma_{\rm inclusive}$  can be calculated from the perturbative amplitudes listed in \S\ref{sec:PerturbativeScattering}  and single-particle scattering amplitudes given in \S \ref{sec:SO3Dyon} respectively, both of which can be written as $\cA=\cM\, \delta(E_f -E_i)$. Using these amplitudes in Fermi's golden rule gives 
\be \label{2d rate}
\exd \Gamma=2\pi |i \cM/(2 L)|^2 \delta(E_f - E_i) \frac{\exd k' \, L}{\pi} =| \cM|^2 \delta(E_f - E_i) \,\exd k'/(2L),
\ee
for both the interaction picture and inclusive Heisenberg picture differential rate $\exd \Gamma_2$ and  $\exd \Gamma_{\rm inclusive}$, respectively. 
The factor of $1/(2L)$ in the above differential rate again comes about because we cannot choose the initial $S$-wave state to have a specific momentum when working in the continuum limit. In principle we could remove this factor in the same way as above, by redefining our initial state in terms of a distribution function in momentum space, however we do not do so here as the definition of the initial $S$-wave  state is purely an intermediate step in the cross section calculation. Exclusive $2$D rates, $ \Gamma_{\rm exclusive}$, can be calculated similarly with \textit{e.g.} differential rates for processes in which no pairs are produced given by \eqref{2d rate} where $\cA$ is an exclusive amplitude such as \eqref{- to -} - \eqref{+ to +1}.

The  $4$D differential rates for the scattering processes described here
are then independent of the system length $2 L$, as advertised, and become
\be
\exd \Gamma_{\scriptscriptstyle 4D}=\frac{\cN^2}{8\pi^2 k^2}\,\delta_{\mfs, \mfc}\,| \cM|^2 \delta(E_f - E_i) \,\exd k' \,\rho(\vec{k})\,\exd^3 k.
\ee
The  differential cross section is obtained from $\exd \Gamma_{\scriptscriptstyle 4 D}$ by factoring out the flux of initial particles, given by
$\mfF_i= \exd n \, v_{\rm rel}= \exd n$, where $v_{\rm rel}=1$ is the relative velocity between the dyon and incident fermions and $\exd n =\cN^2 [\rho(\vec{k})/(2\pi)^3]\exd^3 k$ is the number density of incident fermions. We   define $\exd n$ as the product of  $ \exd \rm n/\cV$, the density of states, and $i\ol{U}_{\mfs, \mfc, k \hat{z}}(x) \gamma^0 U_{\mfs, \mfc, k \hat{z} }(x)=\cN^2$, the contribution of a plane wave state to the fermion number density.
Finally, we get that the $4$D single-particle $S$-wave differential cross sections $\exd \sigma_s$ simplify to
\be
\exd \sigma_s [f_{\mfs, \mfc}(k) \to f_{\mfs'}(k')] =\frac{\exd \Gamma_{\scriptscriptstyle 4D}}{\mfF_i} = \frac{p_s}{\mfF_i}\exd \Gamma = \frac{\pi}{k^2}\delta_{\mfs, \mfc} |\cM|^2 \delta(E_f-E_i) \, \exd k',
\ee
as we claim in the main text.

\section{The \textit{in} and \textit{out} vacuum}
\label{App: In and Out vacuum}

 The Bogoliubov relations imply that the $in$ and $out$ vacua, defined as
\be
\bm{a}^{\ins}_{\mathfrak{s}, k}\ket{0_{\ins}}=\ol{\bm{a}}^{\ins}_{\mathfrak{s}, k}\ket{0_{\ins}}=0, \quad \bm{a}^{\outs}_{\mathfrak{s}, k}\ket{0_{\outs
}}=\ol{\bm{a}}^{\outs}_{\mathfrak{s}, k}\ket{0_{\outs}}=0, 
\ee
do not necessarily coincide due to the possibility of pair production. This appendix shows how to expand  the $in$ vacuum in terms $out$ states and vice versa. We also show how these expansions can be used to evaluate \textit{inclusive} scattering observables, such as the cross sections given in \S \ref{sec:SO3Dyon}.

\subsection*{Expansion of \textit{in} vacuum  in terms of \textit{out} states}

We first note that all $out$ particles and antiparticles with momentum $k>e v$ annihilate the $in$ vacuum, that is
\be
\bm{a}^{\outs}_{\mfs , k}\ket{0_{\ins}}=\ol{\bm{a}}^{\outs}_{\mfs , k}\ket{0_{\ins}}=0, \quad \text{for} \quad k> e v.
\ee
For smaller momenta, the above equation remains satisfied  only for negatively charged particles and antiparticles
\be
\bm{a}^{\outs}_{- , k}\ket{0_{\ins}}=\ol{\bm{a}}^{\outs}_{+ , k}\ket{0_{\ins}}=0, \quad \text{for} \quad 0< k<  e v,
\ee
meaning the $in$ vacuum only contains positively charged particles and antiparticles with momenta in the range  $0<k<e v$.

Since the total vacuum can be written as a tensor product over the vacua for each momentum, $\ket{0_{\ins}}=\underset{k> 0}{\prod}\ket{0^{k}_{\ins}}$, the expansion of the total vacuum in terms of $out$ states can be done on a mode-by-mode basis.
As explained above, for momenta $k>ev$ the single-momentum vacuum must be equal to the corresponding $out$ vacuum up to a phase
\be 
 \ket{0^{k}_{\ins}}= e^{i {\delta}_k}\ket{0^{k}_{\outs}},\quad \text{for} \quad k> e v
\ee
where $\delta_k$ is an arbitrary phase.  For momenta $k < e v$, it is convenient to define the single-mode vacuum $\ket{0^k_{\ins}}$ in the following way
\be
\bm{a}^{\ins}_{+, k}\ket{0^k_{\ins}}=\ol{\bm{a}}^{\ins}_{+, k}\ket{0^k_{\ins}}=0 \quad \text{and} \quad  \bm{a}^{\ins}_{-, -k+ e v}\ket{0^k_{\ins}}=\ol{\bm{a}}^{\ins}_{-, -k+ ev}\ket{0^k_{\ins}}=0,
\ee
as opposed to $\bm{a}^{\ins}_{\mfs, k}\ket{0^k_{\ins}}=\ol{\bm{a}}^{\ins}_{\mfs, k}\ket{0^k_{\ins}}=0$, since the Bogoliubov relations (which impose energy conservation) relate $\mfs=+$ particle operators of momentum $k$ to $\mfs=-$ antiparticle operators of momentum $-k + e v$. With this definition, the most general form of the single-mode vacuum is:
\be\label{single mode in vacuum}
\ket{0^k_{\ins}}=\bigl[B_{0;0}+B_{1;0}(\bm{a}^{\outs}_{+, k})^{\star}+B_{0;1}(\ol{\bm{a}}^{\outs}_{-, -k+ ev})^{\star}+B_{1;1}(\bm{a}^{\outs}_{+, k})^{\star}(\ol{\bm{a}}^{\outs}_{-, -k+ ev})^{\star}\Bigr]\ket{0^k_{\outs}}.
\ee
The action of any $out$ operator on the single-mode vacuum $\ket{0^{k}_{\ins}}$ can be calculated in more than one way: Directly, by using the above expansion in terms of $out$ states, or by rewriting the operator using the Bogoliubov relations and then acting on the state. This can be used to determine the $B_{i;j}$ coefficients in \eqref{single mode in vacuum}, as we now show. Consider the state $\bm{a}^{\outs}_{+, k}\ket{0^{k}_{\ins}}$, which  is equal to:
\be\label{+ k in 1}
\bm{a}^{\outs}_{+, k}\ket{0^k_{\ins}}=\bigl[B_{1;0}+B_{1;1}(\ol{\bm{a}}^{\outs}_{-, -k+ ev})^{\star}\Bigr]\ket{0^k_{\outs}},
\ee
but can also be rewritten as 
{\small
\bea\label{+ k in 2}
&&\bm{a}^{\outs}_{+, k}\ket{0^{k}_{\ins}}=\cT^{+-}_{\ins} e^{i \mfa}\,(\ol{\bm{a}}^{\ins}_{-, -k+ e v})^{\star}\ket{0^{k}_{\ins}}\nn\\&=&\cT^{+-}_{\ins}e^{i \mfa}\,\left(\cT^{--}_{\outs}   (\ol{\bm{a}}^{\outs}_{-, -k+ e v})^{\star}+\cT^{-+}_{\outs} e^{-i \mfa}\bm{a}^{\outs}_{+, k}\right)\nn \\&& \qquad \qquad\times \bigl[B_{0;0}+B_{1;0}(\bm{a}^{\outs}_{+, k})^{\star}+B_{0;1}(\ol{\bm{a}}^{\outs}_{-, -k+ ev})^{\star}+B_{1;1}(\bm{a}^{\outs}_{+, k})^{\star}(\ol{\bm{a}}^{\outs}_{-, -k+ ev})^{\star}\Bigr]\ket{0^k_{\outs}}\nn \\&=&  \cT^{+-}_{\ins}\cT^{--}_{\outs} e^{i \mfa} \Bigl[B_{0;0}(\ol{\bm{a}}^{\outs}_{-, -k+ e v})^{\star} -B_{1;0}(\bm{a}^{\outs}_{+, k})^{\star}(\ol{\bm{a}}^{\outs}_{-, -k+ e v})^{\star}\Bigr]\ket{0^k_{\outs}}\nn \\ && \qquad \qquad \qquad \qquad \qquad \qquad \qquad\qquad +\cT^{+-}_{\ins}\cT^{-+}_{\outs} \Bigl[B_{1;0} +B_{1;1}(\ol{\bm{a}}^{\outs}_{-, -k+ ev})^{\star} \Bigr]\ket{0^k_{\outs}}.
\eea
}%
 In the above, we use the Bogoliubov relations first to rewrite $\bm{a}^{\outs}_{+, k}$ and then again to rewrite $(\ol{\bm{a}}^{\ins}_{-, -k+ e v})^{\star}$. Comparing \eqref{+ k in 1} and \eqref{+ k in 2} shows that, when $\cT^{+-}_{\ins}\cT^{--}_{\outs}\neq 0$\footnote{This condition also implies that $B_{1; 1}\neq 0$, otherwise $B_{1; 0}=0$ would imply $\bm{a}^{\outs}_{+, k}\ket{0^{k}_{\ins}}=0$ which is not consistent since $\bm{a}^{\outs}_{+, k}\ket{0^{k}_{\ins}}=\cT^{+-}_{\ins} e^{i \mfa}\,(\ol{\bm{a}}^{\ins}_{-, -k+ e v})^{\star}\ket{0^{k}_{\ins}}\neq 0$.}
\be\label{B coefficients v2}
B_{1; 0}=0 \quad \text{and} \quad \frac{B_{0;0}}{B_{1; 1}}=\frac{1-\cT^{+-}_{\ins}\cT^{-+}_{\outs}}{\cT^{+-}_{\ins}\cT^{--}_{\outs}}e^{-i \mfa } =\frac{\cT^{--}_{\ins}}{\cT^{+-}_{\ins}}e^{-i \mfa }, 
\ee
where in the last line we rewrite the $\cT_{\outs}$ amplitudes in terms of $\cT_{\ins}$ amplitudes and use \eqref{unitarityT1}. Similarly, the state $\ol{\bm{a}}^{\outs}_{-, -k+ e v}\ket{0^{k}_{\ins}}$ is equal to:
\be\label{- -k+ev in 1}
\ol{\bm{a}}^{\outs}_{-, -k+ e v}\ket{0^k_{\ins}}=\bigl[B_{0;1}-B_{1;1}({\bm{a}}^{\outs}_{+, k})^{\star}\Bigr]\ket{0^k_{\outs}},
\ee
but can also be rewritten as
{\small
\bea\label{- -k+ev in 2}
&&\ol{\bm{a}}^{\outs}_{-, -k+ e v}\ket{0^{k}_{\ins}}=\cT^{-+\,*}_{\ins}e^{i \mfa}\,({\bm{a}}^{\ins}_{+, k})^{\star}\ket{0^{k}_{\ins}}\nn\\&=&\cT^{-+\,*}_{\ins}e^{i \mfa}\,\left(\cT^{++\,*}_{\outs}   ({\bm{a}}^{\outs}_{+, k})^{\star}+\cT^{+-\,*}_{\outs} e^{-i \mfa}\,\ol{\bm{a}}^{\outs}_{-, -k+ e v}\right)\nn \\&& \qquad \qquad\times \bigl[B_{0;0}+B_{0;1}(\ol{\bm{a}}^{\outs}_{-, -k+ ev})^{\star}+B_{1;1}(\bm{a}^{\outs}_{+, k})^{\star}(\ol{\bm{a}}^{\outs}_{-, -k+ ev})^{\star}\Bigr]\ket{0^k_{\outs}}\nn \\&=& \cT^{-+\,*}_{\ins}\cT^{++\,*}_{\outs} e^{i \mfa} \bigl[B_{0;0}({\bm{a}}^{\outs}_{+, k})^{\star}+B_{0;1}({\bm{a}}^{\outs}_{+, k})^{\star}(\ol{\bm{a}}^{\outs}_{-, -k+ ev})^{\star}\Bigr]\ket{0^k_{\outs}} \nn\\&& \qquad \qquad \qquad \qquad \qquad \qquad \qquad +\cT^{-+\,*}_{\ins}\cT^{+-\,*}_{\outs} \,\bigl[B_{0;1}-B_{1;1}(\bm{a}^{\outs}_{+, k})^{\star}\Bigr]\ket{0^k_{\outs}},
\eea
}%
which implies that
\be
B_{0;1}=0,
\ee
when compared to \eqref{- -k+ev in 1}. The remaining coefficients $B_{0; 0}$ and $ B_{1; 1}$ can be determined up to a phase, by imposing the normalization condition $\braket{0^k_{\ins}|0^k_{\ins}}=1$ and using \eqref{B coefficients v2}. For $\cT^{+-}_{\ins}, \cT^{--}_{\ins} \neq 0$, we then get:
\be
\ket{0^{k}_{\ins}}=e^{i \delta_k}\Bigl[\cT^{--}_{\ins}+\cT^{+-}_{\ins}e^{i \mfa} (\bm{a}^{\outs}_{+, k})^{\star}(\ol{\bm{a}}^{\outs}_{-, -k+ e v})^{\star}\Bigr] \ket{0^{k}_{\outs}},
\ee
for momenta $k<e v$, where $\delta_k$ is an arbitrary phase. When $\cT^{+-}_{\ins}=0$ or $\cT^{--}_{\ins}=0$, we define the expansion of $\ket{0^{k}_{\ins}}$ in terms of $out$ states as the $\cT^{+-}_{\ins} \to 0$ and $\cT^{--}_{\ins} \to 0$ limit of the above equation respectively, where the  surviving amplitude is necessarily a phase. 

For general momentum $k$ and choice of $\cT_{\ins}$ amplitudes, the single-mode vacuum is:
\bea
\ket{0^{k}_{\ins}}&=&e^{i \delta_k}\left\{\Theta(k-e v )+\Theta(-k+e v )\Bigl[\cT^{--}_{\ins}+\cT^{+-}_{\ins}e^{i \mfa} (\bm{a}^{\outs}_{+, k})^{\star}(\ol{\bm{a}}^{\outs}_{-, -k+ e v})^{\star}\Bigr]\right\}\ket{0^{k}_{\outs}}.
\eea
Finally, we can write the full $in$ vacuum in the $out$ basis as
\bea
\ket{0_{\ins}}&=&\underset{k> 0}{\prod}\ket{0^{k}_{\ins}}=\left(\underset{k> e v}{\prod}\ket{0^{k}_{\ins}}\right)\left(\overset{k<{ev}}{\underset{k>0}{\prod}}\ket{0^{k}_{\ins}}\right)\nn\\&=&e^{i \delta}\left(\overset{k<{ev}}{\underset{k>0}{\prod}}\left[\cT^{--}_{\ins}+\cT^{+-}_{\ins}e^{i \mfa} (\bm{a}^{\outs}_{+, k})^{\star}(\ol{\bm{a}}^{\outs}_{-, -k+ e v})^{\star}\right]\right)\ket{0_{\outs}},
\eea
where $\delta\coloneqq \underset{k>0}{\sum}\delta_k$ is an arbitrary phase which we set to $0$ in the main text, as it is not observable\footnote{The expansion of $in$ states in terms of the $out$ basis is only relevant when evaluating transition amplitudes. Since all amplitudes  include the $e^{ i\delta}$ factor, no interference experiment can be constructed to measure $\delta$.}. The product over all $k< e v$ modes evaluates to
\bea\label{In vacuum in terms of out states}
\ket{0_{\ins}}&=&e^{i \delta}\,\overset{\frac{ev}{\Delta k}}{\underset{n=0}{\sum}}\,\frac{1}{n!}(\mathcal{T}^{--}_{\ins})^{\frac{ e v }{\Delta k}-n}(\cT^{+-}_{\ins})^n\, e^{  i n \mfa}\nn \\ && \qquad \qquad \times \Bigl[\sum^{e v}_{k_1=0}...\sum^{e v}_{k_n=0}\,(\bm{a}^{\outs}_{+, {k_1}})^{\star}({\ol{\bm{a}}}^{\outs}_{-, {-k_1+ e v}})^{\star} ...(\bm{a}^{\outs}_{+, {k_n}})^{\star}({\ol{\bm{a}}}^{\outs}_{-, {-k_n+ e v}})^{\star}\Bigr] \ket{0_{\outs}}.
\eea
The full $in$ vacuum is normalized, since  the coefficients $B_{i;j}$ were chosen such that the
expansion of each single-mode $in$ vacuum corresponds to a normalized state.

\subsection*{Expansion of \textit{out} vacuum  in terms of \textit{in} states}

The above procedure can be repeated for the $out$ vacuum. We similarly get:
\bea \label{Out vacuum in terms of in states}
\ket{0_{\outs}}&=&e^{-i \delta}\,\overset{\frac{ev}{\Delta k}}{\underset{n=0}{\sum}}\,\frac{1}{n!}(\mathcal{T}^{--}_{\outs})^{\frac{ e v }{\Delta k}-n}(\cT^{+-}_{\outs})^n\, e^{  i n \mfa}\nn \\ && \qquad \qquad \times \Bigl[\sum^{e v}_{k_1=0}...\sum^{e v}_{k_n=0}\,(\bm{a}^{\ins}_{+, {k_1}})^{\star}({\ol{\bm{a}}}^{\ins}_{-, {-k_1+ e v}})^{\star} ...(\bm{a}^{\ins}_{+, {k_n}})^{\star}({\ol{\bm{a}}}^{\ins}_{-, {-k_n+ e v}})^{\star}\Bigr] \ket{0_{\ins}}\nn \\&=&e^{-i \delta}\,\overset{\frac{ev}{\Delta k}}{\underset{n=0}{\sum}}\,\frac{1}{n!}(\mathcal{T}^{--\,*}_{\ins})^{\frac{ e v }{\Delta k}-n}(\cT^{-+\,*}_{\ins})^n\, e^{  i n \mfa}\nn \\ && \qquad \qquad \times \Bigl[\sum^{e v}_{k_1=0}...\sum^{e v}_{k_n=0}\,(\bm{a}^{\ins}_{+, {k_1}})^{\star}({\ol{\bm{a}}}^{\ins}_{-, {-k_1+ e v}})^{\star} ...(\bm{a}^{\ins}_{+, {k_n}})^{\star}({\ol{\bm{a}}}^{\ins}_{-, {-k_n+ e v}})^{\star}\Bigr] \ket{0_{\ins}}.
\eea

\subsection*{Inclusive scattering cross sections in the Born-Oppenheimer approximation}

In the main text, we remark that  the scattering observables of physical interest are often  \textit{inclusive} observables, for which the number of pairs produced by the dyon is unmeasured. We now show how such observables  can be calculated, focusing on the inclusive $S$-wave scattering cross sections of \S \ref{sec:SO3Dyon}. For convenience, we work in discrete normalization as in the rest of this appendix.

To start, we note that in most cases of interest in the main text, \textit{exclusive} amplitudes $\bm \cA_n$ can be written as 
\be\label{UnitaritySum}
\bm \cA_n=\bm \cA_{sc} \,\bm \cA^{n}_{\rm pair},
\ee
that is they factorize into a single-particle scattering amplitude $\bm \cA_{sc}$ and an amplitude to produce $n$ pairs, $\bm \cA^{n}_{\rm pair}$. As a result, the exclusive probability for the process to happen, $P_n$, will be the product
of the probability for the single-particle scattering event, $p_{sc}$, and the probability to produce $n$ pairs, $P^n_{\rm pair}$,
\be
P_n=p_{sc} \,P^{n}_{\rm pair}.
\ee
The inclusive probability for the single-particle scattering process  is given by
\be
p=p_{sc} \,p_{\rm vac}=p_{sc}\,\sum^N_{n=0}P^{n}_{\rm pair},
\ee
where $N\coloneqq\frac{e  v}{\Delta k}$ is the maximum number of $out$ pairs in the $in$ vacuum, while $p_{\rm vac}$ is defined through the above equation  and
is equal to
\bea\label{P pair}
p_{\rm vac}&=& |\braket{0_{\outs}|0_{\ins}}|^2 +\sum^{e v }_{q_1=0}|\braket{0_{\outs}|\ol{\bm a}^{\outs}_{-, -q_1+ e v}\,\bm a^{\outs}_{+, q_1}|0_{\ins}}|^2 +\cdots \nn \\ && \qquad + \frac{1}{n!} \sum^{e v }_{q_1, \cdots q_n=0}|\braket{0_{\outs}|\ol{\bm a}^{\outs}_{-, -q_n+ e v}\,\bm a^{\outs}_{+, q_n}\,\cdots \ol{\bm a}^{\outs}_{-, -q_1+ e v}\,\bm a^{\outs}_{+, q_1}|0_{\ins}}|^2+ \cdots\nn \\ &&\qquad\qquad + \frac{1}{N!} \sum^{e v }_{q_1, \cdots q_N=0}|\braket{0_{\outs}|\ol{\bm a}^{\outs}_{-, -q_N+ e v}\,\bm a^{\outs}_{+, q_N}\,\cdots \ol{\bm a}^{\outs}_{-, -q_1+ e v}\,\bm a^{\outs}_{+, q_1}|0_{\ins}}|^2.
\eea
In the above, the combinatorial prefactors are added to ensure we only sum over distinct final states. Keeping in mind that $out$ states with doubly-occupied pairs do not contribute to the momentum sums  in \eqref{P pair},  $p_{\rm vac}$ simplifies to
\bea
p_{\rm vac}&=& \left(|\cT_{\ins}^{--}|^2\right)^N + N\left(|\cT_{\ins}^{--}|^2\right)^{N-1}|\cT_{\ins}^{+-}|^2+\cdots \nn \\ && \qquad + \frac{N!}{n!(N-n)!} \left(|\cT_{\ins}^{--}|^2\right)^{N-n}\left(|\cT_{\ins}^{+-}|^2\right)^n+ \cdots+  \left(|\cT_{\ins}^{+-}|^2\right)^N \nn \\ &=&\left(|\cT_{\ins}^{--}|^2+|\cT_{\ins}^{+-}|^2\right)^N=1.
\eea
This shows that the inclusive probability for the scattering process described by  $\bm \cA_{sc}$ is
\be
p= p_{sc},
\ee
and so the corresponding inclusive rates and cross sections can be computed in the usual way, using only the single-particle amplitude $\bm \cA_{sc}$ or its continuum-normalization counterpart $\cA_{sc}$.

The previous argument can be used to justify the inclusive cross section formulas given in \S \ref{sec:SO3Dyon},  for all but one of the processes considered in that section. The exception is the process described by the amplitude 
\eqref{+ to +1}, when the momentum of the initial particle, $k$, belongs in the pair production range $k< e v$. In this case, the  exclusive amplitude factorizes as follows
\bea
 &&\bm \cA_n=\bra{0_{\outs}}\overline{\bm a}^{\outs}_{-, -q_n+ e v}{\bm a}^{\outs}_{+, q_n}...\overline{\bm a}^{\outs}_{-, -q_1+ e v}{\bm a}^{\outs}_{+, q_1}\,\bm{a}^{\outs}_{+, k'}(\bm{a}^{\ins}_{+, k})^{\star}\ket{0_{\ins}} \nn\\  
 &&\qquad =\delta_{k k'}\Bigl[\mathcal{T}^{++}_{\ins}\mathcal{T}^{--}_{\ins}-\cT^{-+}_{\ins}\cT^{+-}_{\ins}\Bigr]\left(\cT_{\ins}^{--}\right)^{N- (n+1 )}\left(\cT_{\ins}^{+-}\right)^{ n} e^{i n \mfa} \nn \\ && \qquad =\delta_{k k'}\Bigl[\mathcal{T}^{++}_{\ins}\mathcal{T}^{--}_{\ins}-\cT^{-+}_{\ins}\cT^{+-}_{\ins}\Bigr]\left[\bm \cA^n_{\rm pair}/\cT^{--}_{\ins}\right], 
\eea
when $n< N$ and $k', q_1, \cdots , q_n$ are distinct momenta, and vanishes otherwise. 
The inclusive probability for the relevant single-particle process is given by
\bea
p&=&\sum_{k'}\Bigg[|\braket{0_{\outs}|0_{\ins}}|^2+ \sum^{ev}_{q_1=0}|\braket{0_{\outs}|\ol{\bm a}^{\outs}_{-, -q_1+ e v}\,\bm a^{\outs}_{+, q_1}|0_{\ins}}|^2+ \cdots\nn \\ && \qquad   +\frac{1}{(N-1)!}\sum^{ev}_{q_1, \cdots q_{{\scriptscriptstyle N}-1}=0}|\braket{0_{\outs}|\ol{\bm a}^{\outs}_{-, -q_{N-1}+ e v}\,\bm a^{\outs}_{+, q_{N-1}}\,\cdots \ol{\bm a}^{\outs}_{-, -q_1+ e v}\,\bm a^{\outs}_{+, q_1}|0_{\ins}}|^2\Bigg] \\ && \qquad \qquad \qquad \qquad \qquad \times |\cT_{\ins}^{--}|^{-2}\left|\delta_{k k'}\Bigl[\mathcal{T}^{++}_{\ins}\mathcal{T}^{--}_{\ins}-\cT^{-+}_{\ins}\cT^{+-}_{\ins}\Bigr]\right|^2  \nn.
\eea
To evaluate the above sums, note that the Bogoliubov relations impose that the momentum $k'$ is equal to the momentum of the initial particle, $k$, and so  lies in the pair production range $0<k'< e v$. 
As one of the particle states in the pair production range is already occupied, there are now at most $N-1$ different momenta that each $q_i$ can be equated to. The inclusive probability, $p$, is then equal to
\bea
p&=&\sum_{k'}\Bigg[\left(|\cT_{\ins}^{--}|^{2}\right)^{N - 1}+ (N-1)\left(|\cT_{\ins}^{--}|^{2}\right)^{N - 2}|\cT_{\ins}^{+-}|^{2}+ \cdots   +\left(|\cT_{\ins}^{+-}|^2\right)^{N-1}\Bigg]\nn \\ && \qquad \qquad \qquad \qquad \qquad \qquad \qquad \qquad  \times\left|\delta_{k k'}\Bigl[\mathcal{T}^{++}_{\ins}\mathcal{T}^{--}_{\ins}-\cT^{-+}_{\ins}\cT^{+-}_{\ins}\Bigr]\right|^2 ,
\eea
which simplifies to 
\bea \label{inclusive p}
p&=&\sum_{k'}\left(|\cT_{\ins}^{--}|^{2}+|\cT_{\ins}^{+-}|^2\right)^{N-1}\left|\delta_{k k'}\Bigl[\mathcal{T}^{++}_{\ins}\mathcal{T}^{--}_{\ins}-\cT^{-+}_{\ins}\cT^{+-}_{\ins}\Bigr]\right|^2 \nn \\ &=&\sum_{k'}\left|\delta_{k k'}\Bigl[\mathcal{T}^{++}_{\ins}\mathcal{T}^{--}_{\ins}-\cT^{-+}_{\ins}\cT^{+-}_{\ins}\Bigr]\right|^2.
\eea
The last line of \eqref{inclusive p} shows that the inclusive cross section for this process can be calculated in the usual way, if  the single-particle amplitude 
\be
\bm \cA_{sc}=\delta_{k k'}\Bigl[\mathcal{T}^{++}_{\ins}\mathcal{T}^{--}_{\ins}-\cT^{-+}_{\ins}\cT^{+-}_{\ins}\Bigr],
\ee
or its continuum-normalization counterpart \eqref{ScatAmp05} (which is valid for all $k$), is used instead of exclusive amplitudes.

\section{Bilinear currents}\label{App: Currents}

In this appendix, we show how one can regularize local fermion bilinear operators and compute the $in$ and $out$ vacuum expectation values of the fermion number, electric charge and axial currents, as well as the interaction picture interacting hamiltonian. We further derive the conservation (or non-conservation) equations satisfied by the fermionic currents and show how the boundary condition can be used to directly evaluate their radial components at $r=\epsilon$.

\subsection*{Vacuum expectation values of fermion bilinears}

Writing 
\be
  {\bm \chi}(x) = \sum_{\mfs = \pm}\int^{\infty}_0 \frac{\exd k}{\sqrt{2\pi}} \; \Bigl[ \mfu^{\text{d}}_{\mfs, k}(x) \; { a}^{\text{d}}_{\mfs, k} + \mfv^{\text{d}}_{\mfs, k}(x) \; (\ol{a}^{\text{d}}_{\mfs, k})^\star \Bigr] 
\ee
where  $\text{d}=\{\ins, \outs\}$, we seek the expectation values of fermion bilinears like 
\bea 
   \ol{\bm \chi} M  {\bm \chi} &=& \sum_{\mfs, \mfs' = \pm} \int^{\infty}_0 \frac{\exd k \, \exd k'}{2\pi} \; \Bigl[  \ol \mfu^{\text{d}}_{\mfs, k}  M \, \mfu^{\text{d}}_{\mfs', k'} \left( ({ a}^{\text{d}}_{\mfs, k})^\star \,{ a}^{\text{d}}_{\mfs', k'}\right) +  \ol \mfv^{\text{d}}_{\mfs, k}  M \, \mfv^{\text{d}}_{\mfs', k'}  \left(\ol{ a}^{\text{d}}_{\mfs, k} \,(\ol{ a}^{\text{d}}_{\mfs', k'})^\star \right) \nn\\
   && \qquad\qquad\qquad\qquad + \ol \mfu^{\text{d}}_{\mfs, k}  M \, \mfv^{\text{d}}_{\mfs', k'} \left( ({ a}^{\text{d}}_{\mfs, k})^\star\, (\ol{ a}^{\text{d}}_{\mfs', k'})^\star\right) +  \ol \mfv^{\text{d}}_{\mfs, k}  M \, \mfu^{\text{d}}_{\mfs', k'}  \left(\ol{ a}^{\text{d}}_{\mfs, k} \,{ a}^{\text{d}}_{\mfs', k'} \right) \Bigr]
\eea
where we use the operator-ordering notation for fermions: $(AB) := \frac12 [A,B]$ that ensures classically hermitian expressions remain hermitian. 

The $in$ and $out$ vacuum expectation values of local bilinear operators are formally given by
{\small
\bea 
   \bra{0_{\text{d}}}  \ol{\bm \chi} M  {\bm \chi} \ket{0_{\text{d}}} &=& \sum_{\mfs, \mfs'=\pm} \int^{\infty}_0 \frac{\exd  k \, \exd  k'}{4\pi} \; \Bigl[ -  \ol \mfu^{\text{d}}_{\mfs, k} M \,  \mfu^{\text{d}}_{\mfs', k'} \bra{0_{\text{d}}} { a}^{\text{d}}_{\mfs', k'} ({ a}^{\text{d}}_{\mfs, k})^\star \ket{0_{\text{d}}} +  \ol \mfv^{\text{d}}_{\mfs, k} M \,  \mfv^{\text{d}}_{\mfs', k'}  \bra{0_{\text{d}}}  \ol{ a}^{\text{d}}_{\mfs, k} (\ol{ a}^{\text{d}}_{\mfs', k'})^\star \ket{0_{\text{d}}}   \Bigr]\nn\\
   &=& \sum_{\mfs=\pm}  \int^{\infty}_0 \frac{\exd  k}{4\pi} \; \Bigl[ - \ol \mfu^{\text{d}}_{\mfs, k} M \,  \mfu^{\text{d}}_{\mfs, k}   +  \ol \mfv^{\text{d}}_{\mfs, k}M \,  \mfv^{\text{d}}_{\mfs, k}    \Bigr].
\eea
}
At face value, the above expectation values vanish for any matrix $M$ acting in spin and isospin space, since $\mfu^{\text{d}}_{\mfs, k}(x)=\mfv^{\text{d}}_{\mfs, -k}(x)$ and so
$\ol \mfu^{\text{d}}_{\mfs, k}(x) M \,  \mfu^{\text{d}}_{\mfs, k}(x)=\ol \mfv^{\text{d}}_{\mfs, k}(x) M \,  \mfv^{\text{d}}_{\mfs, k}(x)$ for both $in$ and $out$ modes\footnote{This is true for $M$ such that $\ol \mfu^{\text{d}}_{\mfs, k} M \,  \mfu^{\text{d}}_{\mfs, k}$ is independent of $k$.}. This turns out not to be the case, since the presence of anomalies means more care must be taken when evaluating local operators. To this end, we regularize the fermion bilinear operator $\ol{\bm \chi}(x) M\,  {\bm \chi}(x)$ by  evaluating $\bm{\chi}, \ol{\bm \chi}$ at slightly different points, namely $\ol{\bm \chi}(r+\varepsilon/2, t) M\,  {\bm \chi}(r-\varepsilon/2, t)$\footnote{At $r=\epsilon$, we instead consider the operators $\ol{\bm \chi}(\epsilon+\varepsilon, t) M\,  {\bm \chi}(\epsilon, t)$ since $r\ge \epsilon$. }, and take the limit $\varepsilon \rightarrow 0$ after calculating any matrix elements of interest. Note that the point-split operator $\ol{\bm \chi}(r+\varepsilon/2, t) M\,  {\bm \chi}(r-\varepsilon/2, t)$ is gauge invariant without the addition of a Wilson line, since we point split only in the radial direction along which the gauge field vanishes, $\cA_r(x)=0$. More explicitly, we calculate vacuum expectation values through
\bea 
    \bra{0_{\text{d}}} \ol{\bm \chi} \,M {\bm \chi} \ket{0_{\text{d}}}&=&  \underset{\varepsilon\rightarrow 0}{\lim}\Bigl[\bra{0_{\text{d}}}  \ol{ \chi}_{+}(x_+) \,M\,  {\chi}_{+}(x_-) \ket{0_{\text{d}}}+\bra{0_{\text{d}}}  \ol{ \chi}_{-}(x_+) \,M\,  {\chi}_{-}(x_-) \ket{0_{\text{d}}} \Bigr] ,
\eea
where $x_{\pm}\coloneqq(r_{\pm},t)=(r\pm \frac{1}{2}\varepsilon, t)$. Applying this to the radial component of the axial current (for which $M=i \Gamma^1 \Gamma_\ssA$) gives the following contributions to the $in$ vacuum expectation value:
\bea\label{j0+}
&&\bra{0_{\ins}}\ol{ \chi}_{+}(x_+) \,i \Gamma^1 \Gamma_\ssA\,  {\chi}_{+}(x_-) \ket{0_{\ins}} =   \int^{\infty}_{0} \frac{\exd  k}{4\pi} \; \Bigl[ - \ol \mfu^\ins_{+, k}(x_+) i\Gamma^1 \Gamma_\ssA \,  \mfu^\ins_{+, k}(x_-)   +  \ol \mfv^\ins_{+, k}(x_+) i\Gamma^1 \Gamma_\ssA \,  \mfv^\ins_{+, k}(x_-)    \Bigr]\nn \\ &&\quad= \left[ \left( -1 -  e^{i e v \varepsilon}|\cT^{-+}_{\ins}|^2\right) \left(\frac{r_+}{r_-}\right)^{-\frac{i eQ}{2}} +|\mathcal{T}^{++}_{\ins}|^2 \left(\frac{r_+}{r_-}\right)^{\frac{i eQ}{2}}    \right]\int^{\infty}_{0} \frac{\exd  k}{4\pi} \; (e^{-i k \varepsilon}-e^{i k \varepsilon}), 
\eea
as well as
\bea\label{j0-}
&&\bra{0_{\ins}}\ol{ \chi}_{-}(x_+) \,i \Gamma^1 \Gamma_\ssA\,  {\chi}_{-}(x_-) \ket{0_{\ins}} =   \int^{\infty}_{0} \frac{\exd  k}{4\pi} \; \Bigl[ - \ol \mfu^\ins_{-, k}(x_+) i\Gamma^1 \Gamma_\ssA \,  \mfu^\ins_{-, k}(x_-)   +  \ol \mfv^\ins_{-, k}(x_+) i\Gamma^1 \Gamma_\ssA \,  \mfv^\ins_{-, k}(x_-)    \Bigr]\nn \\ &&\quad= \left[ \left( 1 +  e^{-i e v \varepsilon}|\cT^{+-}_{\ins}|^2\right) \left(\frac{r_+}{r_-}\right)^{\frac{i eQ}{2}} -|\mathcal{T}^{--}_{\ins}|^2 \left(\frac{r_+}{r_-}\right)^{-\frac{i eQ}{2}}    \right]\int^{\infty}_{0} \frac{\exd  k}{4\pi} \; (e^{-i k \varepsilon}-e^{i k \varepsilon}). 
\eea
The momentum integrals in the above equations can be evaluated using the integral representation of the Heaviside theta function
\be\label{Momentum integrals}
\int^{\infty}_{0} \frac{\exd  k}{4\pi} e^{\pm i k \varepsilon}=\underset{\beta \rightarrow 0+}{\lim}\,\frac{i}{2\pi}\int^{\infty}_{-\infty} \frac{\exd  k\, \exd \tau}{4\pi(\tau+ i \beta)} e^{ -i k (\tau \mp \varepsilon)}=\underset{\beta \rightarrow 0+}{\lim}\,\frac{i}{4\pi ( \pm \varepsilon +i \beta)},
\ee
an using \eqref{Momentum integrals} in \eqref{j0+} and \eqref{j0-} shows that the $in$ vacuum expectation value of the radial component of the axial current can be written as 
\bea
\bra{0_{\ins}} \ol{\bm \chi} \,i\Gamma^1 \Gamma_\ssA {\bm \chi} \ket{0_{\ins}} &=&\underset{\varepsilon \rightarrow 0}{\lim}\,\left(-\frac{i}{2\pi \varepsilon}\right)\biggl[ \left( 1 +  e^{-i e v \varepsilon}|\cT^{+-}_{\ins}|^2 +|\mathcal{T}^{++}_{\ins}|^2\right) \left(\frac{r_+}{r_-}\right)^{\frac{i eQ}{2}} \nn \\ &&\qquad \qquad \qquad \qquad -\left( 1 +  e^{i e v \varepsilon}|\cT^{-+}_{\ins}|^2 +|\mathcal{T}^{--}_{\ins}|^2\right)  \left(\frac{r_+}{r_-}\right)^{-\frac{i eQ}{2}}    \biggr].
\eea
Finally, we simplify the above equation by expanding in $\varepsilon$ and  using \eqref{unitarityT1} to get:
\be
\bra{0_{\ins}} j^1_\ssA(x) \ket{0_{\ins}}=\bra{0_{\ins}} \ol{\bm \chi} \,i\Gamma^1 \Gamma_\ssA {\bm \chi} \ket{0_{\ins}}=- \frac{e v}{\pi}|\mathcal{T}^{+-}_{\ins}|^2+\frac{e Q}{\pi r}.
\ee

 The $in$ vacuum expectation value of  other current operators  of interest can similarly be written as
\bea\label{in current vev general}
\bra{0_{\ins}} \ol{\bm \chi} M \,{\bm \chi} \ket{0_{\ins}} &=&\underset{\varepsilon \rightarrow 0}{\lim}\,\left(-\frac{i}{2\pi \varepsilon}\right)\biggl[ \left( q^{\ssM,\,\ins}_- - q^{\ssM,\,\ins}_{+-} e^{-i e v \varepsilon}|\cT^{+-}_{\ins}|^2 -q^{\ssM,\,\ins}_{++}|\mathcal{T}^{++}_{\ins}|^2\right) \left(\frac{r_+}{r_-}\right)^{\frac{i eQ}{2}} \nn \\ &&\qquad \qquad   +\left( q^{\ssM,\,\ins}_+ - q^{\ssM,\,\ins}_{-+} e^{i e v \varepsilon}|\cT^{-+}_{\ins}|^2 -q^{\ssM,\,\ins}_{--}|\mathcal{T}^{--}_{\ins}|^2\right)  \left(\frac{r_+}{r_-}\right)^{-\frac{i eQ}{2}}    \biggr],
\eea
where $q^{\ssM,\, \ins}_{\mfs}, q^{\ssM,\, \ins}_{\mfs \mfs'}$ are listed in table \ref{qM in} for several choices of the matrix $M$. The calculation of current vacuum expectation values goes through in the same way for the $out$ vacuum. We get:
\bea\label{out current vev general}
\bra{0_{\outs}} \ol{\bm \chi} M \,{\bm \chi} \ket{0_{\outs}} &=&\underset{\varepsilon \rightarrow 0}{\lim}\,\frac{i}{2\pi \varepsilon}\biggl[ \left( q^{\ssM,\,\outs}_- - q^{\ssM,\,\outs}_{+-} e^{i e v \varepsilon}|\cT^{+-}_{\outs}|^2 -q^{\ssM,\,\outs}_{++}|\mathcal{T}^{++}_{\outs}|^2\right) \left(\frac{r_+}{r_-}\right)^{-\frac{i eQ}{2}} \nn \\ &&\qquad    +\left( q^{\ssM,\,\outs}_+ - q^{\ssM,\,\outs}_{-+} e^{-i e v \varepsilon}|\cT^{-+}_{\outs}|^2 -q^{\ssM,\,\outs}_{--}|\mathcal{T}^{--}_{\outs}|^2\right)  \left(\frac{r_+}{r_-}\right)^{\frac{i eQ}{2}}    \biggr],
\eea
where $q^{\ssM,\, \outs}_{\mfs}, q^{\ssM,\, \outs}_{\mfs \mfs'}$ are listed in table \ref{qM out}.

\begin{table}[h!]
\centering
\setlength{\arrayrulewidth}{0.3mm}
\setlength{\tabcolsep}{5pt}
\renewcommand{\arraystretch}{1.4}

\begin{tabular}[h!]{c|cccccc}
$M$ & $q^{\ssM, \, \ins}_+$  & $q^{\ssM, \, \ins}_-$ & $q^{\ssM, \, \ins}_{++}$  & $q^{\ssM, \, \ins}_{--}$ & $q^{\ssM, \, \ins}_{-+}$  & $q^{\ssM, \, \ins}_{+-}$  \\
 \hline
 \hline
 $i \Gamma^0$ & 1 & 1 &1  &1 &1&1\\
 \hline
 $i \Gamma^1$ & -1 & -1 &1  &1 &1&1\\
 \hline
 $\frac{i e}{2} \Gamma^0 \tau_3$ & $\frac{e}{2}$ & -$\frac{e}{2}$ &$\frac{e}{2}$  & -$\frac{e}{2}$ & -$\frac{e}{2}$&$\frac{e}{2}$\\
 \hline
 $\frac{i e}{2} \Gamma^1 \tau_3$ & -$\frac{e}{2}$ & $\frac{e}{2}$ &$\frac{e}{2}$  &-$\frac{e}{2}$ & -$\frac{e}{2}$&$\frac{e}{2}$\\
 \hline
 $i \Gamma^0\Gamma_\ssA$ & 1 & -1 &-1  &1 &1&-1\\
 \hline
 $i \Gamma^1\Gamma_\ssA$ &- 1 & 1 &-1  &1 &1&-1\\
\end{tabular}
\caption{\small Numerical values of $q^{\ssM, \, \ins}_{\mfs}, q^{\ssM, \, \ins}_{\mfs \mfs'}$, defined by equation \eqref{in current vev general}, for various choices of the matrix $M$.}
\label{qM in}
\end{table}

We see that the  regularized vacuum matrix elements of the components of the fermion number current are given by
\be\label{jF in,out}
\bra{0_{\ins}}{j}^0_\ssB(x)\ket{0_{\ins}} = \bra{0_{\outs}}{j}^0_\ssB(x)\ket{0_{\outs}}=0 \quad
\hbox{and} \quad \bra{0_{\ins}}{j}^1_\ssB(x)\ket{0_{\ins}} = \bra{0_{\outs}}{j}^1_\ssB(x)\ket{0_{\outs}}= \frac{2 i }{\pi \varepsilon}, 
\ee
while those of the electromagnetic current are  
\bea \label{j in,out}
  \bra{0_{\ins}}{j}^0_{\ssF}(r,t)\ket{0_{\ins}} &=& \bra{0_{\outs}}{j}^0_{\ssF}(r,t)\ket{0_{\outs}}=\frac{e^2 v}{2\pi}|\mathcal{T}^{+-}_{\ins}|^2-\frac{e^2 Q}{2\pi r} \nn\\
  \hbox{and} \quad
   \bra{0_{\ins}}{j}^1_{\ssF}(r,t)\ket{0_{\ins}} &=& -\bra{0_{\outs}}{j}^1_{\ssF}(r,t)\ket{0_{\outs}}=  \frac{e^2 v }{2 \pi}|\mathcal{T}^{+-}_{\ins}|^2 \,.
\eea
Finally, the axial current vacuum matrix elements are
\bea
   \bra{0_{\ins}}{j}^0_\ssA(r,t)\ket{0_{\ins}} &=& -\bra{0_{\outs}}{j}^0_\ssA(r,t)\ket{0_{\outs}}=-\frac{e  v }{\pi}|\mathcal{T}^{+-}_{\ins}|^2 \nn\\
   \hbox{and} \quad \bra{0_{\ins}}{j}^1_\ssA(r,t)\ket{0_{\ins}} &=& \bra{0_{\outs}}{j}^1_\ssA(r,t)\ket{0_{\outs}}=-\frac{ e v }{\pi}|\mathcal{T}^{+-}_{\ins}|^2+\frac{ e Q}{\pi r} \,.
\eea

\begin{table}[h!]
\centering
\setlength{\arrayrulewidth}{0.3mm}
\setlength{\tabcolsep}{5pt}
\renewcommand{\arraystretch}{1.4}

\begin{tabular}[h!]{c|cccccc}
$M$ & $q^{\ssM, \, \outs}_+$  & $q^{\ssM, \, \outs}_-$ & $q^{\ssM, \, \outs}_{++}$  & $q^{\ssM, \, \outs}_{--}$ & $q^{\ssM, \, \outs}_{-+}$  & $q^{\ssM, \, \outs}_{+-}$  \\
 \hline
 \hline
 $i \Gamma^0$ & 1 & 1 &1  &1 &1&1\\
 \hline
 $i \Gamma^1$ & 1 & 1 &-1  &-1 &-1&-1\\
 \hline
 $\frac{i e}{2} \Gamma^0 \tau_3$ & $\frac{e}{2}$ & -$\frac{e}{2}$ &$\frac{e}{2}$  & -$\frac{e}{2}$ & -$\frac{e}{2}$&$\frac{e}{2}$\\
 \hline
 $\frac{i e}{2} \Gamma^1 \tau_3$ & $\frac{e}{2}$ & -$\frac{e}{2}$ &-$\frac{e}{2}$  &$\frac{e}{2}$ & $\frac{e}{2}$&-$\frac{e}{2}$\\
 \hline
 $i \Gamma^0\Gamma_\ssA$ & -1 & 1 &1  &-1 &-1&1\\
 \hline
 $i \Gamma^1\Gamma_\ssA$ &- 1 & 1 &-1  &1 &1&-1\\
\end{tabular}
\caption{\small  Numerical values of $q^{\ssM, \, \outs}_{\mfs}, q^{\ssM, \, \outs}_{\mfs \mfs'}$, defined by equation \eqref{out current vev general}, for various choices of $M$.}
\label{qM out}
\end{table}

 The vacuum expectation value of the interaction hamiltonian in the interaction picture \eqref{HintFieldBasis} (with $\delta \cC^{pv}=\delta \cC^{ps}=0$) can be caclulated similarly. This expectation value is defined as
\bea
\bra{0}H_{\rm int}\ket{0}&=& \underset{\varepsilon \to 0}{\lim}\sum_{\mfs=\pm}\int^{\infty}_0\frac{\exd k}{8\pi}\Bigl[- \overline{\mfu}^{\ins 0}_{\mfs, k}(\epsilon+ \varepsilon,t) \Bigl(\delta\mathcal{C}^s_{\mfs\mfs'}-i \delta\mathcal{C}^{v}_{\mfs\mfs}\Gamma^{0}\Bigr)\,\mfu^{\ins 0}_{\mfs, k}(\epsilon,t) \nn \\ &&\qquad \qquad \qquad \qquad \qquad+ \overline{\mfv}^{\ins 0}_{\mfs, k}(\epsilon+ \varepsilon,t) \Bigl(\delta\mathcal{C}^s_{\mfs\mfs'}-i \delta\mathcal{C}^{v}_{\mfs\mfs}\Gamma^{0}\Bigr)\,\mfv^{\ins 0}_{\mfs, k}(\epsilon,t) \Bigr],
\eea
where $\ket{0}=\ket{0_{\ins}}=\ket{0_{\outs}}$ in the interaction picture and is equal to
\bea
\bra{0}H_{\rm int}\ket{0}&=&\underset{\varepsilon \to 0}{\lim}\sum_{\mathfrak{s}=\pm}(\delta\mathcal{C}^s_{\mathfrak{s} \mathfrak{s}}-\delta\mathcal{C}^v_{\mathfrak{s} \mathfrak{s}}) \left[\left(\frac{\epsilon+\varepsilon}{\epsilon}\right)^{{i \mfs  e Q}/{2}}-\left(\frac{\epsilon+\varepsilon}{\epsilon}\right)^{{-i \mfs e Q}/{2}}\right]\int^{\infty}_0\frac{\mathrm{d}k}{8\pi}\,(e^{i k \varepsilon}-e^{-i k \varepsilon})\nn \\ &=&- \frac{e Q}{4\pi\epsilon}\sum_{\mathfrak{s}=\pm}\mfs\,(\delta\mathcal{C}^s_{\mathfrak{s} \mathfrak{s}}-\delta\mathcal{C}^v_{\mathfrak{s} \mathfrak{s}}),
\eea
which matches \eqref{Hint vev}.

\subsection{Current conservation equations}

The fermionic currents satisfy conservation (or non-conservation) equations which can be derived directly from the $2$D Dirac equation.  That is, the current $j^{\alpha}_\ssM(x)\coloneqq i \ol{\bm \chi}(x)\Gamma^{\alpha} M \bm\chi(x) $ 
 satisfies the equation
\bea
\partial_t\, j^t_\ssM(x)&=&  i \, \underset{\varepsilon \rightarrow 0}{\lim} \left[\left(\partial_t \overline{\bm{\chi}}(r_+,t)\right)\Gamma^0\,M\bm{\chi}(r_-,t)+\overline{\bm{\chi}}(r_+ ,t)\Gamma^0\,M\left(\partial_t\bm{\chi}(r_-,t)\right)\right]\nn\\&=&i \underset{\varepsilon \rightarrow 0}{\lim}\Big[ -\left(\partial_r\overline{\bm{\chi}}(r_+,t)\right)\Gamma_c\,\Gamma^0\,M\bm{\chi}(r_-, t)+\overline{\bm{\chi}}(r_+,t)\Gamma^0\,M \,\Gamma_c \left(\partial_r\bm{\chi}(r_-,t)\right) \nn\\ &&-\frac{i e}{2}\Big(\mathcal{A}^3_0(r_+, t)\,\overline{\bm{\chi}}(r_+, t)\Gamma^0\tau_3\, M\bm{\chi}(r_-, t)- \mathcal{A}^3_0(r_-, t)\,\overline{\bm{\chi}}(r_+, t)\Gamma^0\, M \tau_3\bm{\chi}(r_-, t)\Big)\Big]\nonumber\\&=&i\underset{\varepsilon \rightarrow 0}{\lim}\Big[\partial_r \left(\overline{\bm{\chi}}(r_+, t)\Gamma^0\Gamma_c\,M\,\bm{\chi}(r_-, t)\right)- \frac{i e \varepsilon}{2}(\partial_r\mathcal{A}^3_0(r))\overline{\bm{\chi}}(r_+, t)\Gamma^0\, M\tau_3\bm{\chi}(r_-, t)\Big]\nn\\&=&-\partial_r{j}^r_\ssM(x)+\underset{\varepsilon \rightarrow 0}{\lim}\,\frac{ \varepsilon e Q}{2 r^2} \,\overline{\bm{\chi}}(r_+, t)\Gamma^0\, M\tau_3\bm{\chi}(r_-, t),
\eea
where  $M \in\{1, \frac{e}{2}  \tau_3, \Gamma_\ssA\}$ and we use $[M, \Gamma_c]=[M, \tau_3]=0$. This can be equivalently rewritten as
\bea
    \partial_{\alpha}j^{\alpha}_\ssB(x)&=&  \frac{e Q}{2 r^2}\underset{\varepsilon \rightarrow 0}{\lim}  \left( \varepsilon\,\overline{\bm{\chi}}(r_+, t)\Gamma^0\, \tau_3\bm{\chi}(r_-, t)\right)=-i \frac{e Q}{e r^2}\underset{\varepsilon \rightarrow 0}{\lim}\left(\varepsilon j^{0}_\ssF(x)\right) , \nn\\ \partial_{\alpha}j^{\alpha}_\ssF(x)&=& \frac{e (e Q)}{4 r^2}\underset{\varepsilon \rightarrow 0}{\lim}  \left( \varepsilon\,\overline{\bm{\chi}}(r_+, t)\Gamma^0\bm{\chi}(r_-, t)\right)=-i \frac{e (e Q)}{4 r^2}\underset{\varepsilon \rightarrow 0}{\lim}\left(\varepsilon j^{0}_\ssB(x)\right) , \nonumber\\ \partial_{\alpha}j^{\alpha}_\ssA(x)&=& -\frac{e Q}{2 r^2}\underset{\varepsilon \rightarrow 0}{\lim}  \left( \varepsilon\,\overline{\bm{\chi}}(r_+, t)\Gamma^1\bm{\chi}(r_-, t)\right) =i \frac{e Q}{2 r^2}\underset{\varepsilon \rightarrow 0}{\lim}\left(\varepsilon j^{1}_\ssB(x)\right),
\eea
for each choice of $M$, which shows that the conservation of the fermionic currents hinges on whether $j^{\alpha}_\ssB, j^0_\ssF$ are singular in the small $\varepsilon$ limit, or not. The dominant contributions to the source terms in the above conservation equations come from the vacuum expectation values of $j^{\alpha}_\ssB, j^0_\ssF$. Equations \eqref{jF in,out}, \eqref{j in,out} then imply that the fermion number and electric charge current are conserved, while the axial current satisfies the anomaly equation
\be
  \partial_{\alpha}j^{\alpha}_\ssA(x)=-\frac{e Q}{ \pi r^2}= \left[\frac{e}{2}-\left(-\frac{e}{2}\right)\right]\frac{1}{2\pi}\epsilon^{\alpha \beta} \cF^3_{\alpha \beta}(x),
\ee
which we rewrite in the last line to emphasize the fact that the top and bottom components of the doublet contribute the usual factor of $\frac{q_{\mfs}}{2\pi}\epsilon^{\alpha \beta} \cF^3_{\alpha \beta}(x)$ to the $4$-divergence of the axial current, where $q_{\mfs}= \frac{1}{2} \mfs e$. We take the difference, as opposed to the sum, of these two contributions since $j^{\alpha}_\ssA(x)$ is defined with an extra $\tau_3$ matrix compared to the standard definition of an axial current in $2$D.

\subsection{Boundary currents}

At $r=\epsilon$ the radial components of fermionic currents can be evaluated by using the boundary condition $\Gamma^1 \bm{\chi}(\epsilon,t)=-O_{\cB}(\mfa) \,\bm{\chi}(\epsilon,t)$,  which can also be rewritten as $ \ol{\bm{\chi}}(\epsilon,t)\Gamma^1= \ol{\bm{\chi}}(\epsilon,t) O_{\cB}(\mfa)$. We get:
{\small
\bea\label{boundary currents}
j^1_\ssB(\epsilon, t)&=&\frac{i}{2}\left(\ol{\bm{\chi}}(\epsilon,t) \Gamma^1\bm{\chi}(\epsilon,t)+\ol{\bm{\chi}}(\epsilon,t) \Gamma^1\bm{\chi}(\epsilon,t)\right)=\frac{i}{2}\left(\ol{\bm{\chi}}(\epsilon,t) O_{\cB}(\mfa)\bm{\chi}(\epsilon,t)-\ol{\bm{\chi}}(\epsilon,t) O_{\cB}(\mfa)\bm{\chi}(\epsilon,t)\right)=0,\nn \\ j^1_\ssF(\epsilon, t)&=&\frac{i e}{4}\left(\ol{\bm{\chi}}(\epsilon,t) \Gamma^1 \tau_3\bm{\chi}(\epsilon,t)+\ol{\bm{\chi}}(\epsilon,t) \tau_3\Gamma^1\bm{\chi}(\epsilon,t)\right)=\frac{i}{2}\ol{\bm{\chi}}(\epsilon,t) \left[ O_{\cB}(\mfa), \frac{e}{2} \tau_3\right]\bm{\chi}(\epsilon,t), \\ j^1_\ssA(\epsilon, t)&=&\frac{i}{2}\left(\ol{\bm{\chi}}(\epsilon,t) \Gamma^1 \Gamma_\ssA\bm{\chi}(\epsilon,t)-\ol{\bm{\chi}}(\epsilon,t) \Gamma_\ssA\Gamma^1\bm{\chi}(\epsilon,t)\right)=\frac{i}{2}\ol{\bm{\chi}}(\epsilon,t) \left\{ O_{\cB}(\mfa), \Gamma_\ssA\right\}\bm{\chi}(\epsilon,t) \nn.
\eea
}%
The above currents vanish when the boundary action \eqref{dyonEFT22D} is invariant under field transformations of the form $\delta \bm{\chi}(\epsilon, t)= i \theta M \bm{\chi}(\epsilon, t)$, where $\theta$ is a constant parameter and $M=1$ gives the transformation corresponding to the fermion number current, $M=\frac{e}{2}\tau_3$ to the electric charge current  and $M=\Gamma_\ssA$ to the axial current.

As argued earlier in this appendix, the fermion bilinears appearing in  \eqref{boundary currents} should  be regulated by \textit{e.g.} a point-splitting procedure.  Such a regularization procedure generally need not preserve the equality in \eqref{boundary currents}. However, if we regularize the boundary currents  in the following way
\bea
(j^1_\ssB(\epsilon, t))_{\rm reg}&:=&\frac{i}{2}\,\left(\ol{\bm{\chi}}(\epsilon,t) \Gamma^1\bm{\chi}(\epsilon+ \varepsilon,t)+\ol{\bm{\chi}}(\epsilon+\varepsilon,t) \Gamma^1\bm{\chi}(\epsilon,t)\right),\nn \\ (j^1_\ssF(\epsilon, t))_{\rm reg}&:=&\,\frac{i e}{4}\left(\ol{\bm{\chi}}(\epsilon,t) \Gamma^1 \tau_3\bm{\chi}(\epsilon+\varepsilon,t)+\ol{\bm{\chi}}(\epsilon+\varepsilon,t) \tau_3\Gamma^1\bm{\chi}(\epsilon,t)\right), \nn\\ (j^1_\ssA(\epsilon, t))_{\rm reg}&:=&\,\frac{i}{2}\left(\ol{\bm{\chi}}(\epsilon,t) \Gamma^1 \Gamma_\ssA\bm{\chi}(\epsilon+ \varepsilon,t)-\ol{\bm{\chi}}(\epsilon+\varepsilon,t) \Gamma_\ssA\Gamma^1\bm{\chi}(\epsilon,t)\right),
\eea
and the remaining  fermion bilinears  in  \eqref{boundary currents} as
{\small
\bea
\left(\frac{i}{2}\ol{\bm{\chi}}(\epsilon,t)\, [O_{\cB}(\mfa),1]\,\bm{\chi}(\epsilon,t)\right)_{\rm reg}&:=&\frac{i}{2}\,\left(\ol{\bm{\chi}}(\epsilon,t) O_{\cB}(\mfa)\bm{\chi}(\epsilon+ \varepsilon,t)-\ol{\bm{\chi}}(\epsilon+\varepsilon,t) O_{\cB}(\mfa)\bm{\chi}(\epsilon,t)\right),\nn \\ \,\left(\frac{i}{2}\ol{\bm{\chi}}(\epsilon,t) \left[ O_{\cB}(\mfa), \frac{e}{2} \tau_3\right]\bm{\chi}(\epsilon,t)\right)_{\rm reg}&:=&\,\frac{i e}{4}\left(\ol{\bm{\chi}}(\epsilon,t) O_{\cB}(\mfa) \tau_3\bm{\chi}(\epsilon+\varepsilon,t)-\ol{\bm{\chi}}(\epsilon+\varepsilon,t) \tau_3O_{\cB}(\mfa)\bm{\chi}(\epsilon,t)\right), \nn\\ \left(\frac{i}{2}\ol{\bm{\chi}}(\epsilon,t) \left\{ O_{\cB}(\mfa), \Gamma_\ssA\right\}\bm{\chi}(\epsilon,t)\right)_{\rm reg}&:=&\,\frac{i}{2}\left(\ol{\bm{\chi}}(\epsilon,t) O_{\cB}(\mfa)\Gamma_\ssA\bm{\chi}(\epsilon+ \varepsilon,t)+\ol{\bm{\chi}}(\epsilon+\varepsilon,t) \Gamma_\ssA O_{\cB}(\mfa) \bm{\chi}(\epsilon,t)\right),\nn\\
\eea
}%
then the regularization scheme does preserve the equality in \eqref{boundary currents}, which can be seen by using the boundary conditions satisfied by $\bmchi(\epsilon), \ol{\bm \chi}(\epsilon) $.

\end{appendix}

\end{document}